\documentclass{article}
\usepackage{graphicx,amsmath,epsfig,fancyheadings,wasysym,psfrag}
\usepackage{times}
\usepackage[hang]{subfigure}

\usepackage{kniga}
\usepackage[comma,sort&compress]{natbib}
\begin{document}           

\renewcommand{\baselinestretch}{1.5}
\normalsize

\title{Dynamics of alliance formation and the egalitarian revolution}
\author{Sergey Gavrilets$^{*\dag}$, Edgar A. Duenez-Guzman$^{\ddag}$ and Michael D. Vose$^{\ddag}$\\
Departments of $^*$Ecology and Evolutionary Biology,
$^{\dag}$Mathematics,\\ and 
$^{\ddag}$Computer Science, University of Tennessee,\
Knoxville, TN 37996, USA.\\ 
$^*$corresponding author.
Phone: 865-974-8136,\
fax: 865-974-3067,\\
email: sergey@tiem.utk.edu}

\maketitle

\begin{abstract}

{\bf Background.} Arguably the most influential force in human history is the formation of social coalitions and alliances (i.e., long-lasting coalitions) and their impact on individual power. Understanding the dynamics of alliance formation and its consequences for biological, social, and cultural evolution is a formidable theoretical challenge. In most great ape species, coalitions occur at individual and group levels and among both kin and non-kin. Nonetheless, ape societies remain essentially hierarchical, and coalitions rarely weaken social inequality. In contrast, human hunter-gatherers show a remarkable tendency to egalitarianism, and human coalitions and alliances occur not only among individuals and groups, but also among groups of groups. These observations suggest that the evolutionary dynamics of human coalitions can only be understood in the context of social networks and cognitive evolution. 

{\bf Methodology/Principal Findings.} Here, we develop a stochastic model describing the emergence of networks of allies resulting from within-group competition for status or mates between individuals utilizing dyadic information. The model shows that alliances often emerge in a phase transition-like fashion if the group size, awareness, aggressiveness, and persuasiveness of individuals are large and the decay rate of individual affinities is small. With cultural inheritance of social networks, a single leveling alliance including all group members can emerge in several generations. 

{\bf Conclusions/Significance.}  We propose a simple and flexible theoretical approach for studying the dynamics of alliance emergence applicable where game-theoretic methods are not practical. Our approach is both scalable and expandable. It is scalable in that it can be generalized
to larger groups, or groups of groups. It is expandable in that it allows for inclusion of additional factors such as behavioral, genetic, social, and cultural features.  Our results suggest that a rapid transition from a hierarchical society of great apes to an egalitarian society of hunter-gatherers (often referred to as ``egalitarian revolution'') could indeed follow an increase in human cognitive abilities. The establishment of stable group-wide egalitarian alliances creates conditions promoting the origin of cultural norms favoring the group interests over those of individuals.
\end{abstract}

keywords: coalition | alliance | social | network | egalitarian


\section*{Introduction}

Coalitions and alliances (i.e., long-lasting coalitions) are often observed in a number of mammals 
including hyenas, wolves, lions, cheetahs, coatis, meerkats, and dolphins \cite{har92}. 
In primates, both kin and non-kin, and both within-group and group-level coalitions are  
a very powerful means of achieving increased reproductive success via increased dominance
status and access to mates and other resources 
\cite{goo86,deW00,har92,wid00,ver00,mit03,new04}.
In humans, coalitions occurs at many different levels
(ranging from within-family to between-nation states)
and represent probably the most dominant factor in social interactions that has shaped
human history 
\cite{joh87,kna91,boe99,car70,rub02,tur03,tur05,wri77}.


The evolutionary forces emerging from coalitionary interactions may have been extremely important
for the origin of our ``uniquely unique'' species \citep{ale90,fli05}. For example,
it has been argued that the evolution of human brain size and intelligence during Pleistocene
was largely driven by selective forces arising from intense competition between individuals
for increased social and reproductive success
(the ``social brain'' hypothesis, also known as the ``Machiavellian intelligence'' hypothesis;
\cite{jol66,hum76,byr88,ale90,whi97,dun98,dun03,fli05,str05,gea05,rot05,gav06a}). 
Coalition formation is one of the most powerful strategies in competitive interactions
and thus it should have been an important ingredient of selective forces acting in early
humans.
Moreover, one can view language as a tool that originally emerged for simplifying the 
formation and improving the efficiency of coalitions and alliances. It has also been argued 
that the establishment of stable group-wide egalitarian alliances
in early human groups should have created conditions promoting
the origin of conscience, moralistic aggression, altruism, and other norms favoring the 
group interests over those of individuals \citep{boe07}. 
Increasing within-group cohesion should also promote the group efficiency in between-group 
conflicts \citep{wra99,bow07} and intensify cultural group selection \citep{ric05}.

In spite of their importance for biological, social and cultural evolution, our 
understanding of how coalitions and alliances are formed, maintained and break down is limited. 
Existing theoretical approaches for studying coalitions in animals 
are deeply rooted in cooperative game theory, economics, and operations 
research 
\cite{kah84,mye91,klu02,kon03}.
These approaches are usually limited by consideration of coalitions of two individuals against one,
focus on conditions under which certain coalitions are successful and/or profitable
and assume (implicitly or explicitly) that individuals are able to evaluate these conditions
and join freely coalitions that maximize their success \cite{noe94,dug98,joh00,pan03,scha04,whi05a,whi05b,scha06,mes07}.
As such, they typically do not capture the dynamic nature of coalitions and/or are not 
directly applicable to individuals lacking the abilities to enter into binding agreements and 
to obtain, process, and use complex 
information on costs, benefits, and consequences of different actions involving multiple parties \cite{ste05}.
These approaches do not account for the effects of friendship and the memory of past events and
acts which all are important in coalition formation and maintenance.
Other studies emphasize the importance of Prisoner's Dilemma as a paradigm for the emergence of 
cooperative behavior in groups engaged in the public goods game \cite{boy88,bac06}.
These studies have been highly successful in identifying conditions that favor the evolution of cooperation among unrelated individuals in the face of incentives to cheat. 
Prisoner's Dilemma however is often not appropriate for studying coalitionary 
behavior \cite{noe92,ham03} especially when individuals cooperate to compete directly with other individuals or coalitions \cite{ale90,fli05} and within-coalition interactions are mutualistic rather than altruistic and the benefit of cooperation is immediate. 
The social network dynamics that result from coalition formation remain largely unexplored.
Here, we propose a simple and flexible theoretical approach for studying the dynamics of alliance 
emergence applicable where game-theoretic methods are not practical. Our method is related to
recent models of social network formation and games on graphs with dynamic linking
\cite{sky00,pem04a,pem04b,pac06,san06,hru06}.
In our novel approach, alliances are defined in a natural way (via affinity matrices; see below)
and emerge from low-level processes. The approach is both
scalable and expandable. It is scalable in that it can be generalized to larger groups, 
or groups of groups, and potentially applied to modeling the origin and evolution of 
states \citep{car70,wri77,tur03,tur05,mar92,ian02,rub02}. It is expandable 
in that it allows for inclusion of additional factors such as behavioral, 
genetic, social, and cultural features. 
One particular application of our approach is an analysis of conditions under which
intense competition for a limiting resource between individuals with intrinsically different 
fighting abilities could lead 
to the emergence of a single leveling alliance including all members of the group.
This application is relevant with regard to recent discussions of ``egalitarian 
revolution'' (i.e. a rapid transition from a hierarchical society of great 
apes to an egalitarian society of human hunter-gatherers, \cite{boe99}),  
and whether it could have been triggered by an increase in human cognitive abilities \cite{ale90,fli05}.

%
%
%

\section*{Model}

We consider a group of $N$ individuals continuously engaged in
competition for status and/or access to a limited resource. Individuals differ
with regard to their fighting abilities $s_i$ ($1 \le i \le N$).
The attitude of individual $i$ to individual $j$ is described by a variable 
$x_{ij}$  which we call affinity. We allow for both positive and negative 
affinities. Individual affinities control the probabilities of getting coalitionary 
support (see below). The group state is characterized by an $N \times N $  matrix with elements 
$x_{ij}$  which we will call the affinity matrix. 


Time is continuous. Below we say that an event occurs at rate $r$ if 
the probability of this event during a short time interval $dt$ is $rdt$.

We assume that each individual gets engaged in a conflict with another 
randomly chosen individual at rate $\ga$ which we treat as a constant 
for simplicity. 
Each other member of the group is
aware of the conflict with a constant probability $\go$.  Each
individual, say individual $k$, aware of a conflict between
individuals $i$ and $j$ (``initiators''), evaluates a randomly chosen
initiator of a conflict, say, individual $i$, and helps him or not
with probabilities $h_{ki}$ and $1-h_{ki}$, respectively. In the
latter case, individual $k$ then evaluates the other initiator of the
conflict and helps him or not with probabilities $h_{kj}$ and
$1-h_{kj}$, respectively. 
We note 
that the coalitionary support may be vocal rather than physical \cite{wit07}.
Below we will graphically illustrate the group state using matrices with 
elements $h_{ij}$  which we will call interference matrices.

The interference probabilities $h_{ij}$ are
given by an S-shaped function of affinity $x_{ij}$ and are scaled by
two parameters: $\gb$ and $\eta$.  A baseline interference rate $\gb$ controls the
probability of interference on behalf of an individual the affinity
towards whom is zero; $\gb$ can be viewed as a measure of individual
aggressiveness (i.e., the readiness to interfere in a conflict) or
persuasiveness (i.e., the ability to attract help).  A slope parameter
$\eta$ controls how rapidly the probability of interference increases
with affinity.  In numerical simulations we will use function
\[
h_{ki}=\left[ 1+ \frac{1-\gb}{\gb}\ \exp(-\eta x_{ki}) \right]^{-1}.
\]
Note that the probability of help $h_{ij}$ changes from $0$ to $\gb$ to $1$
as affinity $x_{ij}$ changes from large negative values to zero to large positive values.

For simplicity, we assume that interference decisions
are not affected by who else is interfering and on which side.  We
also assume that individuals join coalitions without regard to their
probability of winning. This
assumption is sensible as a first step because predicting the outcomes
of conflicts involving multiple participants and changing alliances
would be very challenging for apes and hunter-gatherers. 

As a result of interference, an initially dyadic conflict may transform
into a conflict between two coalitions. 
[Here, coalition is a group of individuals on the same side of a particular conflict.]
The fighting ability
$S_I$ of a coalition $I$ with $n$ participants is defined as 
$\ov{s}_n n^2$, where $\ov{s}_n$ is the average fighting ability of the
participants. This formulation follows the classical Lanchester-Osipov square
law 
\cite{kin02,hel93,wil02}
which captures a larger importance of the size 
of the coalition over the individual strengths of its
participants.  The probability that coalition $I$ prevails over
coalition $J$ is $S_{I}/(S_{I}+S_{J})$.  

Following a conflict resolution we
update the affinities of all parties involved by a process analogous
to reinforcement learning \cite{mac02}.  
The affinities of winners are changed by $\gd_{ww}$, of the losers by $\gd_{ll}$, the affinities
of winners to losers by $\gd_{wl}$, and those of losers to winners by $\gd_{lw}$.
The $\gd$-values reflect the effects of the costs and benefits of interference on future actions. 
It is natural to assume that the affinities of winners increase ($\gd_{ww}>0$) and 
those of antagonists decrease ($\gd_{wl} < 0, \gd_{lw}<0$). The change in the affinities of
losers $\gd_{ll}$ can be of either sign or zero. 
Parameters $\gd_{ww}, \gd_{wl}, \gd_{lw}$ and $\gd_{ll}$  are considered to be constant. We note that a negative impact of costs of interfering in a conflict on the probability of future interferences can be captured by additionally reducing all affinities between members of a coalition by a fixed value $\gd$.


We assume that coalitions
are formed and conflicts are resolved on a time-scale much faster than that of conflict initiation.
Finally, to reflect a reduced importance of past events relative to more recent events in controlling
one's affinities, affinities decay towards $0$ at a constant rate $\gm$ \cite{whi01}.
Table 1 summarizes our notaion.

\begin{table}[t]
\centering
\caption{Main dinamic variables, parameters, and other variables, functions, and statistics.}
\begin{tabular}{|c|c|} 
\hline
$x_{ij}$ 	&	affinity of individual $i$ to individual $j$\\
\hline
$N$		&	group size\\
$s_i$		&	fighting ability of individual $i$\\
$\ga$		&	conflict initiation rate\\
$\go$		&	awareness\\
$\gb$		&	baseline interference rate\\
$\eta$		&	slope parameter\\
$\gd_{ww},\gd_{wl},\gd_{lw},\gd_{ll}$	&	changes in affinity after conflict resolution\\
$\mu$		&	affinity decay rate\\
$\gk$		&	strength of social network inheritance\\
$\gga$		&	birth rate\\
\hline
$h_{ij}$	&	probability that individual $i$ helps individual $j$; is given by \\
&  an $S$-shaped function of affinity with parameters $\gb$ and $\eta$\\
$S_I=\ov{s}n^2$	&	strength of coalition $I$ with $n$ members and average fighting ability $\ov{s}$\\
$S_I/(S_I+S_J)$	&	probability that coalition $I$ wins a conflict with coalition $J$\\
$X_i$		&	proportion of conflicts won by individual $i$ since birth\\
$Y_i=\sum_k b_k/A_i$	&	expected social success of individual $i$; $A_i$ is the age \\
& of individual 
$i$ and $b_k$ is the benefit of the $k$th conflict\\
$H_X,H_Y$	&	standard deviations of $X_i$ and $Y_i$  in the group (measures of inequality)\\
$C^{(1)},C^{(2)},\ov{h}$	&	clustering coefficients and the average probability of help in an alliance\\
\hline
\end{tabular}
\end{table}

\section*{Results and Their Biological Interpretation}

To gain intuition about the model's behavior we ran numerical simulations
with all affinities initially zero.
We analyzed the structure of the interference matrix $h_{ij}$, looking for emerging alliances.
We say individuals $i$ and $j$ are allies if their interference probabilities
$h_{ij}$ and $h_{ji}$ both exceed the baseline interference rate $\gb$ by at least 50\%.
An alliance is a connected network of allies.

We also measured a number of statistics including
the average and variance of affinities,
the proportion of individuals who belong to an alliance,
the number and sizes of alliances,
the clustering coefficients $C^{(1)}$ and $C^{(2)}$ \cite{new03},
related to the probability that two allies of an individual are themselves allies.  The average interference probability and the clustering coefficients can be interpreted as measuring the ``strength'' of alliances.

To make interpretation of model dynamics easier,  we computed
the proportion $X_i$ of conflicts won since birth, and the
expected social success $Y_i=\sum_k b_k/A_i$, where $A_i$ is the age of individual $i$,
the sum is over all conflicts $k$ he has participated in,
the benefit $b_k$ is $1/n_k$ if $i$ was a member of a winning coalition of $n_k$ individuals,
and $b_k$ is $0$ if $i$ was on the losing side.
Although in our model the probability of winning always increases with the coalition size, 
the benefit $b_k$ always decreases with the coalition size. The net effect of the alliance size
on the expected benefits of its members will depend on the sizes and composition of all alliances in
the group. 
Note that our interpretation of $Y_i$ as a measure of expected social success makes sense both 
if all members on the winning side share equally the reward or 
if the spoils of each particular conflict goes to a randomly chosen member of the winning coalition.  
The former may be the case when the reward is an increase in status or rank. 
The latter may correspond to situations similar to those in
baboons fighting over females, where members of the winning coalition may 
race to the female and whoever reaches her first becomes the undisputed consort for some time \cite{noe92}. 
Nonequal sharing of benefits can be incorporated in the model in a straightforward way.
Note also that being a member of a losing coalition always reduces relative social success.

We also calculated 
the standard deviations $H_X$ and $H_Y$ of $X_i$ and $Y_i$ values.
These statistics measure the degree of ``social inequality'' in the group.

\begin{figure}[t]
\centering
\subfigure{ 	
		\includegraphics[width=0.5in]{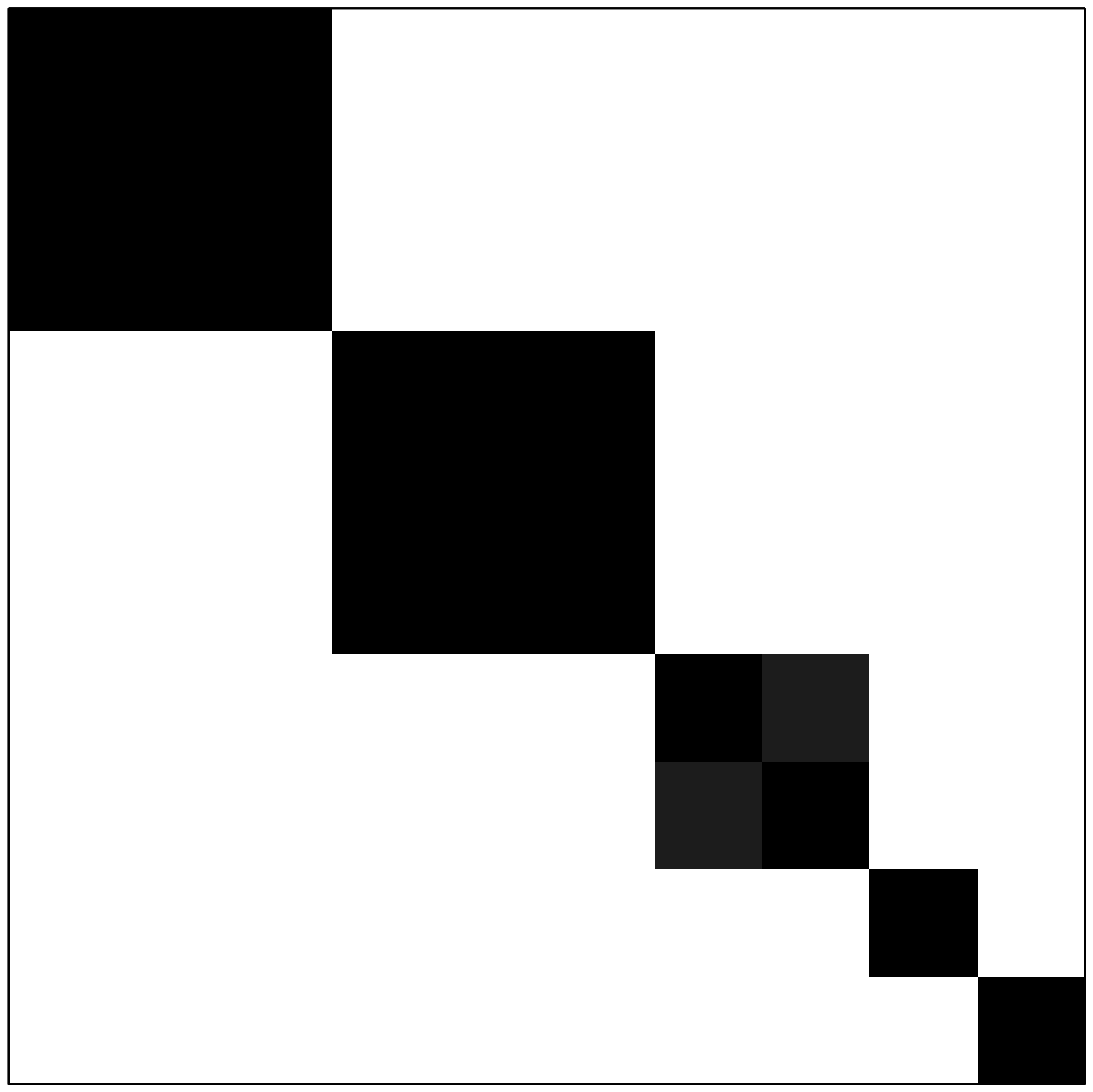}}%
 		\hspace{.05cm}%
\subfigure{	\includegraphics[width=0.5in]{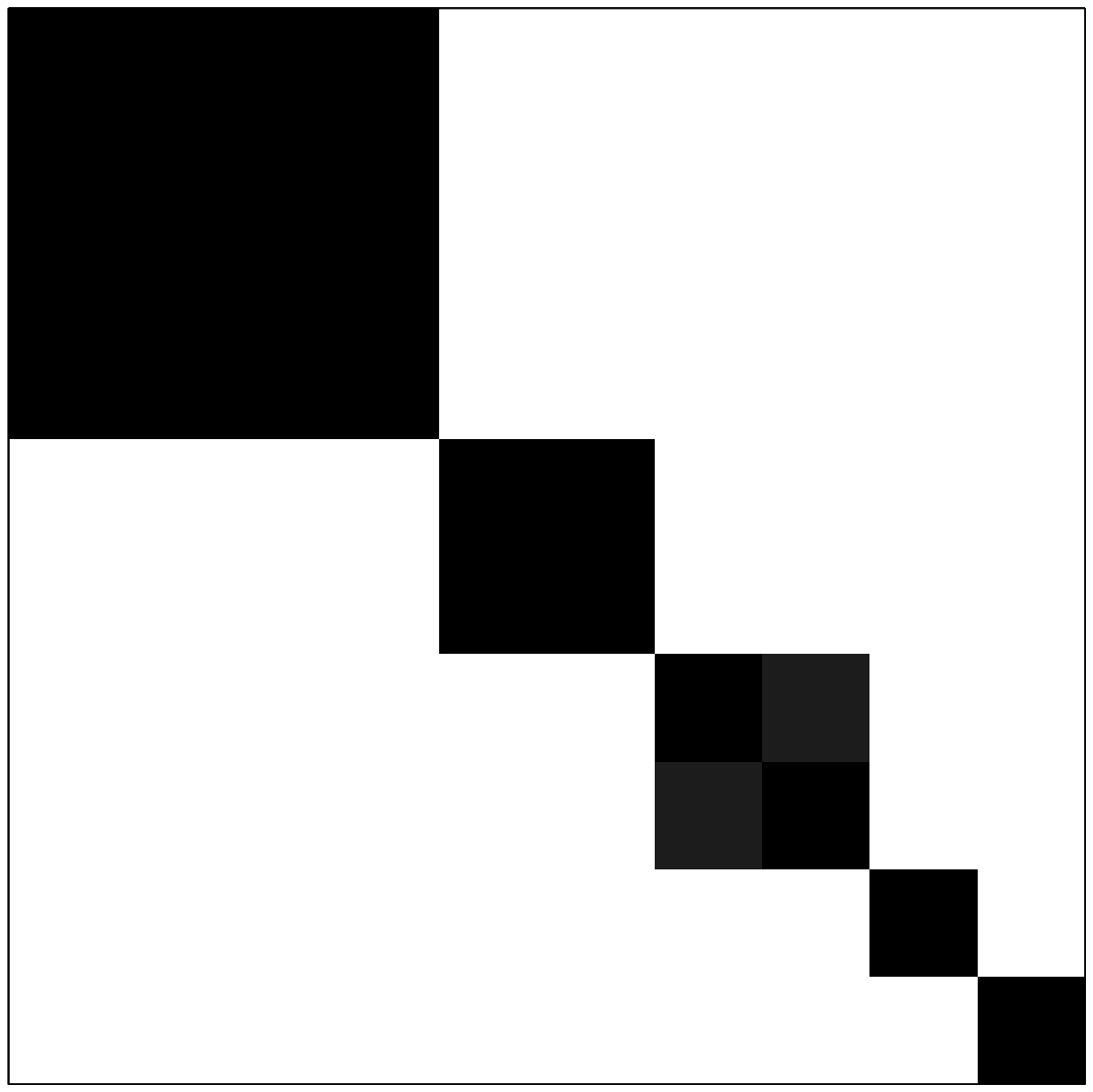}}%
 		\hspace{.05cm}%
\setcounter{subfigure}{0}
\subfigure[]{	\includegraphics[width=0.5in]{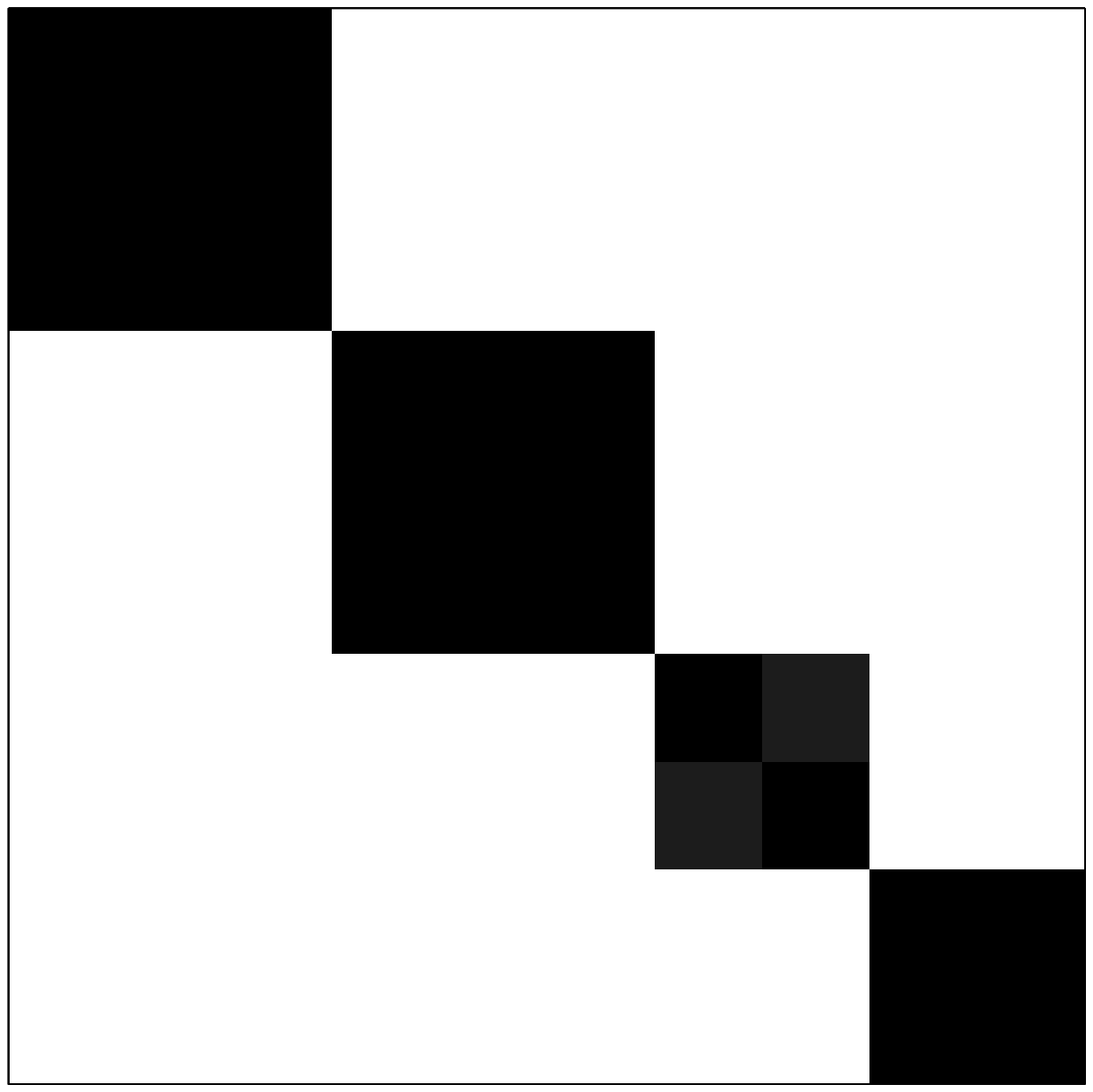}}%
 		\hspace{.05cm}%
\subfigure{	\includegraphics[width=0.5in]{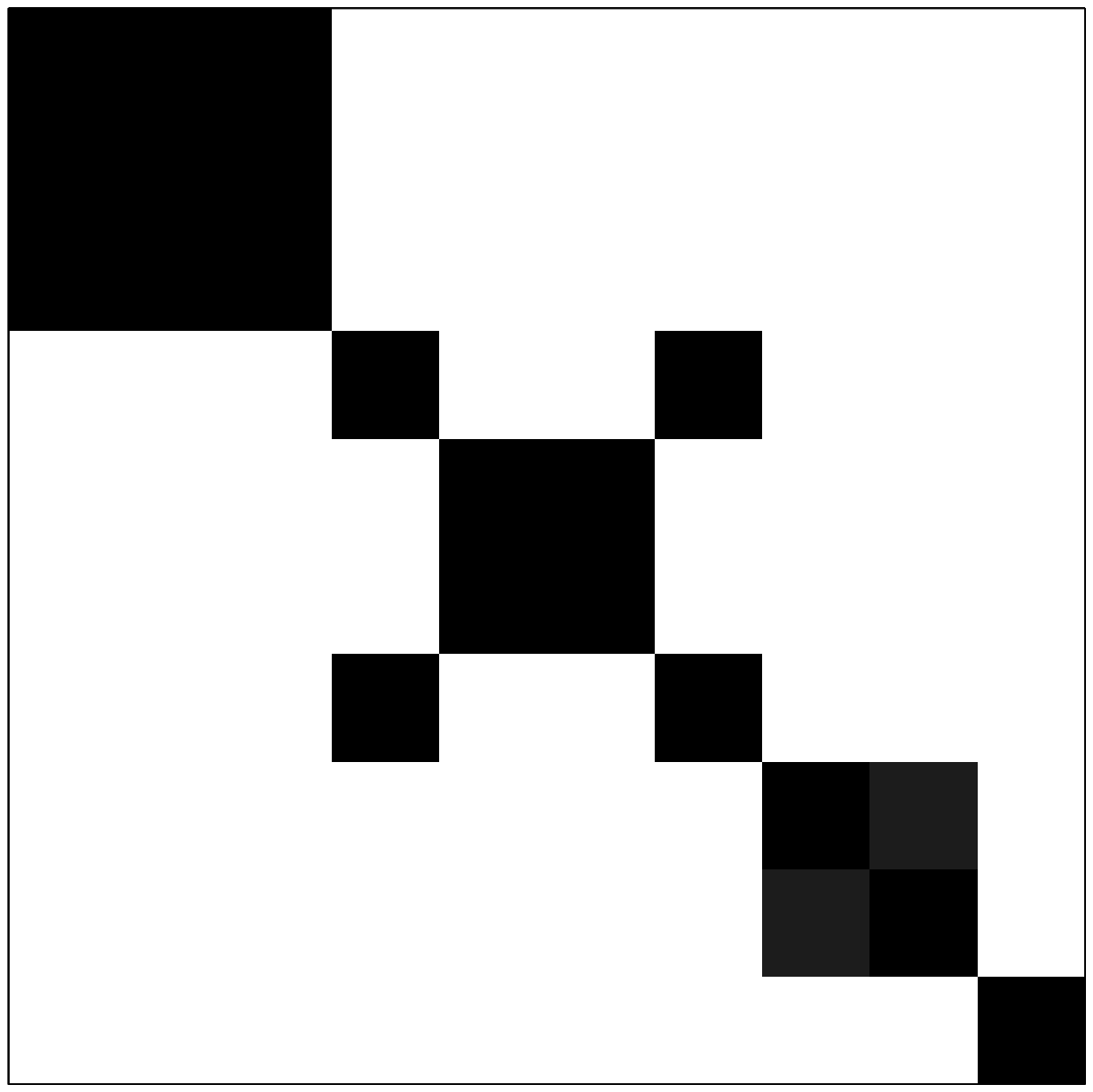}}%
 		\hspace{.05cm}%
\subfigure{	\includegraphics[width=0.5in]{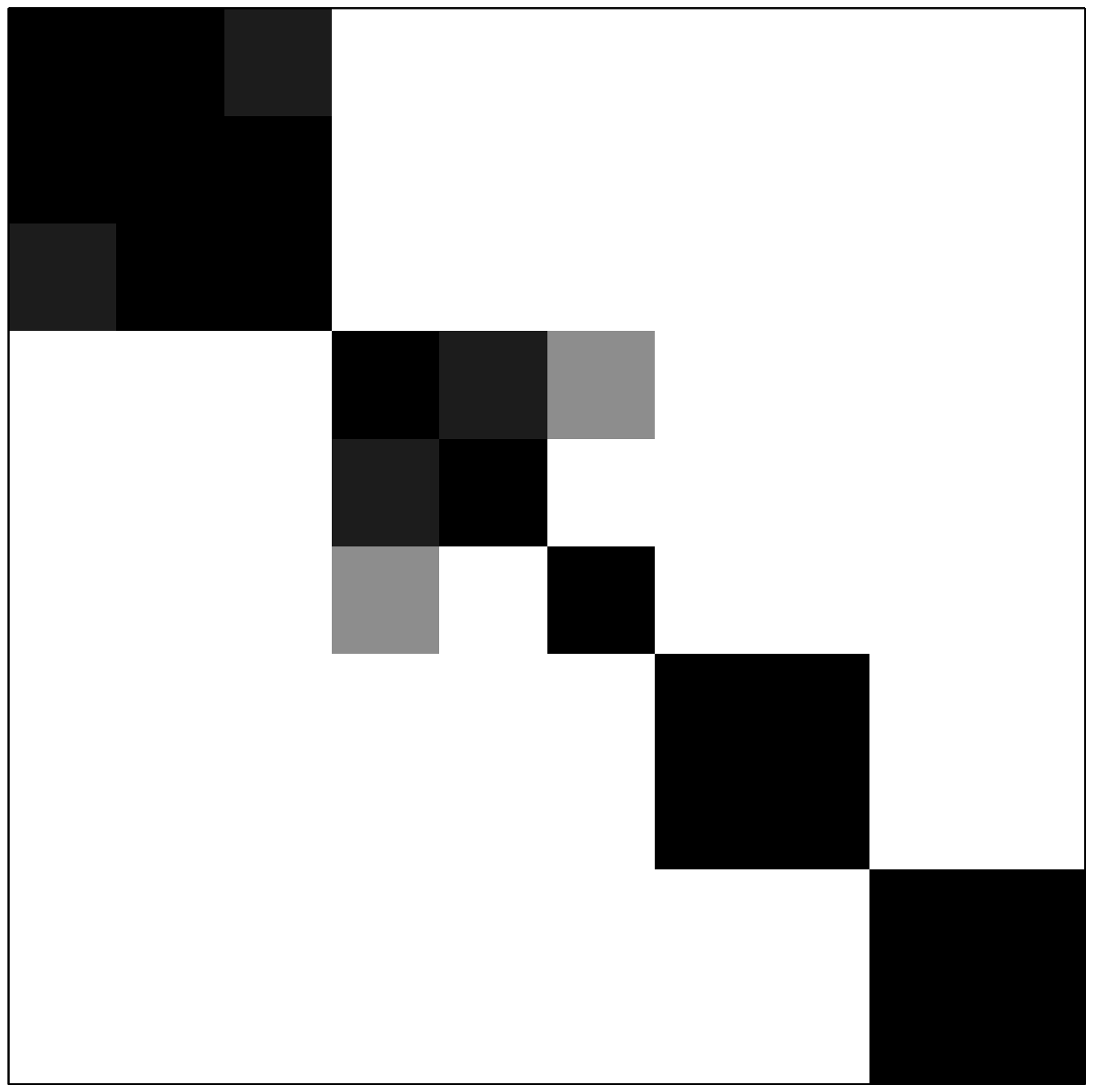}}%
 		\hspace{.5cm}%
\subfigure{ 	
		\includegraphics[width=0.5in]{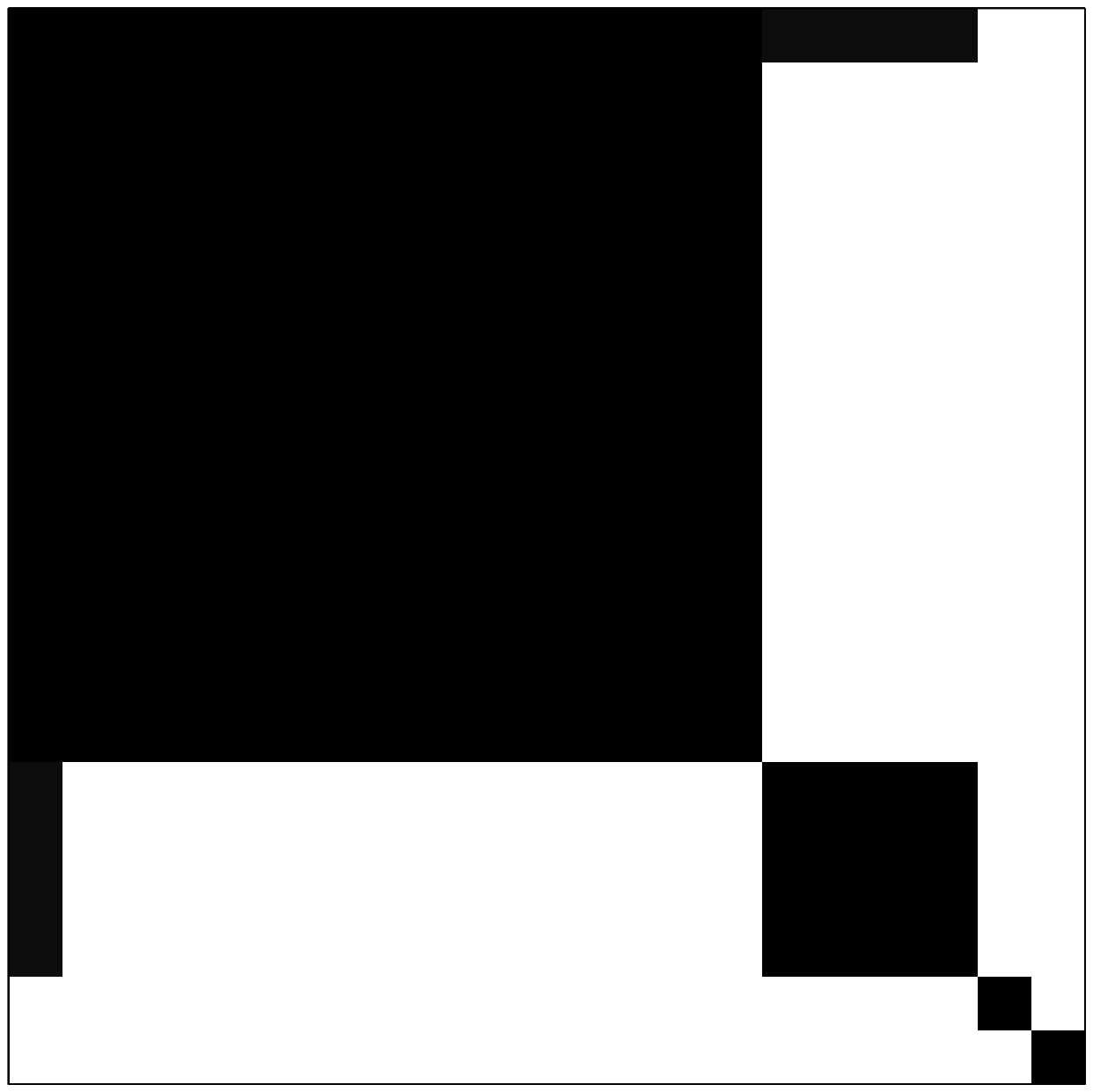}}%
 		\hspace{.05cm}%
\subfigure{	\includegraphics[width=0.5in]{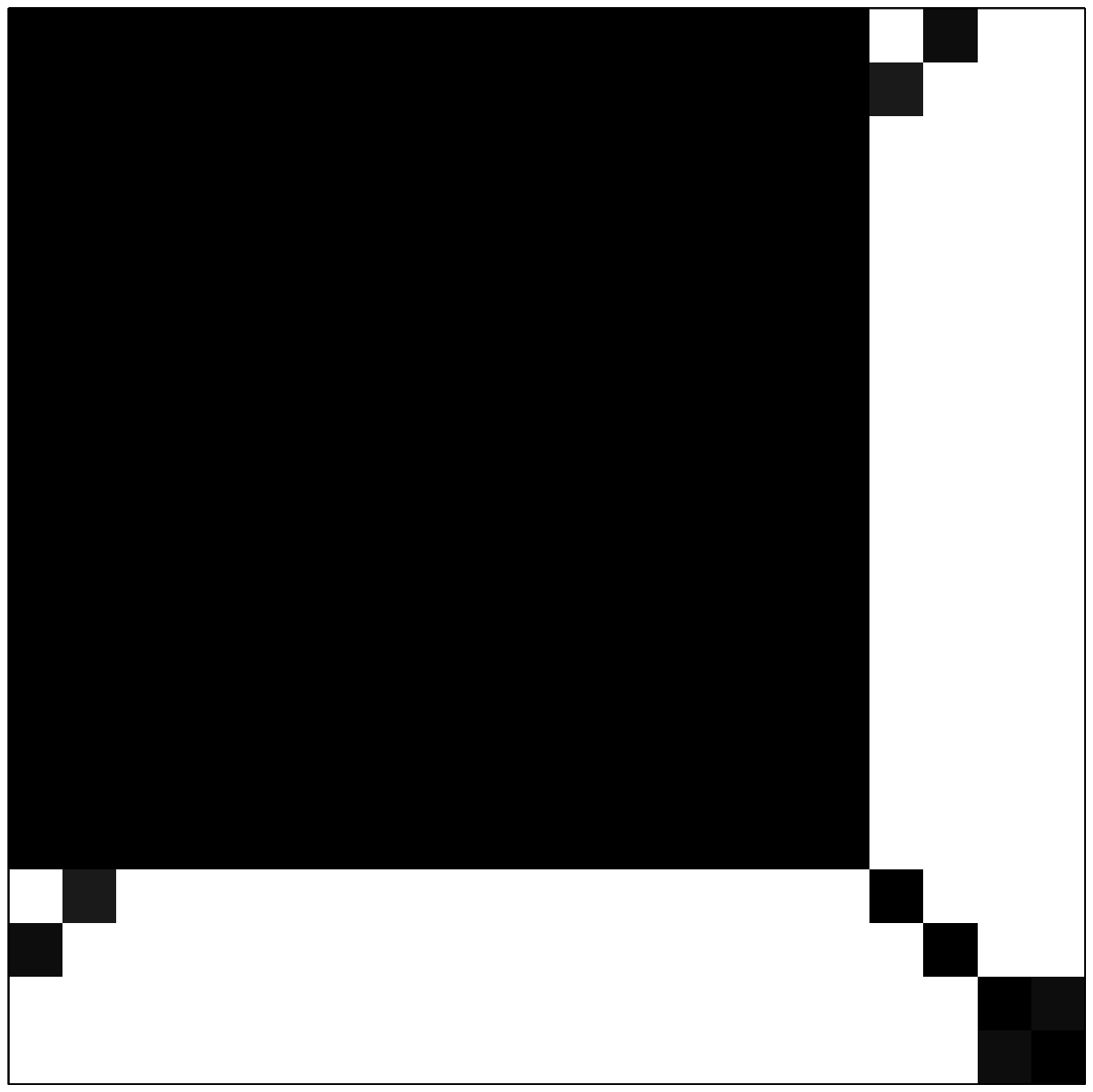}}%
 		\hspace{.05cm}%
\setcounter{subfigure}{1}
\subfigure[]{	\includegraphics[width=0.5in]{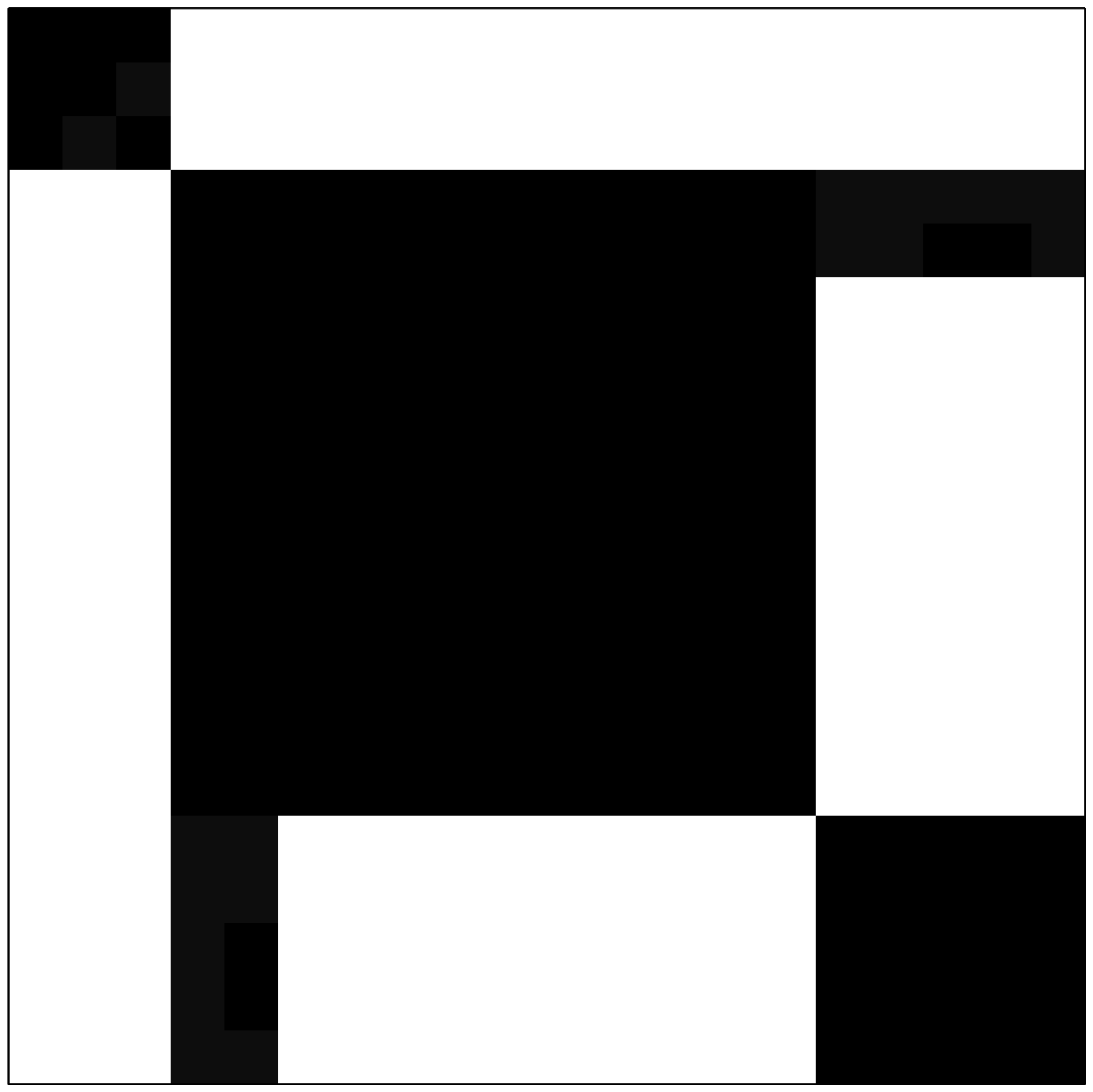}}%
 		\hspace{.05cm}%
\subfigure{	\includegraphics[width=0.5in]{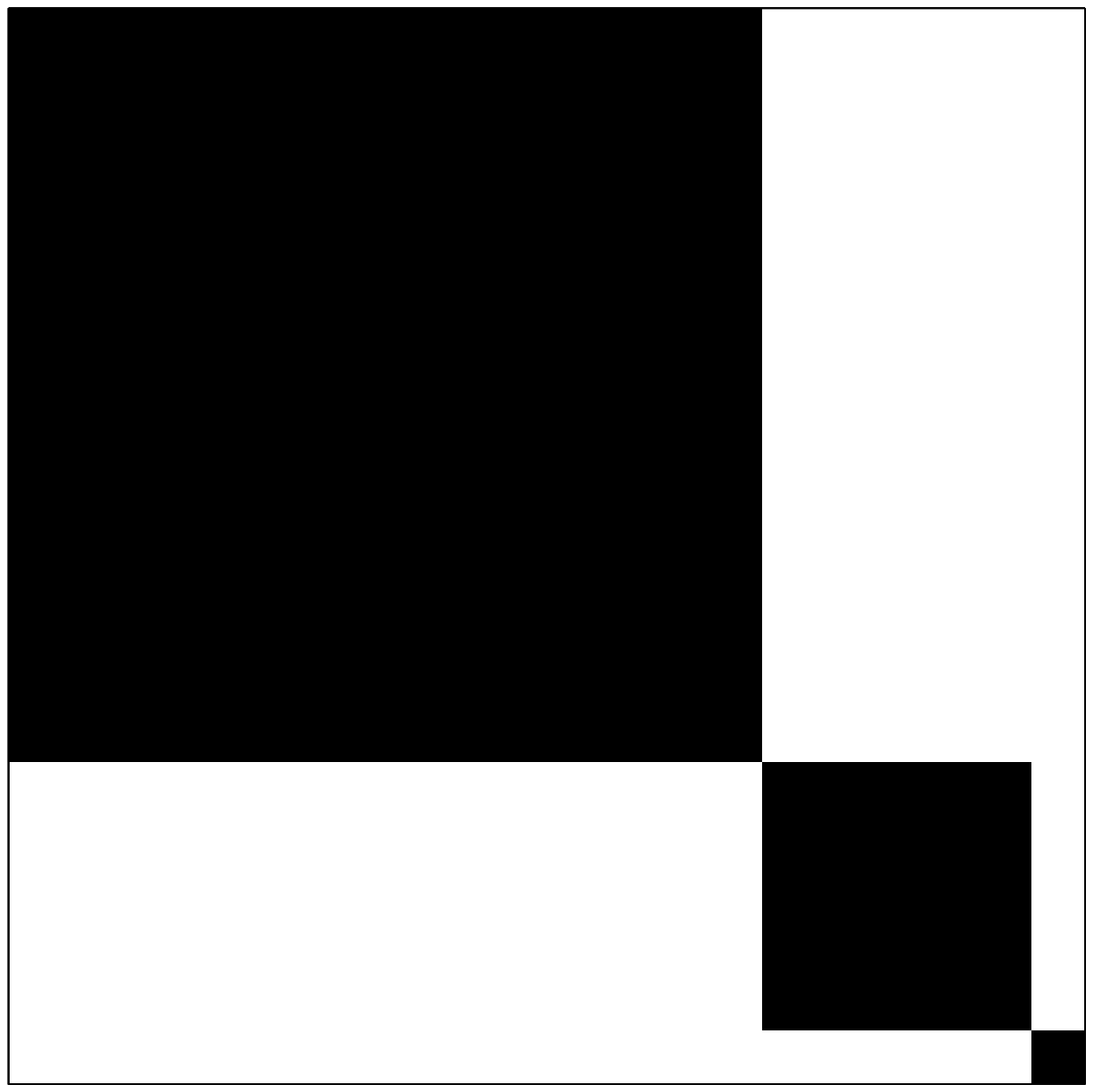}}%
 		\hspace{.05cm}%
\subfigure{	\includegraphics[width=0.5in]{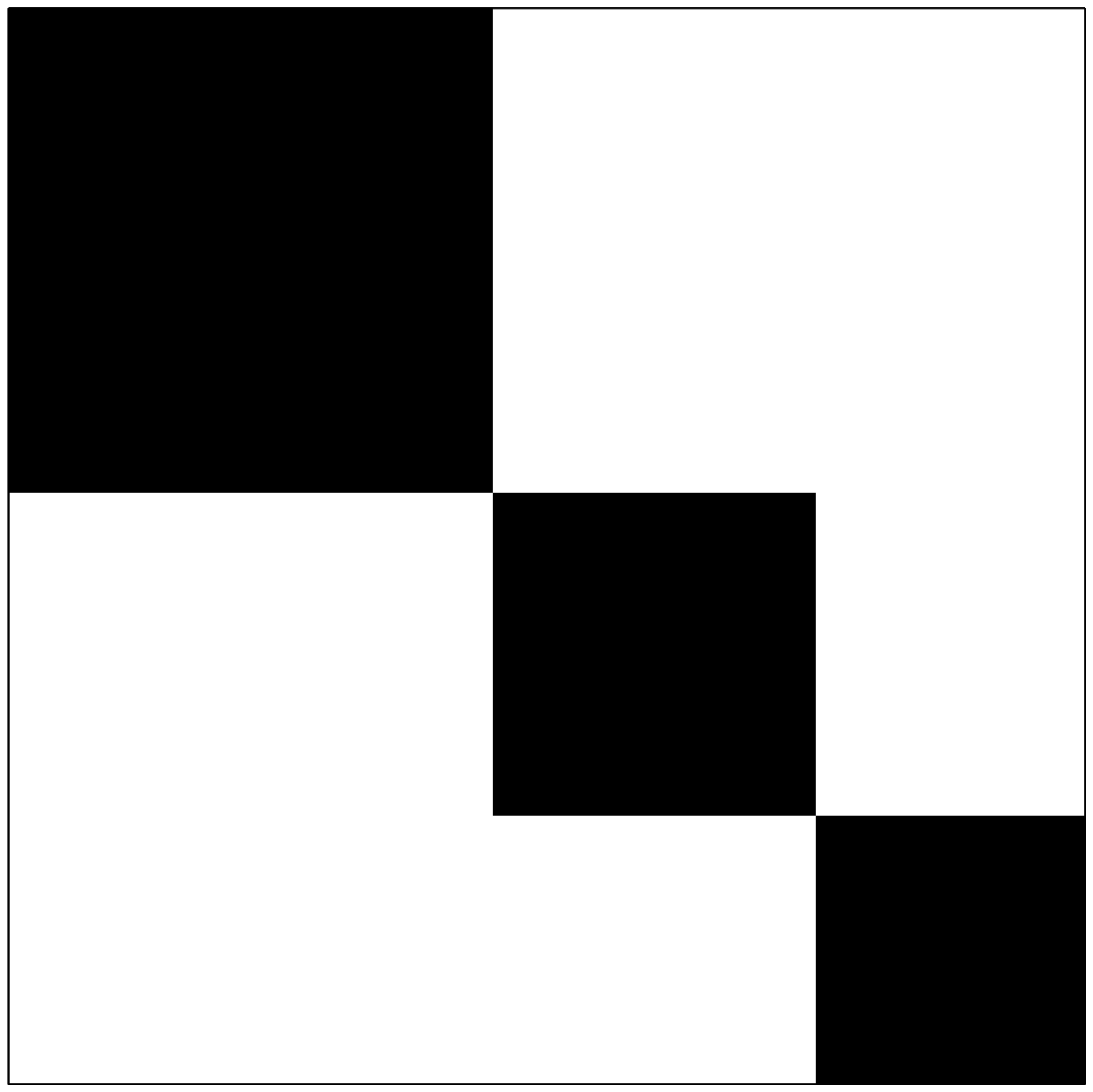}}\\

\subfigure{ 	
		\includegraphics[width=0.5in]{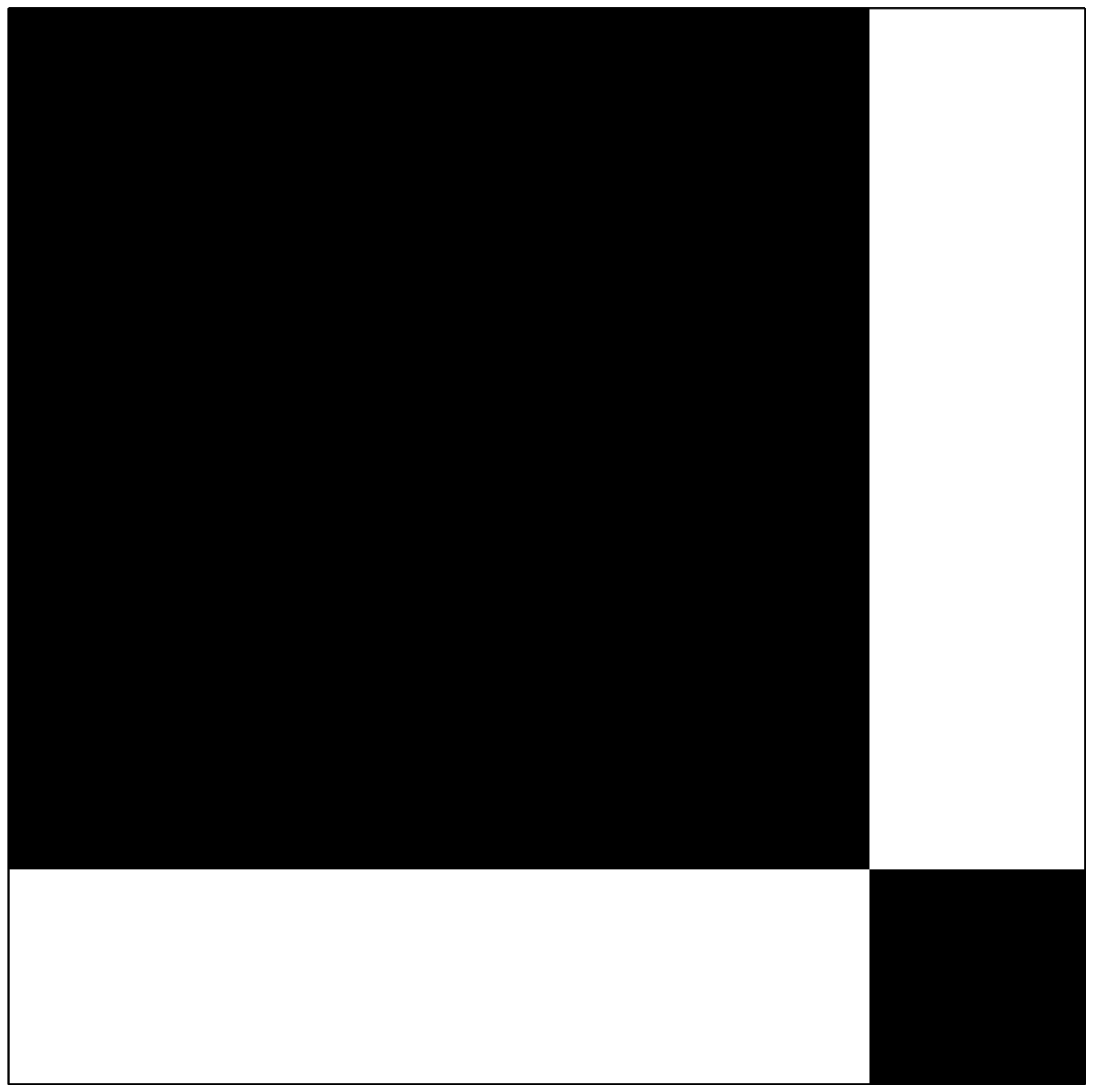}}%
 		\hspace{.05cm}%
\subfigure{	\includegraphics[width=0.5in]{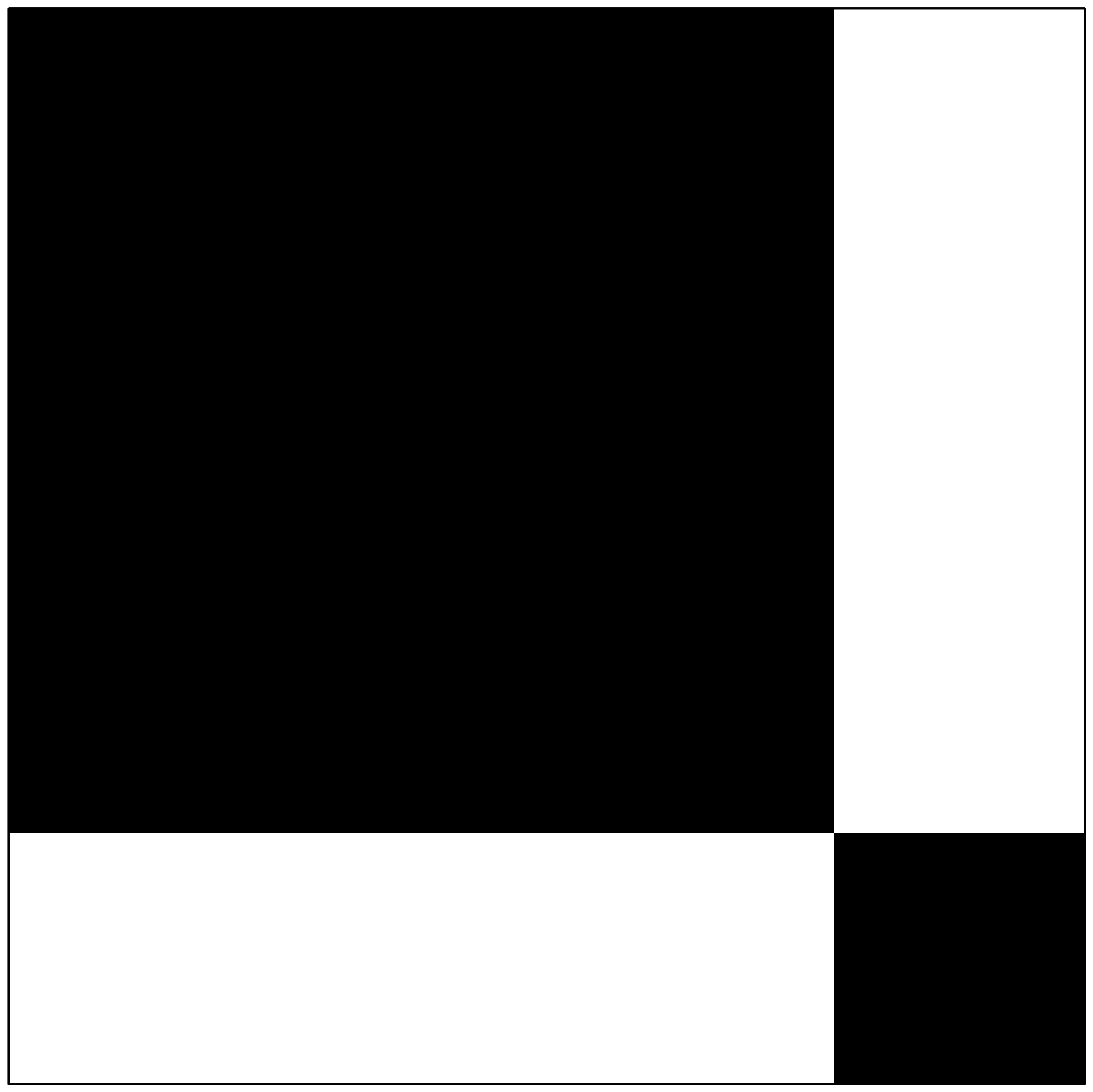}}%
 		\hspace{.05cm}%
\setcounter{subfigure}{2}
\subfigure[]{	\includegraphics[width=0.5in]{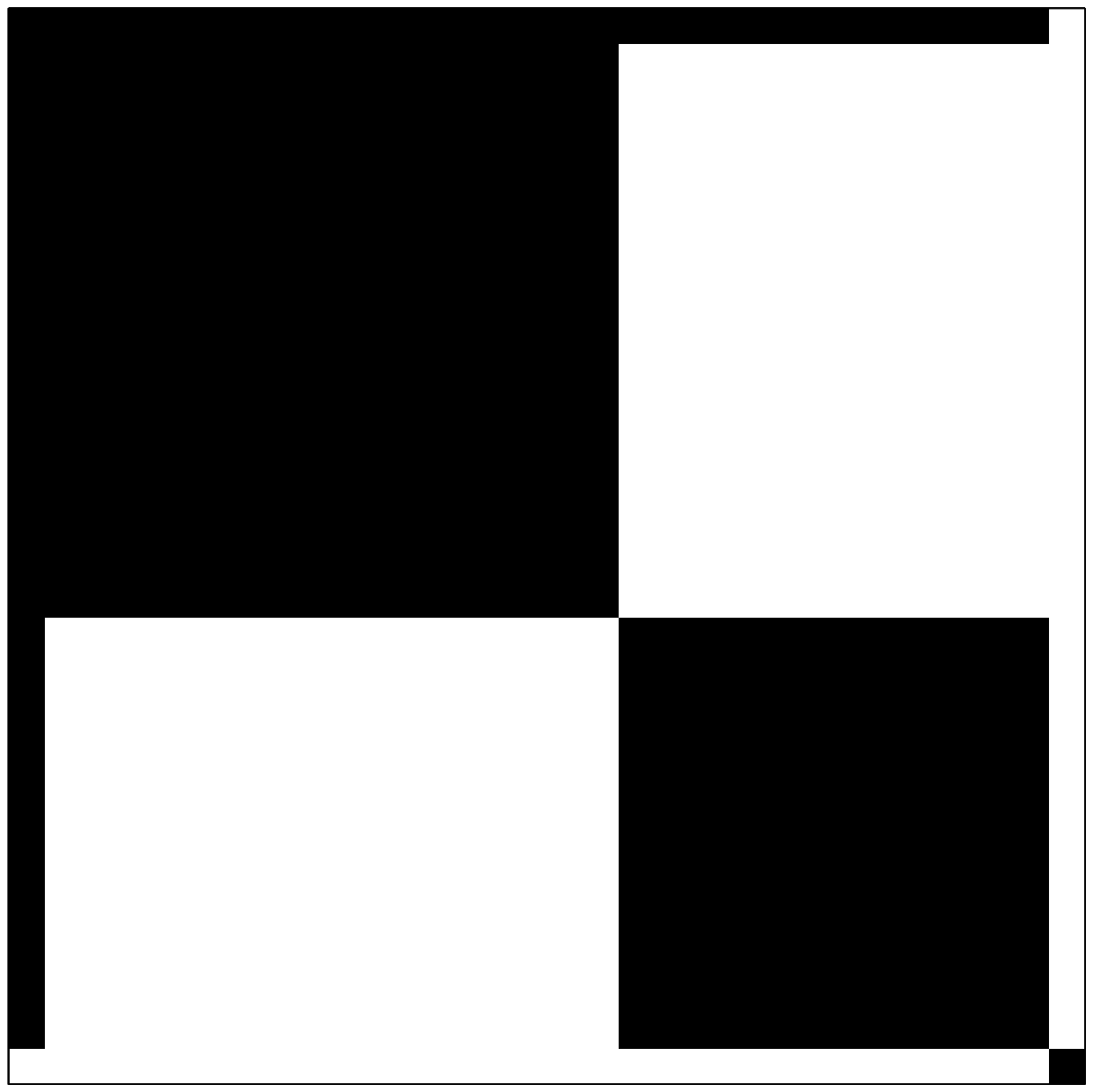}}%
 		\hspace{.05cm}%
\subfigure{	\includegraphics[width=0.5in]{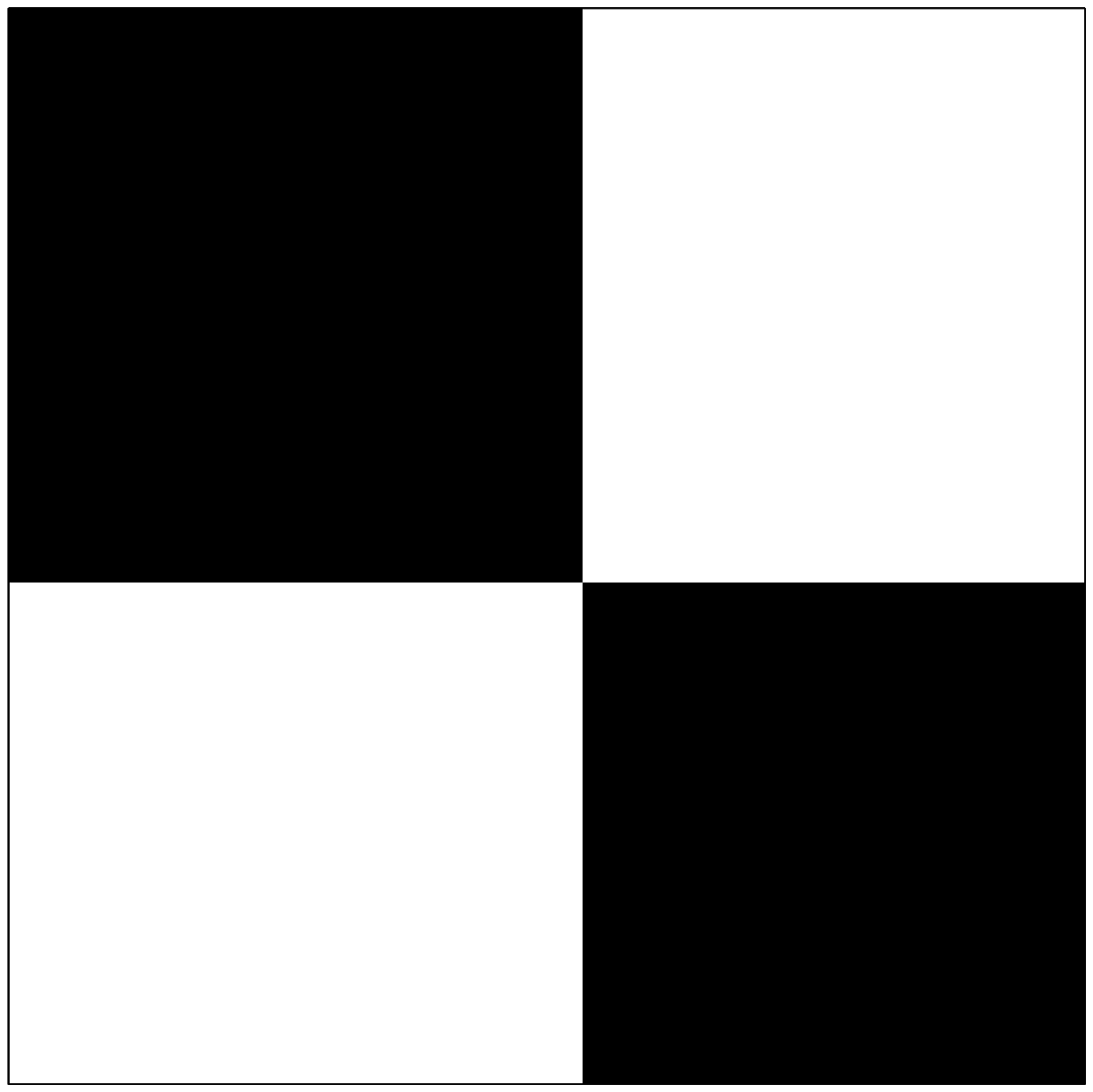}}%
 		\hspace{.05cm}%
\subfigure{	\includegraphics[width=0.5in]{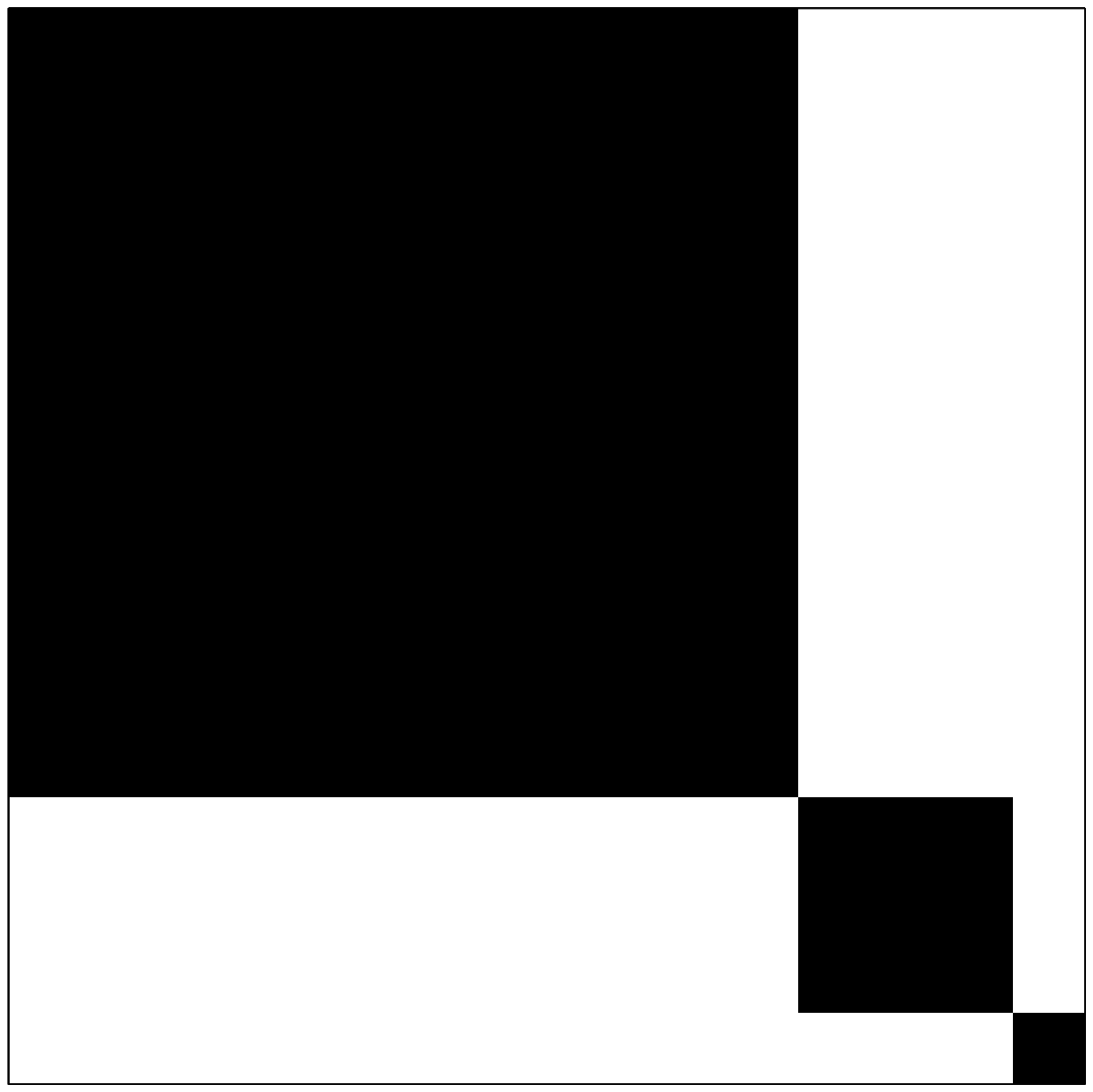}}%
 		\hspace{.5cm}%
\subfigure{ 	
		\includegraphics[width=0.5in]{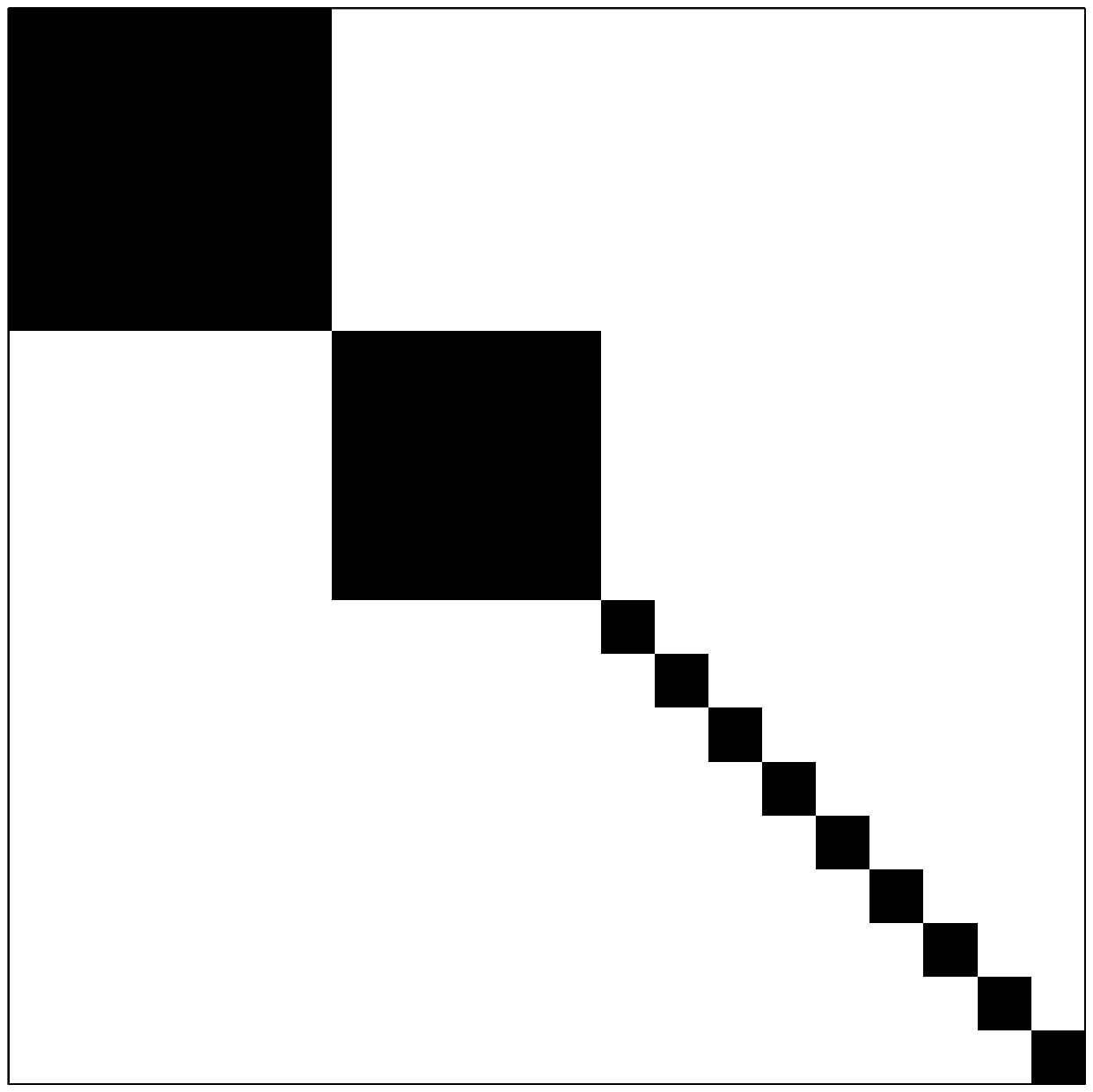}}%
 		\hspace{.05cm}%
\subfigure{	\includegraphics[width=0.5in]{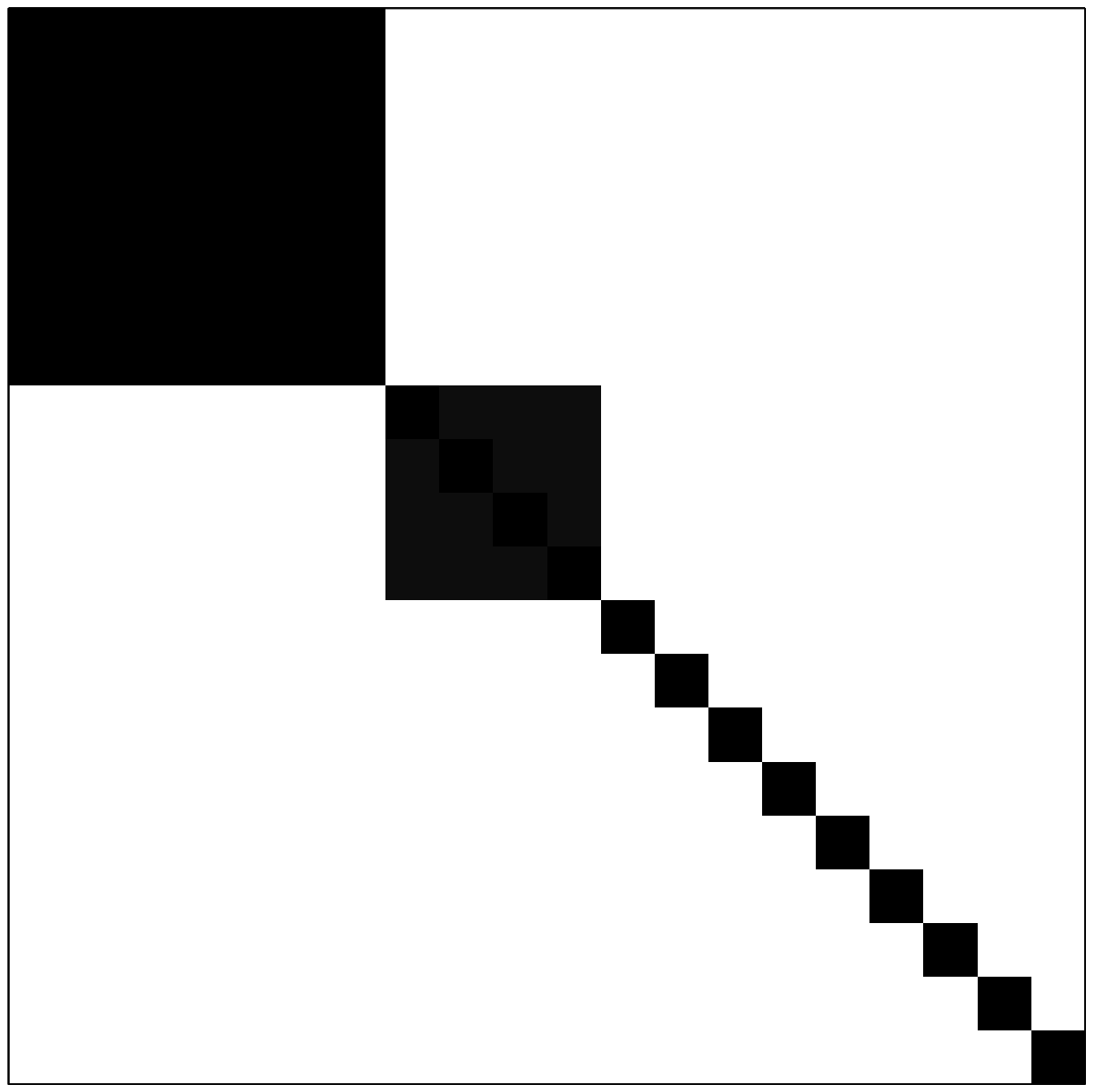}}%
 		\hspace{.05cm}%
\setcounter{subfigure}{3}
\subfigure[]{	\includegraphics[width=0.5in]{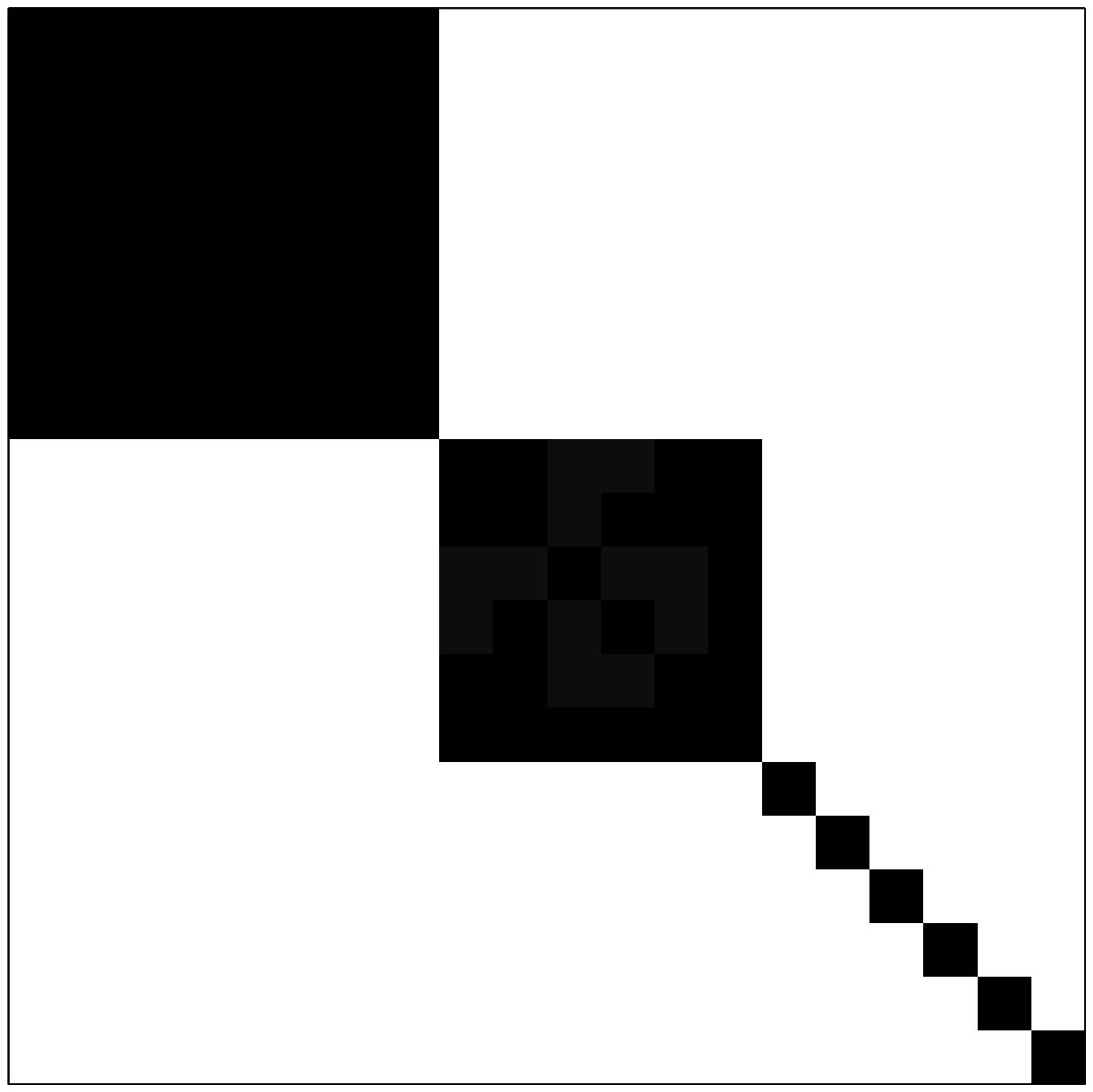}}%
 		\hspace{.05cm}%
\subfigure{	\includegraphics[width=0.5in]{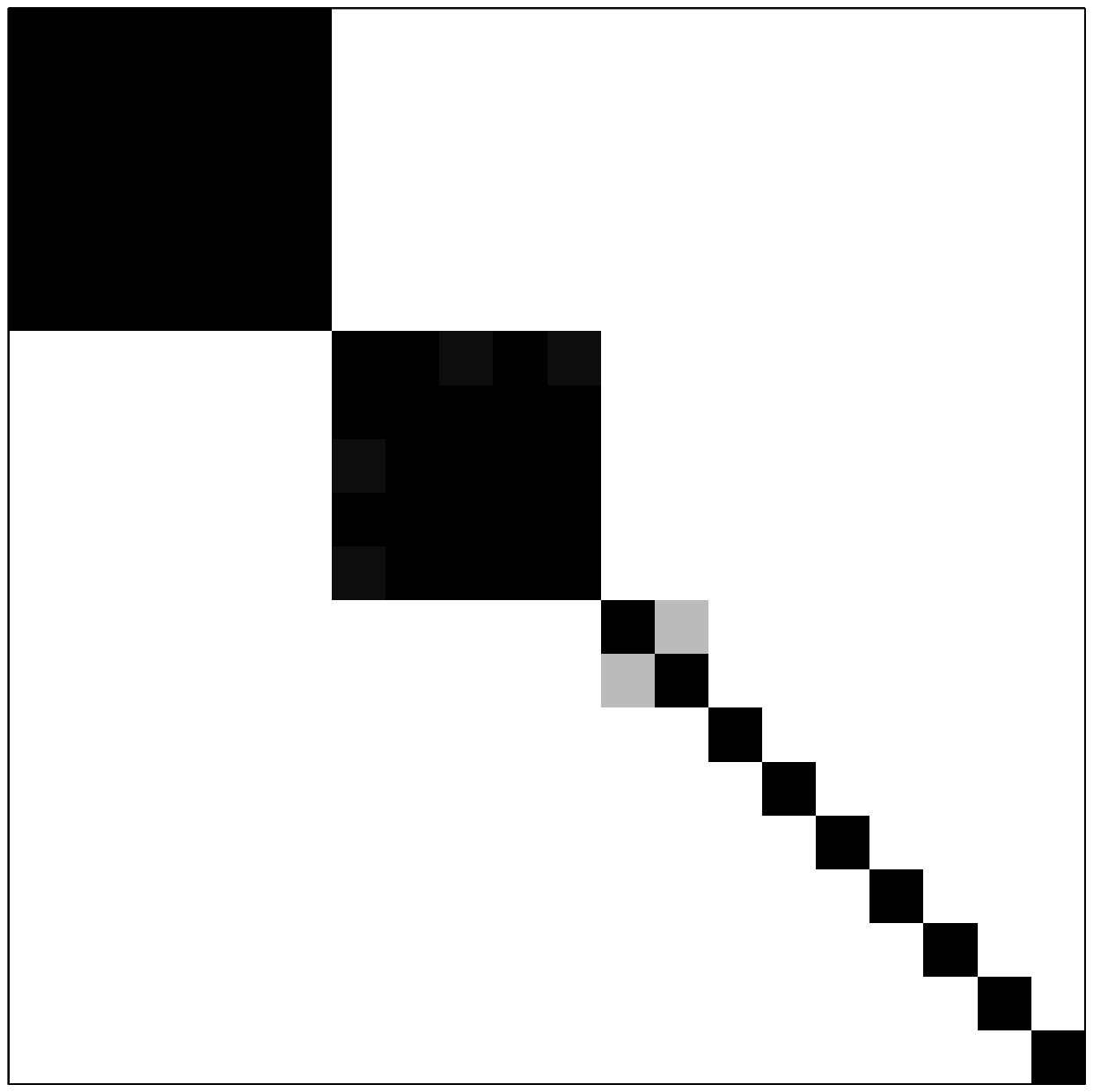}}%
 		\hspace{.05cm}%
\subfigure{	\includegraphics[width=0.5in]{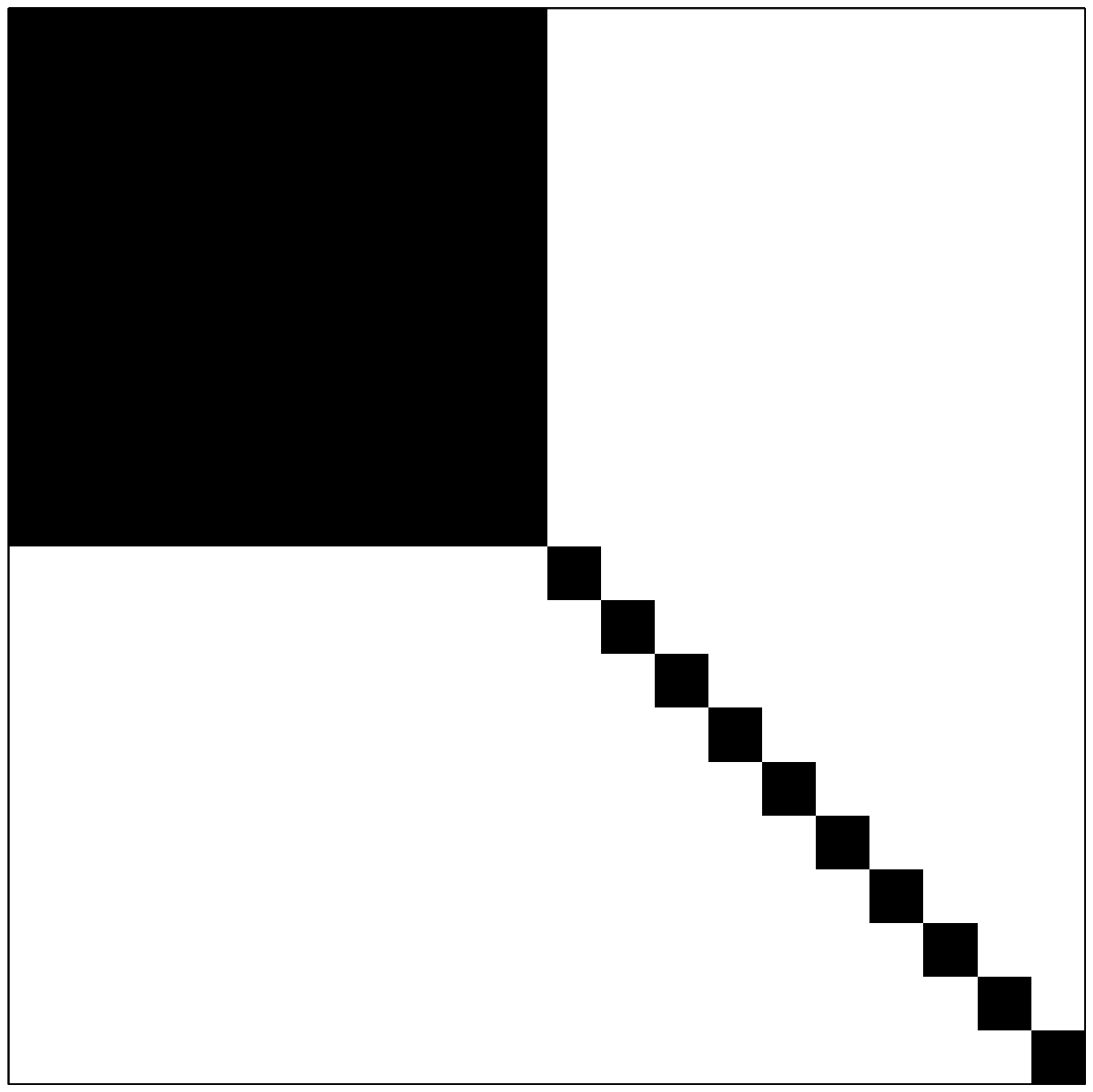}}\\

\subfigure{ 	
		\includegraphics[width=0.5in]{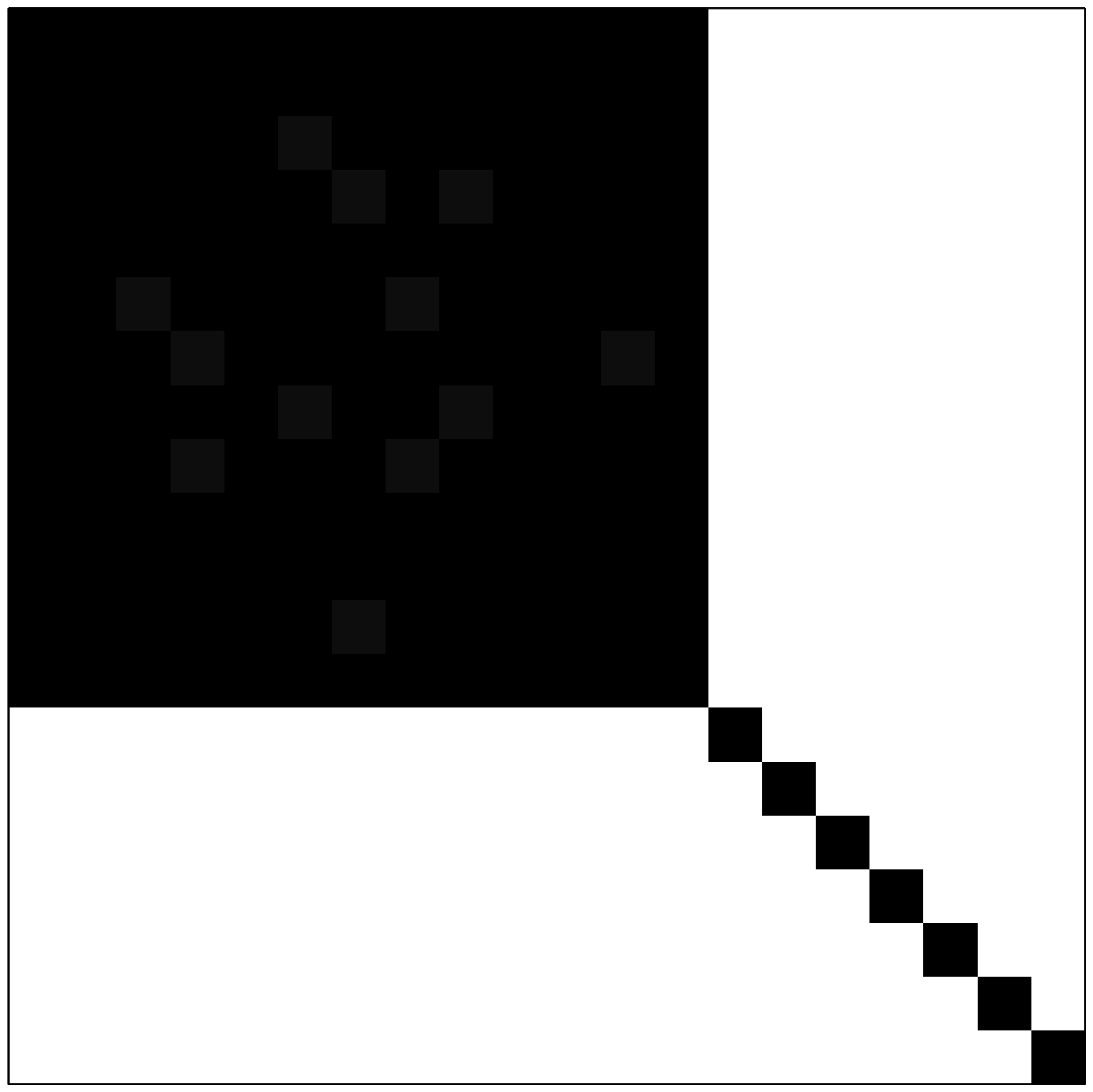}}%
 		\hspace{.05cm}%
\subfigure{	\includegraphics[width=0.5in]{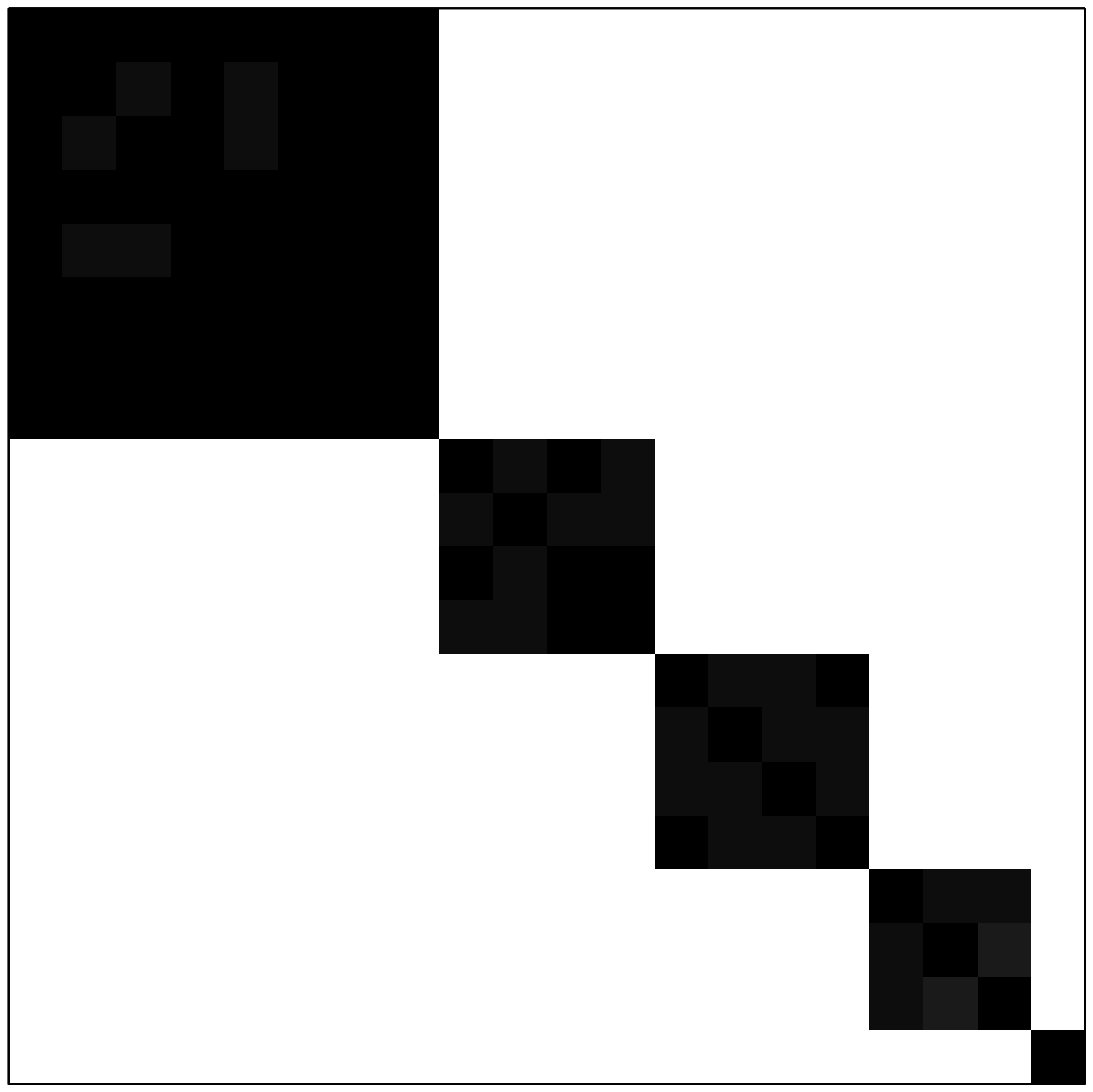}}%
 		\hspace{.05cm}%
\setcounter{subfigure}{4}
\subfigure[]{	\includegraphics[width=0.5in]{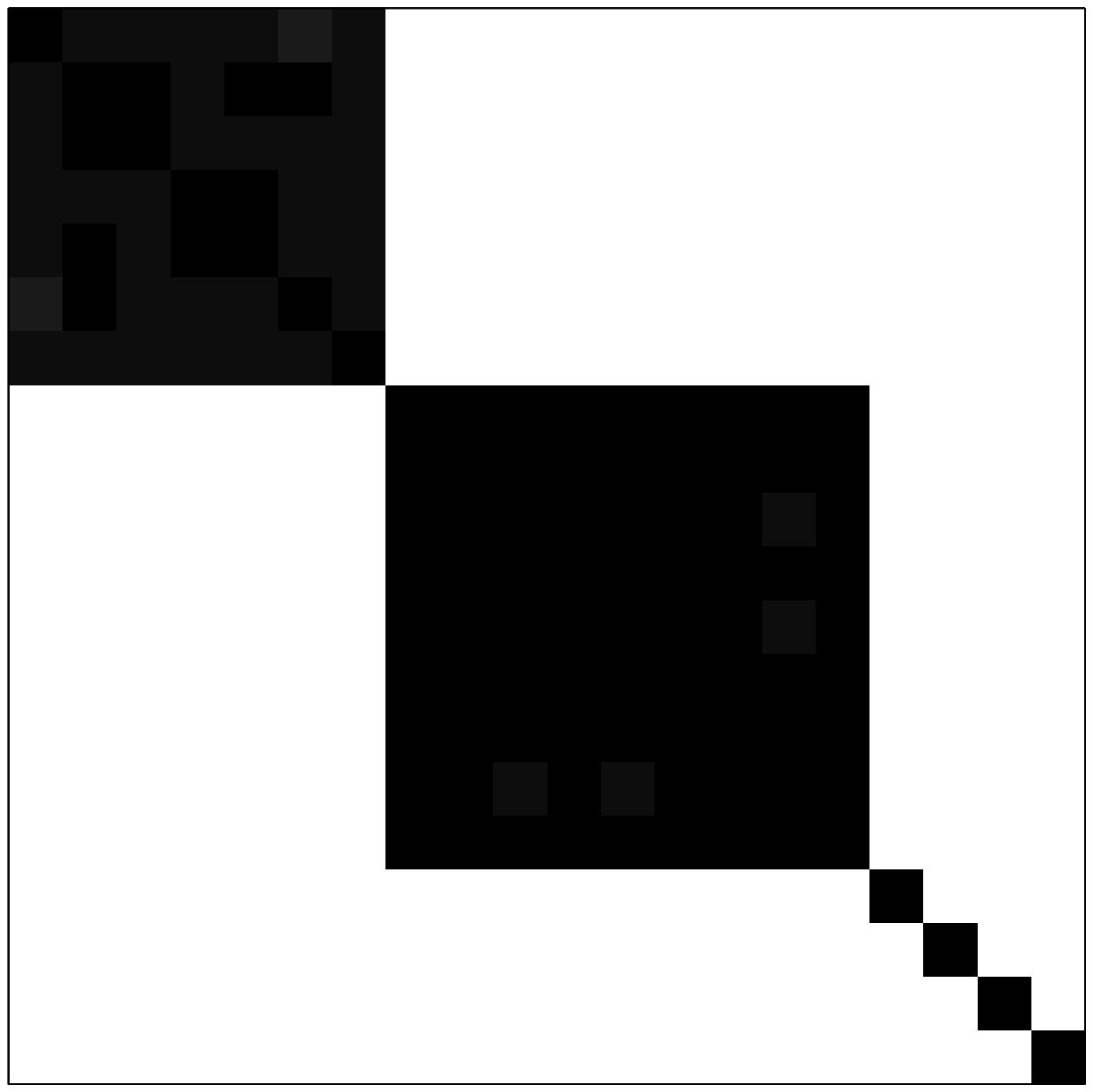}}%
 		\hspace{.05cm}%
\subfigure{	\includegraphics[width=0.5in]{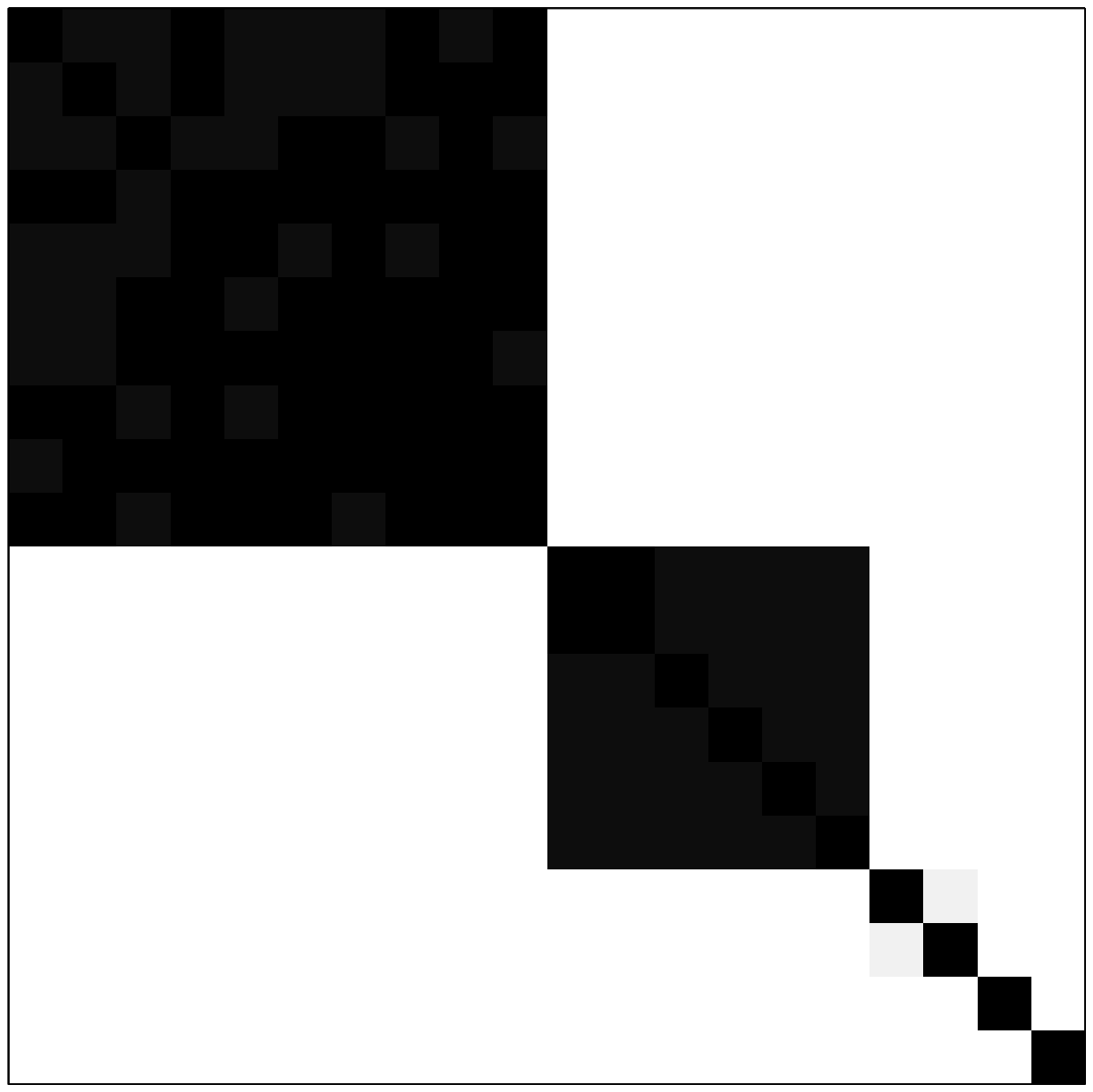}}%
 		\hspace{.05cm}%
\subfigure{	\includegraphics[width=0.5in]{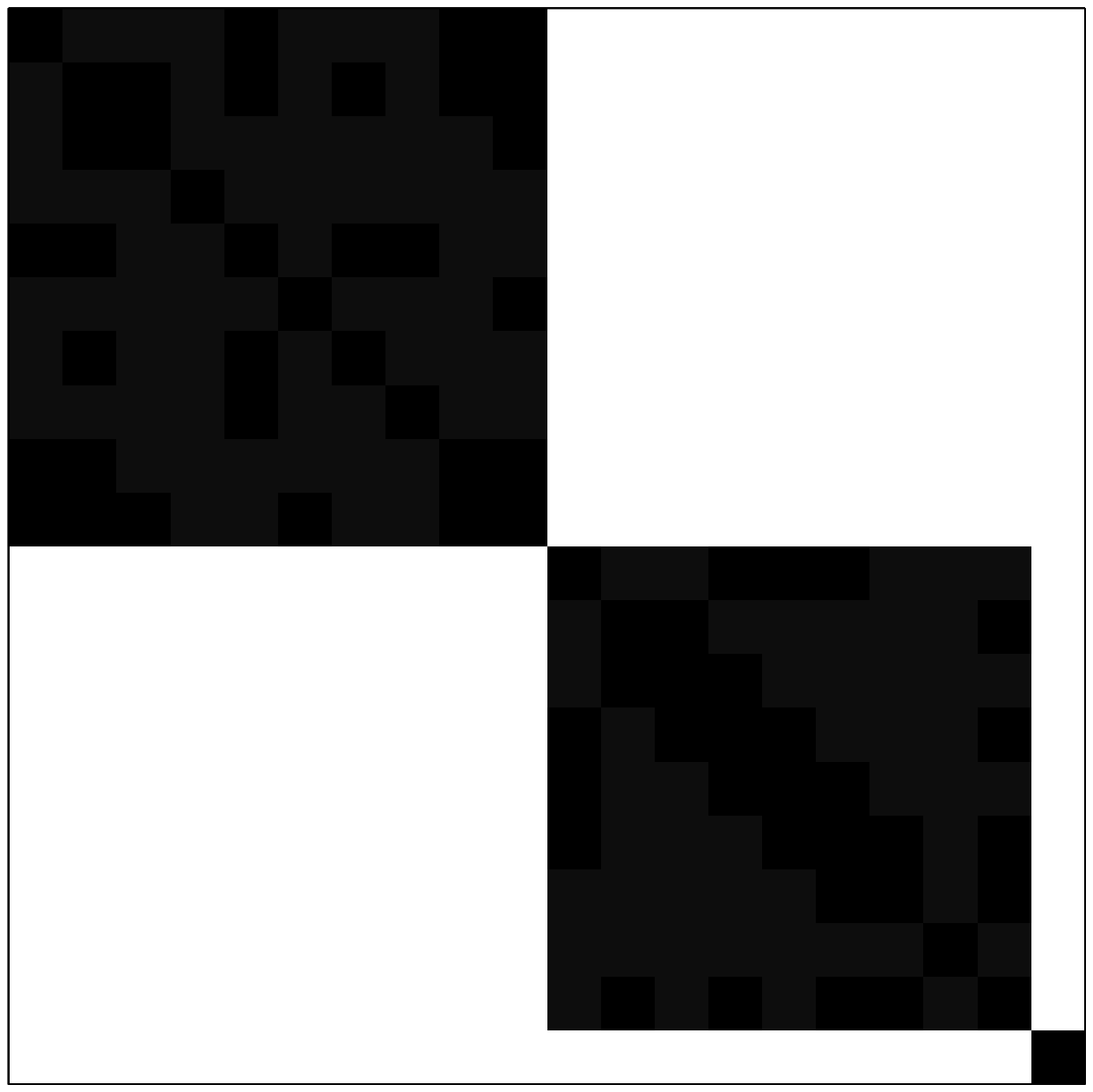}}%
 		\hspace{.5cm}%
\subfigure{ 	
		\includegraphics[width=0.5in]{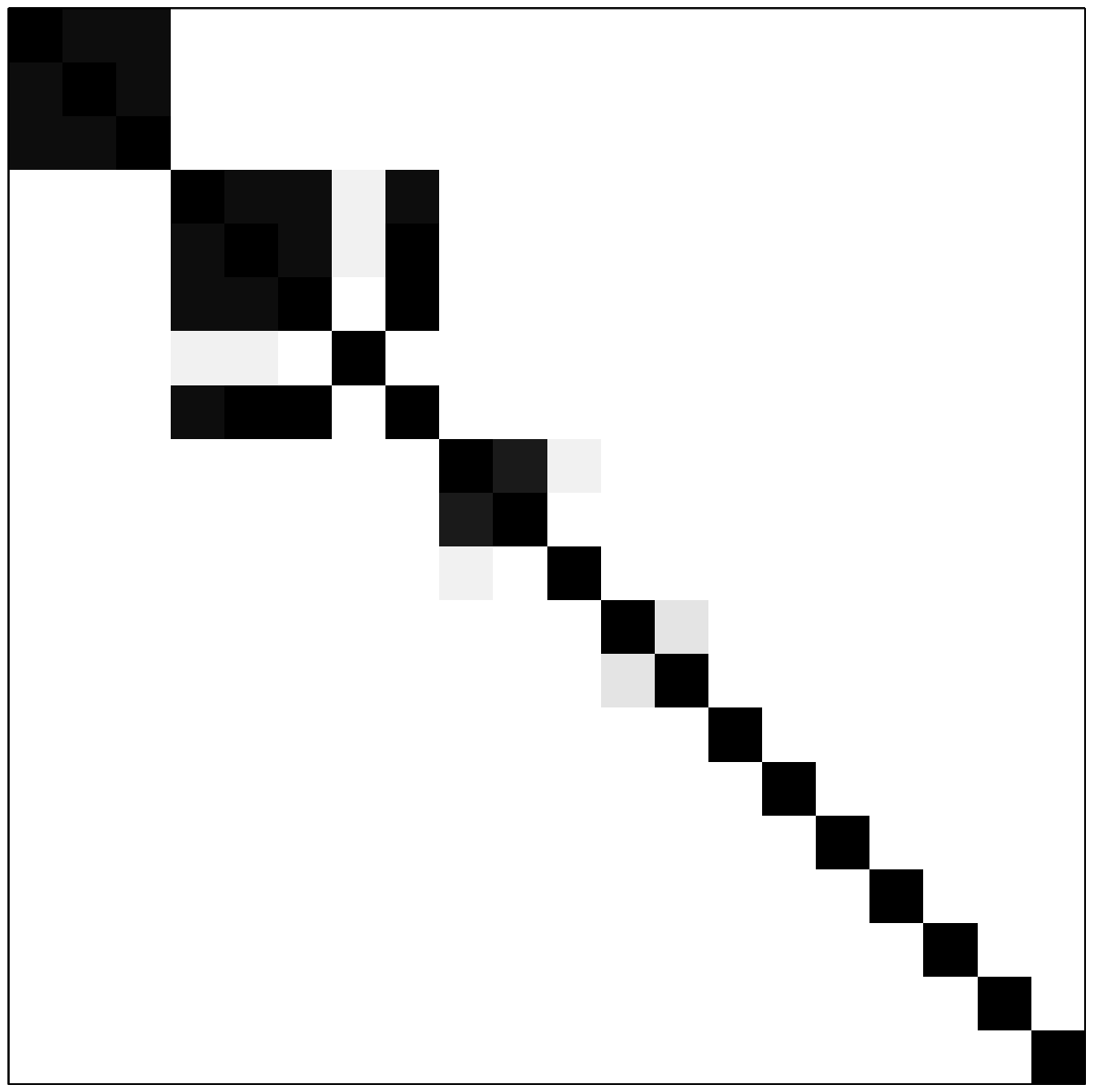}}%
 		\hspace{.05cm}%
\subfigure{	\includegraphics[width=0.5in]{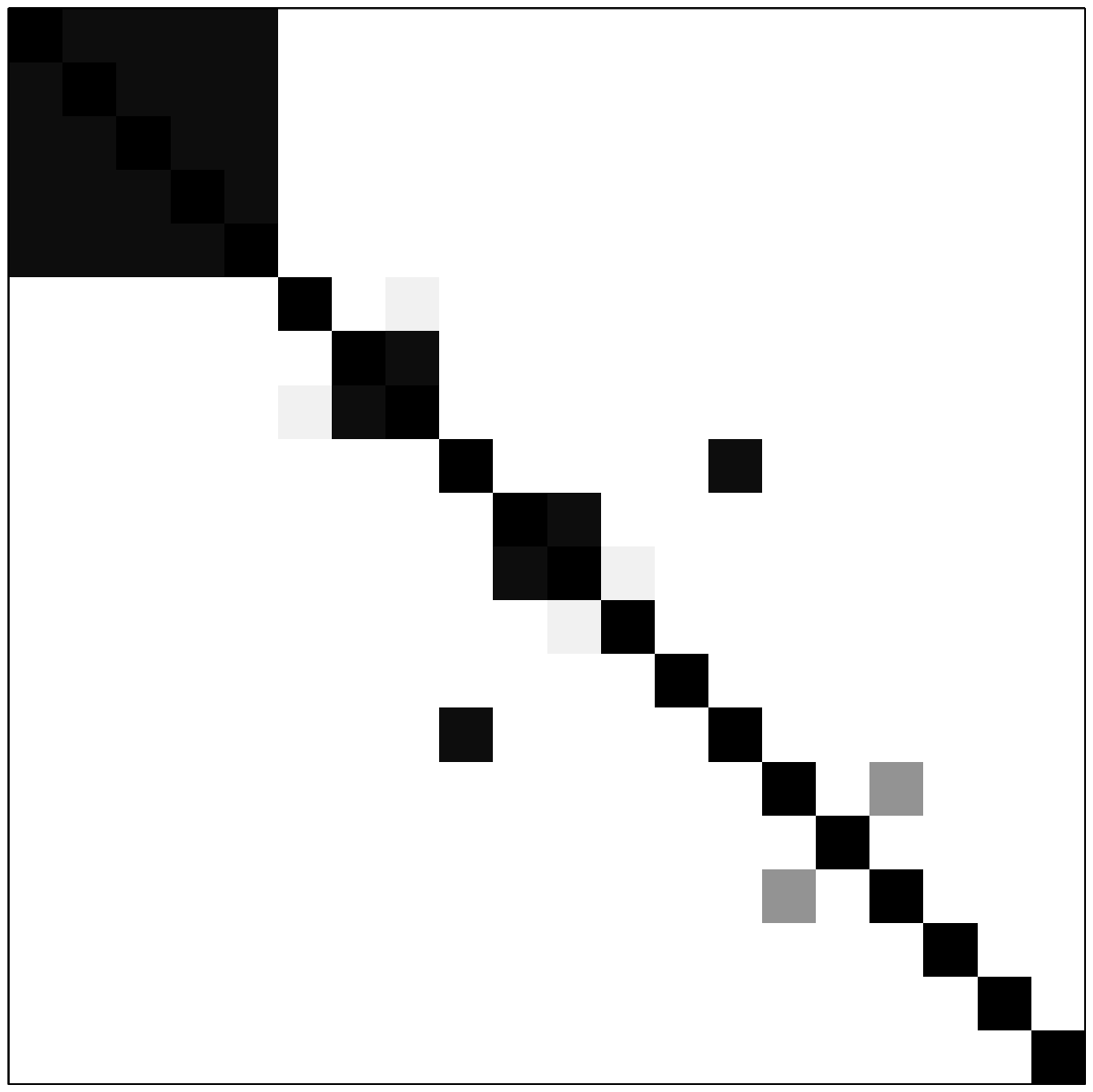}}%
 		\hspace{.05cm}%
\setcounter{subfigure}{5}
\subfigure[]{	\includegraphics[width=0.5in]{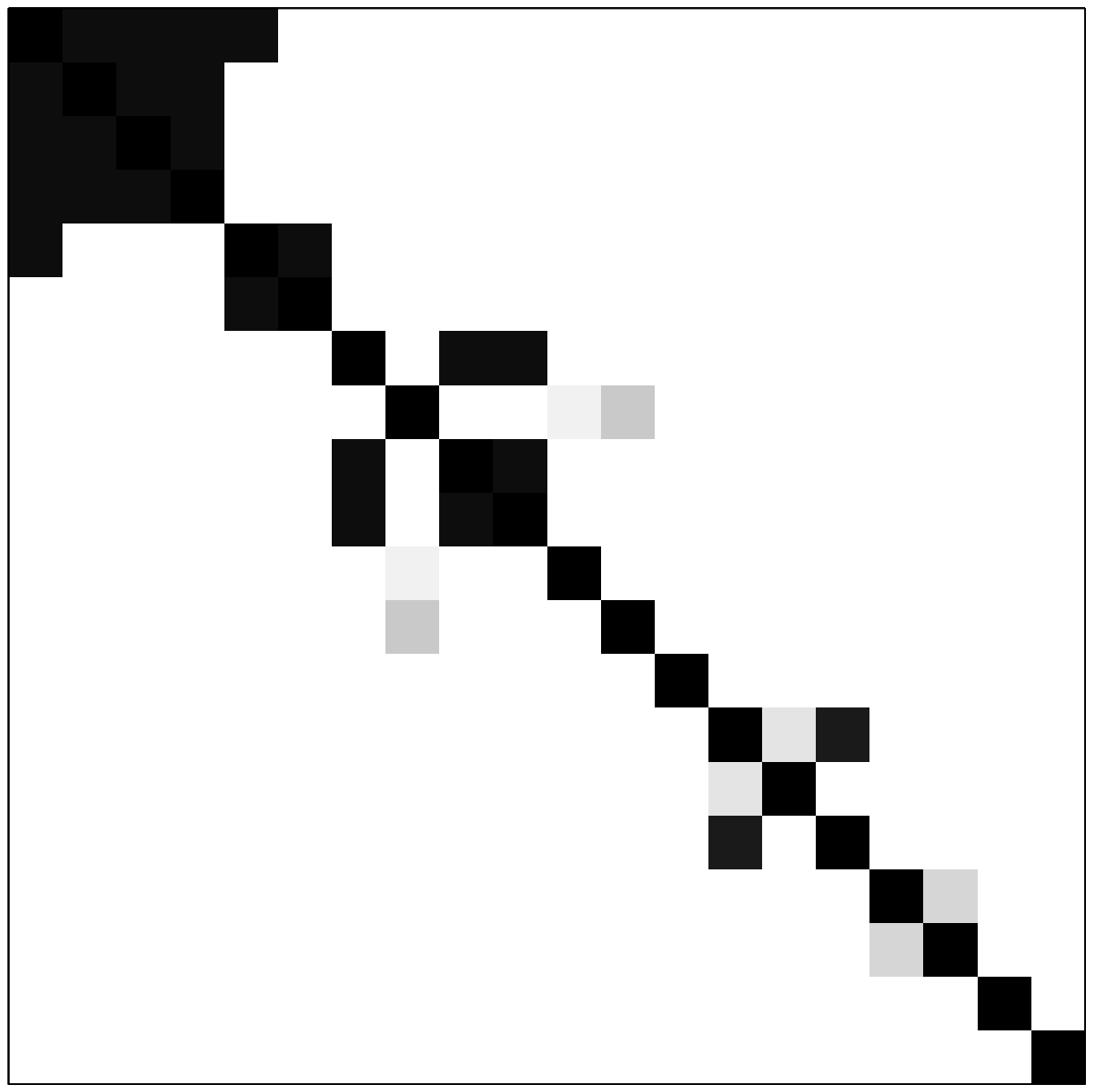}}%
 		\hspace{.05cm}%
\subfigure{	\includegraphics[width=0.5in]{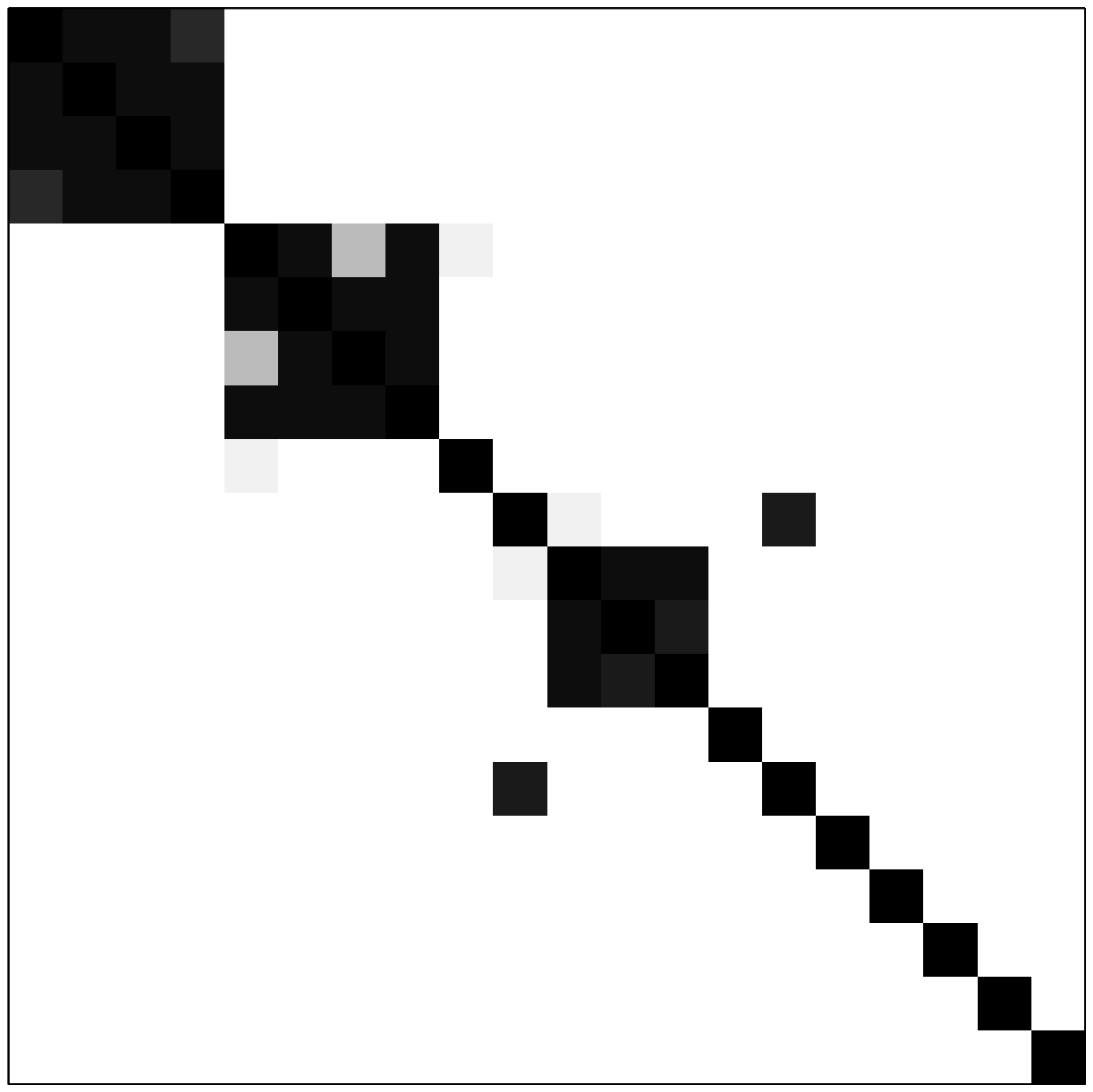}}%
 		\hspace{.125cm}%
\subfigure{	\includegraphics[width=0.5in]{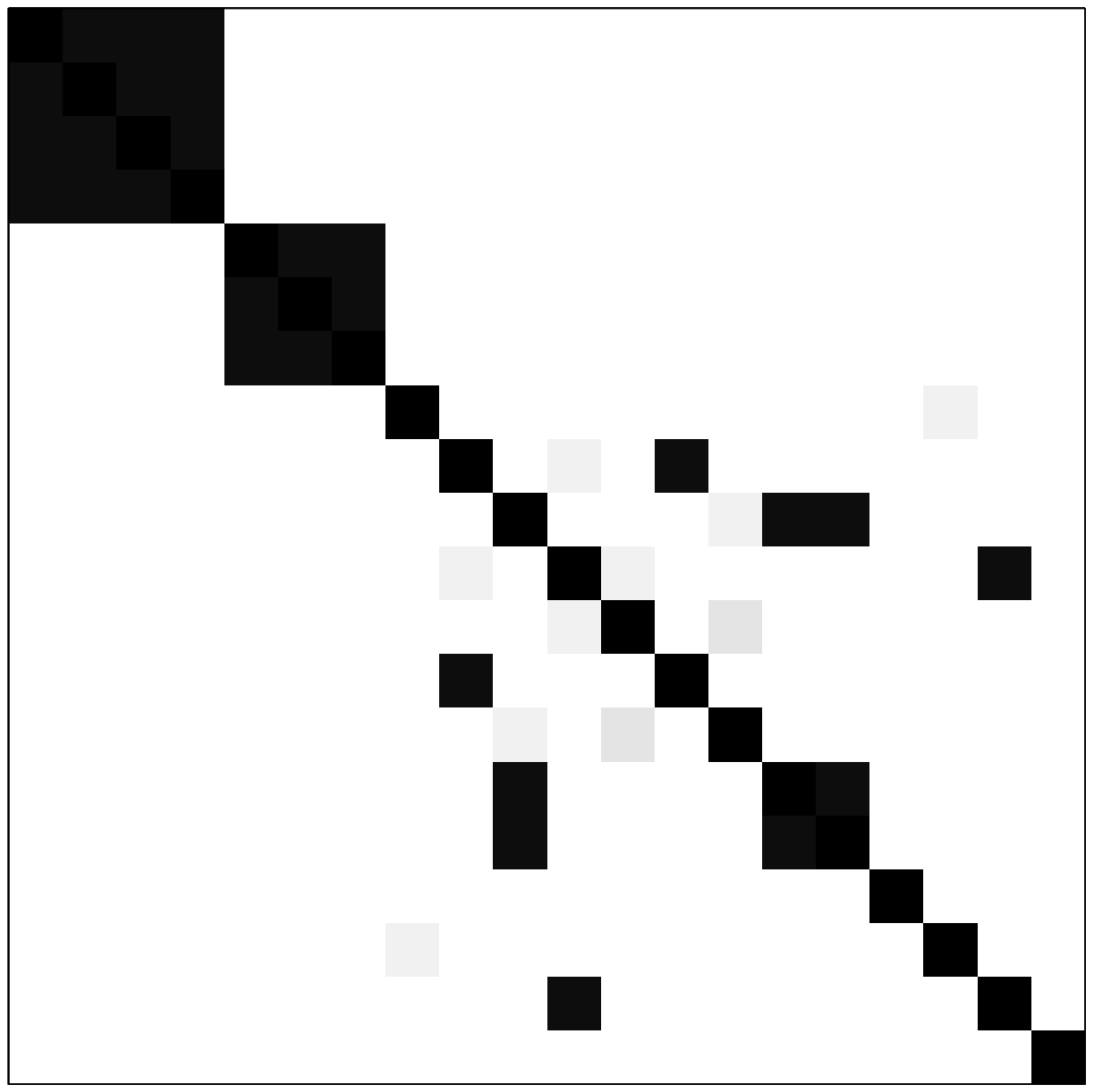}}\\
\caption{{\small {\bf Interference matrices at time 1000.}
Values of $h_{ij}$ are gray-scale coded from 0 (white) to 1 (black), with diagonal elements set to black. 
The smallest squares on the diagonal represent unaffiliated individuals.
For display purposes, alliances are ordered according to their clustering coefficients $C^{(1)}$
so that stronger alliances occur first along the diagonal. Parameters have
default values except where noted.
(a)~$N=10$.
(b)~$N=20$.
(c)~$N=30$.
(d)~$N=20, \,\gd_{ll}=-0.5$.
(e)~$N=20, \,\mu=0.1$.
(f)~$N=20, \,\go=0.25$.
}}
\label{sample}
\end{figure}

Figure~\ref{sample} 
illustrates some coalitionary regimes observed in simulations using
a default set of parameters ($\ga=1, \gb=0.05$, $\gd_{ww}=1, \gd_{ll}=0.5, \gd_{wl}=-0.5, \gd_{lw}=-0.5,
\eta=0.5, \go=0.5$, $\mu_a=0.05$) unless noted otherwise. 
This figure shows the $N\times N$  interference matrices using small squares arranged in an $N\times N$ array with each of the squares color-coding for the corresponding value of $h_{ij}$  using the gray scale. The squares on the diagonal are painted black for convenience. 
In all examples,
individual strengths $s_i$ are chosen randomly and
independently from a uniform distribution on $[0,10]$ resulting in strong between-individual
variation.

{\em Emergence of alliances.}\quad
In our model, the affinity between any two individuals is reinforced if they are on a winning side of a conflict and is decreased if they are on the opposite sides; all affinities also decay to zero at a constant rate. The resulting state represents a balance between factors increasing and decreasing affinities. Although the emergence of alliances is in no way automatic, simulations show that under certain conditions they do emerge. The size, strength, and temporal stability of alliances depend on parameters and may vary dramatically from one run to another even with the same parameters. However, once one or more alliances with high values of $C^{(1)}, C^{(2)}$ and 
$\ov{h}$ are formed, they are typically stable. Individuals belonging to the same alliance have very similar social success which is only weakly correlated with their fighting abilities. That is, the social success is now defined not by the individual?s fighting ability but by the size and strength of the alliance he belongs to. Individuals from different alliances can have vastly different social success, so that the formation of coalitions and alliances does not necessarily reduce social inequality in the group as a whole.

\begin{figure}[tb]
\centering
\subfigure[]{ 	
		\includegraphics[width=2in]{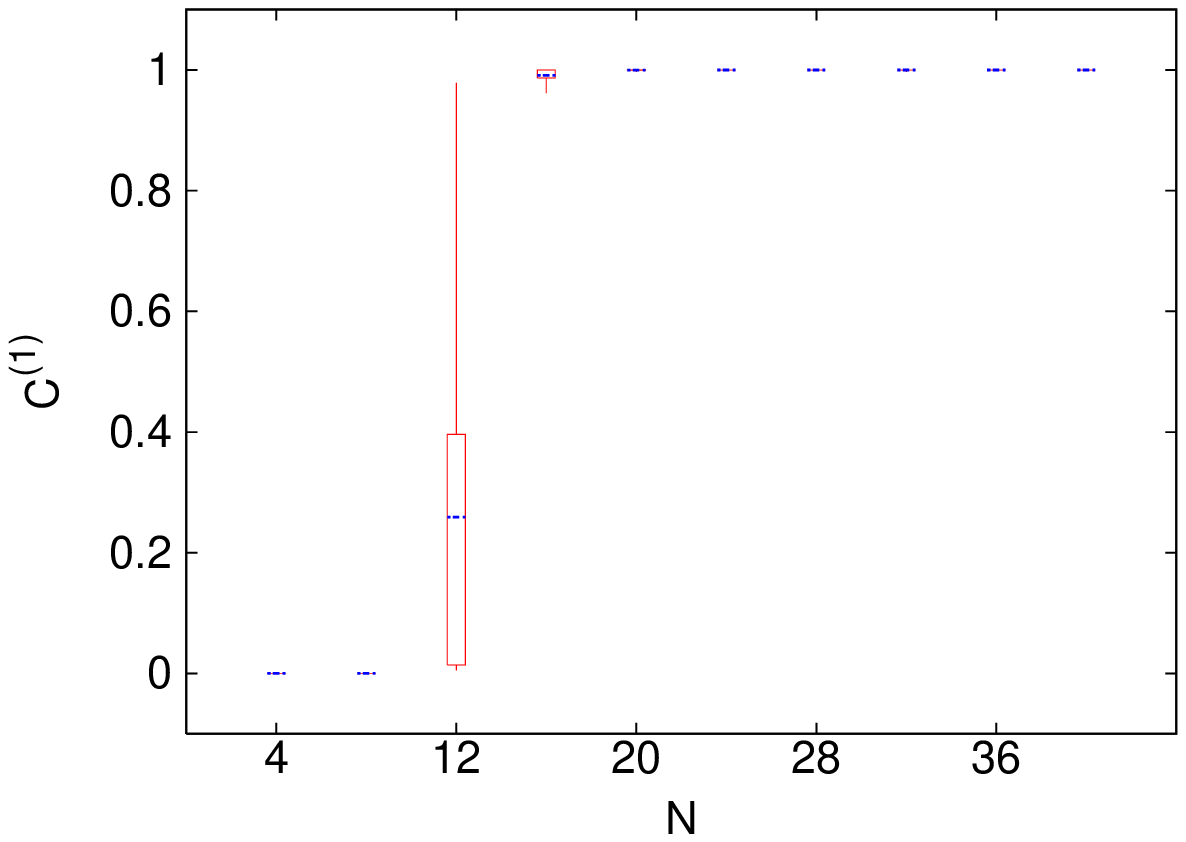}}%
 		\hspace{.125cm}%
\subfigure[]{	\includegraphics[width=2in]{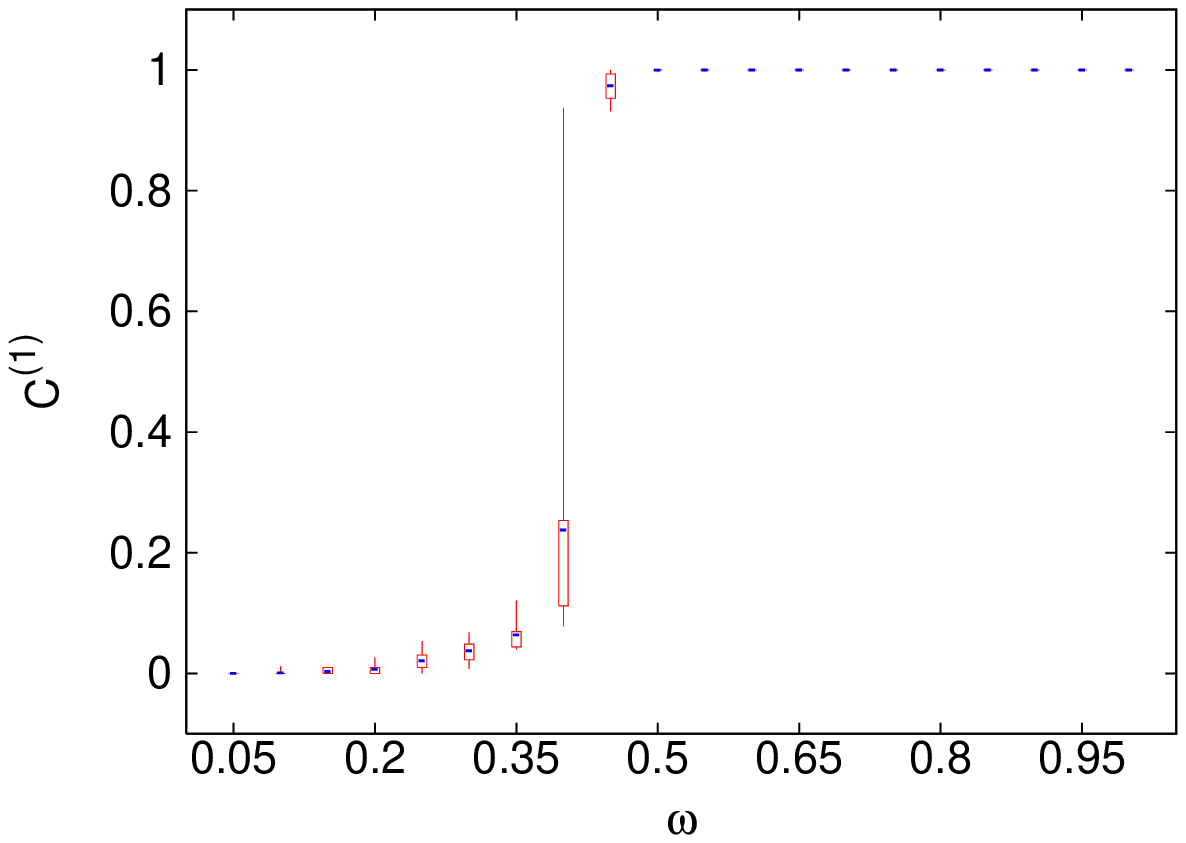}}%
 		\hspace{.125cm}%
\subfigure[]{	
		\includegraphics[width=2in]{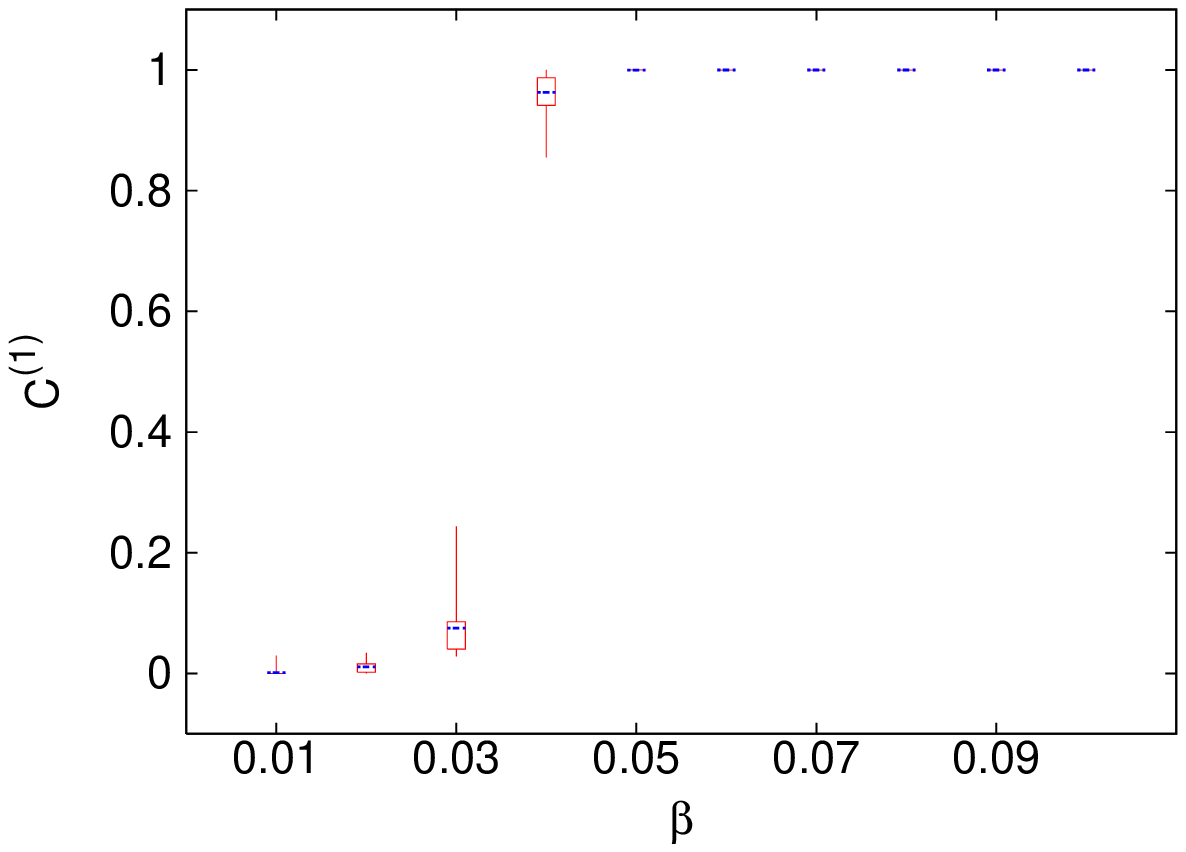}}\\
\subfigure[]{	
		\includegraphics[width=2in]{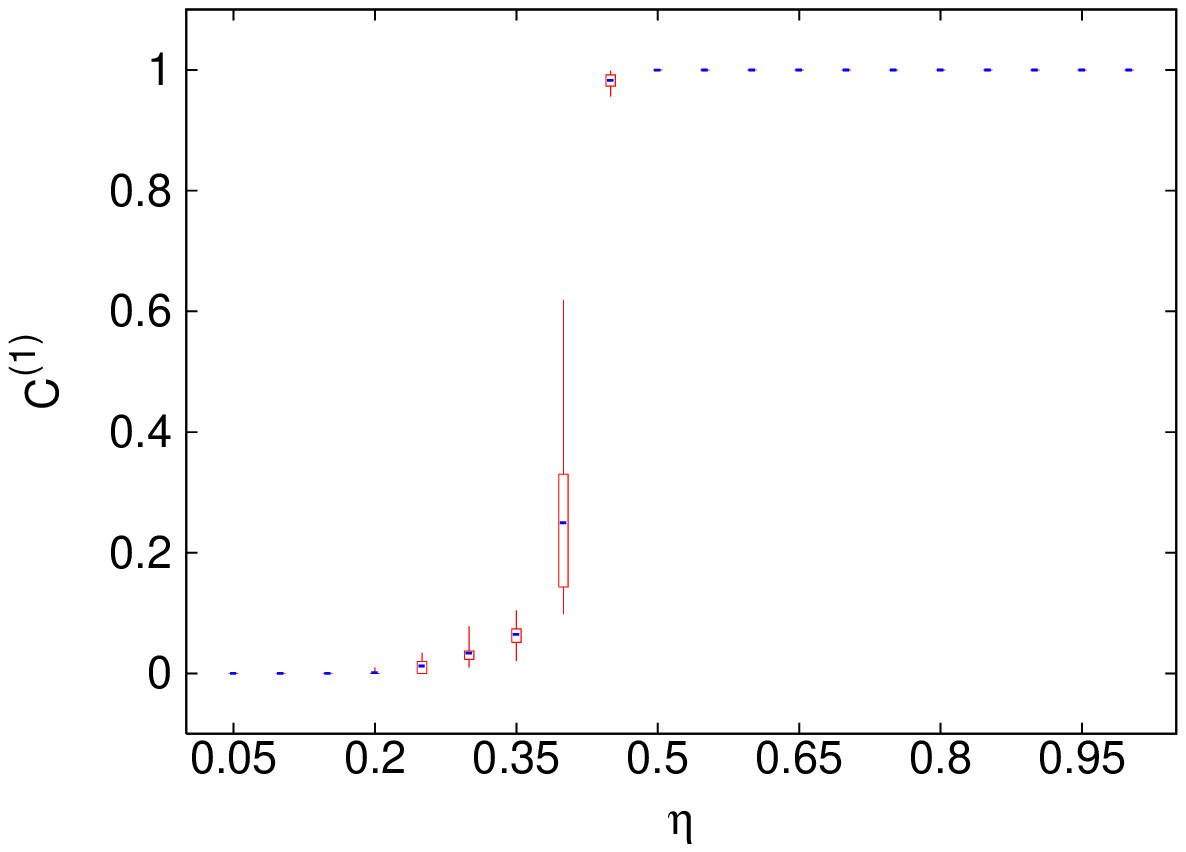}}%
		\hspace{.125cm}%
\subfigure[]{	\includegraphics[width=2in]{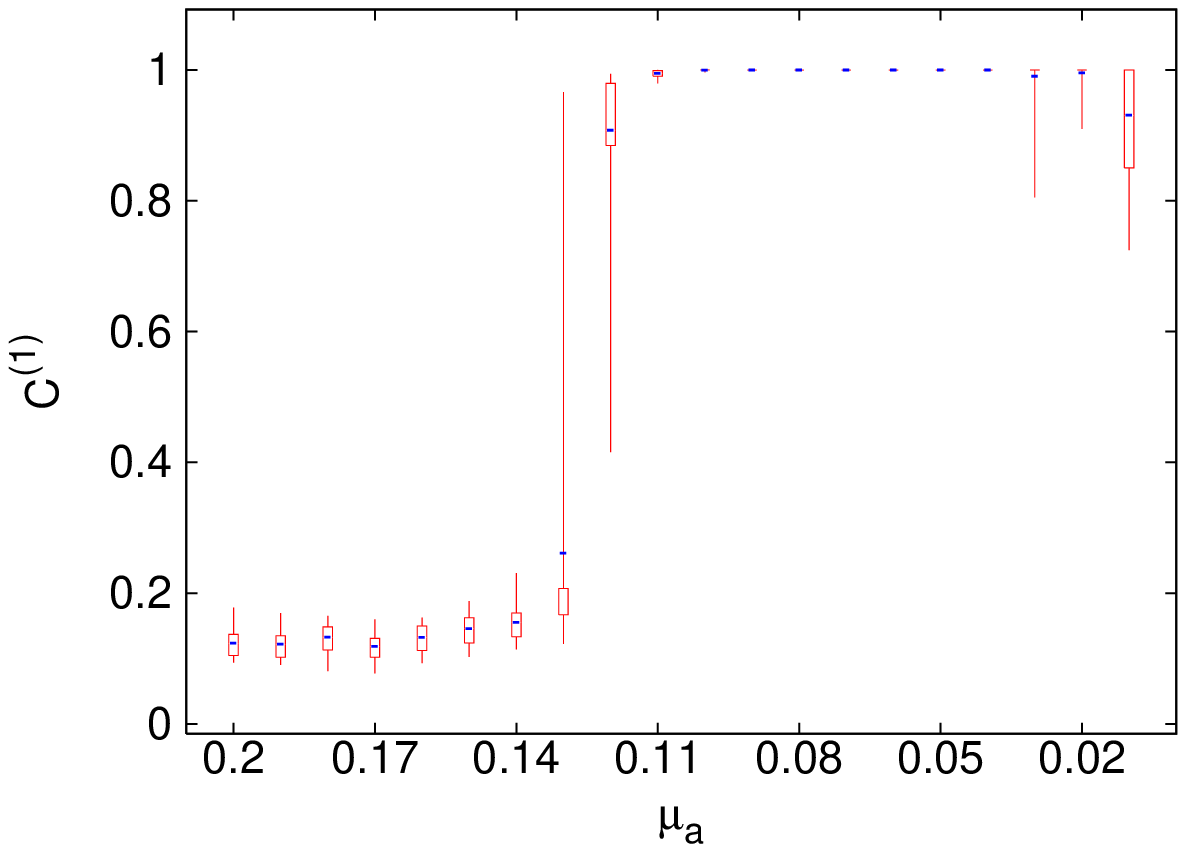}}%
 		\hspace{.125cm}%
\subfigure[]{	
		\includegraphics[width=2in]{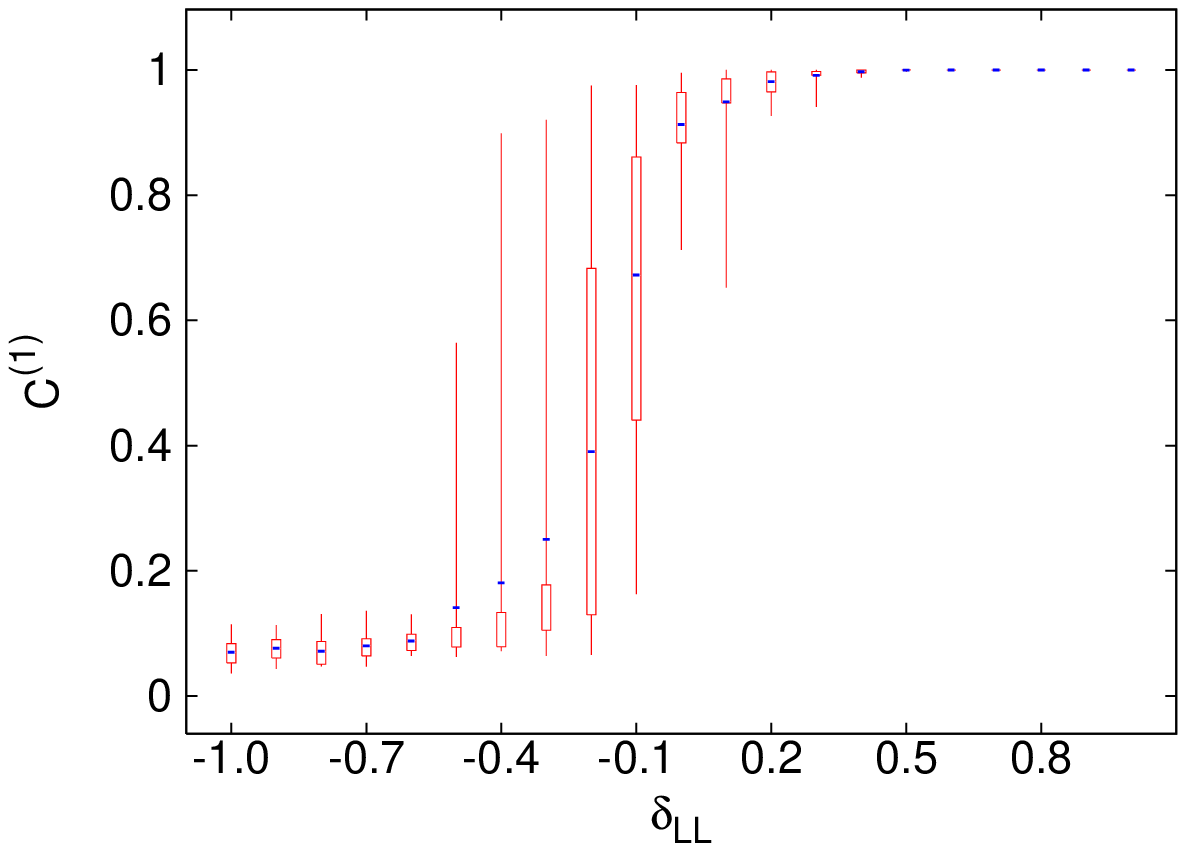}}\\
\caption{{\small {\bf Tukey plots for the effects of
$N,\go,\gb,\eta,\mu,\gd_{ll}$ on the $C^{(1)}$ measure of the largest
alliance.} Each graph shows the effect of changing a single parameter
from its default value (results for each parameter value are averaged
over 20 runs, using data from time 1000 to 2000). The vertical lines
extend from minimum to maximum observations, the dashed lines
depict averages, and the boxes extend from lower to upper
quartiles.}}
\label{phase_transition}
\end{figure}

{\em Phase transition.}\quad
We performed a detailed numerical study of the effects of individual parameters of the properties of 
the system. As expected, increasing the frequency of interactions (which can be achieved
by increasing the group size $N$, 
the awareness probability $\go$, baseline interference rate $\gb$, or the slope parameter $\eta$) 
and reducing the affinity decay rate $\mu$ all promote alliance formation.
Most interestingly, some characteristics 
change in a phase transition-like pattern as some parameters
undergo small changes. For example, Figure~\ref{phase_transition} 
show that increasing $N, \go, \gb, \eta$, or decreasing $\mu$ result
in a sudden transition from no alliances to at least one very strong alliance
with all members always supporting each other. Parameter $\gd_{ll}$ has a similar
but less extreme effect, whereas parameters $\gd_{wl}$ and $\gd_{lw}$ have relatively weak effects
(Supplementary Information).
Similar threshold-like behavior is exhibited by
the $C^{(2)}$-measure, the average probability of help $\ov{h}$ within the largest alliance, 
the number of alliances, and the numbers 
of alliances with $C^{(1)}>0.5$ and with $\ov{h}>0.5$
(Supplementary Information).
Interestingly, formation of multiple alliances is hindered when affinities
between individuals fighting 
on the same side decrease as a result of losing (i.e., if $\gd_{ll}<0$).


{\em Cultural inheritance of social networks.}\quad
Next, we extended the model to larger temporal scales by allowing for birth/death
events, and the cultural inheritance of social networks.
New individuals are born at a constant rate $\gga$.
Each birth causes the death of a different randomly chosen individual.
We explored two rather different scenarios of cultural inheritance.
In the first, the offspring inherits the social network of its parent who is chosen among all 
individuals with a probability proportional to the rate of social success $Y_i$. 
This scenario requires special social bonds between parents and offspring.
In the second, each new individual inherits affinities of its ``role model'' (chosen 
from the whole group either with a uniform probability or
with a probability proportional to the rate of social success $Y_i$). 
Under both scenarios, if individual $i^*$ is an offspring (biological in the first scenario or cultural 
in the second scenario) of individual $i$, then we set $x_{i^*j}=\gk x_{ij}$
for each other individual $j$ in the group (parameter $0 \leq \gk \leq 1$ controls
the strength of social network inheritance).
In the parent-offspring case, the affinities of other individuals to the son
are proportional to those to the father: $x_{ji^*}=\gk x_{ji}$ 
and $x_{i^*i} = x_{ii^*}$ is set to $\gk$ times the maximum existing affinity in the group.
In the role model case, other individuals initially have zero affinities to the 
new member of the group: $x_{ji^*}=0$.

%

\begin{figure}[t]
\centering
\subfigure{ 	
		\includegraphics[width=0.5in]{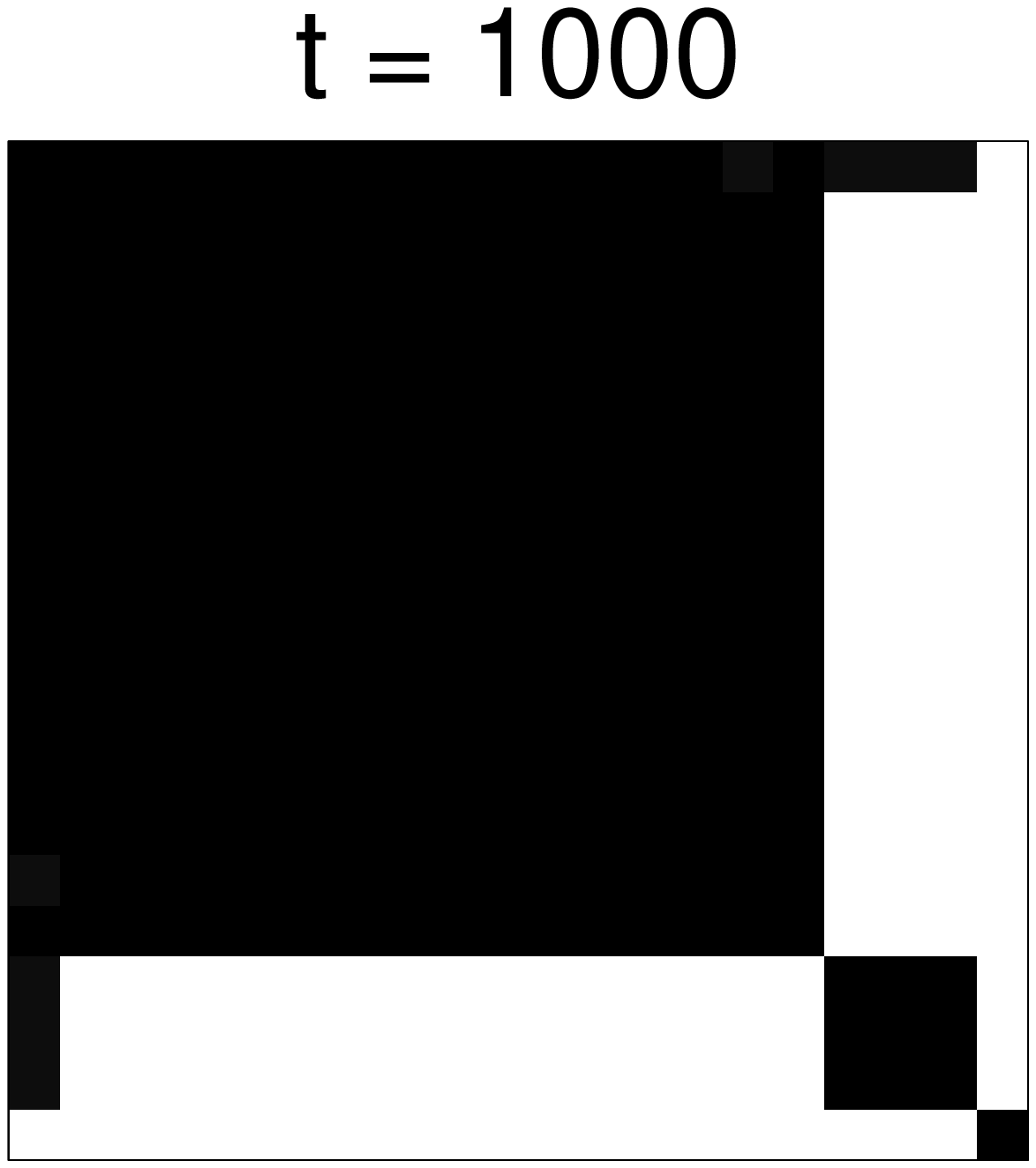}}%
 		\hspace{.125cm}%
\subfigure{	
		\includegraphics[width=0.5in]{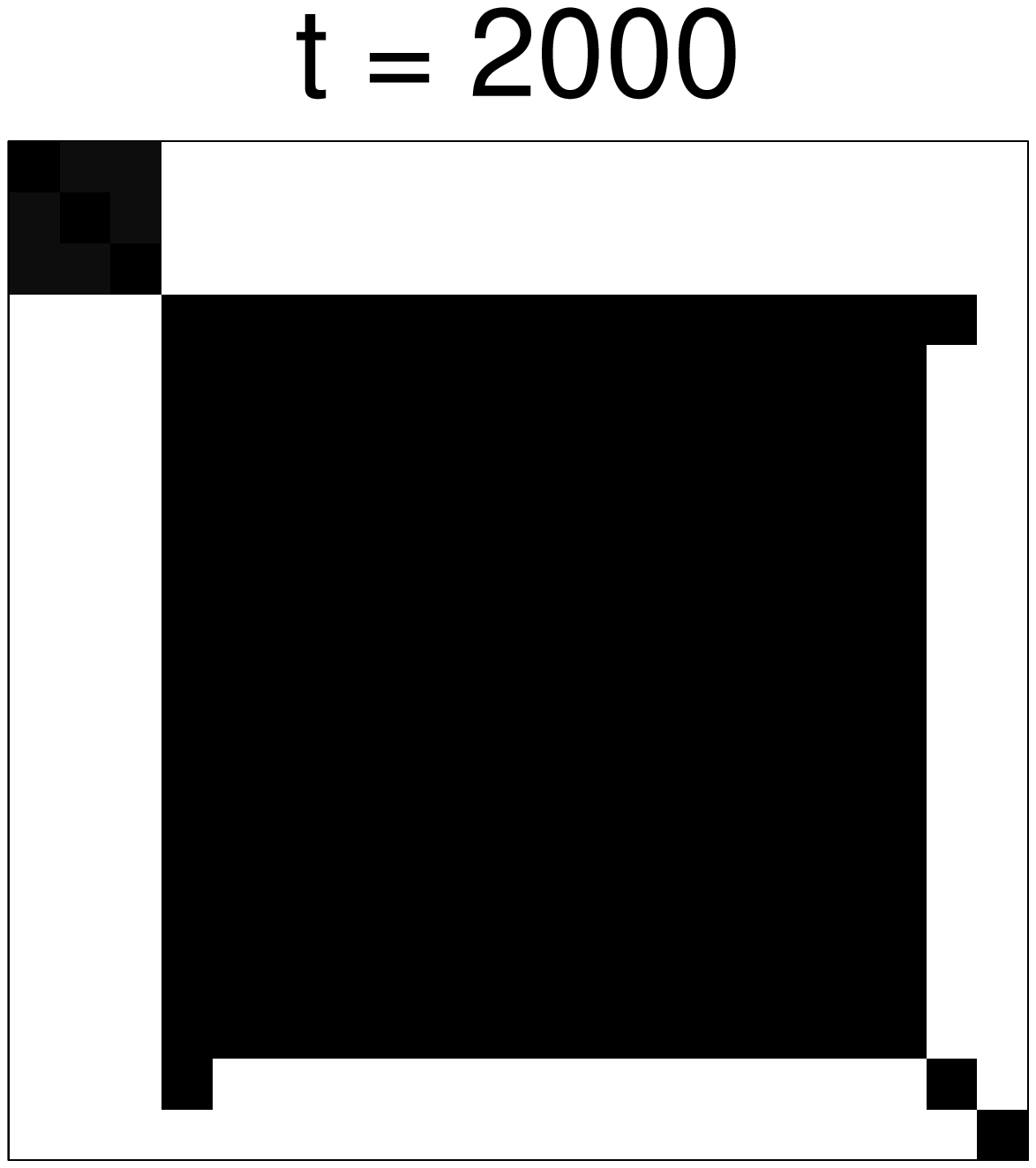}}%
		 \hspace{.125cm}%
\subfigure{	
		\includegraphics[width=0.5in]{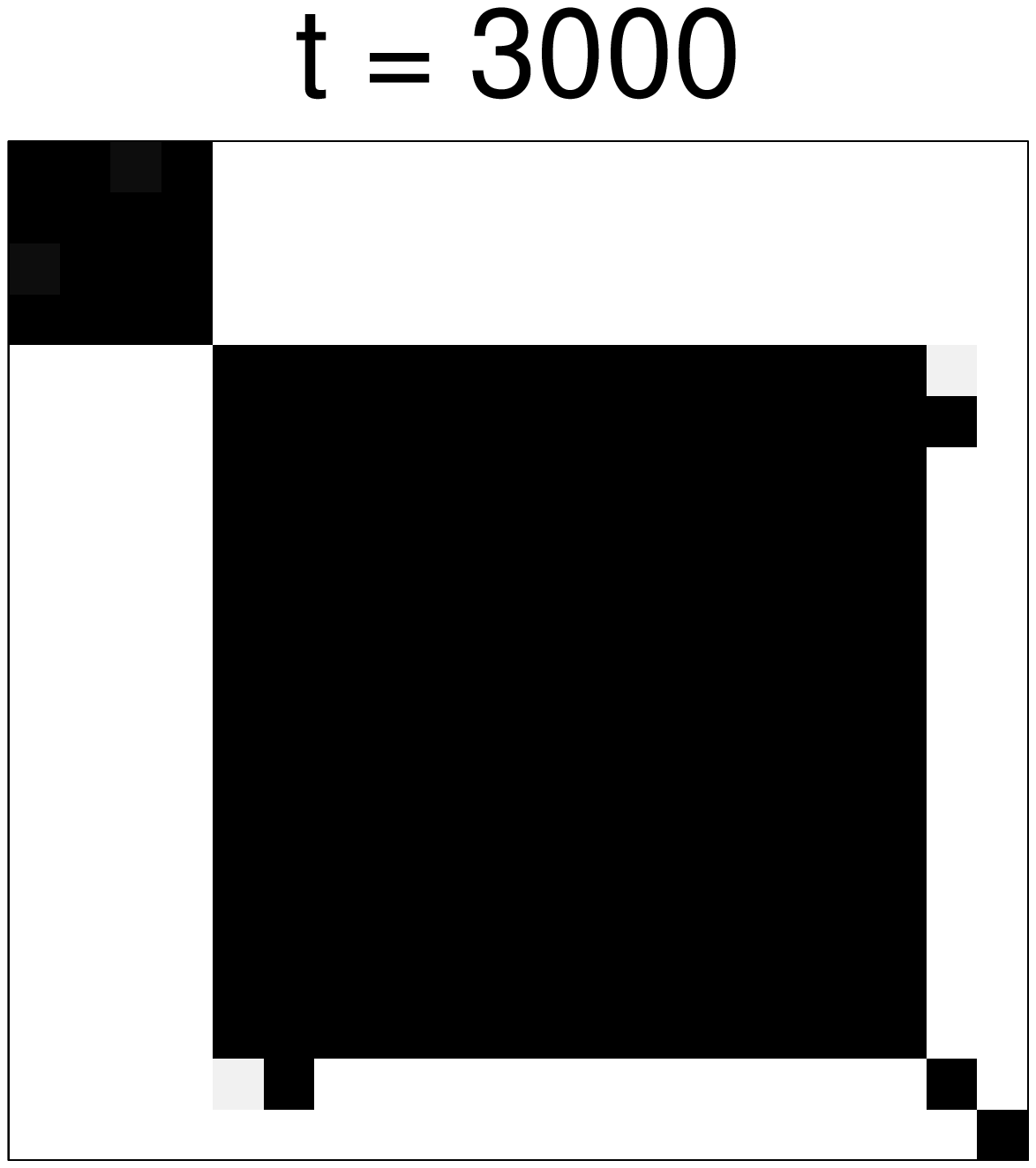}}%
 		\hspace{.125cm}%
\subfigure{	
		\includegraphics[width=0.5in]{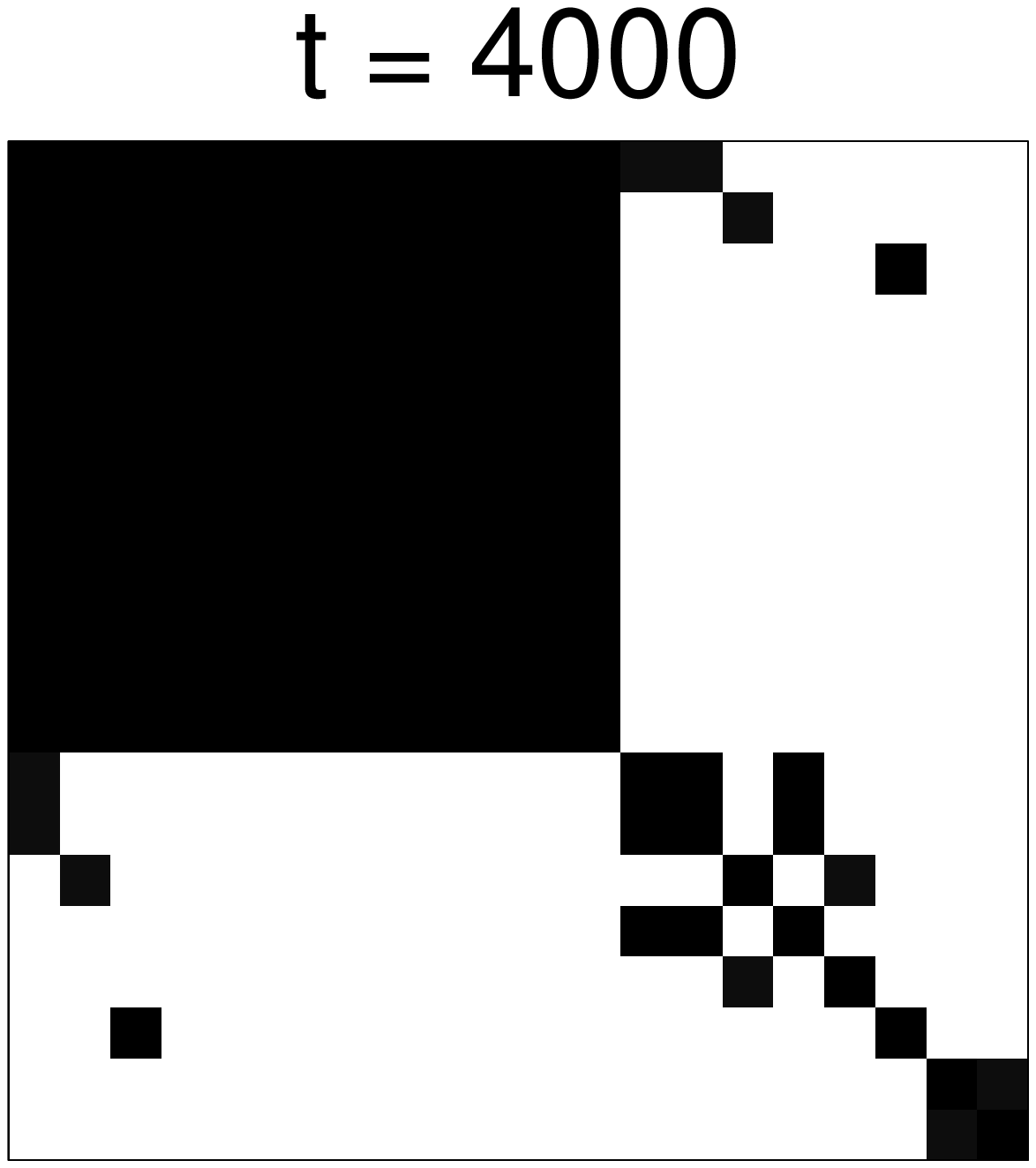}}%
		\hspace{.125cm}%
\subfigure{	
		\includegraphics[width=0.5in]{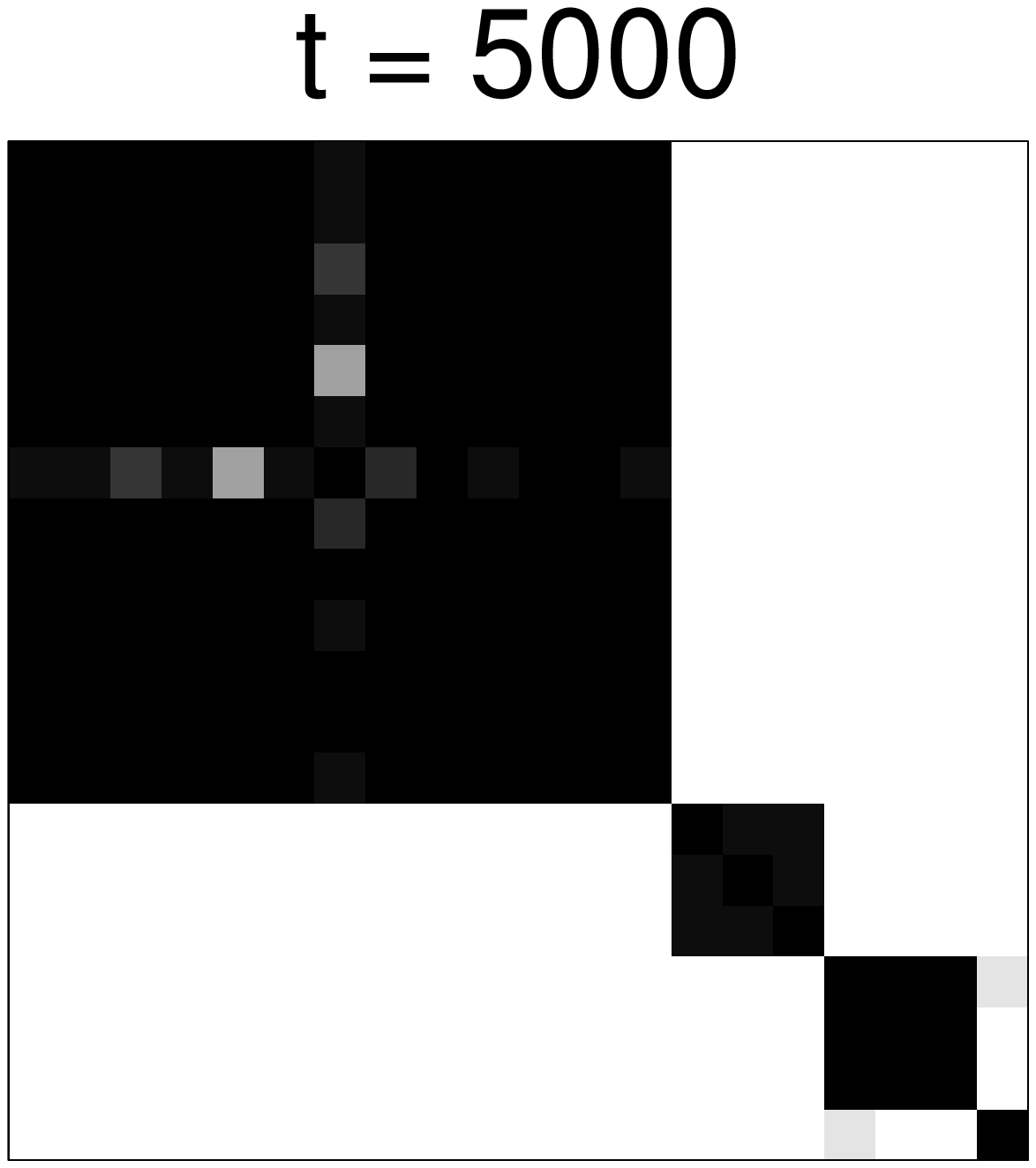}}%
		\hspace{.125cm}%
\subfigure{	
		\includegraphics[width=0.5in]{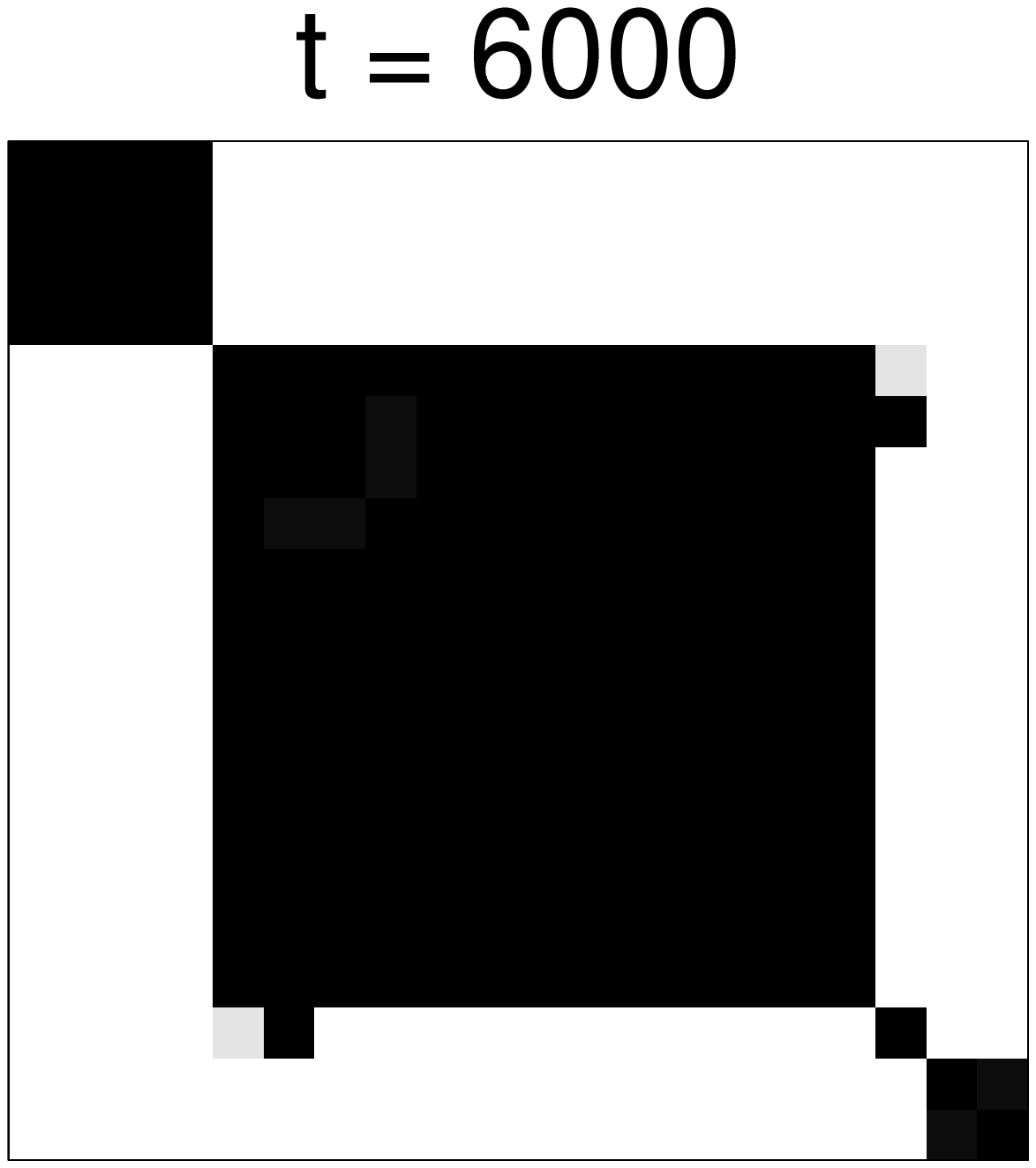}}%
		\hspace{.125cm}%
\subfigure{	
		\includegraphics[width=0.5in]{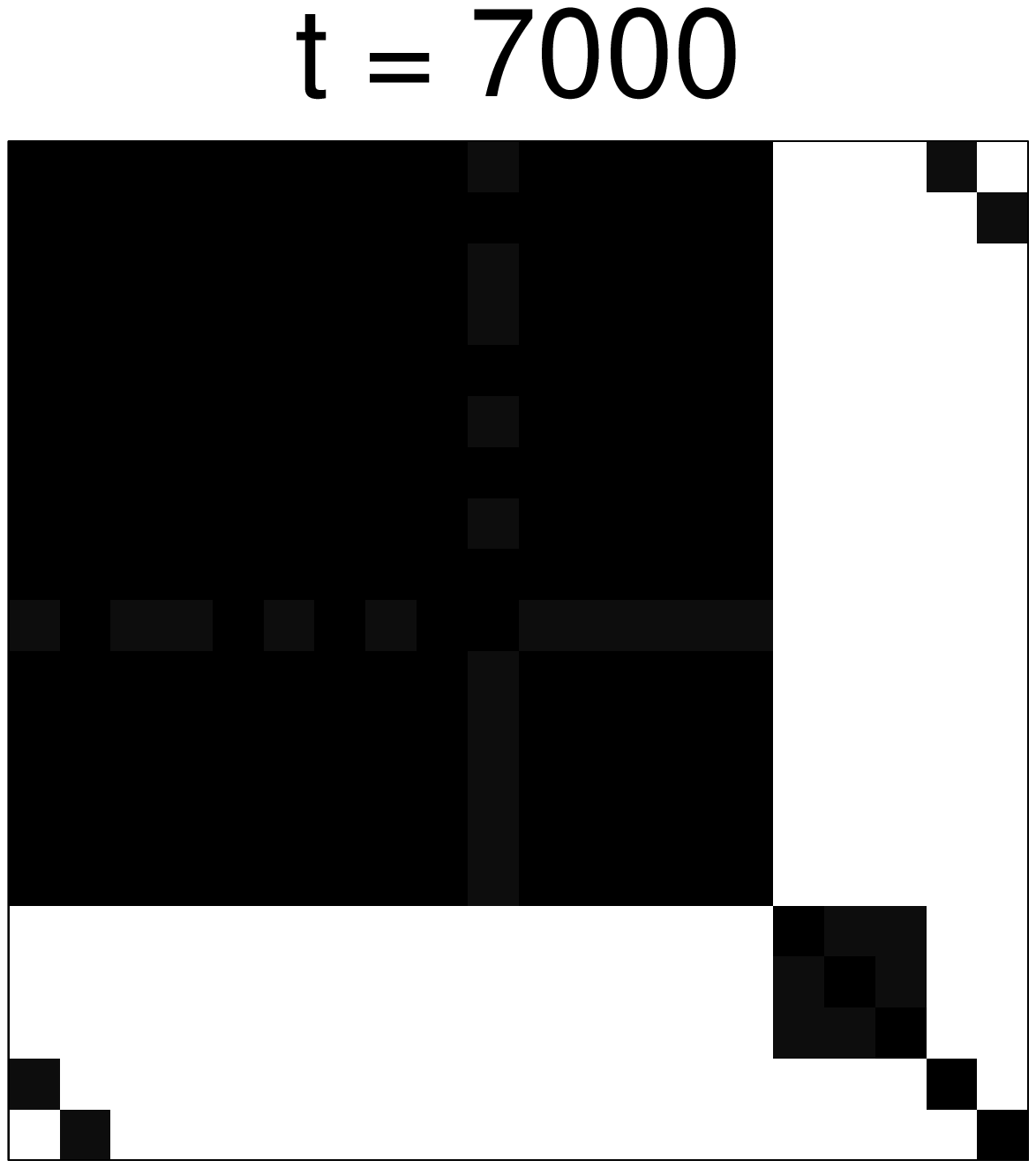}}%
		\hspace{.125cm}%
\subfigure{	
		\includegraphics[width=0.5in]{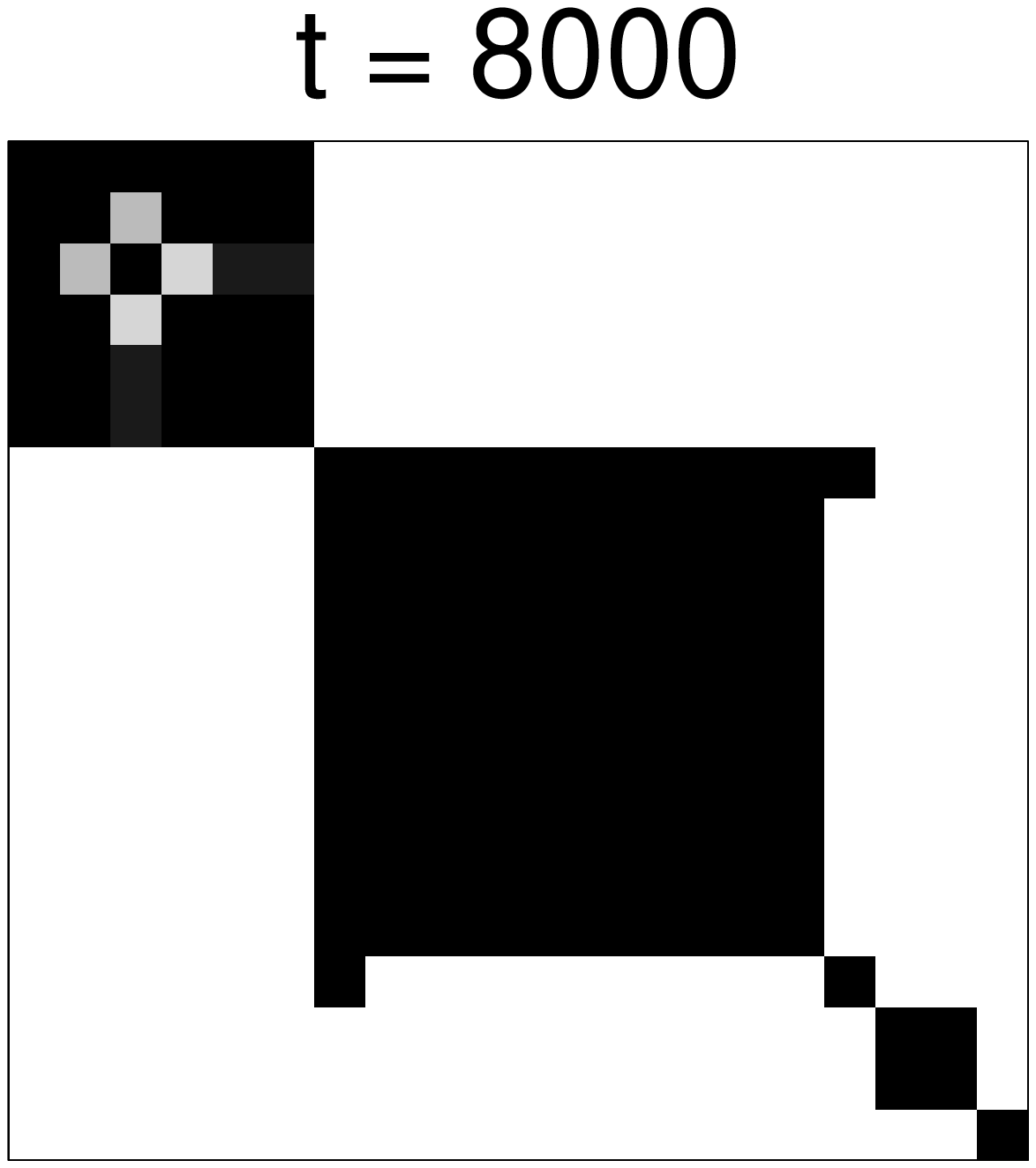}}%
		\hspace{.125cm}%
\subfigure{	
		\includegraphics[width=0.5in]{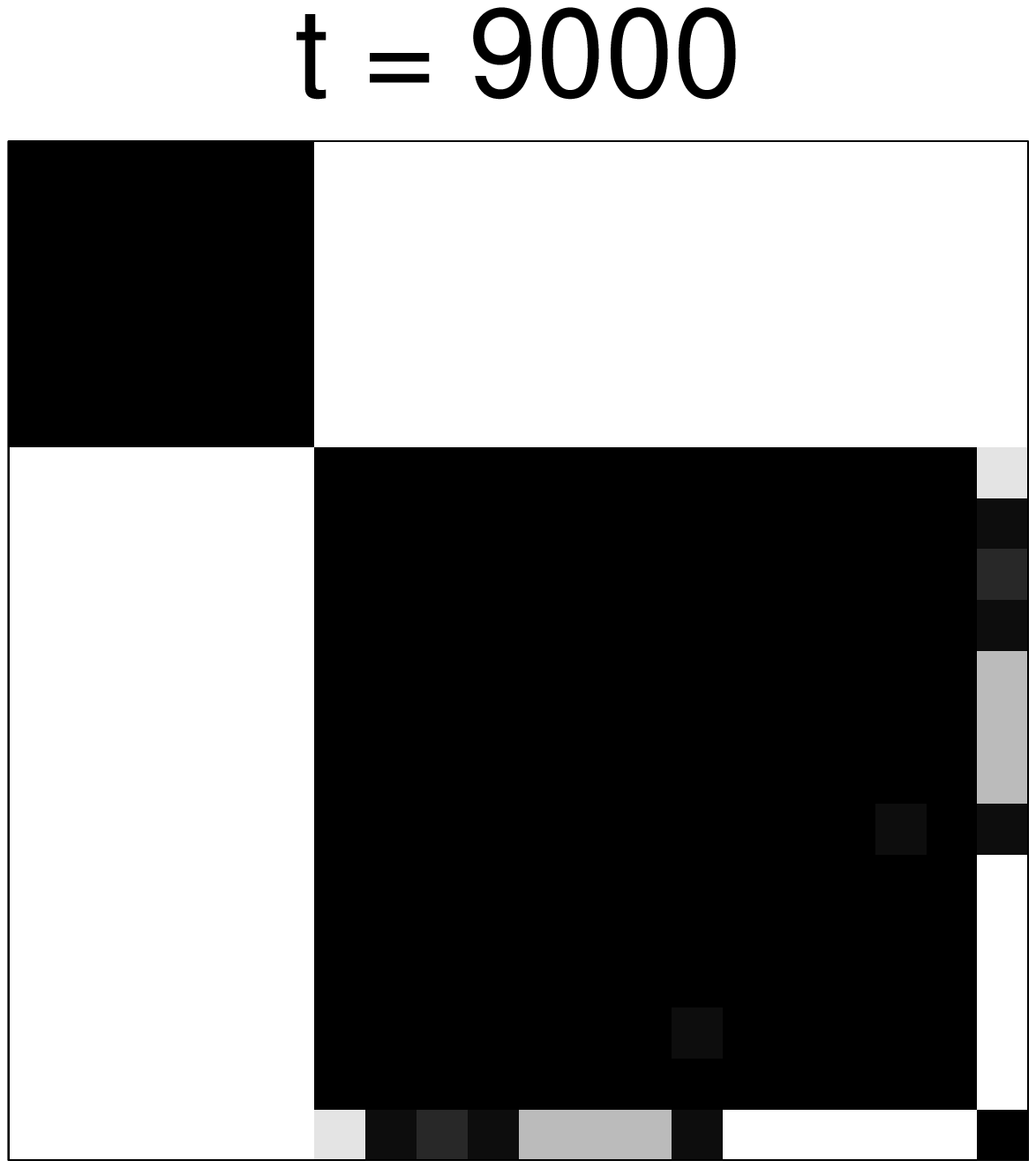}}%
		\hspace{.125cm}%
\subfigure{	
		\includegraphics[width=0.5in]{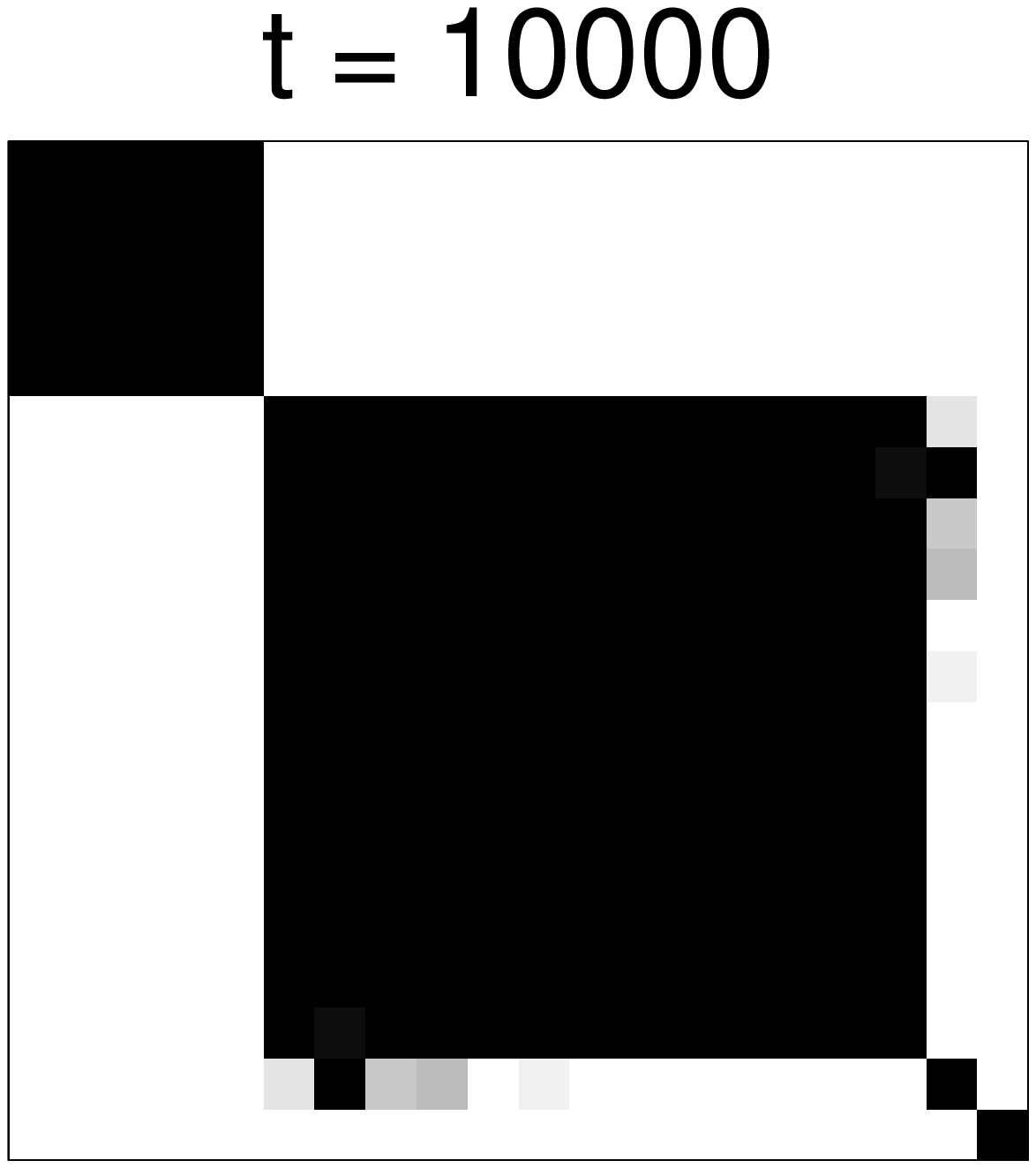}}\\
		\hspace{.125cm}%
\subfigure{	
		\includegraphics[width=0.5in]{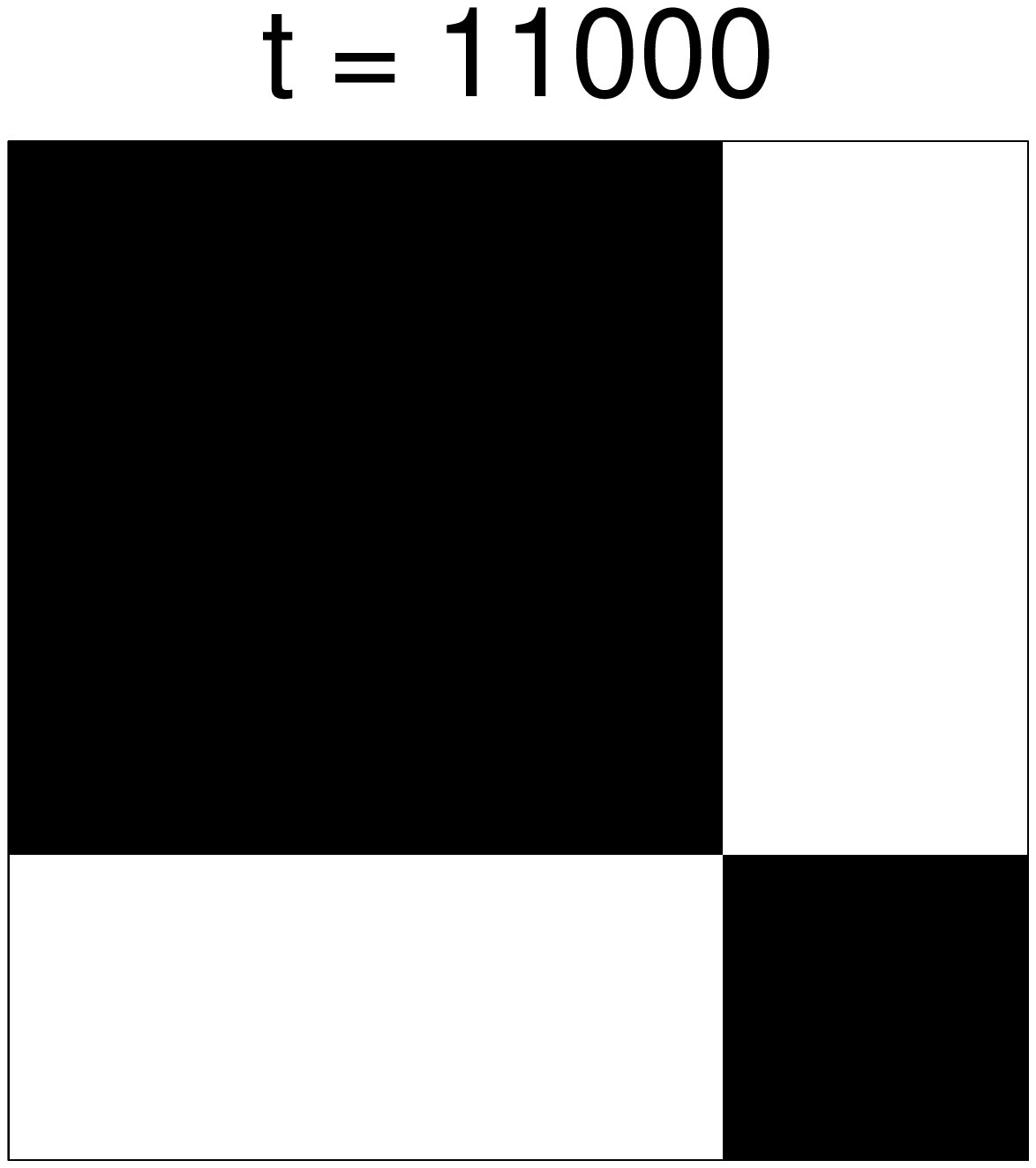}}%
		\hspace{.125cm}%
\subfigure{	
		\includegraphics[width=0.5in]{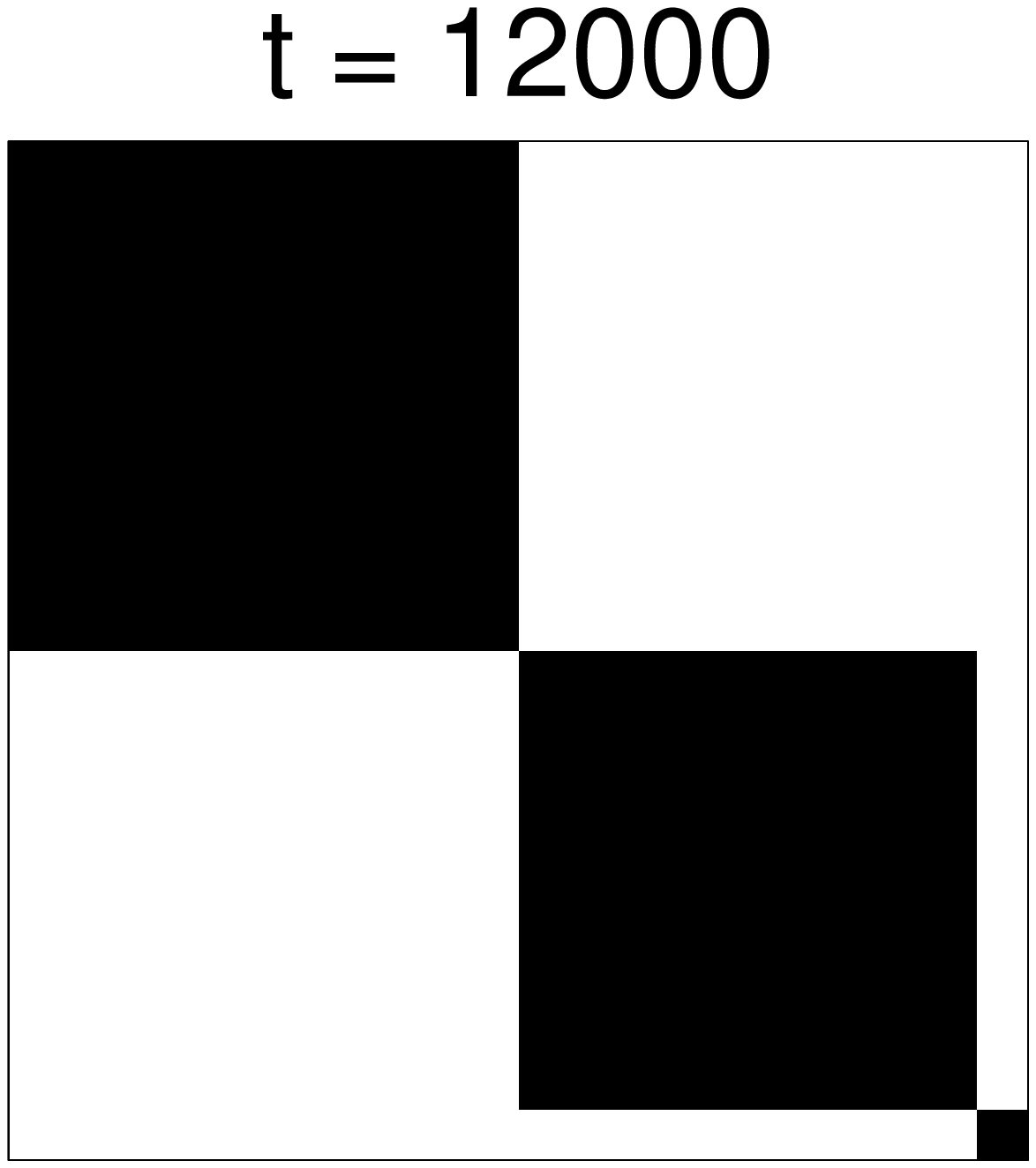}}%
		\hspace{.125cm}%
\subfigure{	
		\includegraphics[width=0.5in]{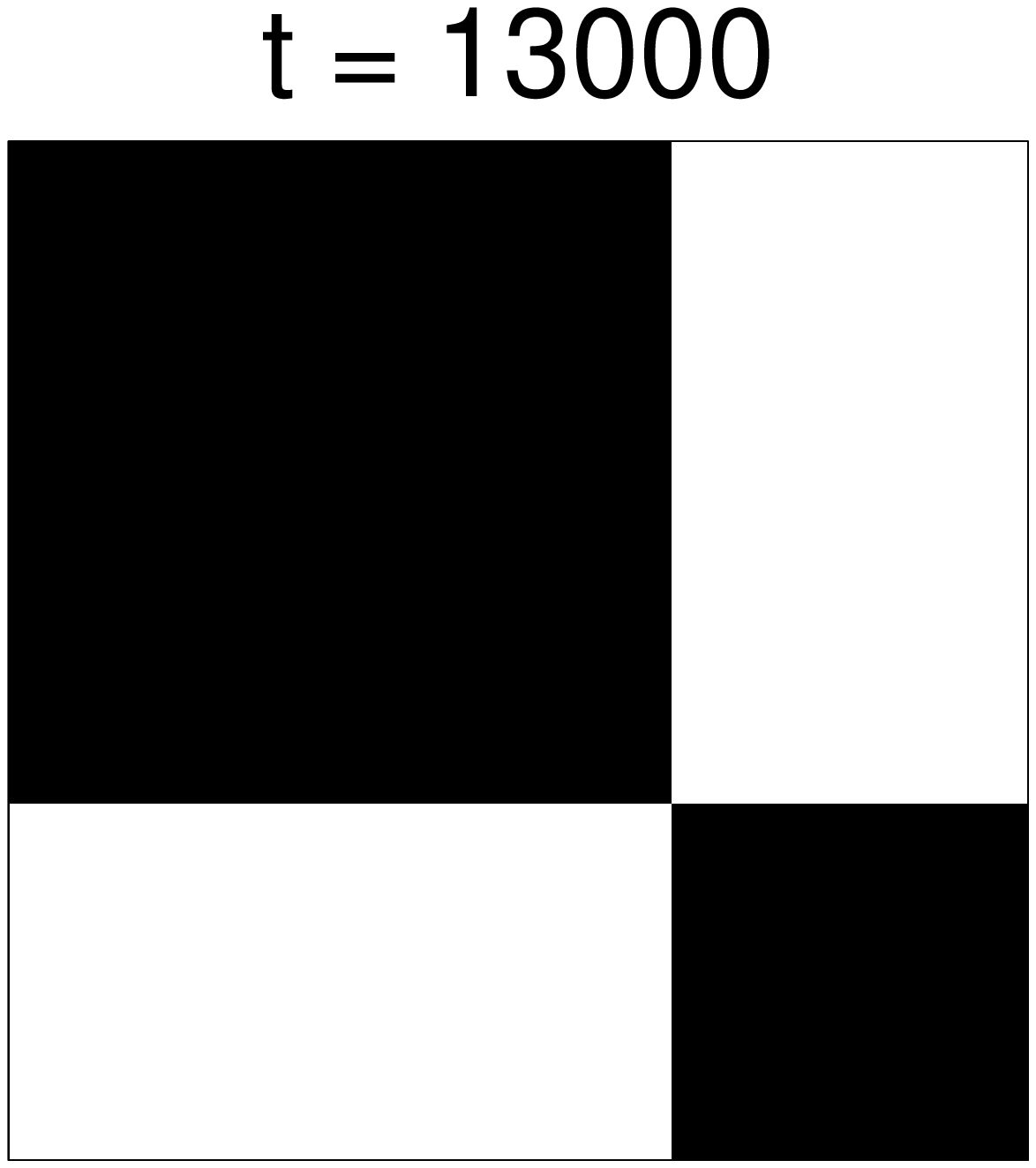}}%
		\hspace{.125cm}%
\subfigure{	
		\includegraphics[width=0.5in]{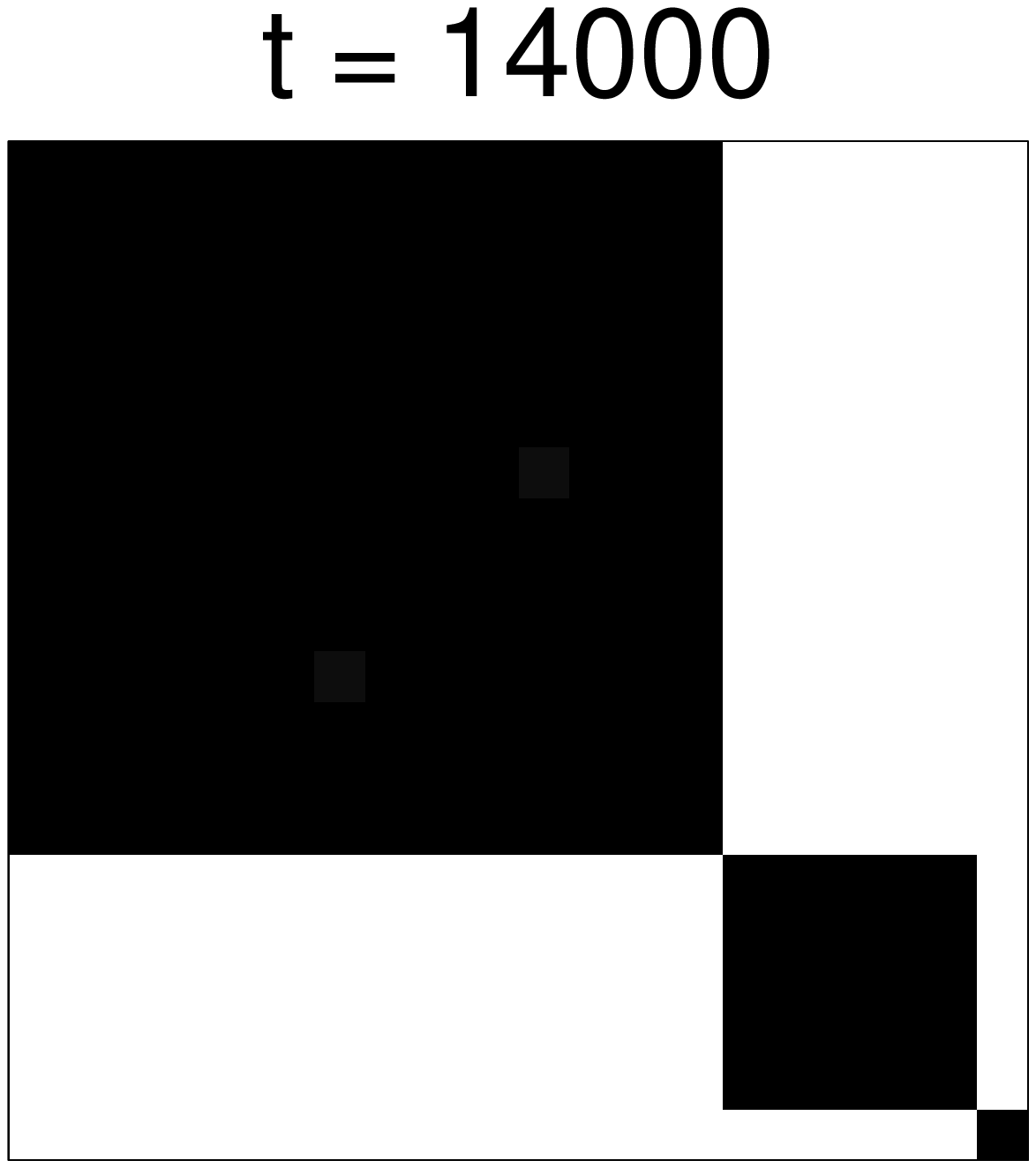}}%
		\hspace{.125cm}%
\subfigure{	
		\includegraphics[width=0.5in]{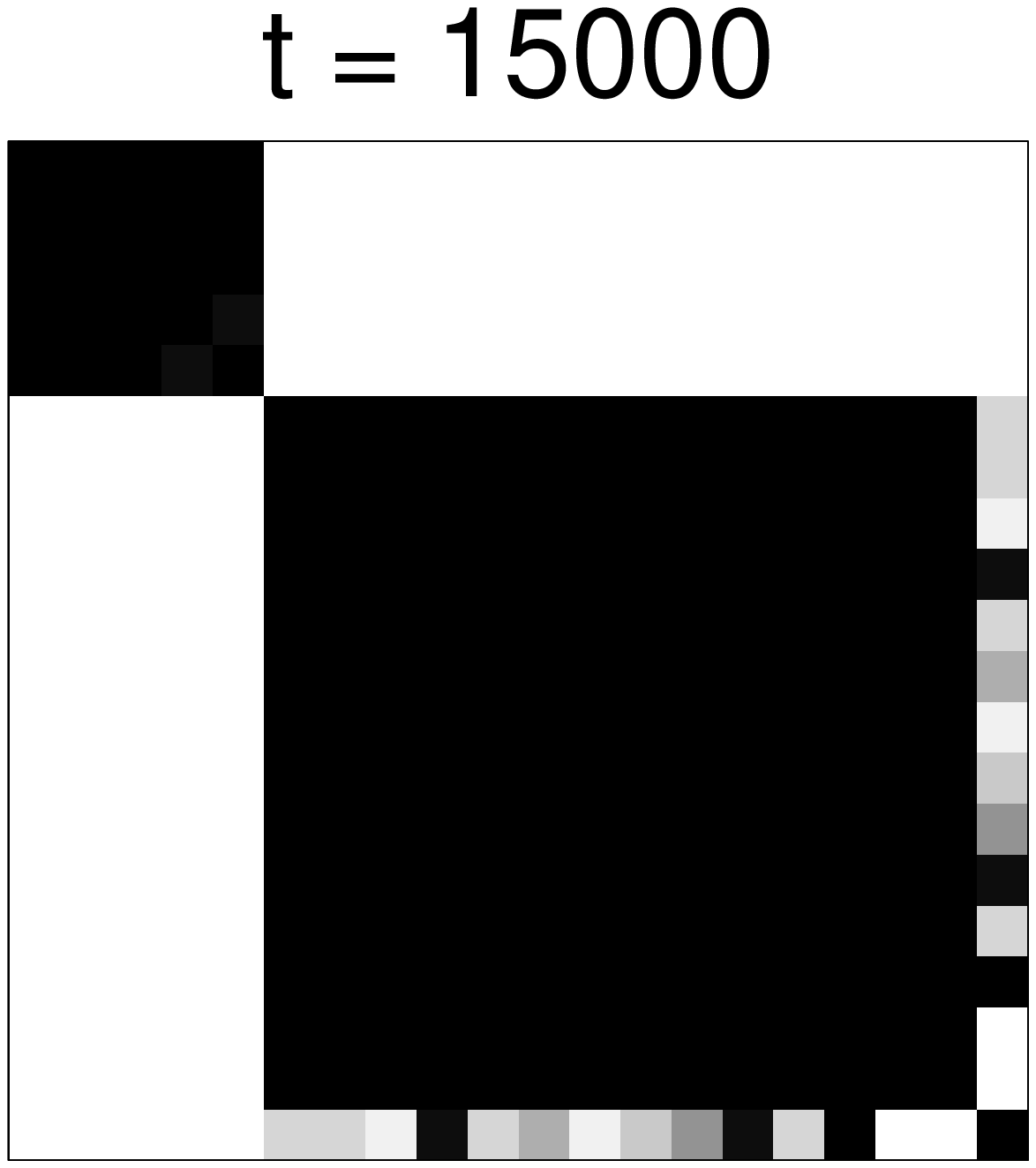}}%
		\hspace{.125cm}%
\subfigure{	
		\includegraphics[width=0.5in]{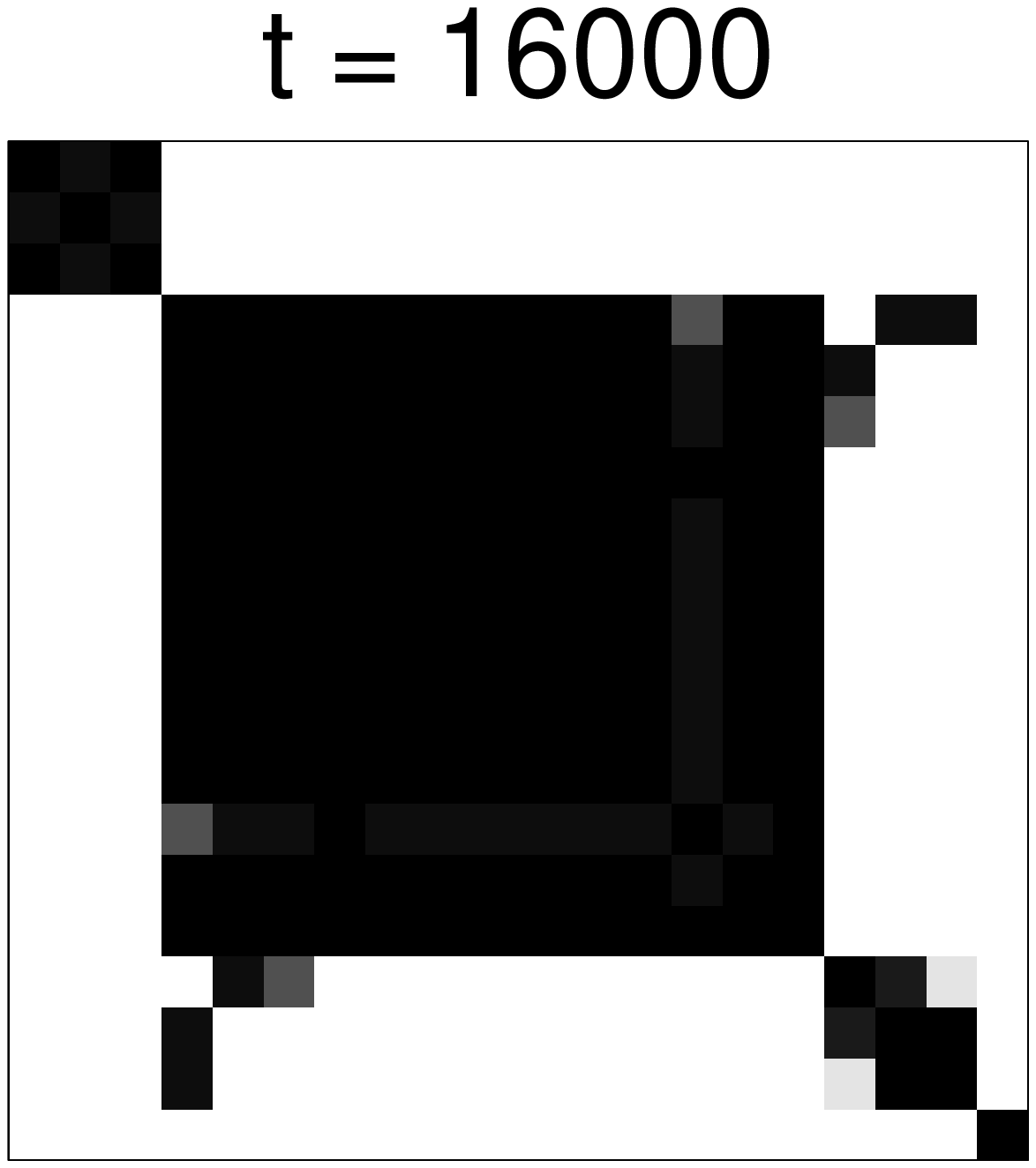}}%
		\hspace{.125cm}%
\subfigure{	
		\includegraphics[width=0.5in]{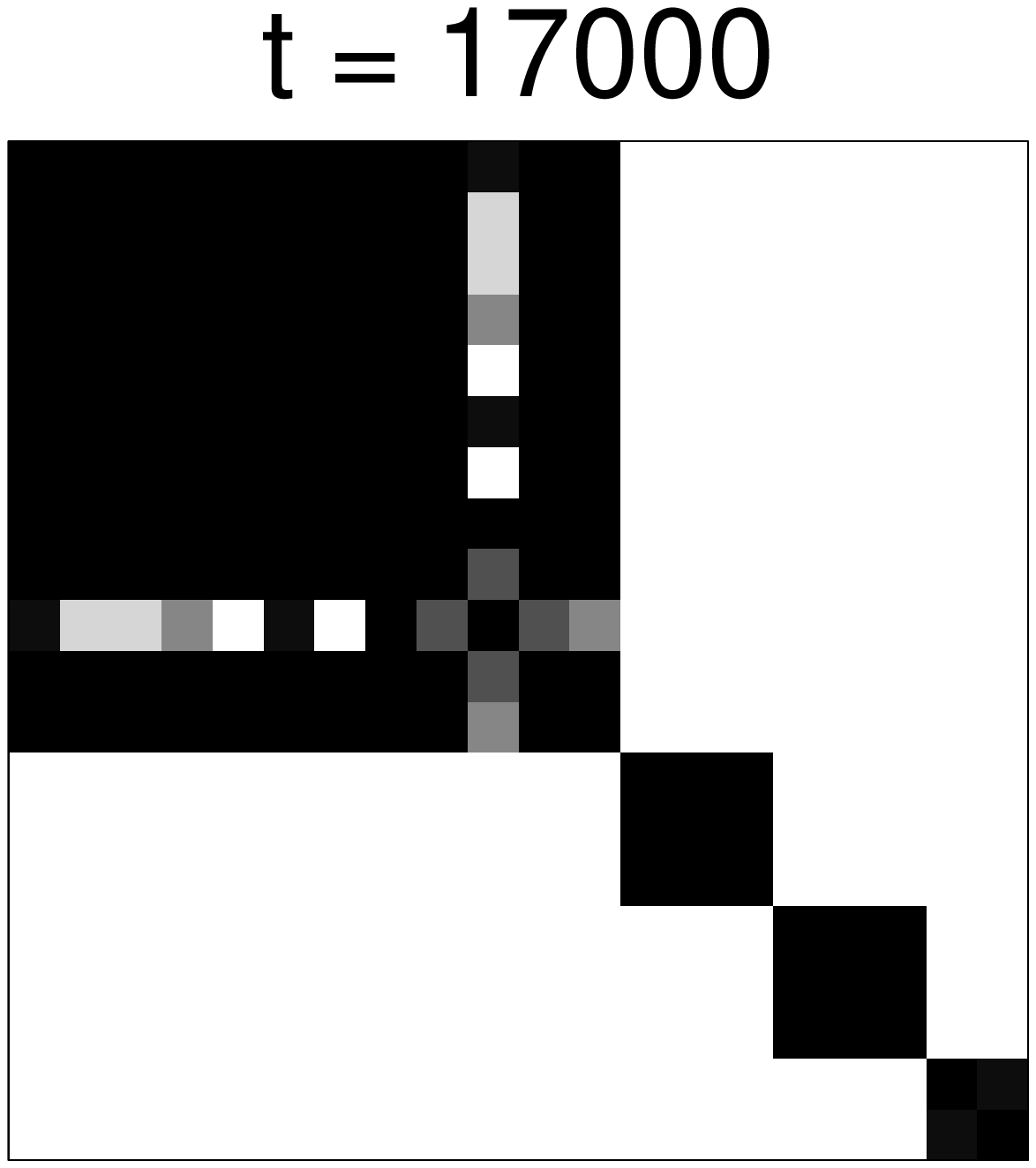}}%
		 \hspace{.125cm}%
\subfigure{	
		\includegraphics[width=0.5in]{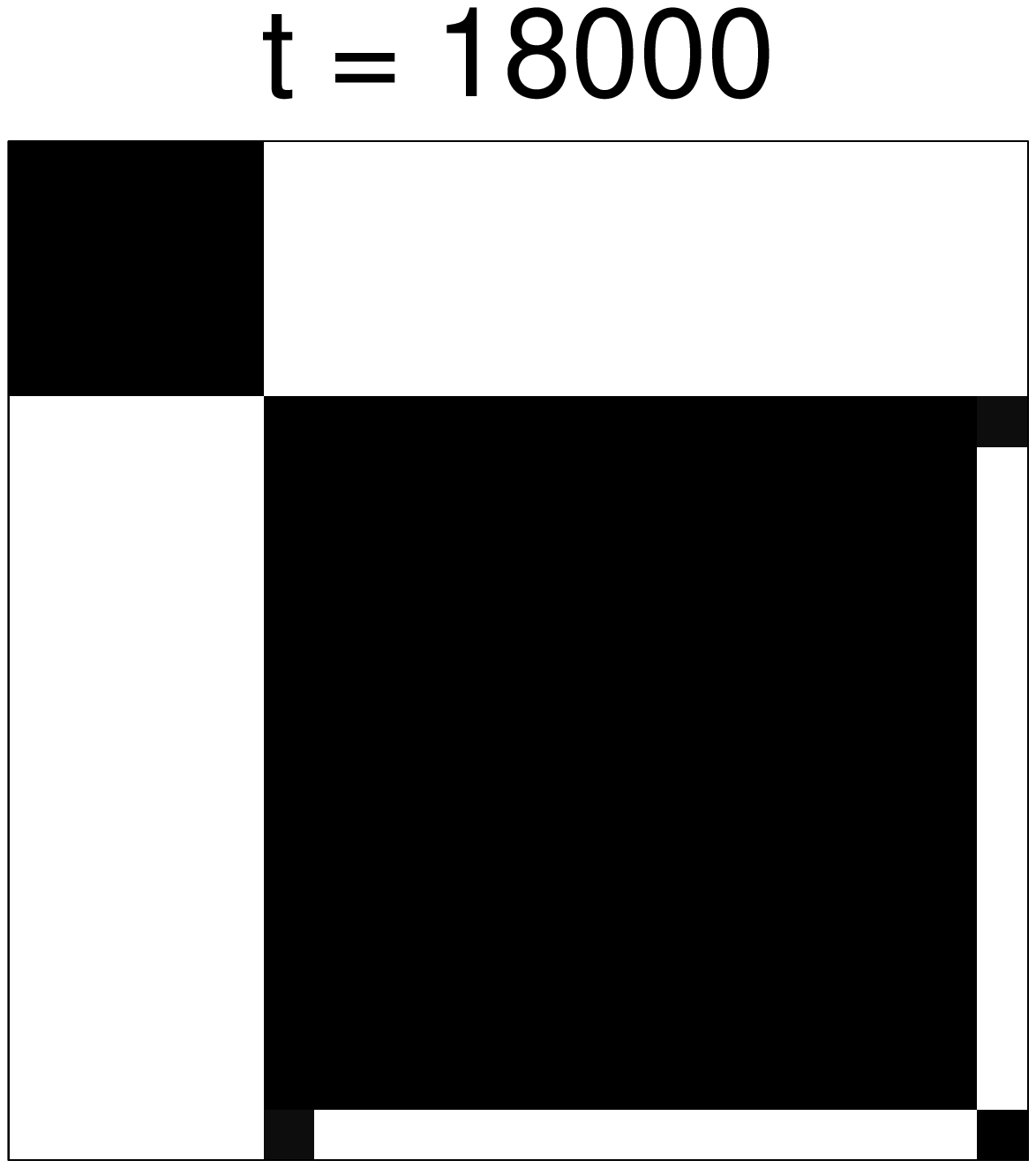}}
		\hspace{.125cm}%
\subfigure{	
		\includegraphics[width=0.5in]{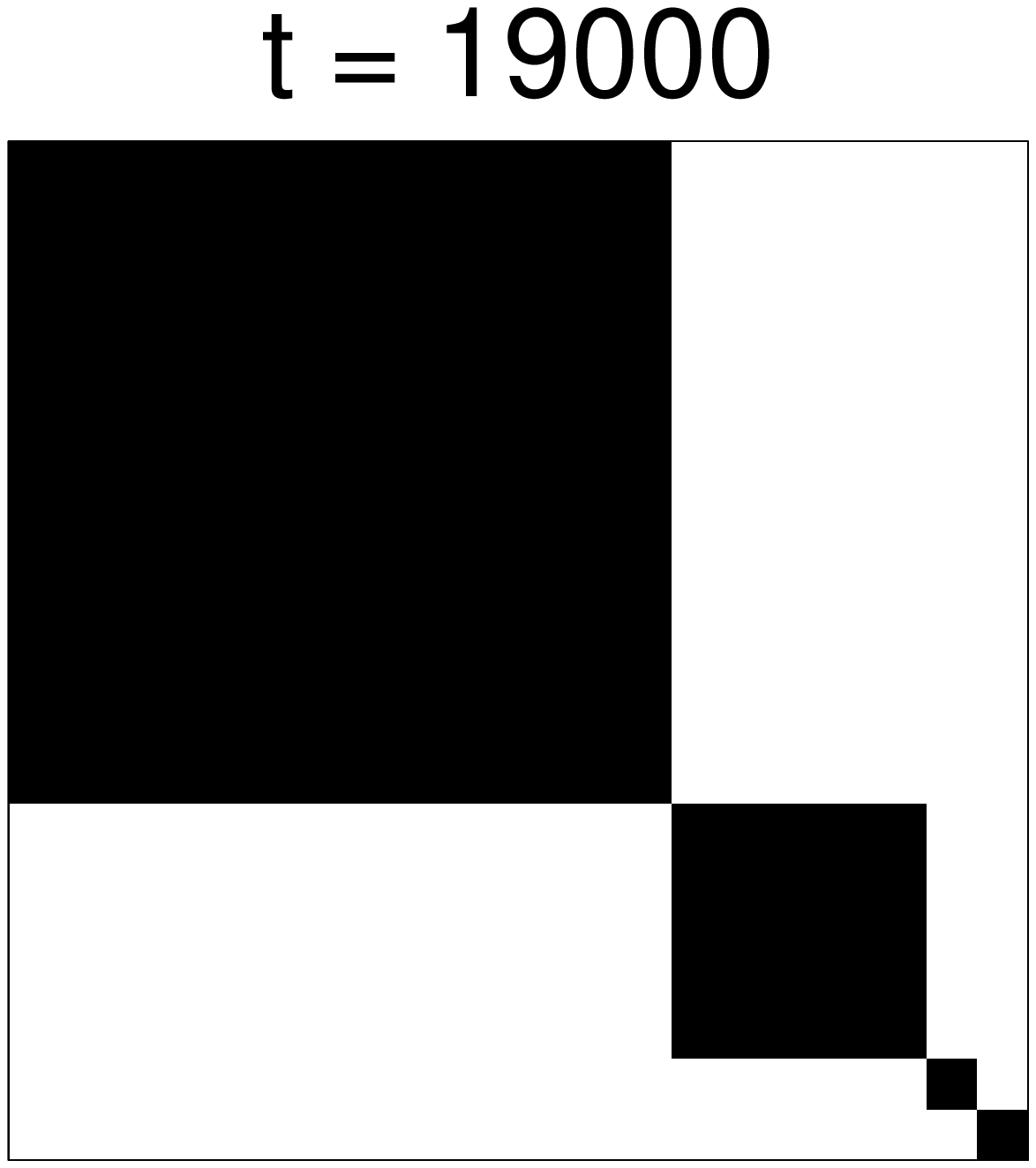}}%
		 \hspace{.125cm}%
\subfigure{	
		\includegraphics[width=0.5in]{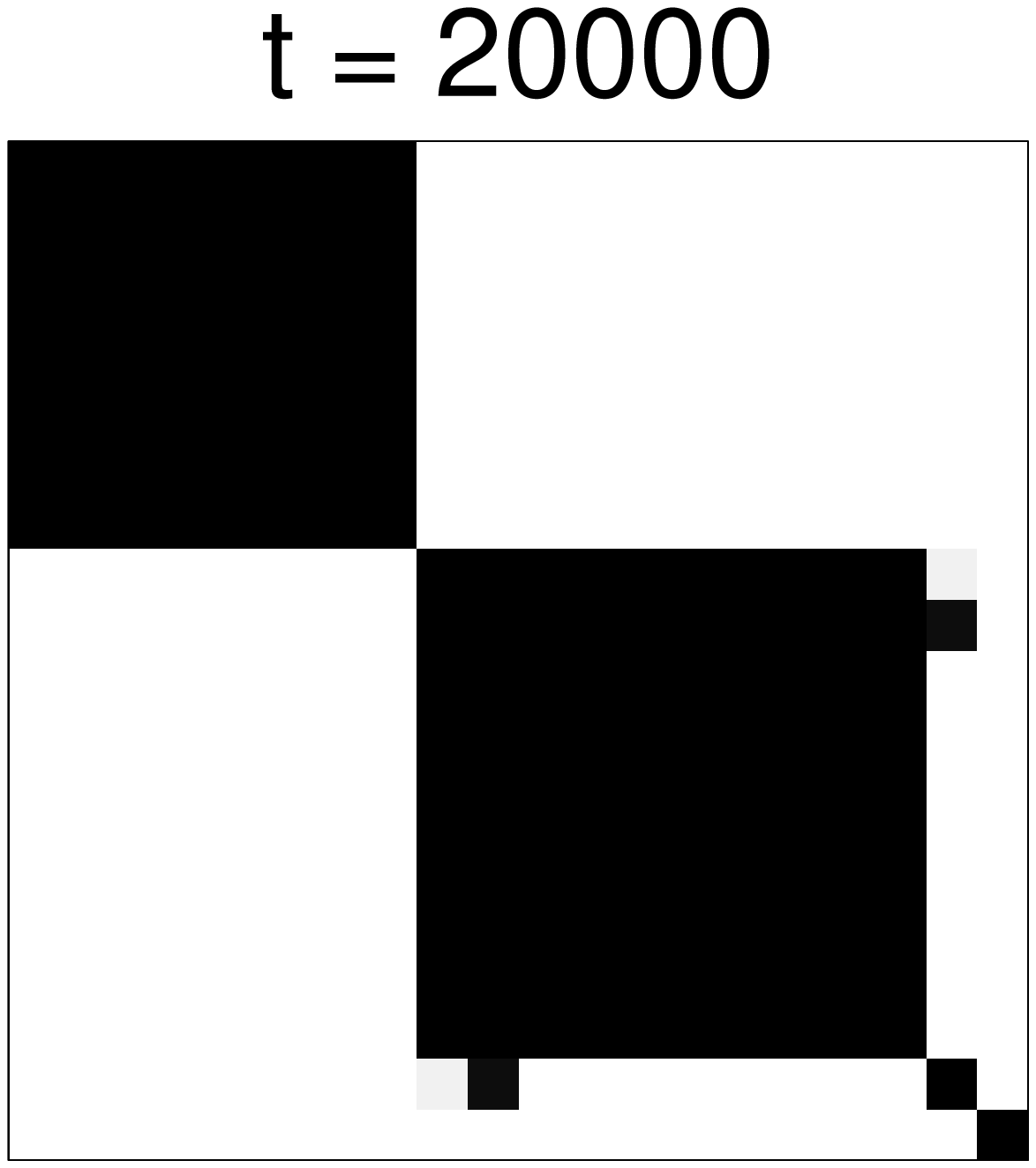}}
\caption{{\small {\bf Dynamics of interference matrix; no 
cultural inheritance ($\gk=0$)}.
Parameters values are default with $N=20$ and $\gga=0.001$ 
(average life span 1000).
See the legend of Fig.1.
}}
\label{orphands}
\end{figure}

{\em Stochastic equilibrium.}\quad
If cultural inheritance of social networks is weak ($\gk$ is small), 
a small number of alliances are maintained across generations in
stochastic equilibrium (see 
Figure~\ref{orphands}). 
This happens because the death of individuals tends to decrease the size of existing alliances while new individuals are initially unaffiliated and may form new affinities. 
This regime is similar to coalitionary structures recently identified in
a community of wild chimpanzees in Uganda \cite{mit03} and in populations of bottlenose
dolphins in coastal waters of Western Australia \cite{con01}  
and eastern Scotland \cite{lus06}.

{\em Egalitarian state.}\quad
If cultural inheritance of social networks is faithful ($\gk$ is large), 
the dynamics become dramatically different due to intense
selection between different alliances. 
Now the turnover of individuals creates conditions for growth of alliances. 
Larger alliances
increase in size as a result of their members winning more conflicts, 
achieving higher social success, and parenting (biologically or culturally)
more offspring who themselves become members of the paternal alliance. 
As a result of this positive feedback loop (analogous to that of positive frequency-dependent selection), the system exhibits a strong tendency towards approaching a state in which all members of the group belong to the same alliance and have very similar social success in spite of strong variation in their fighting abilities.
Figure 4 contrasts the egalitarian state with the stochastic equilibrium illustrated in Figure 3 above. One can see that at the egalitarian state, the average affinity is increased while the standard deviation of affinity and the hierarchy measures are decreased. Although at the egalitarian state the correlation of individual strength and social success can be substantial,  it does not result in social inequality.
This ``egalitarian'' state can be reached in several generations.


\begin{figure}
\centering
\subfigure{ 	
		\includegraphics[width=2.75in]{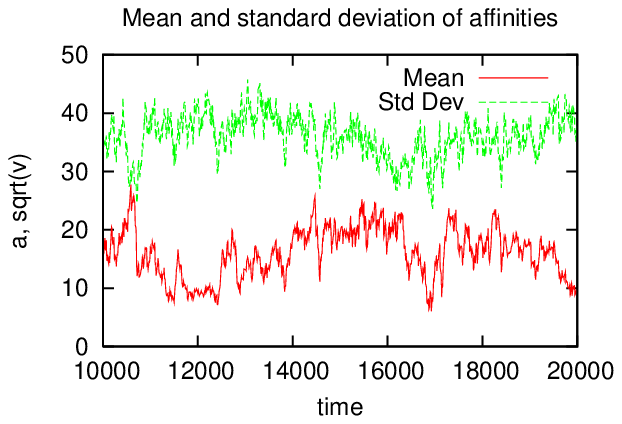}}%
\subfigure{	
		\includegraphics[width=2.75in]{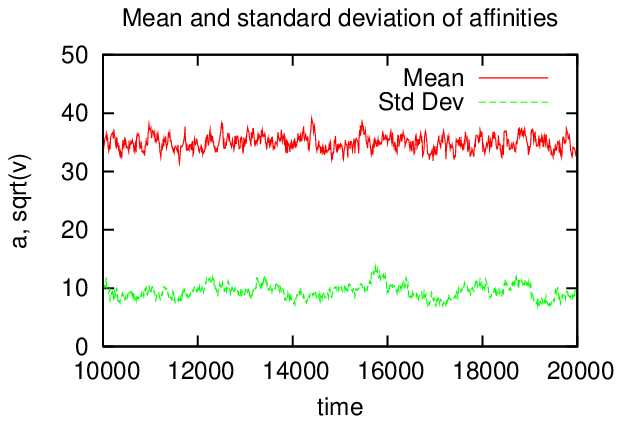}}\\
\subfigure{	
		\includegraphics[width=2.75in]{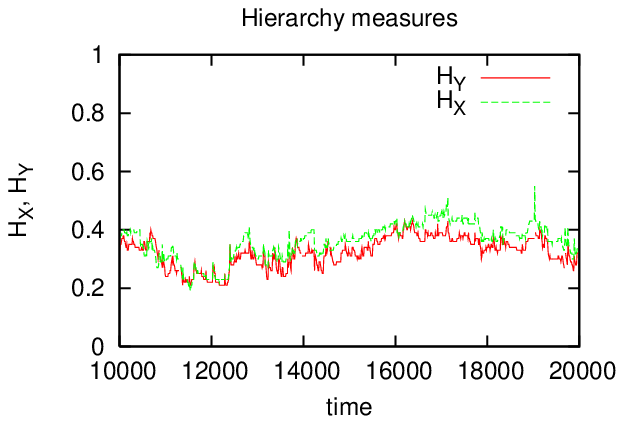}}%
\subfigure{	
		\includegraphics[width=2.75in]{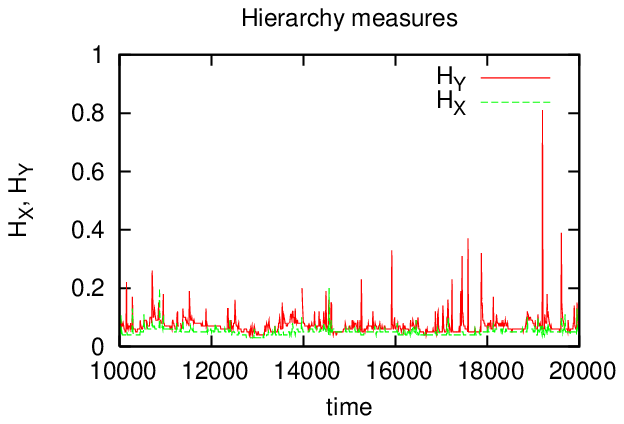}}\\
\subfigure{	
		\includegraphics[width=2.75in]{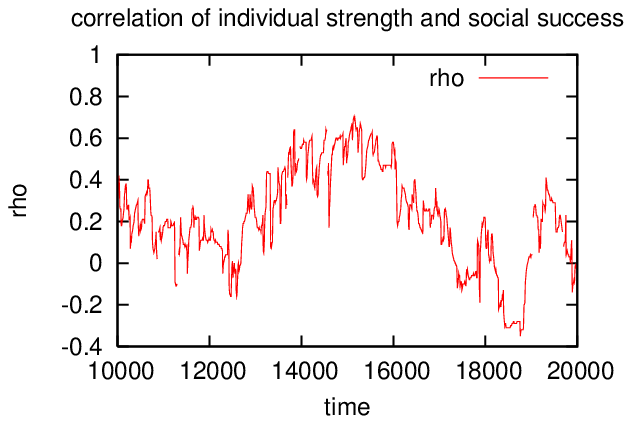}}%
\subfigure{	
		\includegraphics[width=2.75in]{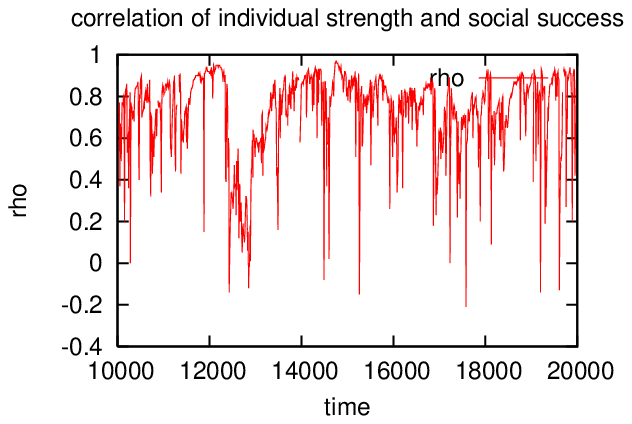}}
\caption{
Contrasting a state with a small number of alliances maintained in
stochastic equilibrium with an egalitarian state. 
The left column of graphs correspond to the run shown in Figure~\ref{orphands} which 
resulted in a small number of alliances maintained in
stochastic equilibrium. The right column of graphs correspond 
to a run with $\gk=1$ (complete cultural inheritance of social network) and $\mu=0.025$
(increased memory of past events) which resulted in an egalitarian regime.
With several alliances present simultaneously (left graphs), the average affinity 
is small, the variance of affinities is large, the measures of social inequality
$H_x$ and $H_y$ are large, and the correlation between social success $Y_i$ and individual
fighting ability $s_i$ is small.
In the egalitarian state (right graphs), the average affinity 
is large, the variance of affinities is small, the measures of social inequality
$H_x$ and $H_y$ are small, and the correlation between social success $Y_i$ and individual
fighting ability $s_i$ is large.
}
\label{correlation}
\end{figure}

\begin{figure}[t]
\centering
\subfigure[]{ 	
		\includegraphics[width=2.in]{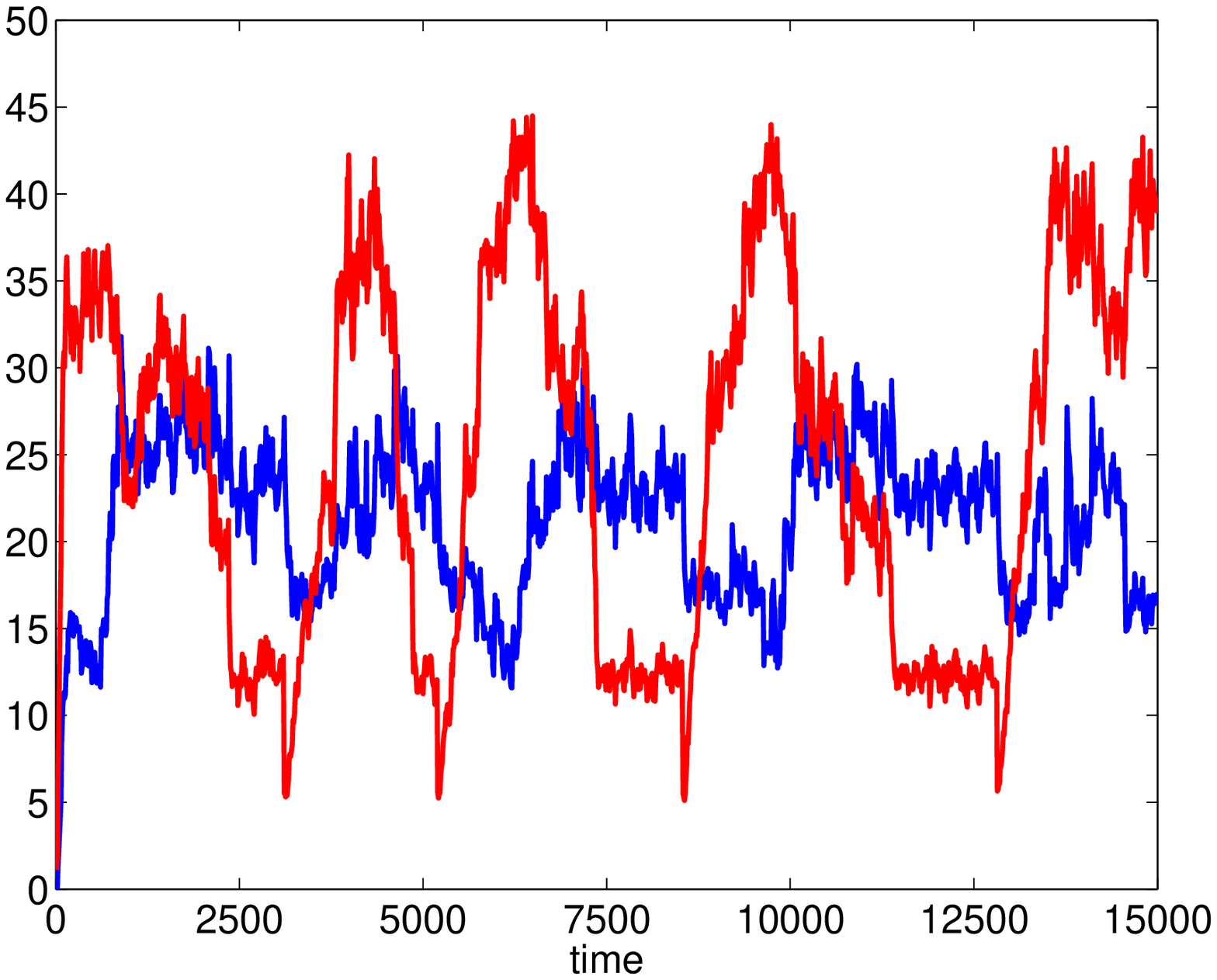}}%
 		\hspace{.125cm}%
\subfigure[]{	
		\includegraphics[width=2.in]{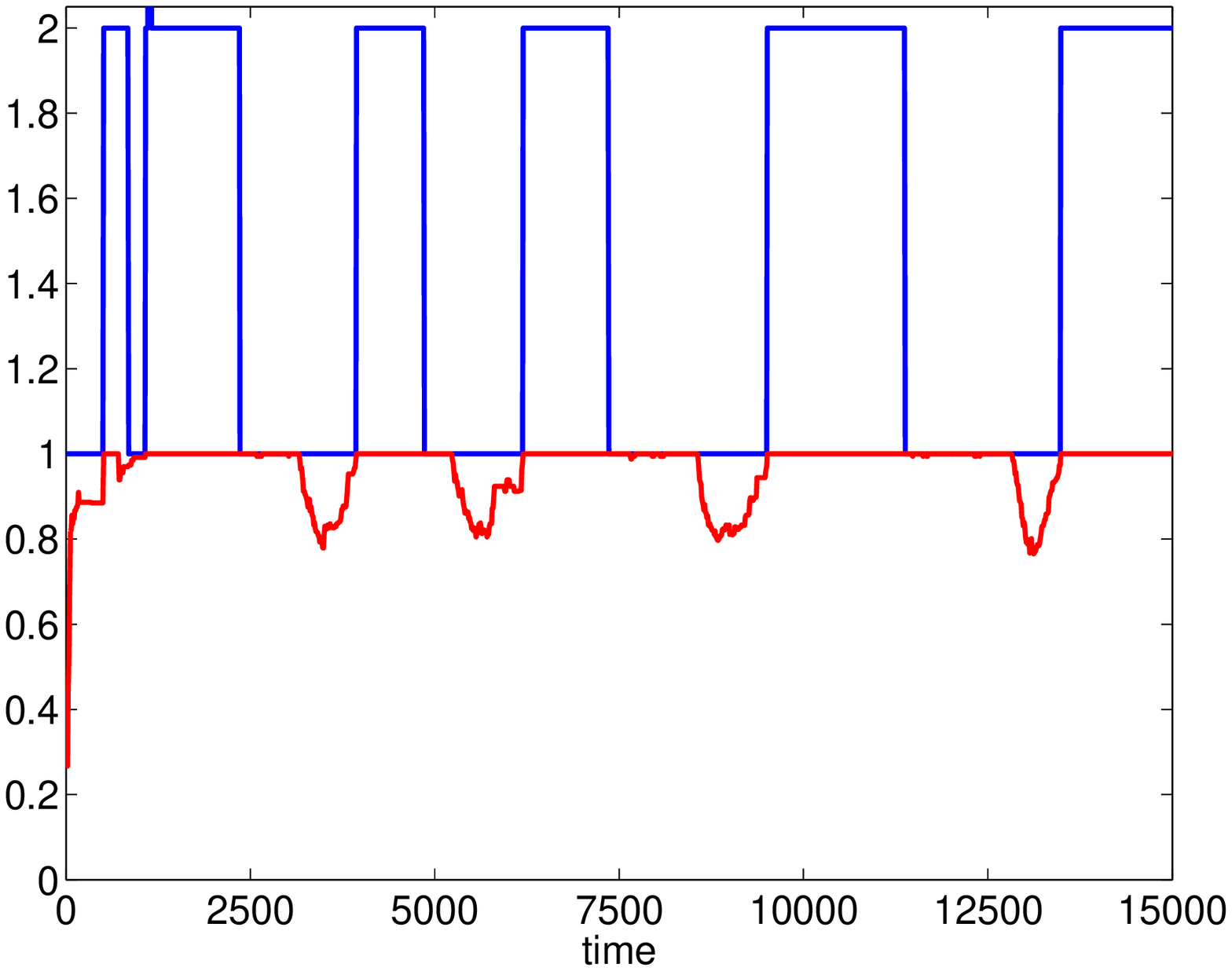}}%
 		\hspace{.125cm}%
\subfigure[]{	
		\includegraphics[width=2.in]{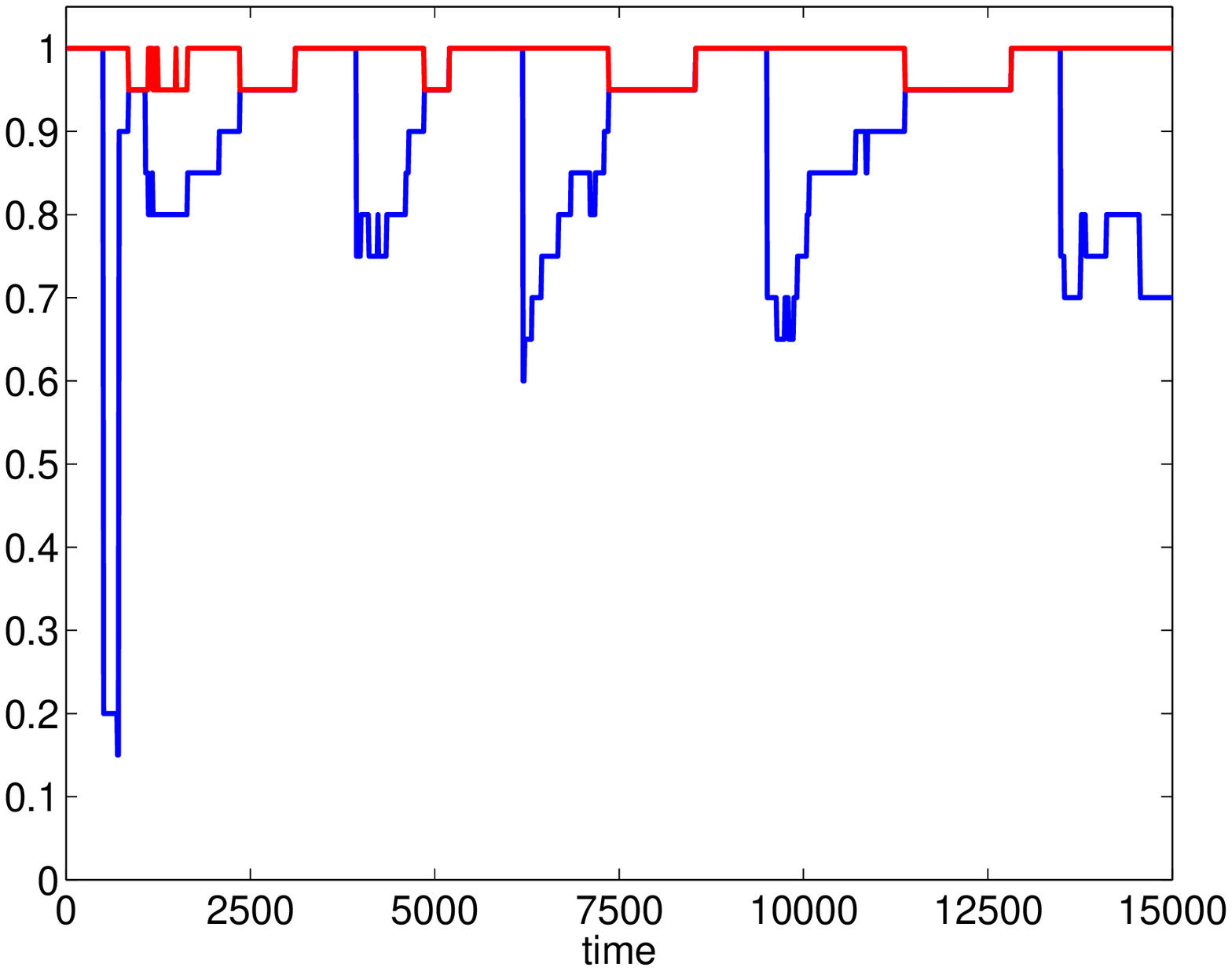}}\\ 
		\hspace{.125cm}%
\subfigure{ 	
		\includegraphics[width=0.5in]{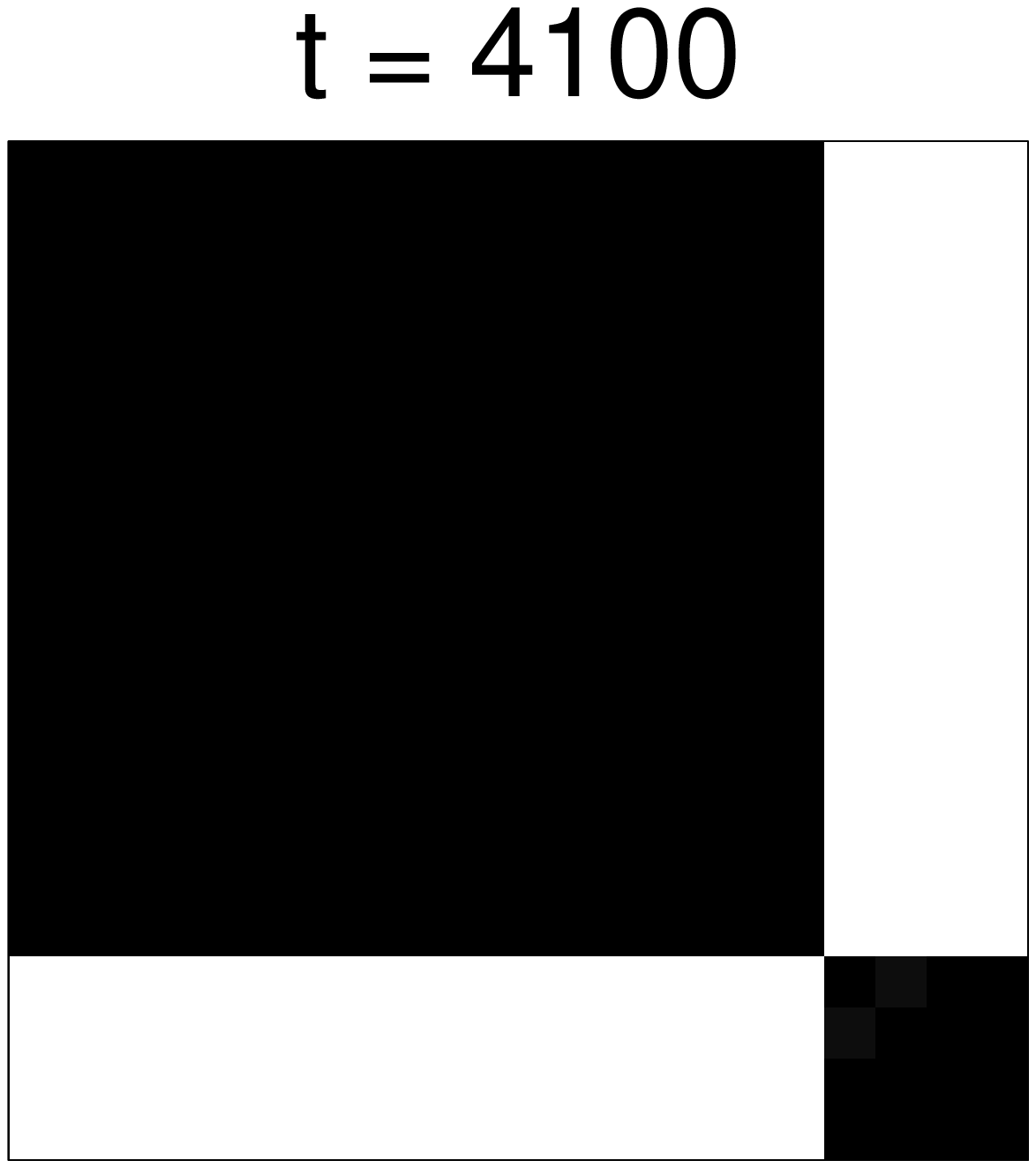}}%
 		\hspace{.125cm}%
\subfigure{	
		\includegraphics[width=0.5in]{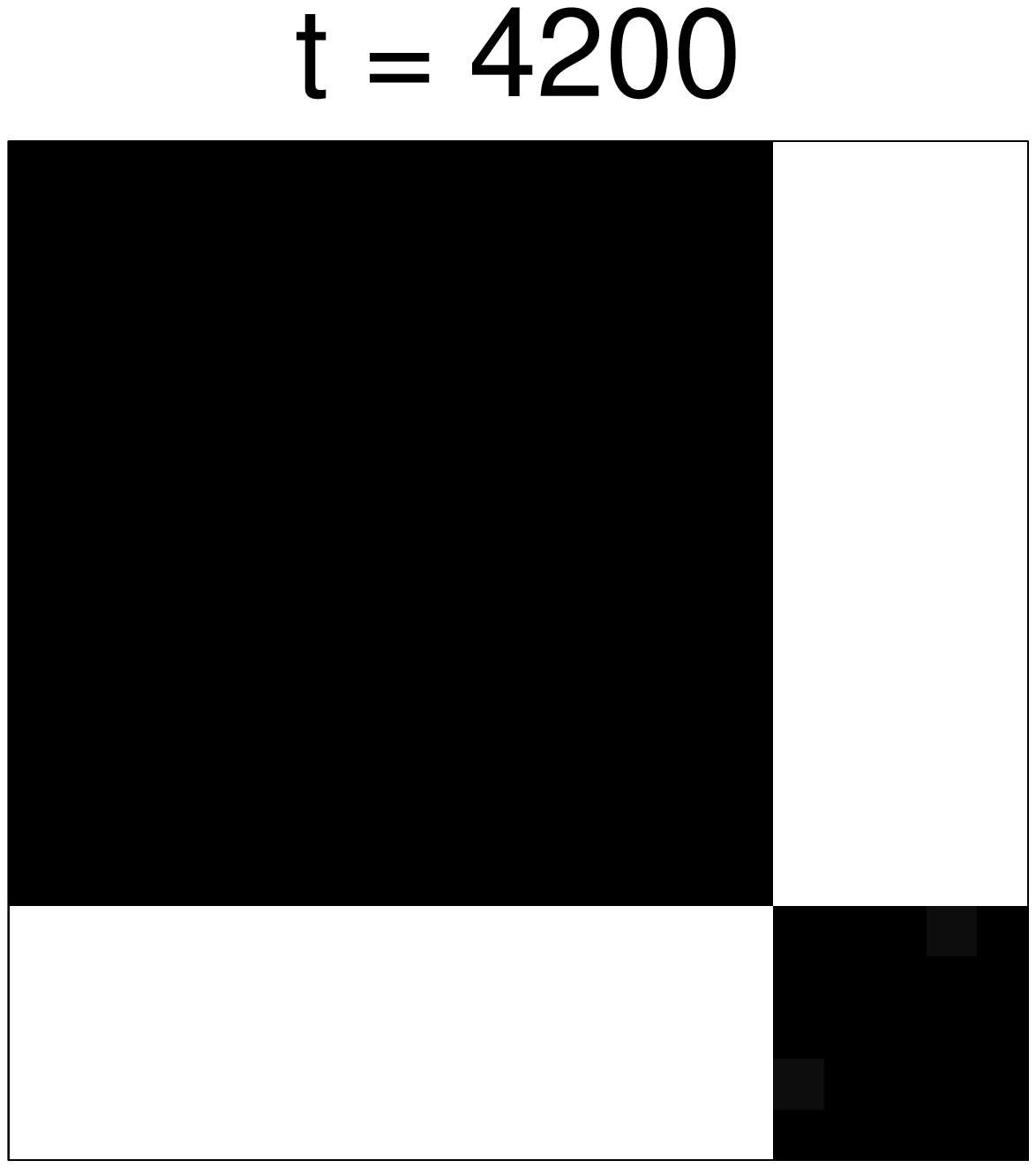}}%
 		\hspace{.125cm}%
\subfigure{	
		\includegraphics[width=0.5in]{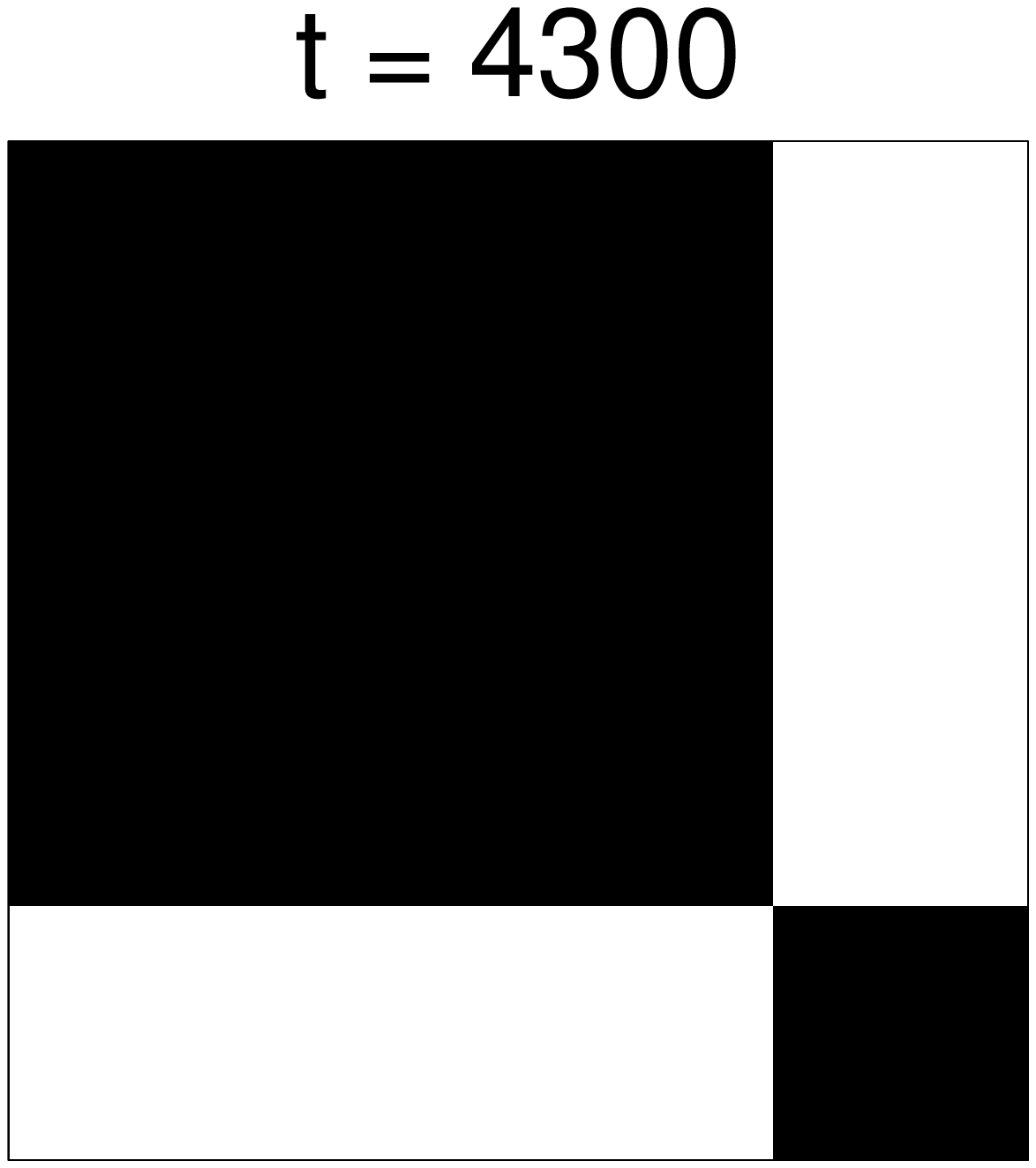}}%
		\hspace{.125cm}%
\subfigure{	
		\includegraphics[width=0.5in]{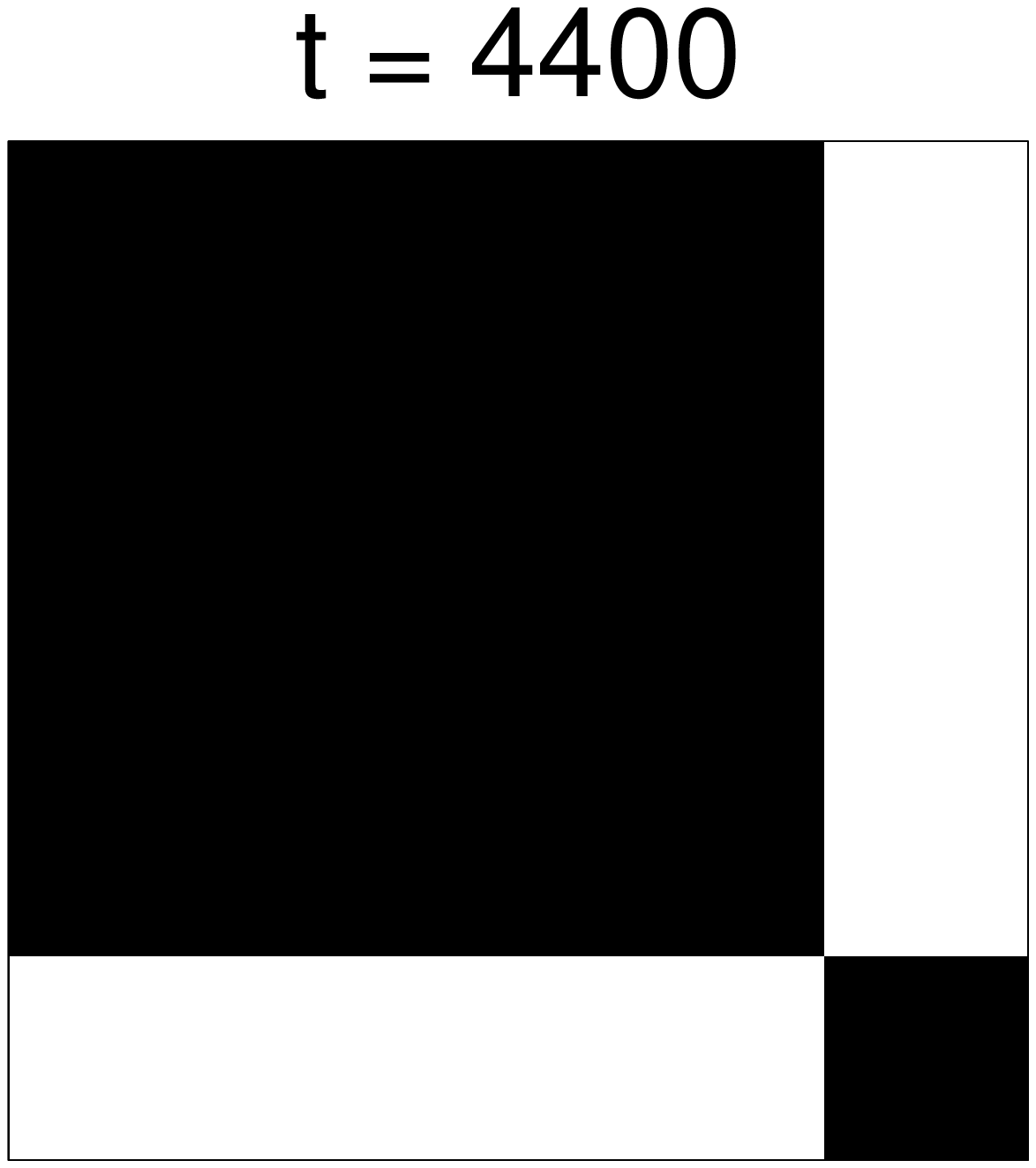}}%
		\hspace{.125cm}%
\subfigure{	
		\includegraphics[width=0.5in]{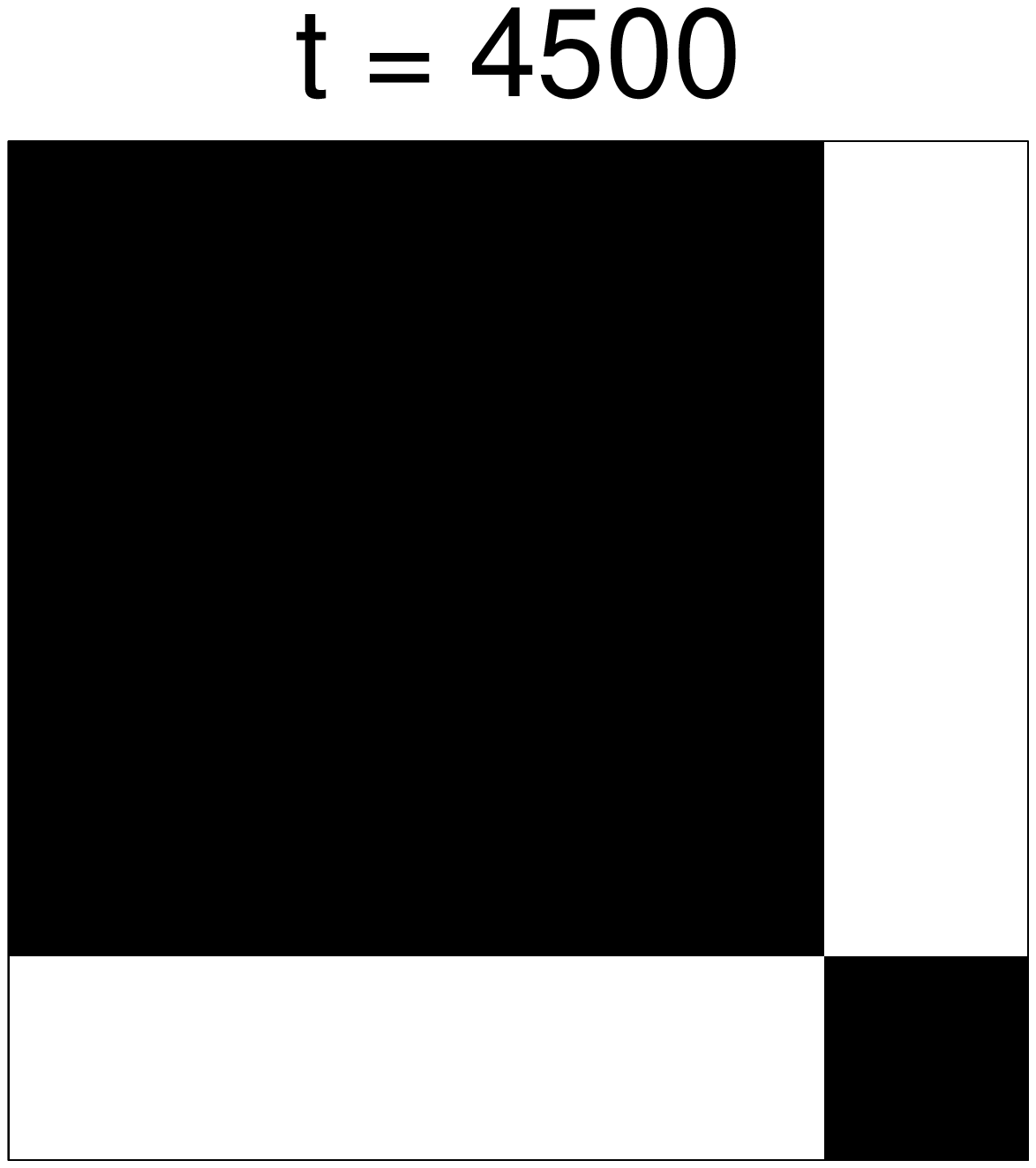}}%
		\hspace{.125cm}%
\subfigure{	
		\includegraphics[width=0.5in]{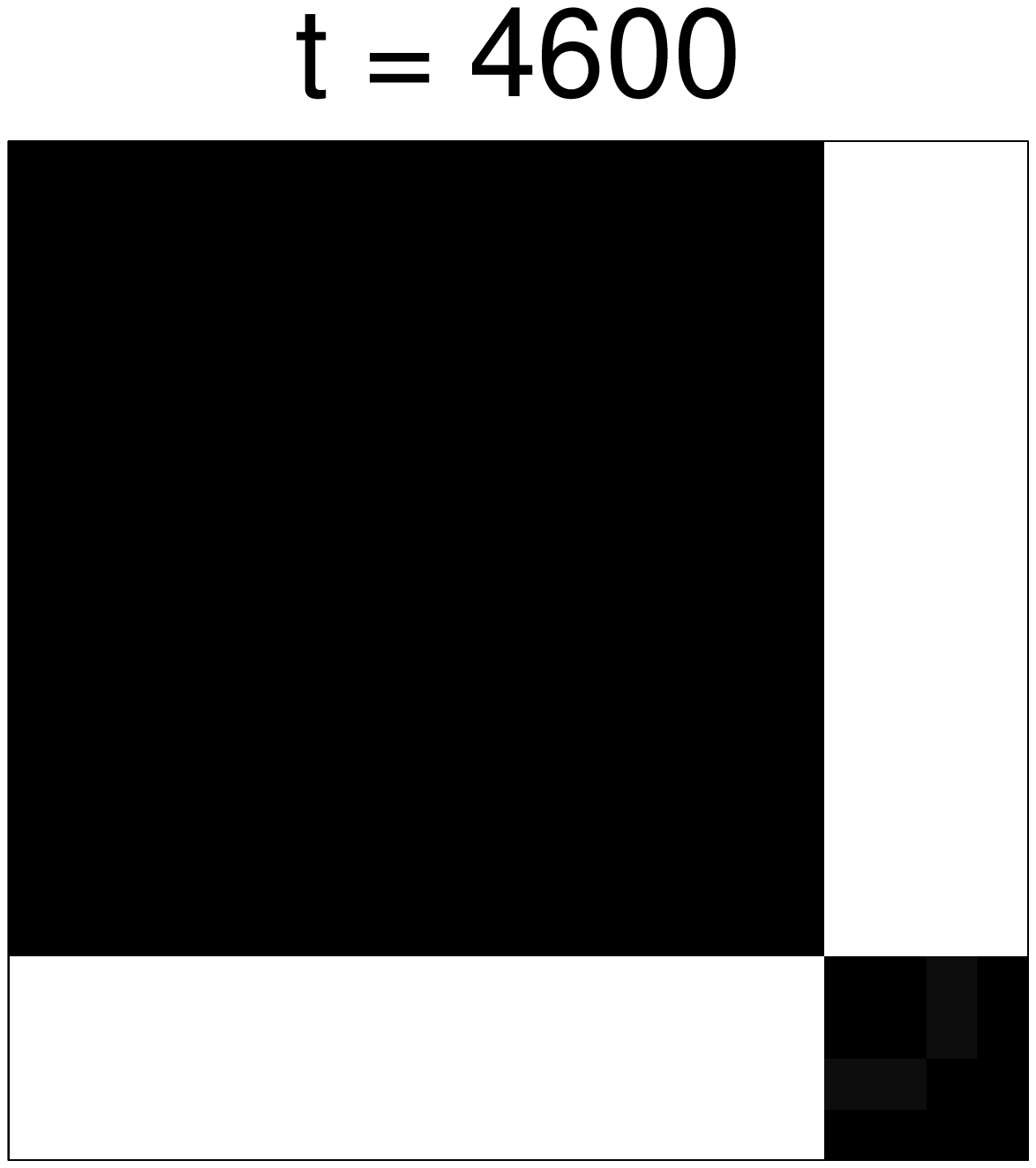}}%
		\hspace{.125cm}%
\subfigure{	
		\includegraphics[width=0.5in]{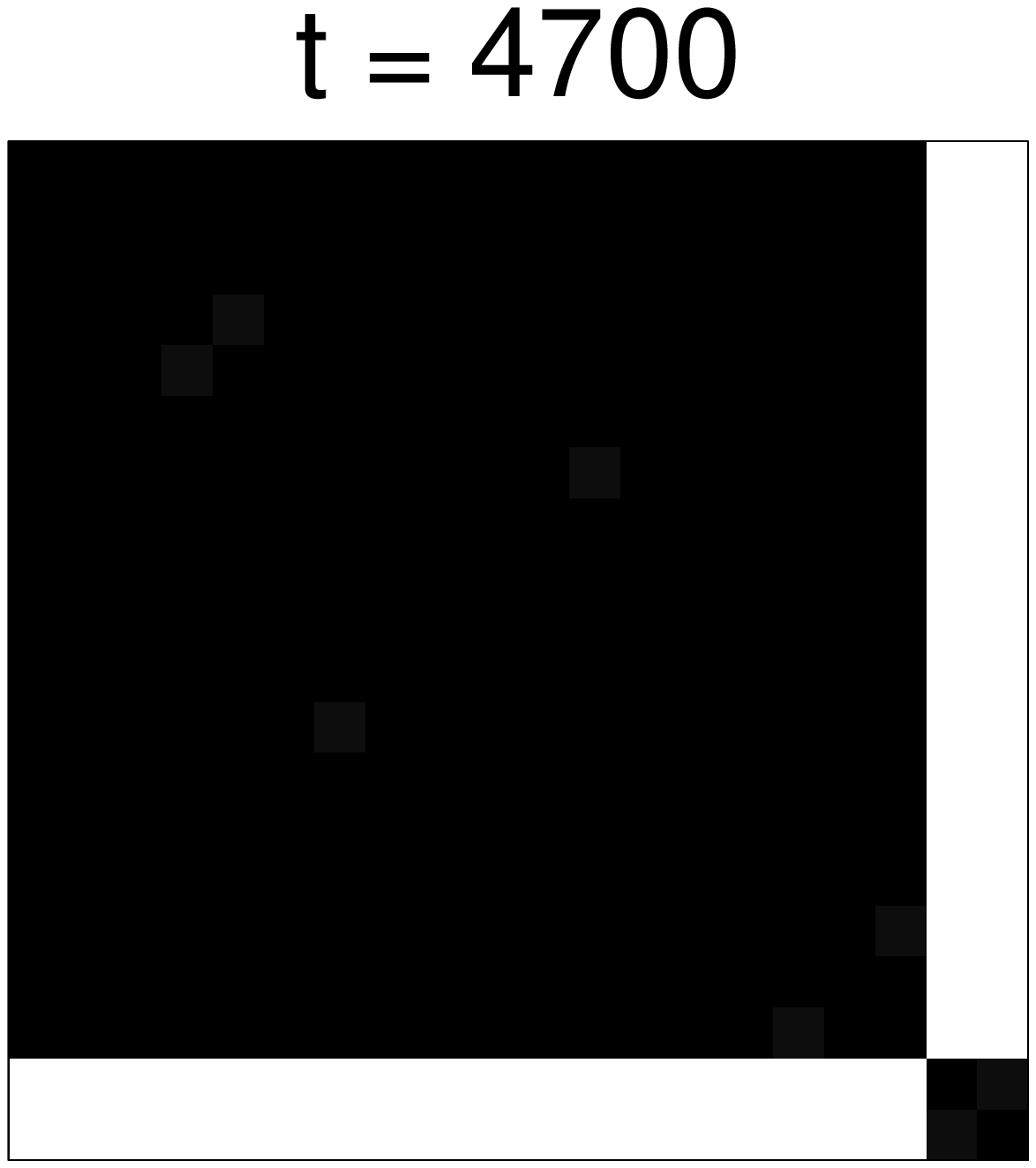}}%
		\hspace{.125cm}%
\subfigure{	
		\includegraphics[width=0.5in]{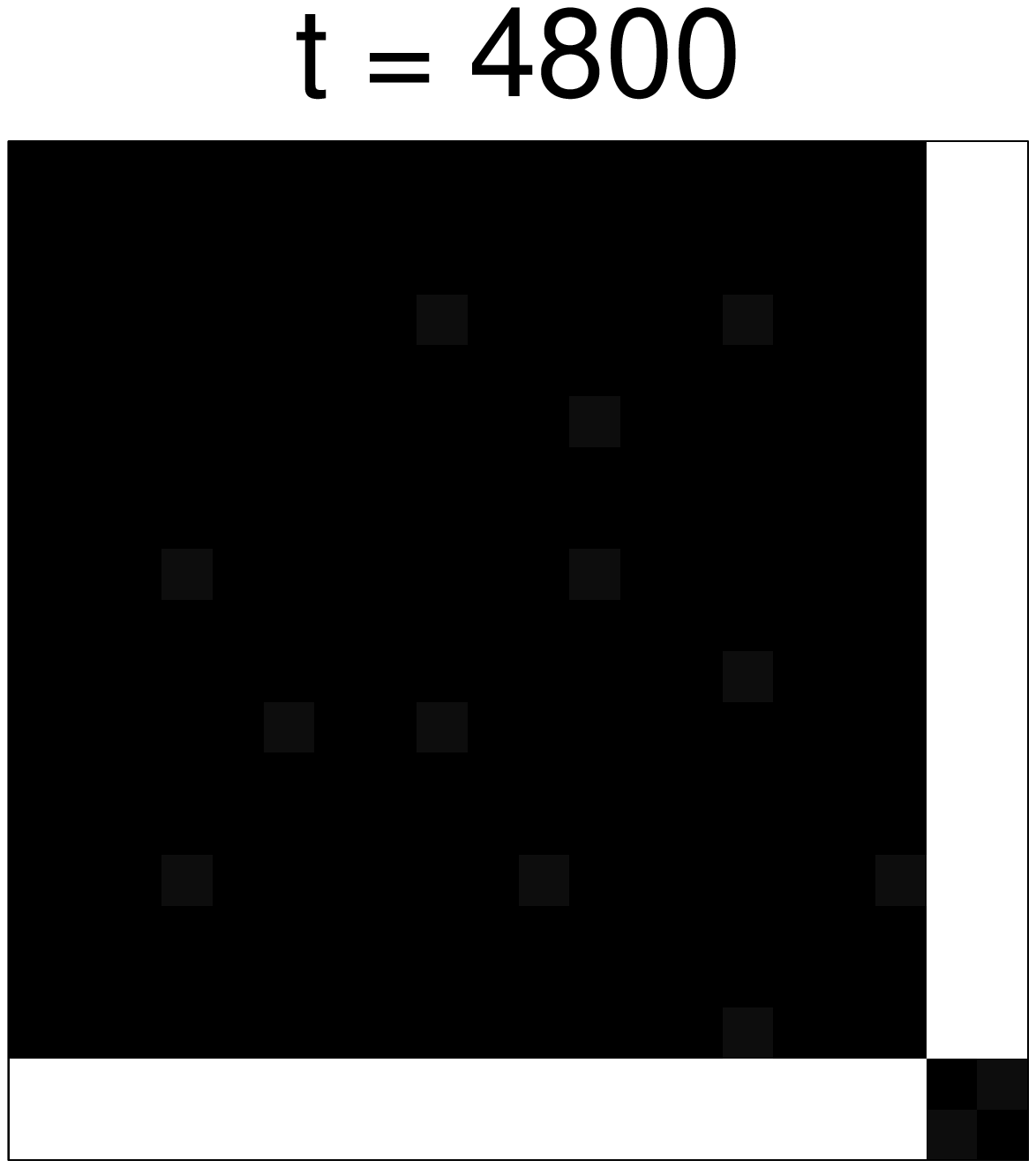}}%
		\hspace{.125cm}%
\subfigure{	
		\includegraphics[width=0.5in]{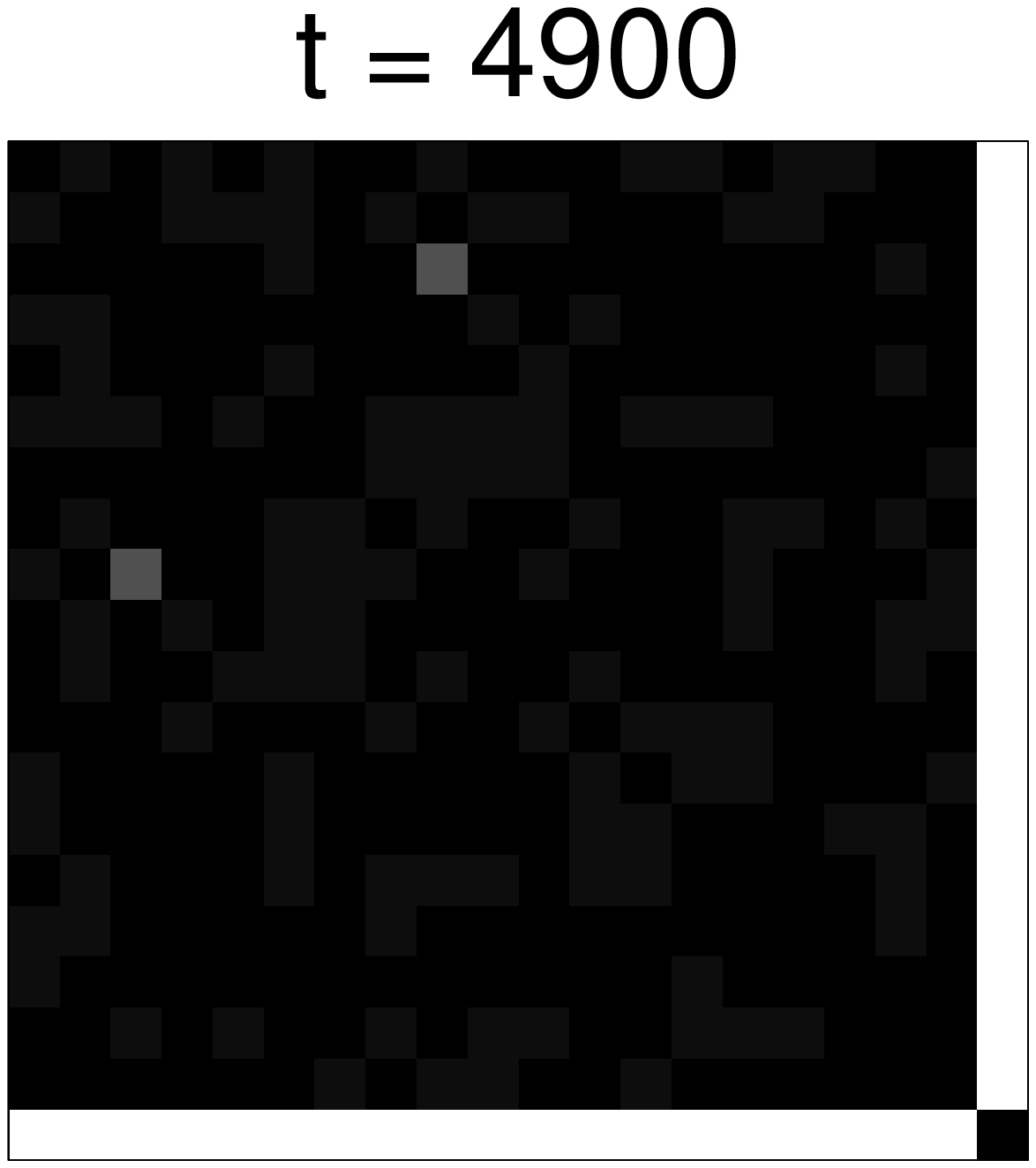}}%
		\hspace{.125cm}%
\subfigure{	
		\includegraphics[width=0.5in]{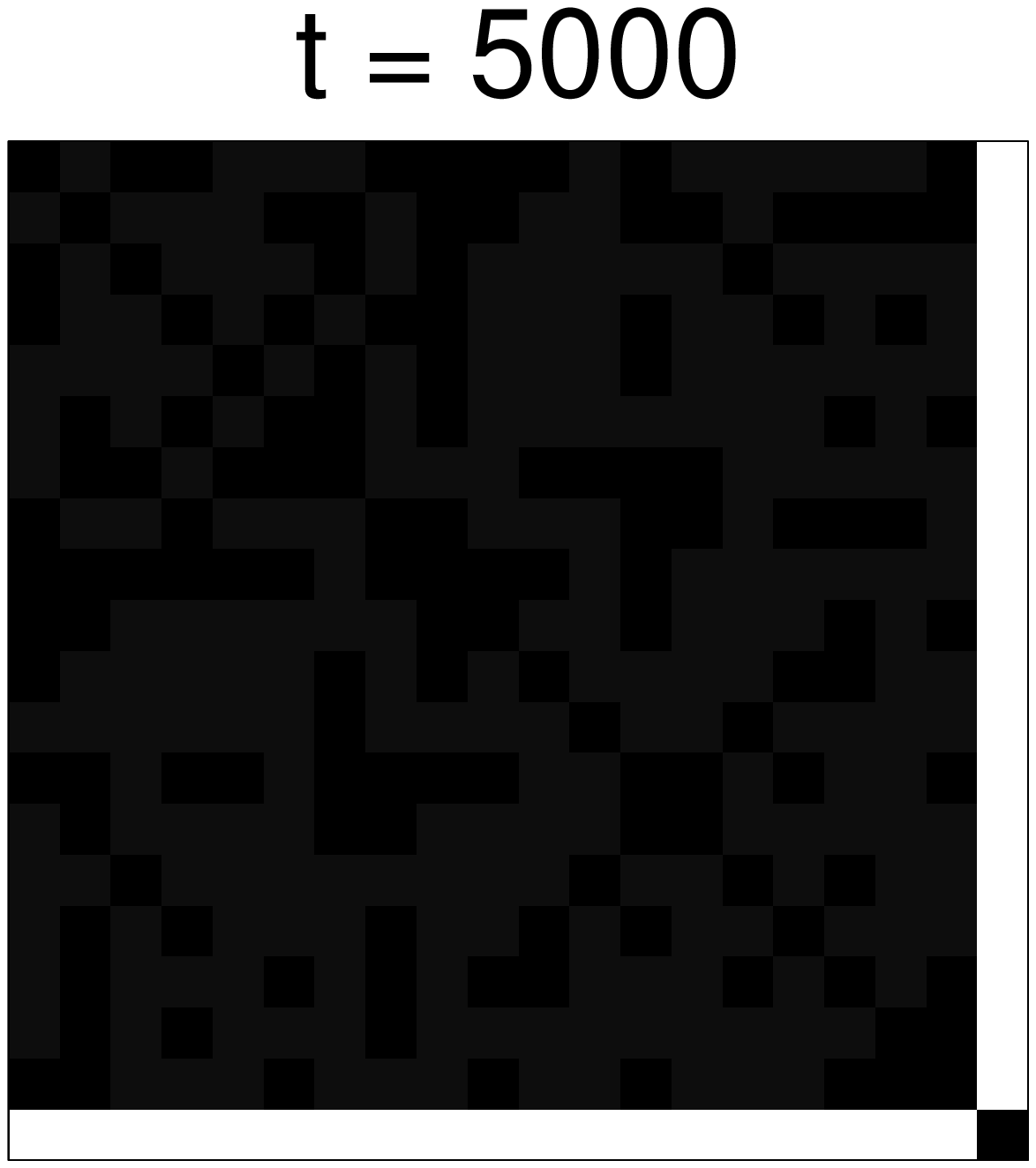}}\\
\subfigure{ 	
		\includegraphics[width=0.5in]{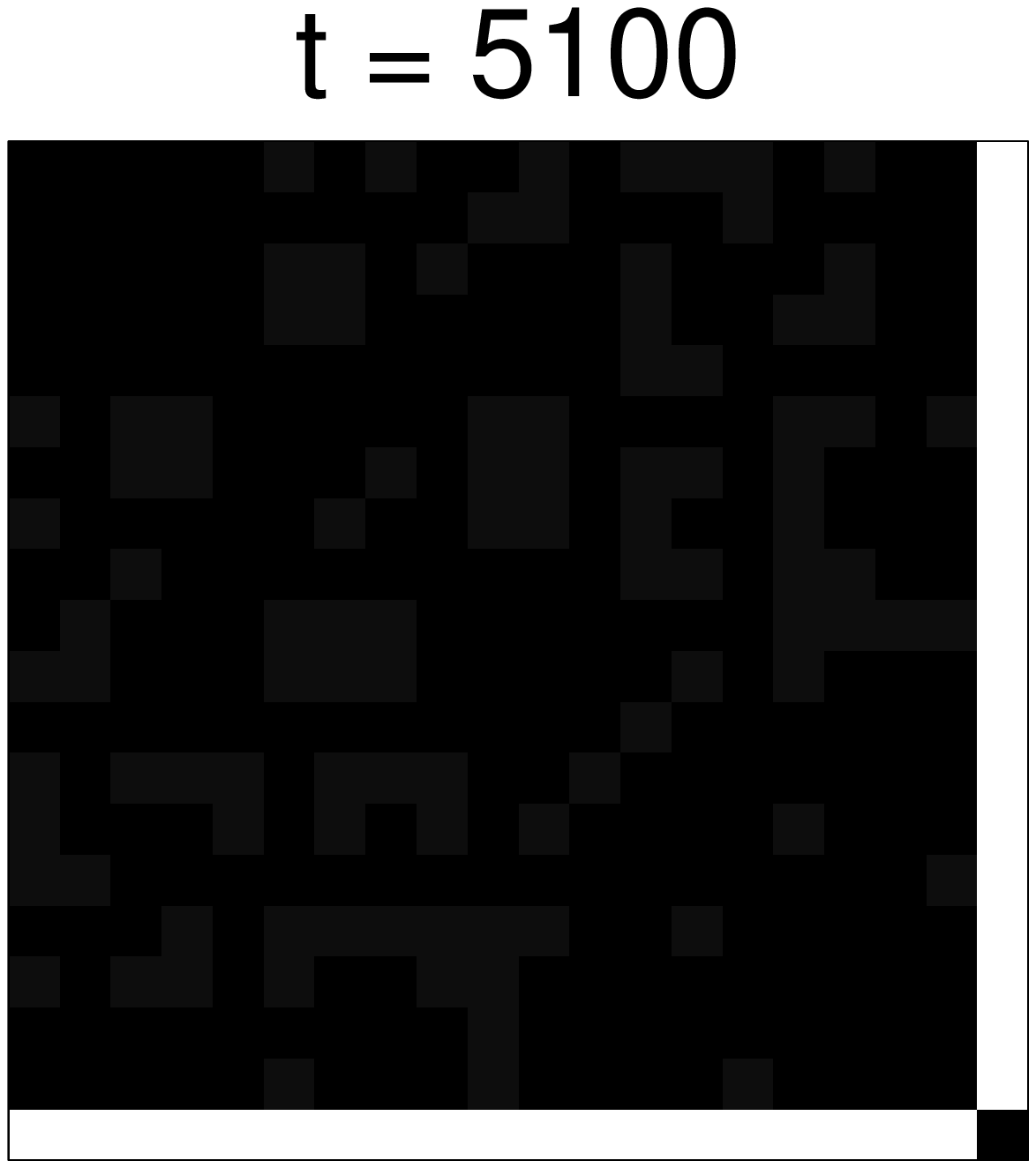}}%
 		\hspace{.125cm}%
\subfigure{	
		\includegraphics[width=0.5in]{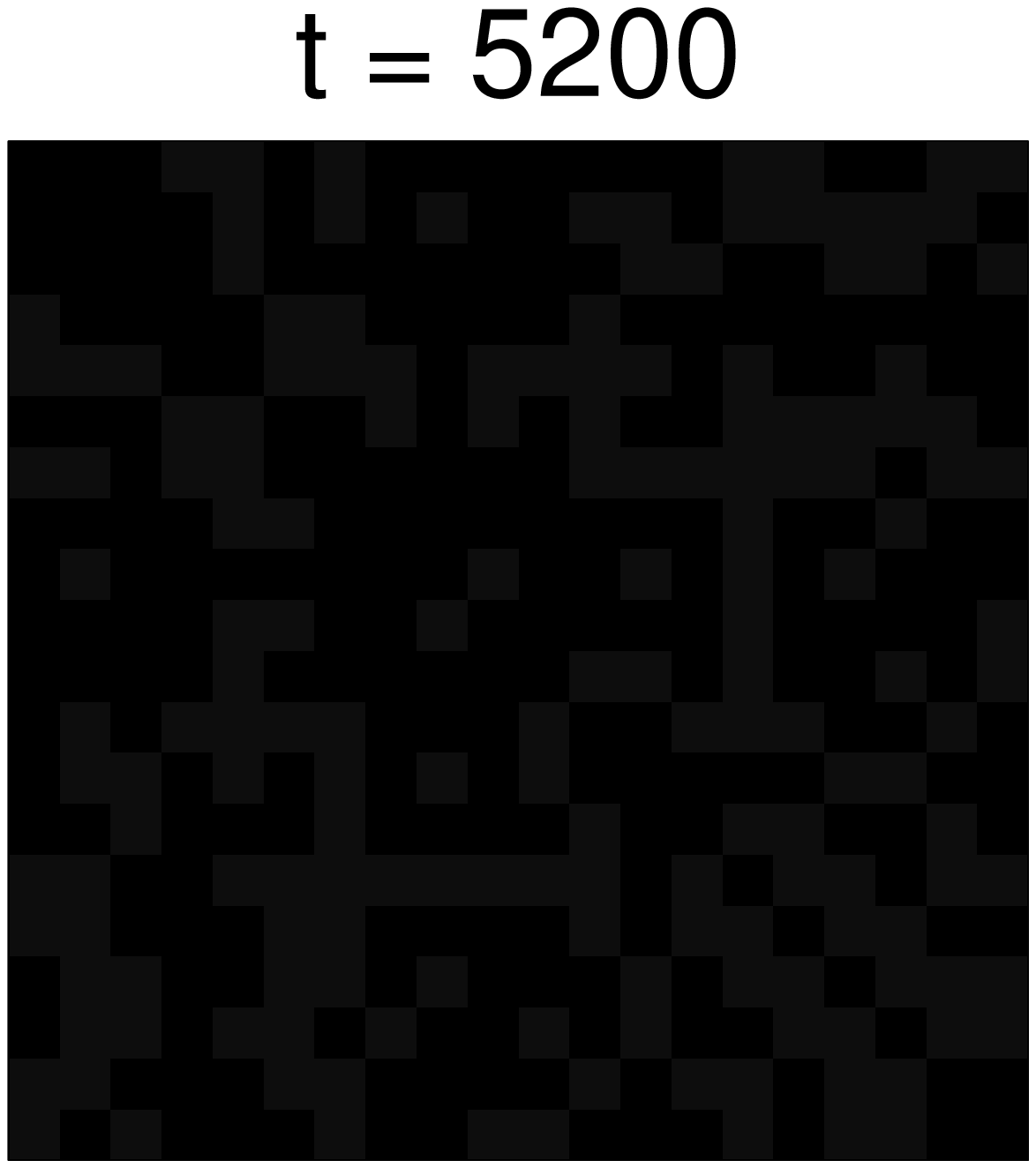}}%
 		\hspace{.125cm}%
\subfigure{	
		\includegraphics[width=0.5in]{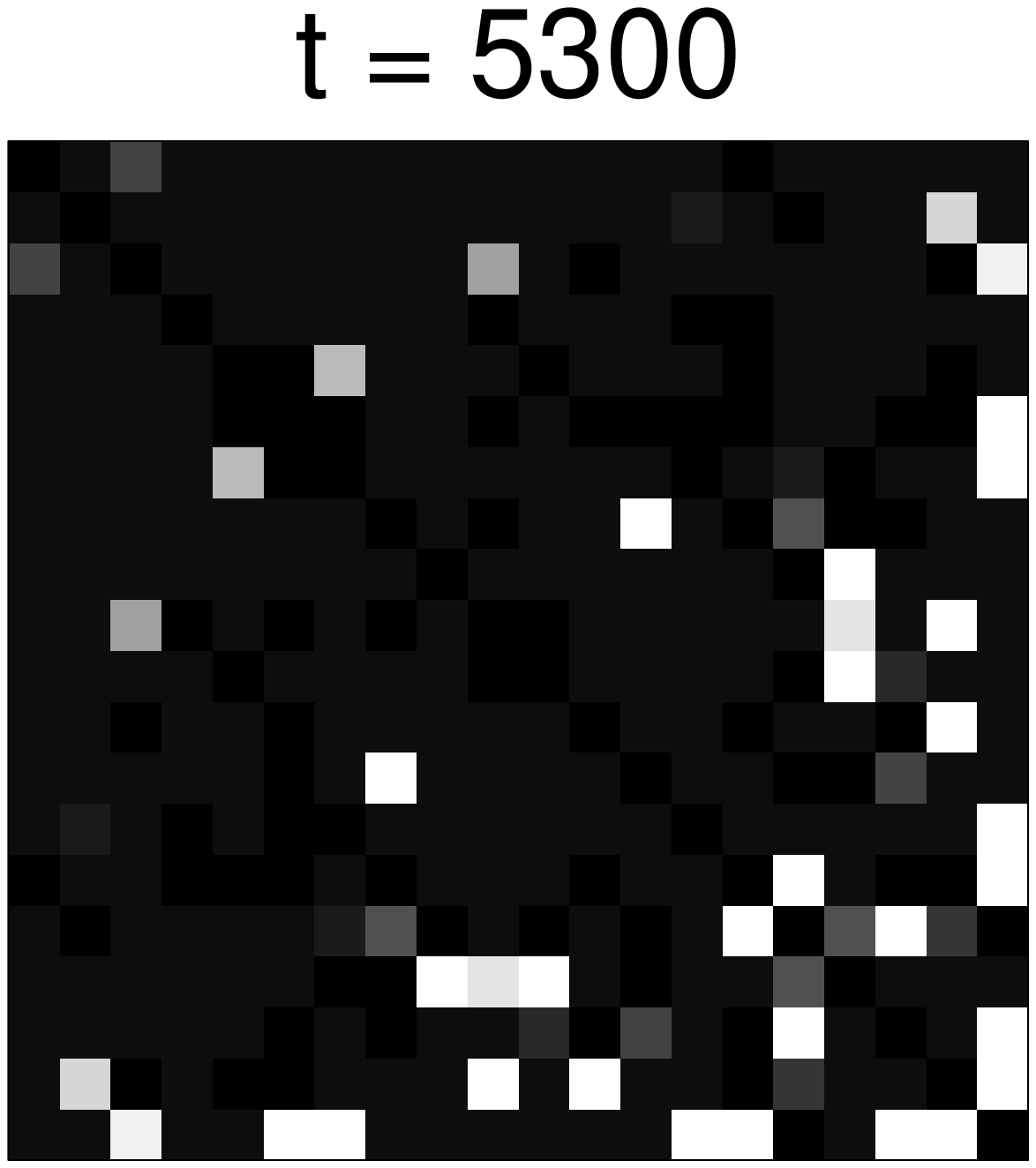}}%
		\hspace{.125cm}%
\subfigure{	
		\includegraphics[width=0.5in]{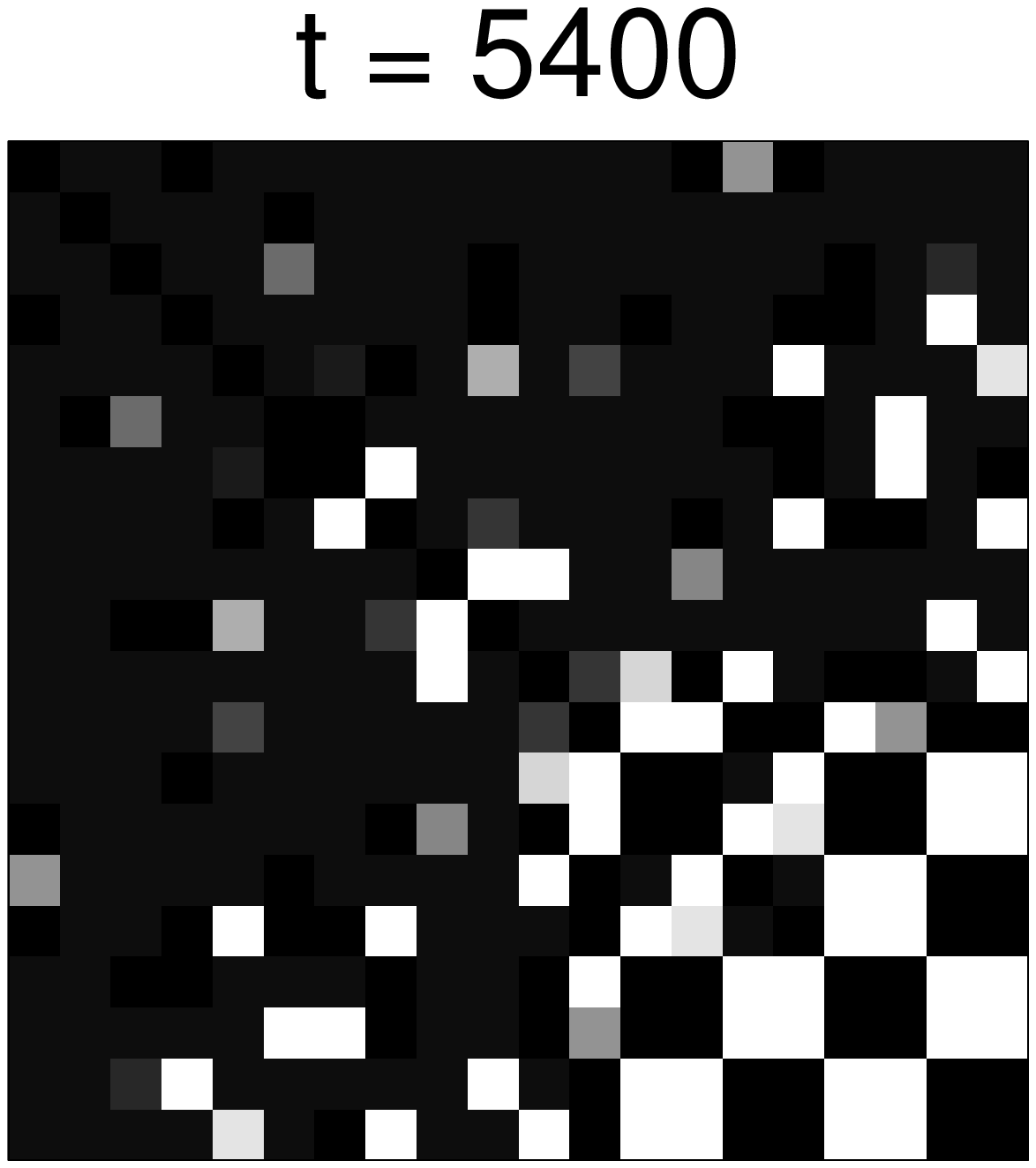}}%
		\hspace{.125cm}%
\subfigure{	
		\includegraphics[width=0.5in]{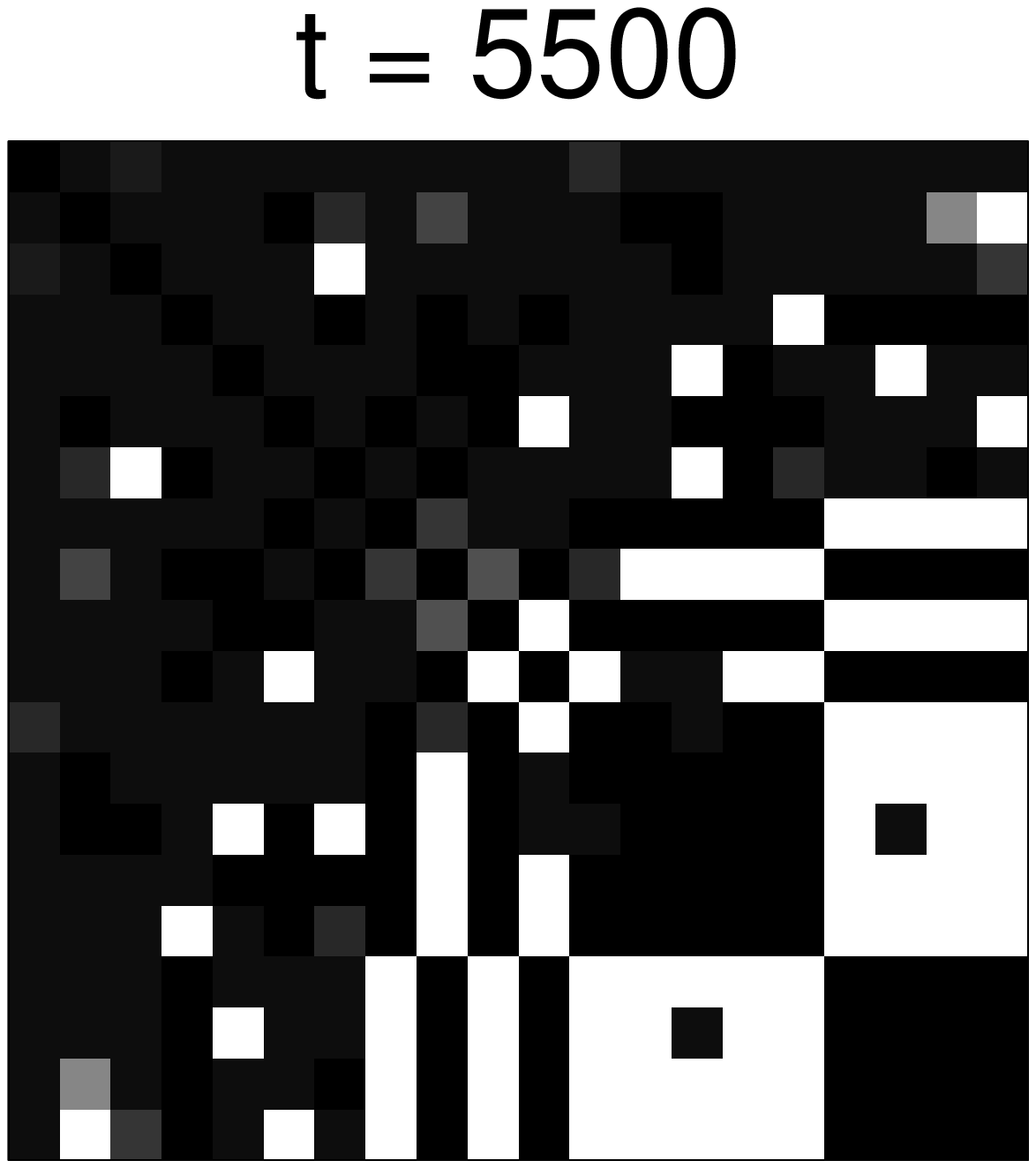}}%
		\hspace{.125cm}%
\subfigure{	
		\includegraphics[width=0.5in]{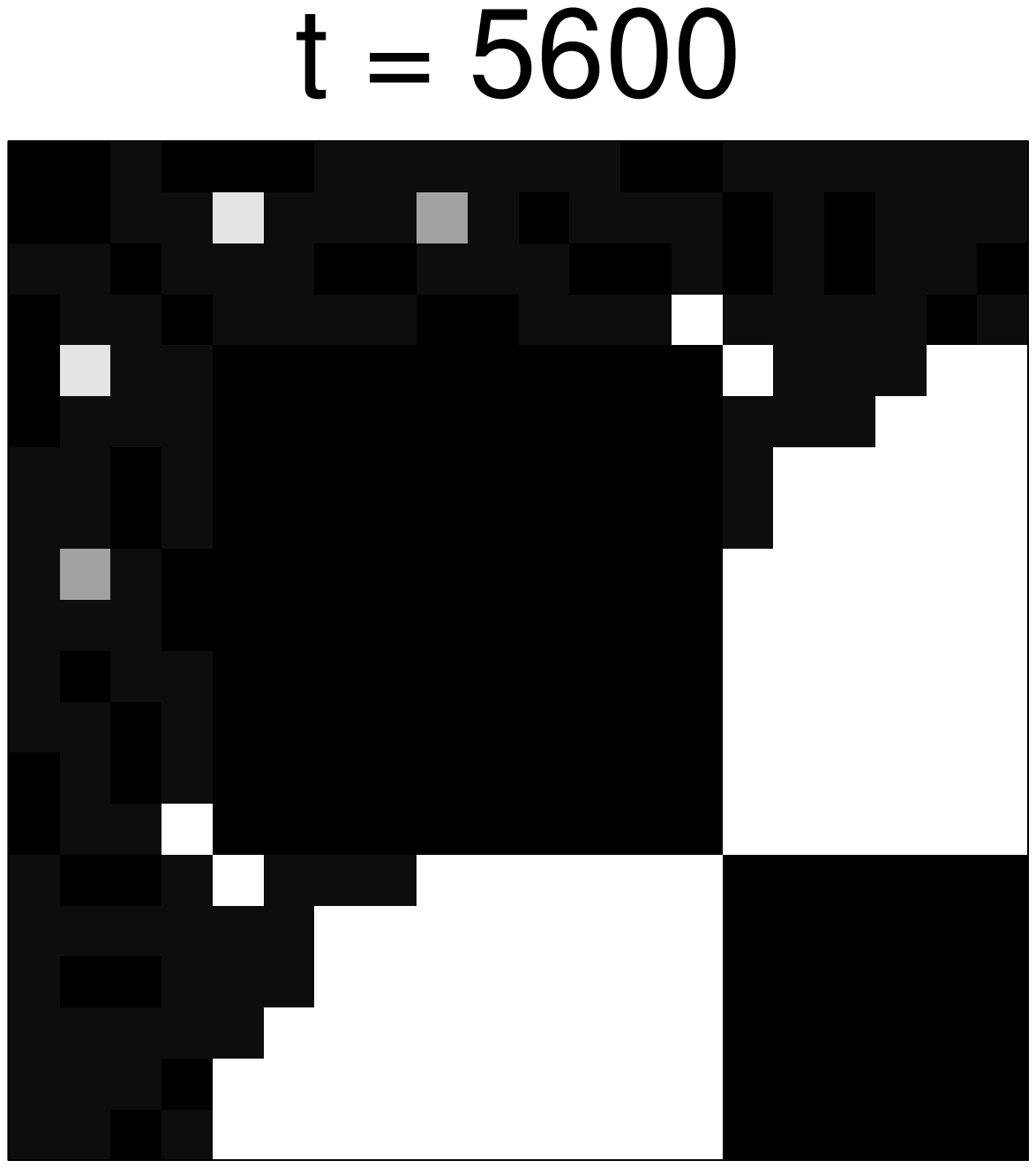}}%
		\hspace{.125cm}%
\subfigure{	
		\includegraphics[width=0.5in]{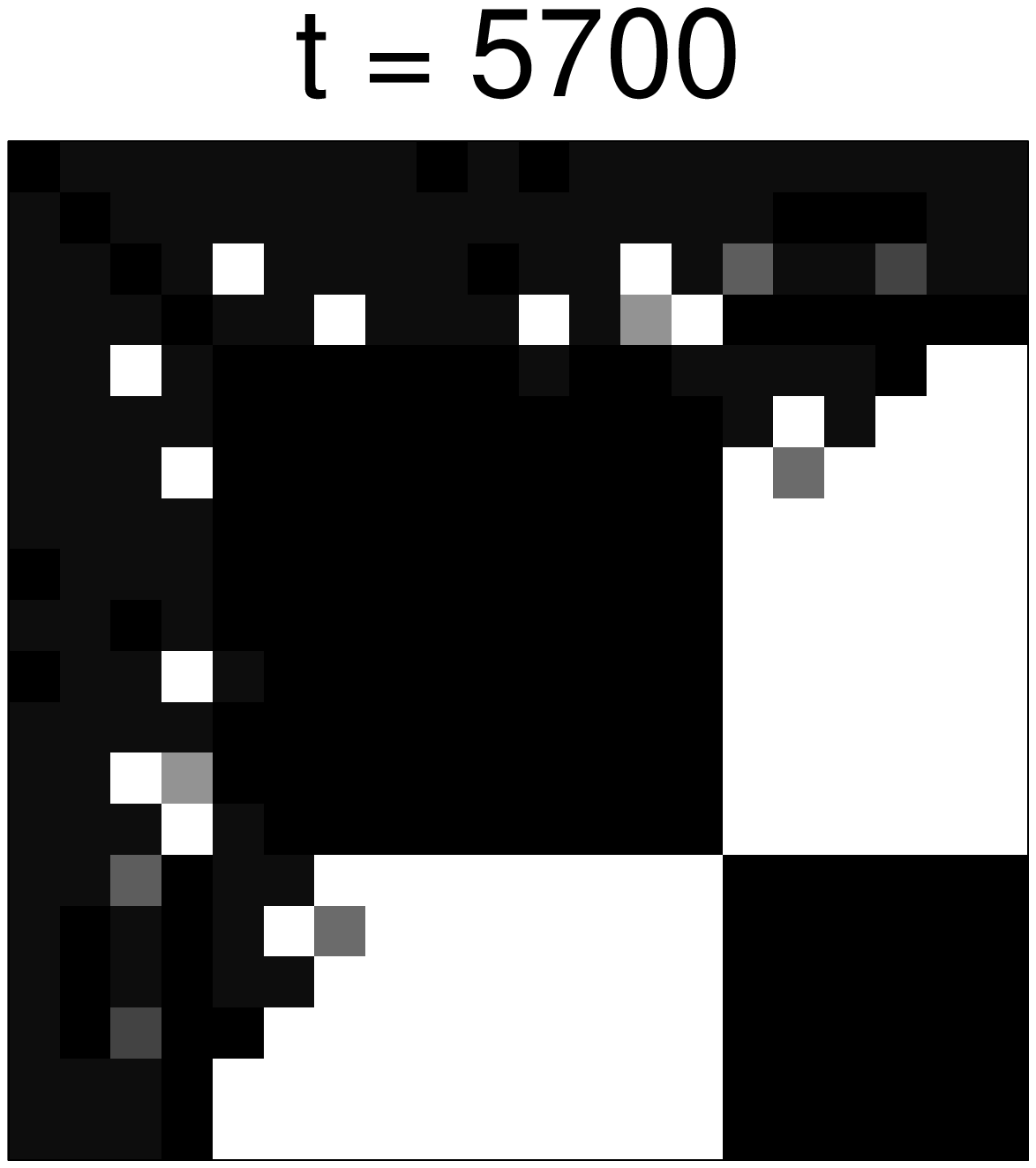}}%
		\hspace{.125cm}%
\subfigure{	
		\includegraphics[width=0.5in]{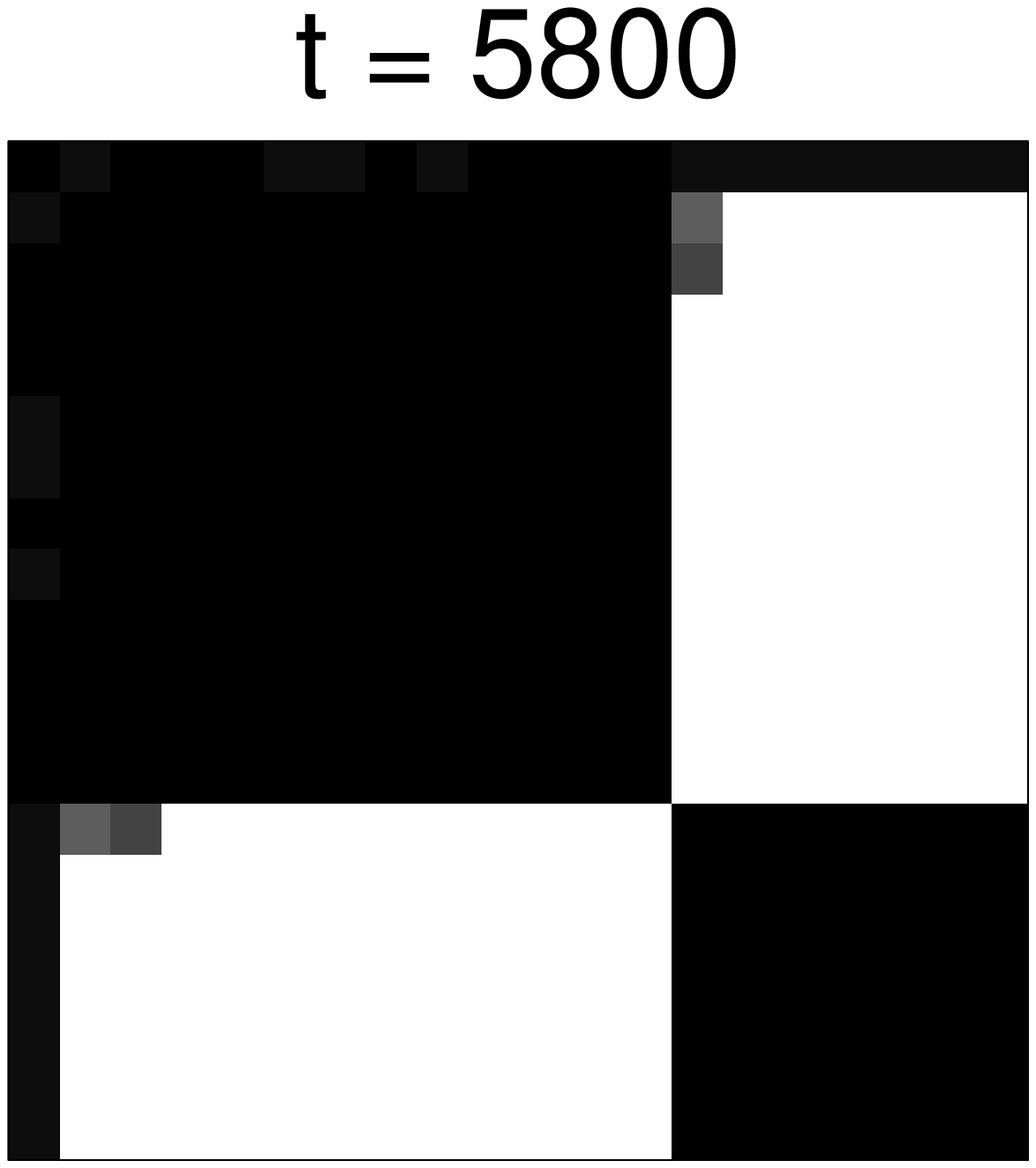}}%
		\hspace{.125cm}%
\subfigure{	
		\includegraphics[width=0.5in]{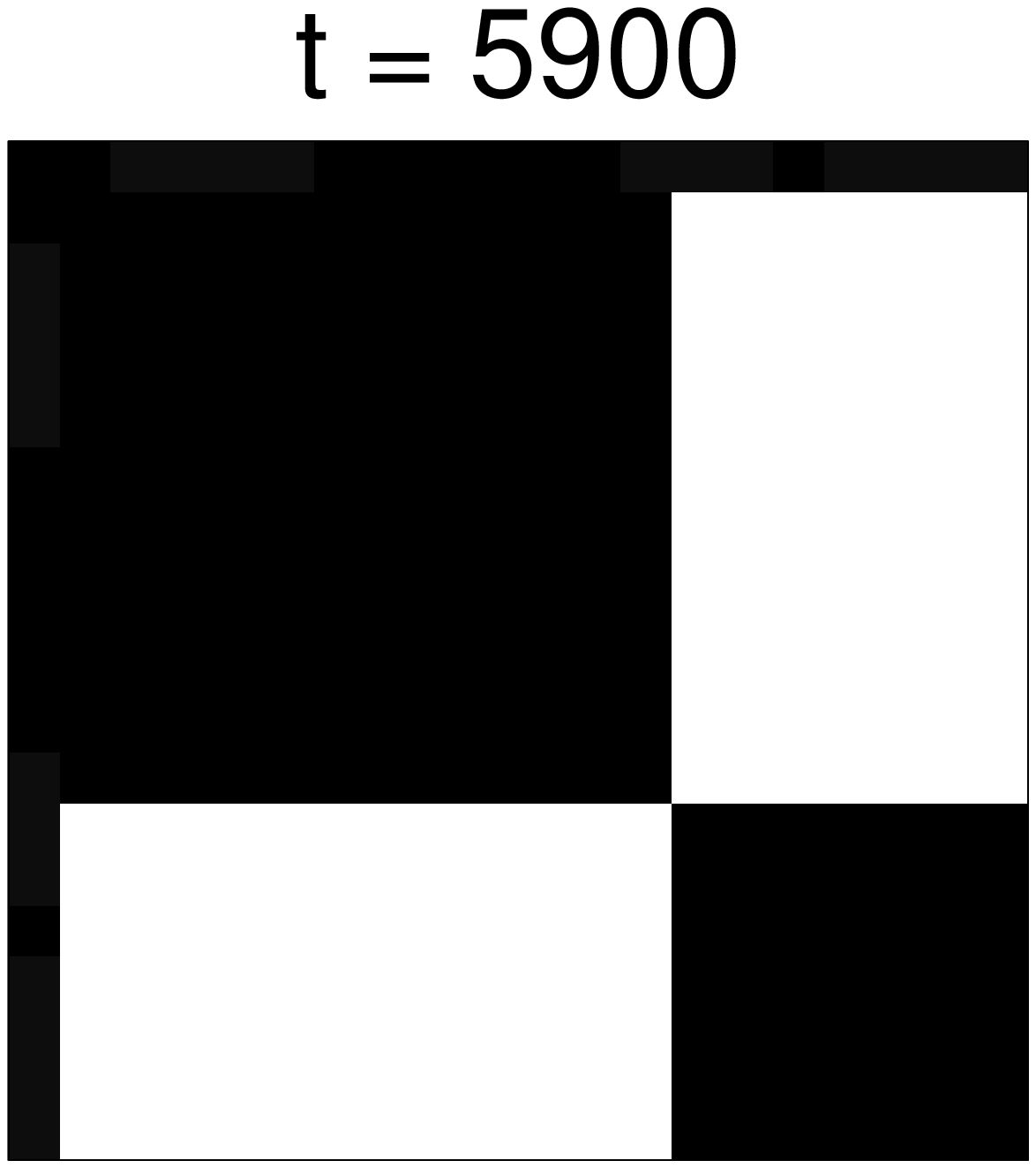}}%
		\hspace{.125cm}%
\subfigure{	
		\includegraphics[width=0.5in]{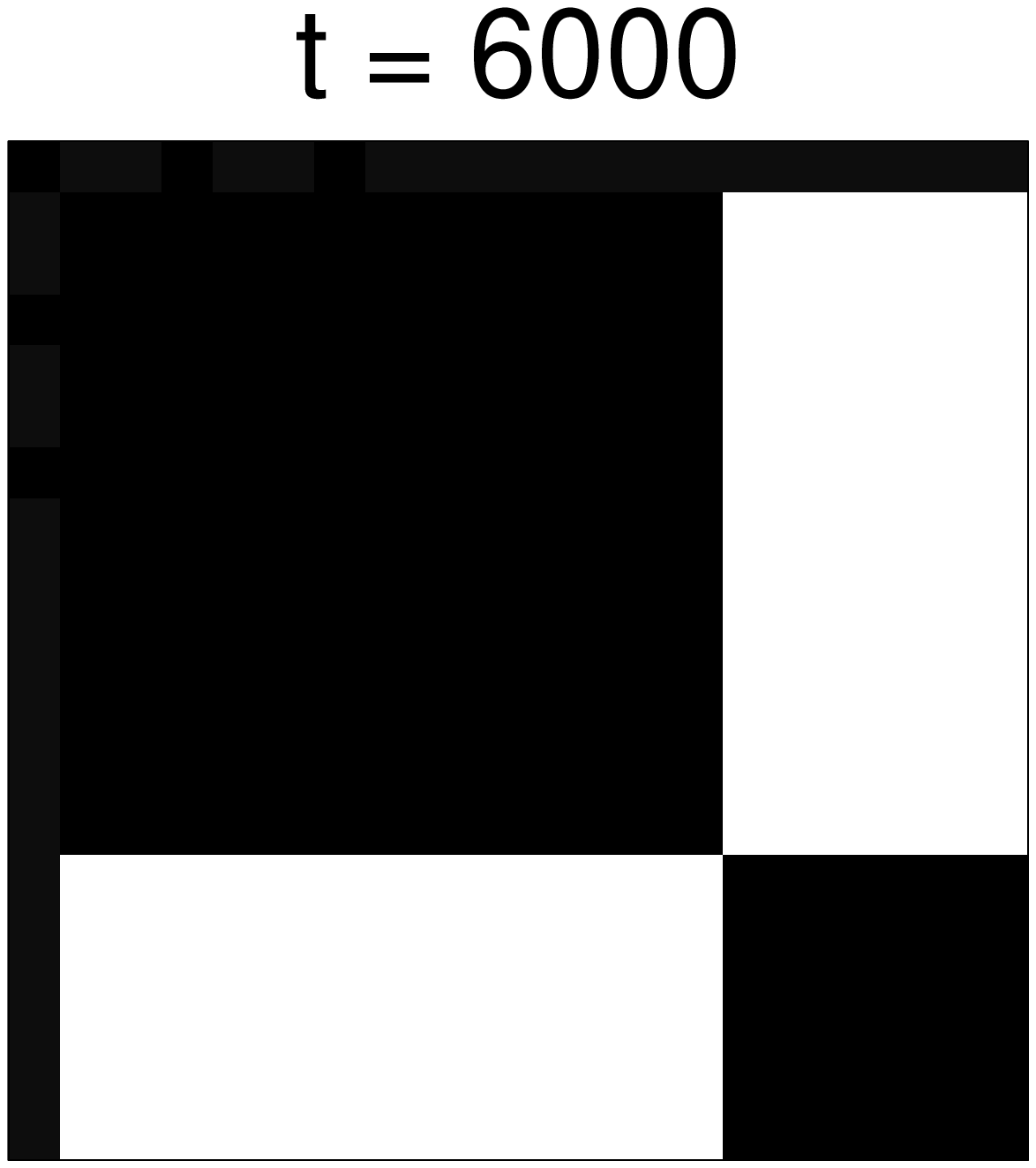}}\\
\subfigure{ 	
		\includegraphics[width=0.5in]{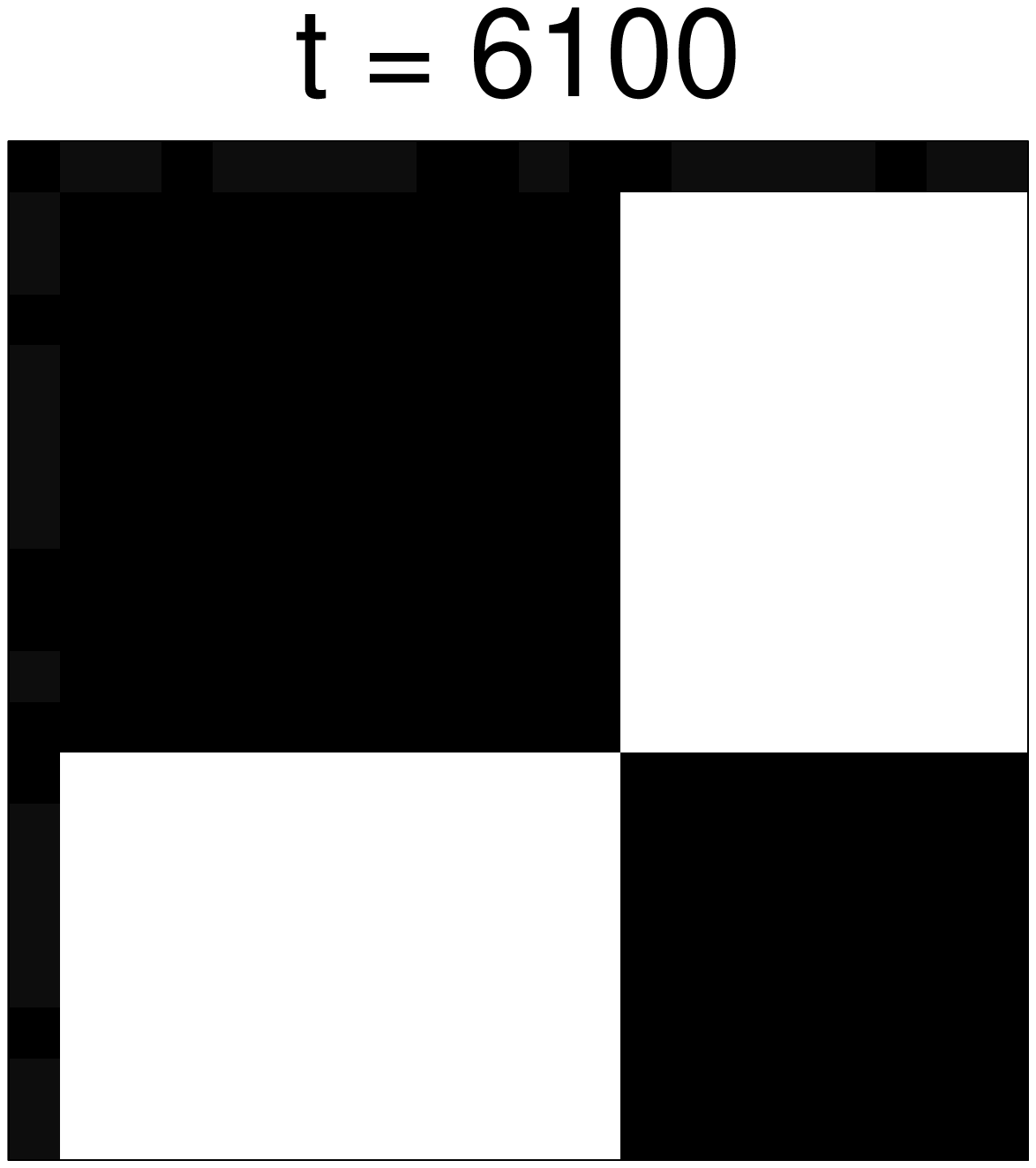}}%
 		\hspace{.125cm}%
\subfigure{	
		\includegraphics[width=0.5in]{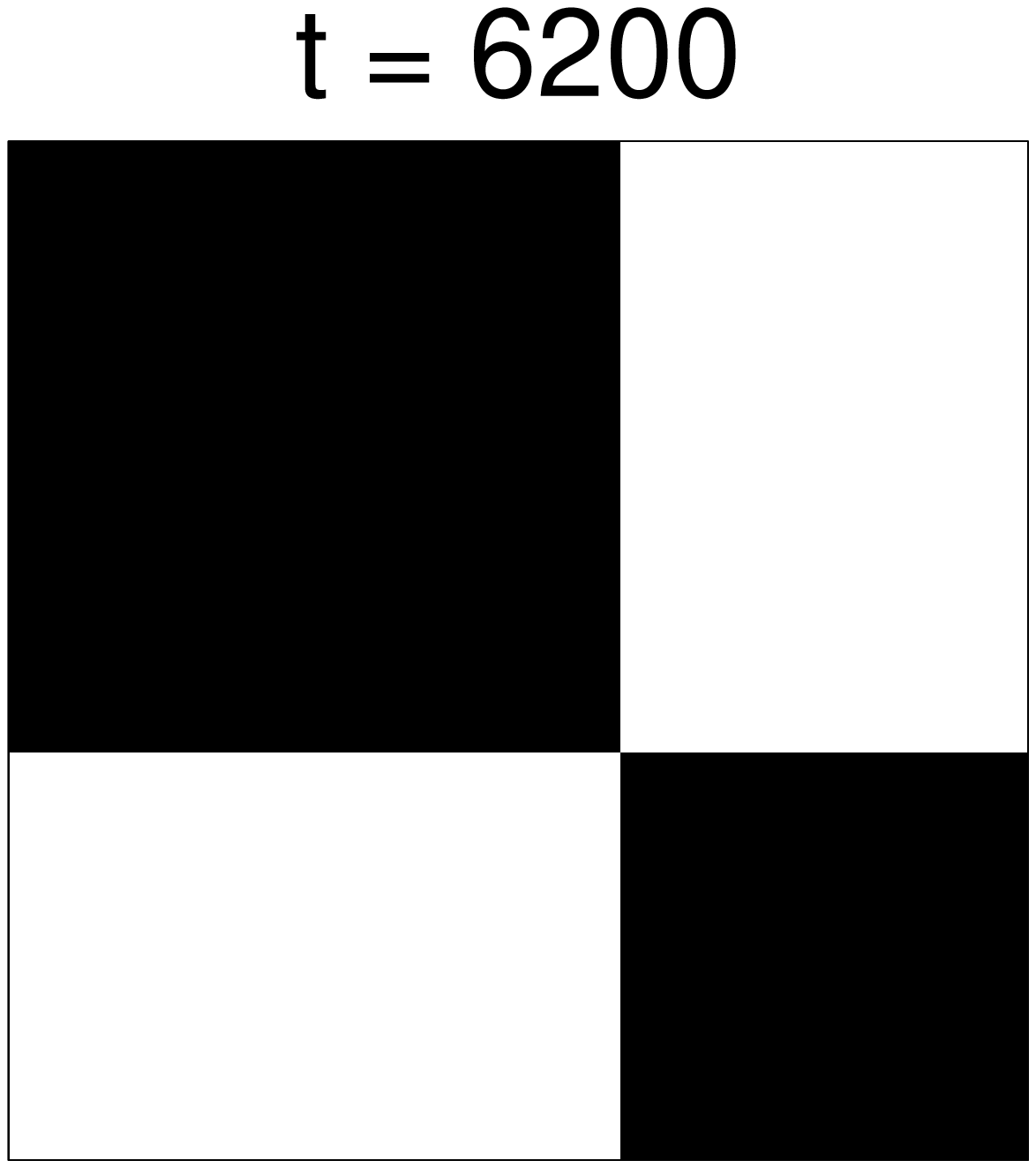}}%
 		\hspace{.125cm}%
\subfigure{	
		\includegraphics[width=0.5in]{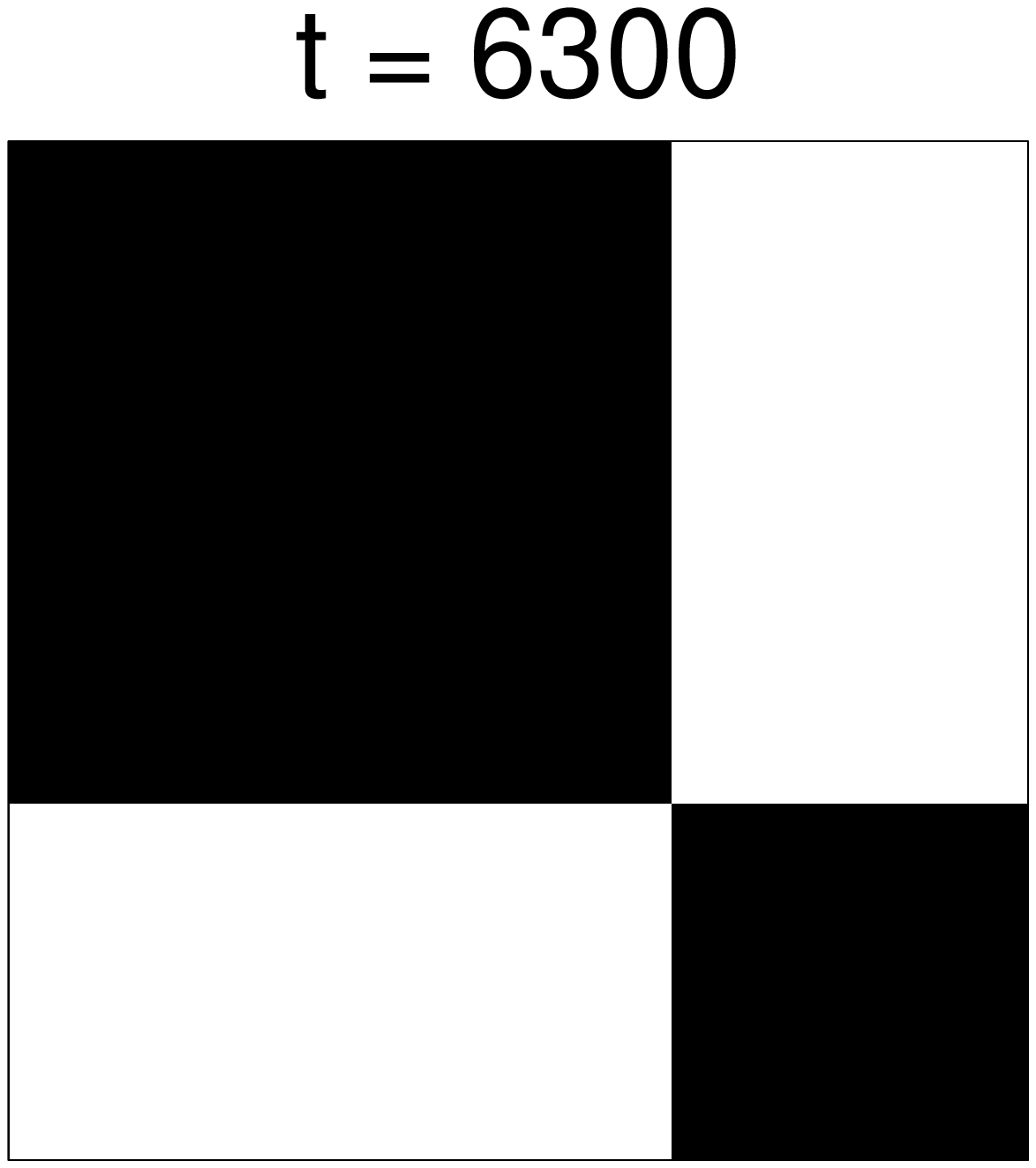}}%
		\hspace{.125cm}%
\subfigure{	
		\includegraphics[width=0.5in]{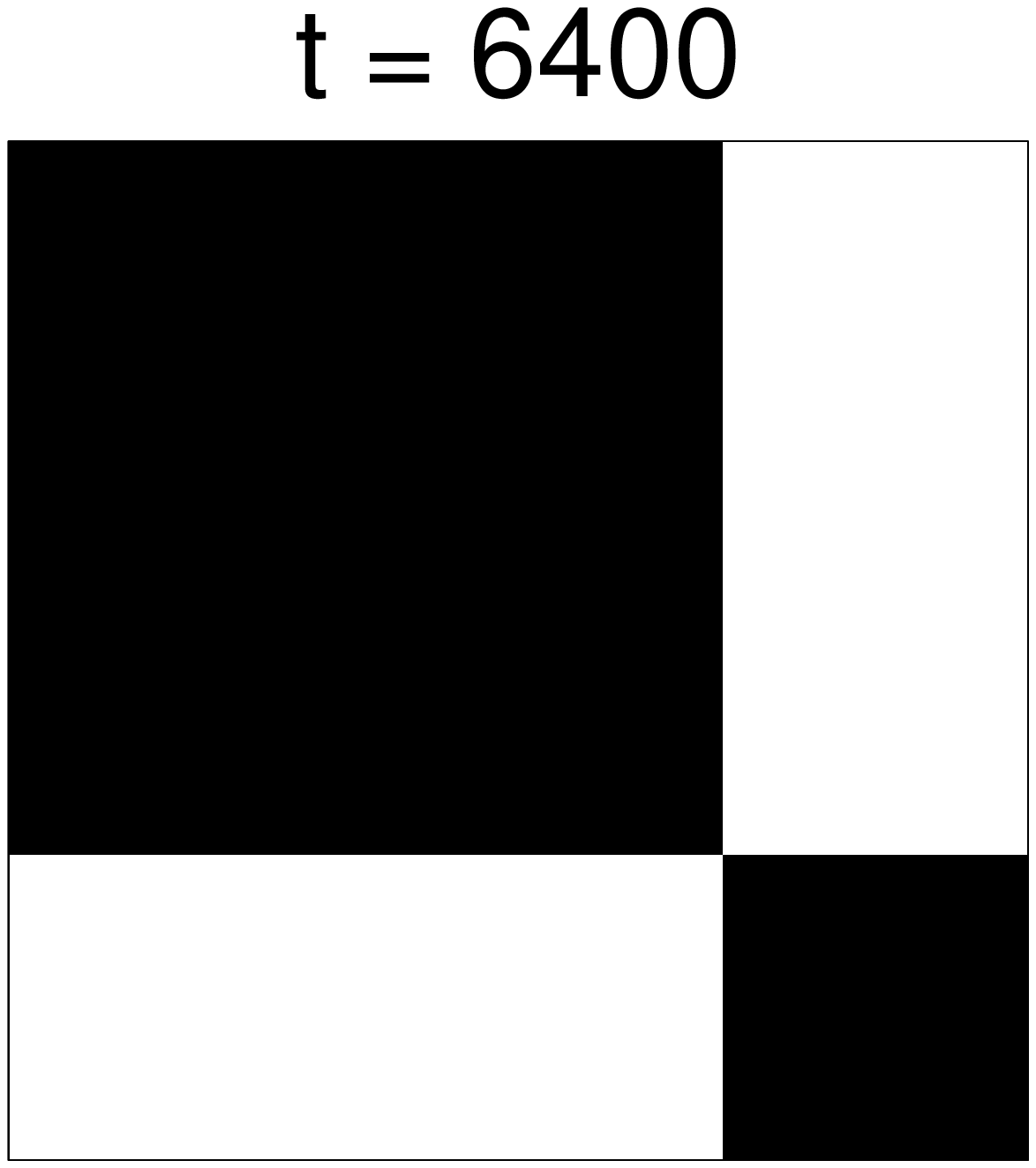}}%
		\hspace{.125cm}%
\subfigure{	
		\includegraphics[width=0.5in]{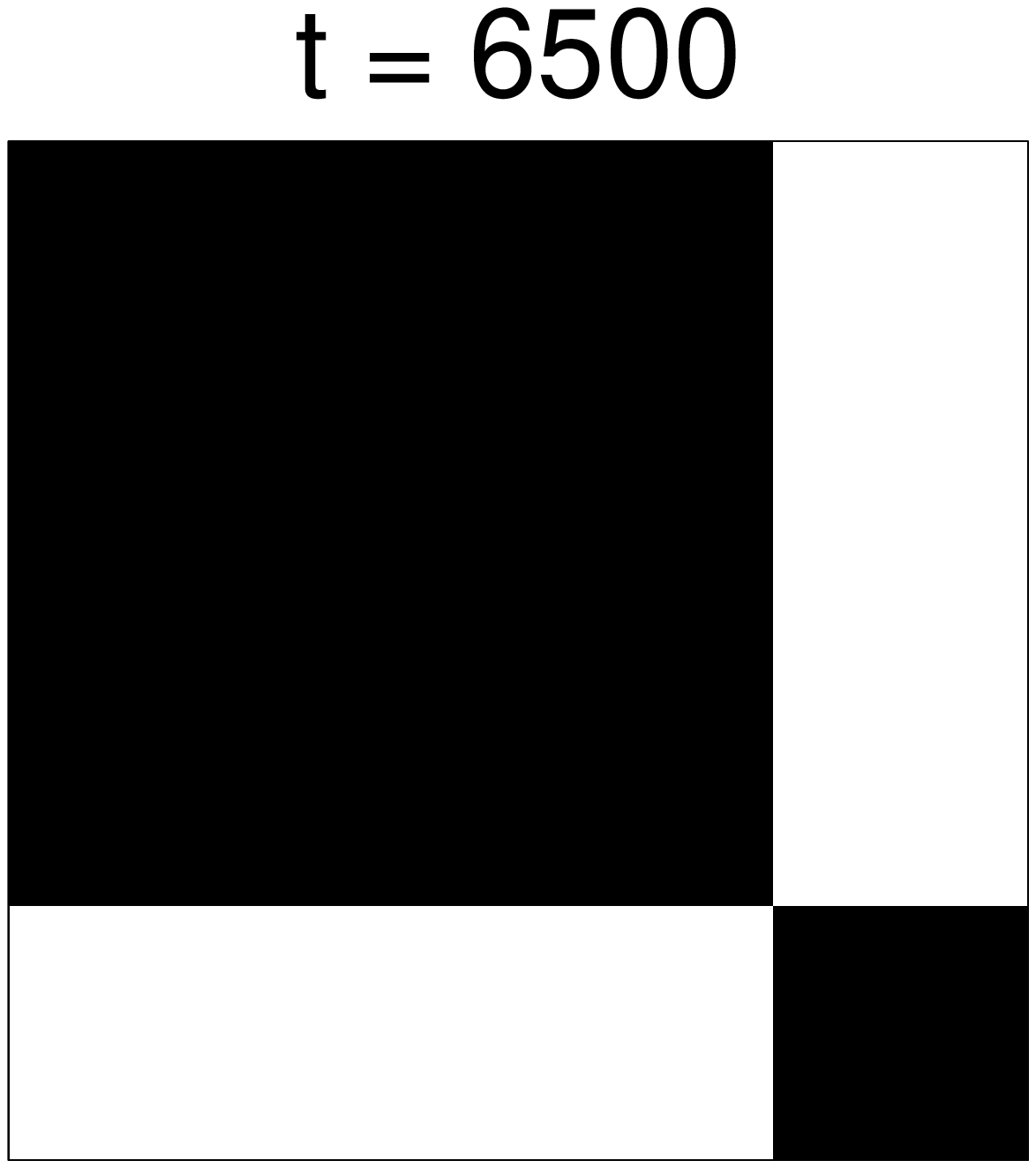}}%
		\hspace{.125cm}%
\setcounter{subfigure}{3}
\subfigure[]{	
		\includegraphics[width=0.5in]{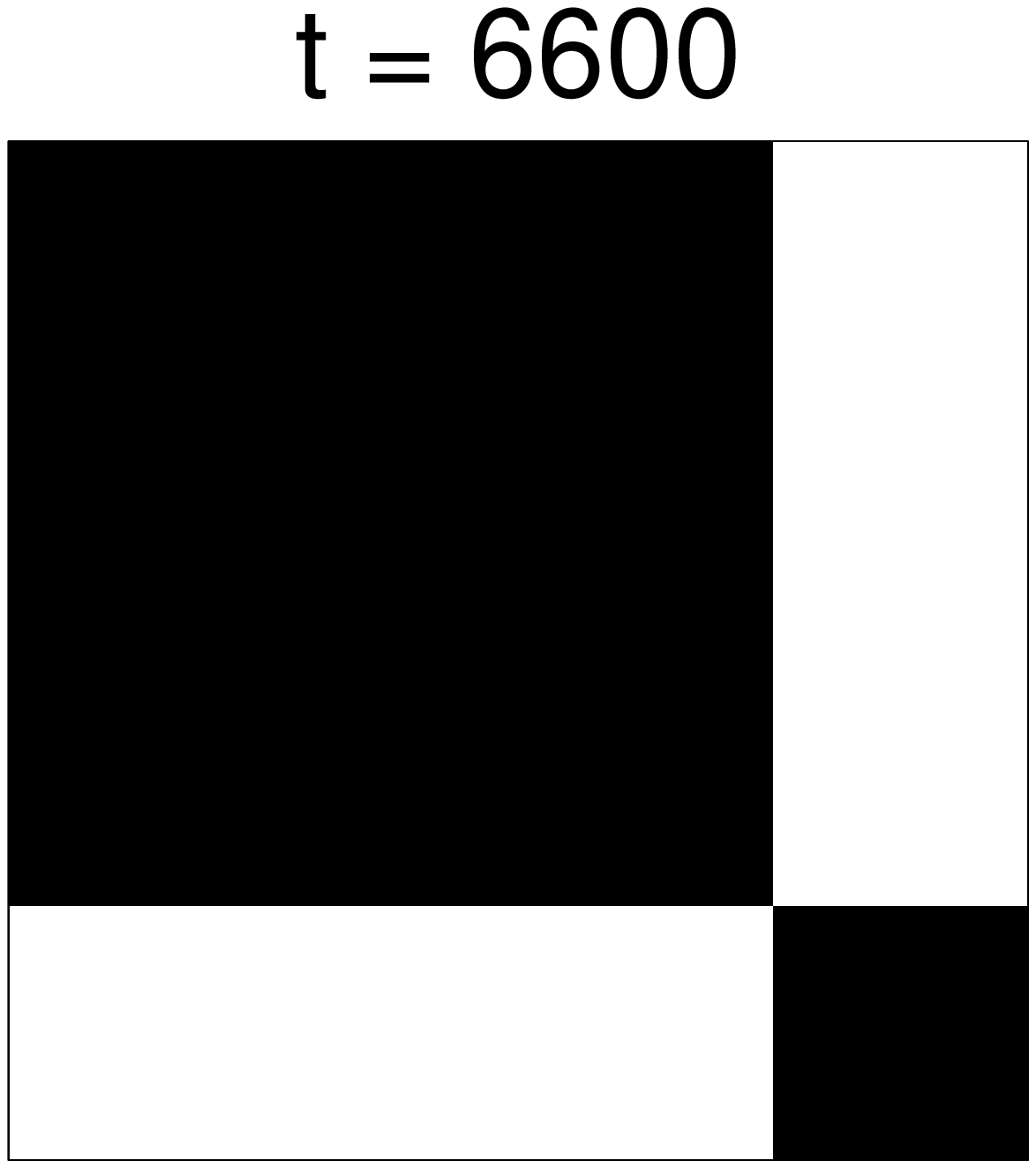}}%
		\hspace{.125cm}%
\subfigure{	
		\includegraphics[width=0.5in]{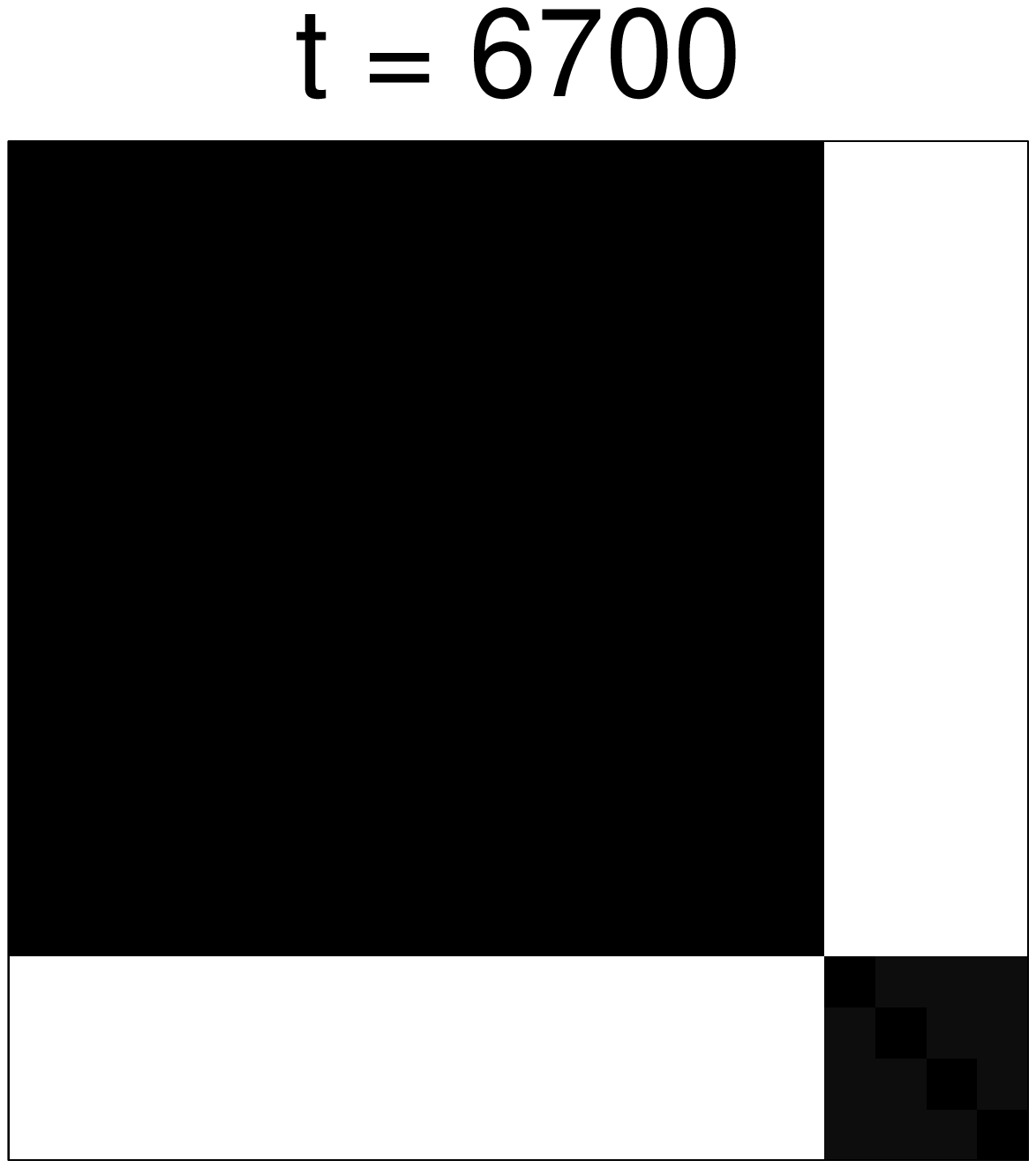}}%
		\hspace{.125cm}%
\subfigure{	
		\includegraphics[width=0.5in]{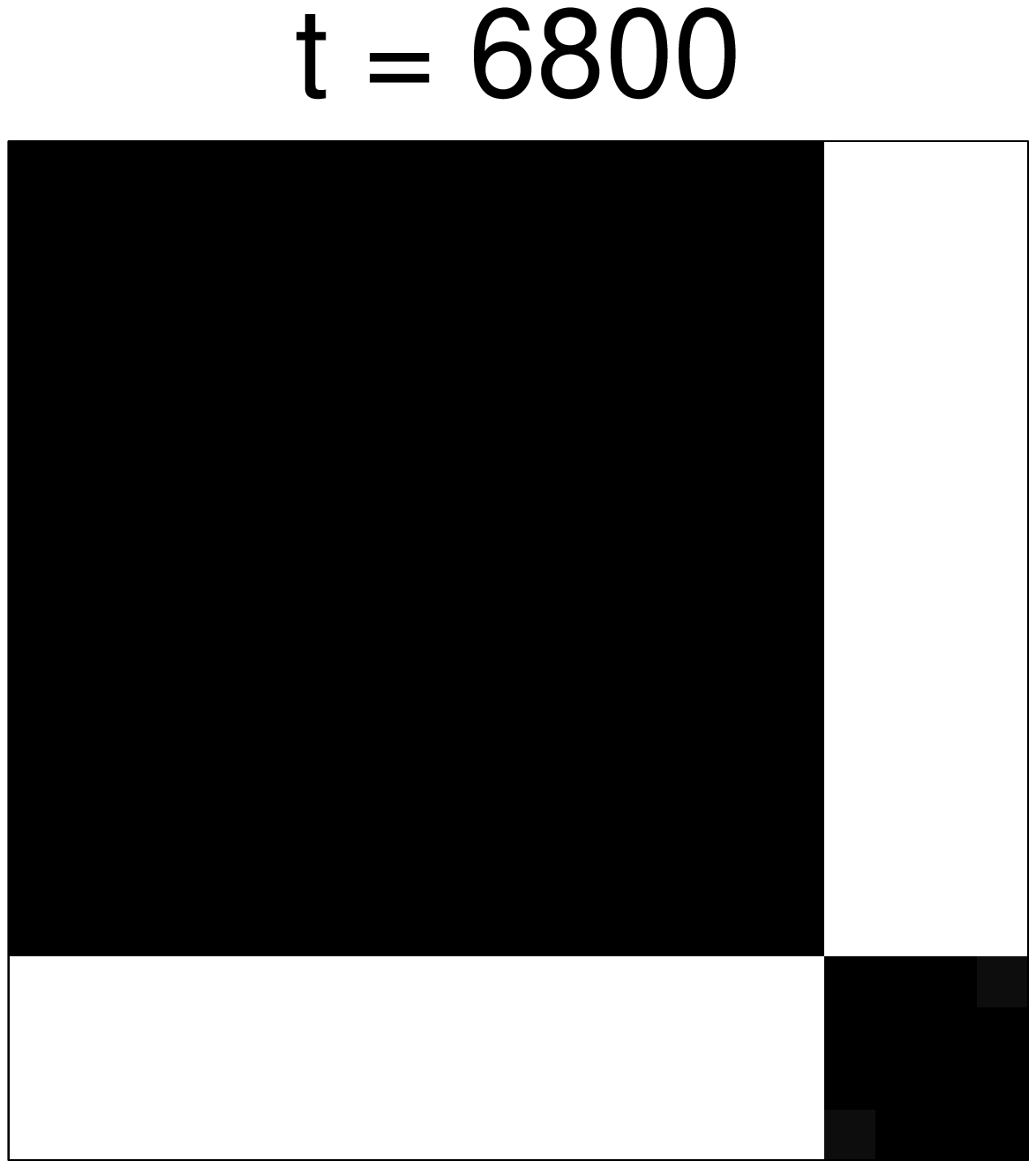}}%
		\hspace{.125cm}%
\subfigure{	
		\includegraphics[width=0.5in]{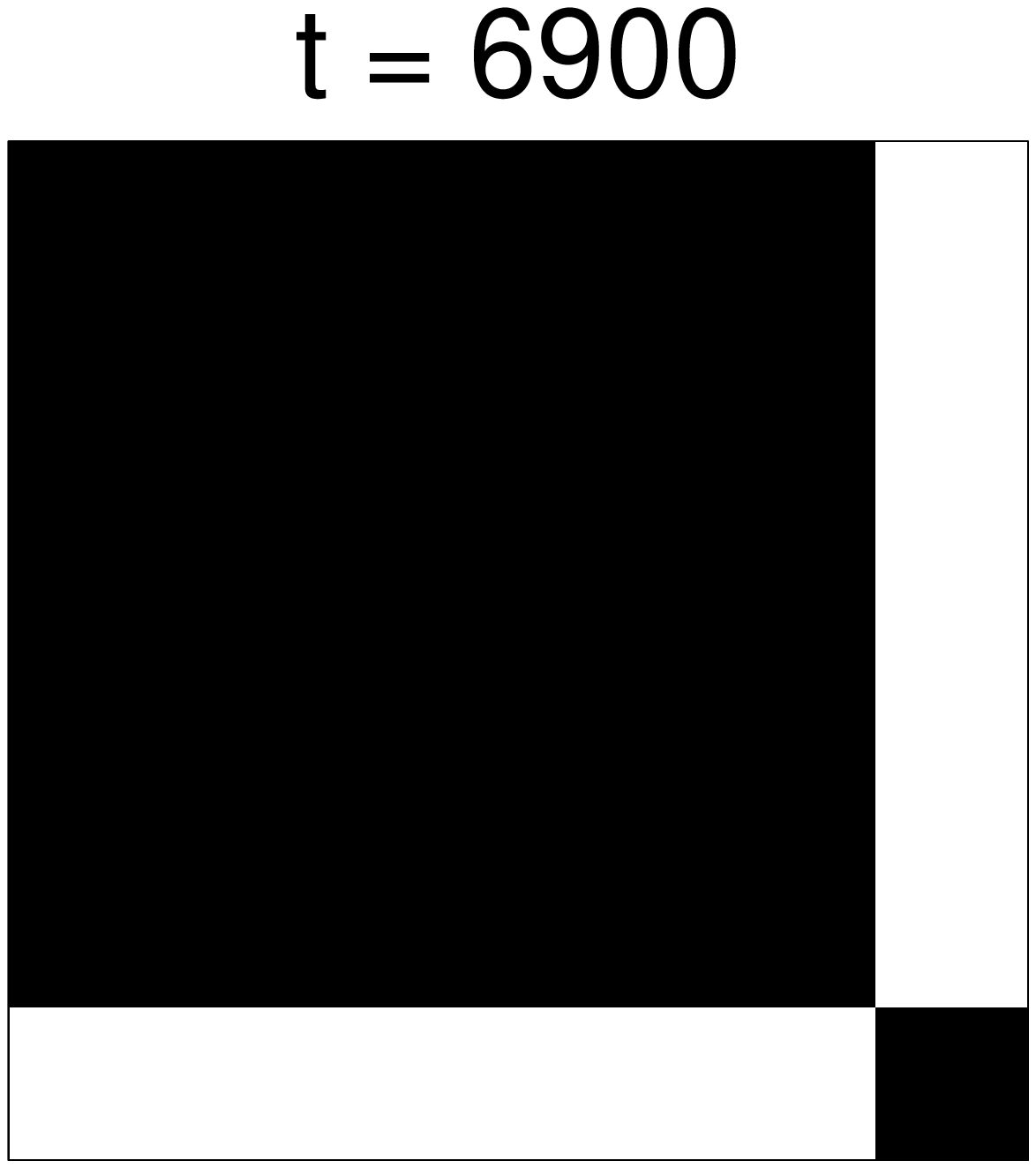}}%
		\hspace{.125cm}%
\subfigure{	
		\includegraphics[width=0.5in]{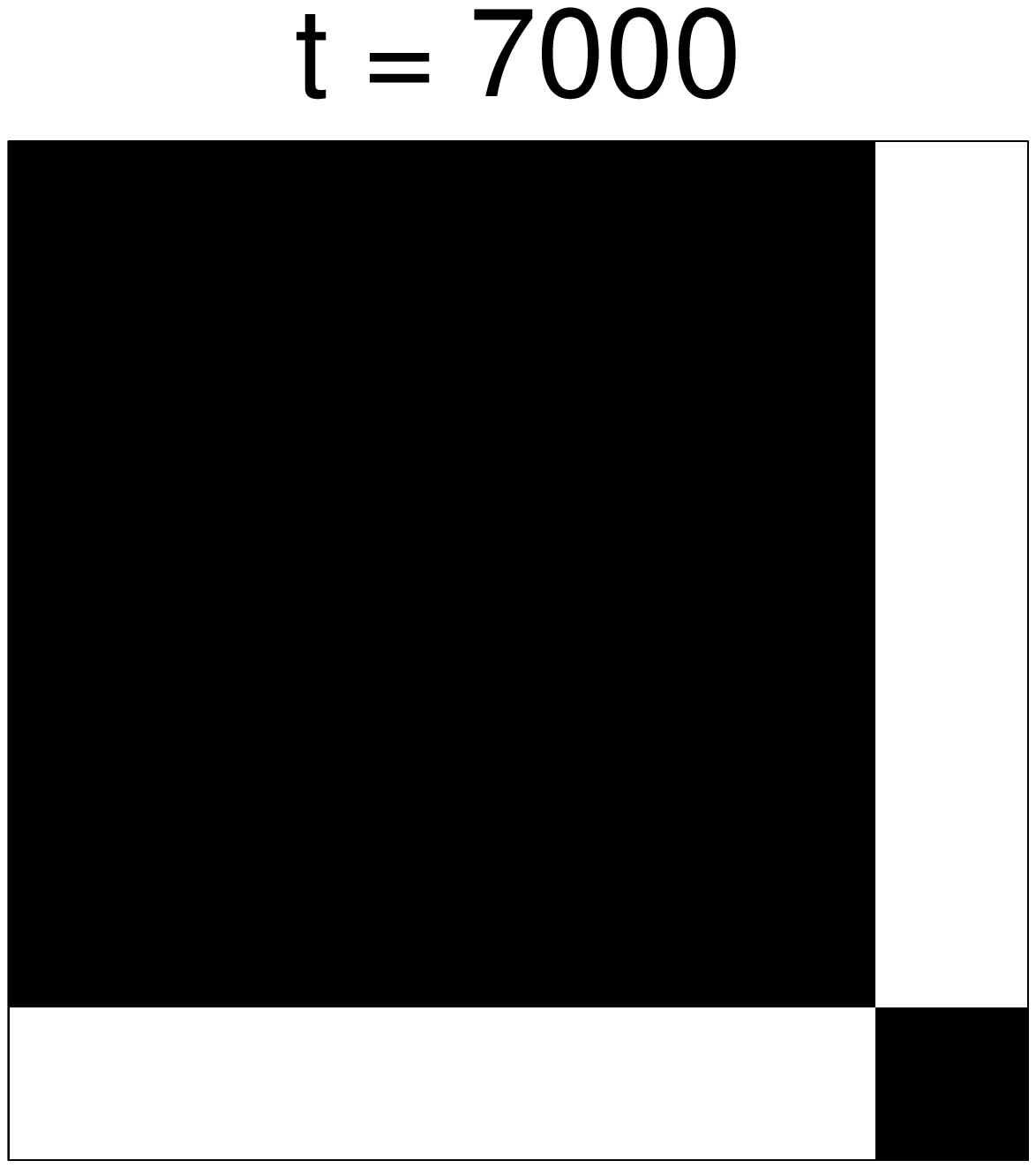}}\\

\caption{{\small {\bf A coalitionary cycle; complete cultural inheritance ($\gk=1$)}. 
Other parameters are as in Figure~\ref{orphands}.
(a)~Average (blue) and standard deviation (red) of affinities. 
(b)~Number of alliances (blue) and clustering coefficient $C^{1}$ for the largest alliance (red).
(c)~Proportions of individuals belonging to an alliance (red) and to the largest alliance (blue).
(d)~Dynamics of the interference matrix between time 4100 and 7200.
}}
\label{cycles}
\end{figure}

{\em Cycling.}\quad
However, 
the egalitarian state is not always stable. Under certain conditions the system 
continuously goes through cycles of increased and decreased cohesion 
(Figure~\ref{cycles}a-c)
in which the egalitarian state is gradually approached  as one alliance 
eventually excludes all others. But once the egalitarian state is established
(in Figure~\ref{cycles}d,
around time 5200), it quickly
disintegrates because of internal conflicts between members of the winning alliance.
Figure~\ref{cycles}d
illustrates one such cycle, showing
that the dominant alliance remains relatively stable
as long as the group excludes at least one member (``outsider'').

{\em Analytical approaches.}\quad
Simple ``mean-field'' approximations help to understand model dynamics. 
These approximations focus on the average $a$ and variance $v$ of affinities
computed over particular coalitions (Supplementary Information). For example,
at an egalitarian state when all individuals have very high affinity to each other,
the dynamics of $a$ and $v$ are predicted to evolve to particular stochastic equilibrium 
values, $a^*$ and $v^*$. 
The egalitarian state is stable if the fluctuations of pairwise affinities
around $a^*$ do not result in negative affinities.
We conjecture that the egalitarian state is stable if $a^*>3 \sqrt{v^*}$, which is roughly
equivalent to $(a^*)^2>10 v^*$, which can be rewritten as
	\[
	   \frac{2 \ga N \go^2}{\mu}>10 \left( \frac{\var \gd}{\ov{\gd}^2}+1-\go^2 \right).
	\]
Here  the mean $\ov{\gd}$ and variance $\var \gd$ are computed over the four $\gd$-coefficients.
Both the approximations and numerical simulations suggest that the egalitarian state cannot be stable
with negative $\ov{\gd}$. Increasing the population size $N$, awareness $\go$,
average $\ov{\gd}$, and decreasing the affinity decay rate $\mu$ and variance $\var \gd$
all promote stability of the egalitarian state.
The agreement of numerical simulations with analytical approximations is very good given 
the stochastic nature of the process.  
Similar approximations can be developed for other regimes. 
In particular, one can show  (Supplementary Information) that the stabilizing effect of ``outsiders'' on 
the persistence of alliances is especially strong in small groups. This happens because
successful conflicts against outsiders simultaneously increase the average $a$
and decrease the variance $v$ of the within-alliance affinities,
with both effects being proportional to $1/N$.


\section*{Discussion}

The overall goal of this paper was to develop a flexible theoretical framework for 
describing the emergence of alliances of individuals able to overcome the power of alpha-types 
in a population and to study the dynamics and consequences of these processes.
We considered a group of individuals competing for rank and/or some limiting resource (e.g.,
mates). We assumed that individuals varied strongly in their fighting abilities. If all conflicts 
were exclusively dyadic, a hierarchy would emerge with a few strongest individuals getting most 
of the resource 
\cite{lan51a,lan51b,bon96,bon99}. 
However there is also a tendency (very small initially) for individuals to interfere 
in an ongoing dyadic conflict thus biasing its outcome one way or another. Positive outcomes of such
interferences increase the affinities between individuals while negative outcomes decrease them. Using a 
minimum set of assumptions about cognitive abilities of individuals, we looked for conditions under 
which long-lasting coalitions (i.e. alliances) emerge in the group. We showed that such an outcome
is promoted by increasing the frequency of interactions (which can be achieved in a number of ways)
and decreasing the affinity decay rate. Most interestingly, the model shows that the shift from a 
state with no alliances to one or more alliances typically occurs in a phase-transition like fashion. 
Even more surprisingly, under certain conditions (that include some cultural inheritance of social networks) 
a single alliance comprising all members of the group can emerge in which the resource is divided
evenly. That is, the competition among nonequal individuals can paradoxically result in their eventual equality.
We emphasize that in our model, egalitarianism emerges from political dynamics of intense competition between individuals for higher social and reproductive success rather than by environmental constraints, social structure, or cultural processes. In other words, within-group conflicts promote the buildup of a group-level alliance. In a sense, once alliances start to form, there is no other reasonable strategy but to join one, and once social networks become highly heritable, a single alliance including all group members is destined to emerge.

Few clarifications are in order. First, in our model coalitionary interactions are mutualistic
in nature rather than altruistic. We note that there are not many examples of truly altruistic behavior 
outside of humans \cite{ste05} with some of those that were initially suggested to be altruistic under 
closer examination turning out to be kin-directed or mutualistic \cite{ste05,kap06}.
Even in humans certain behaviors that are viewed as altruistic may have a rather different origin.
For example, food sharing may have originated as a way to avoid harassment, e.g. in the form of begging 
\cite{ste05}. In any case, modern human behavior is strongly shaped by evolved culture \cite{ric05} 
and might not be a good indicator of factors acting during its origin. 
Second, in our model we avoided the crucial step of the dominant game-theoretic paradigm which is an
explicit evaluation of costs and benefits of certain actions in controlling one's decisions.
In our model, coalitions and alliances emerge from simple processes based on individuals using only
limited ``local'' information (i.e., information on own affinities but not on other individuals' affinities)
rather than as a solution to an optimization task. Our approach is justified not only by its mathematical 
simplicity but by biological realism as well. Indeed, solving the cost-benefit optimization tasks (which 
require rather sophisticated algebra in modern game-theoretic models) would be very difficult for apes and 
early humans \cite{ste05} especially given the multiplicity of behavioral choices
and the dynamic nature of coalitions. Therefore treating coalitions
and alliances in early human groups as an emergent property rather than an optimization task solution
appears to be a much more realistic approach. We note that costs and benefits can be incorporated in our 
approach in a straightforward manner.
Third, one should be careful in applying our model to contemporary humans (whether members of modern societies 
or hunter-gathers). In contemporary humans, an individual's decision on joining coalitions will be strongly 
affected by his/her estimates of costs, benefits, and risks associated as well as by cultural 
beliefs and traditions. These are the factors explicitly left outside of our framework.
%
%
%
%
%

Our results have implications for a number of questions related to human social evolution.
The great apes' societies are very hierarchical; their social system is based on sharp status
rivalry and depends on specific dispositions for dominance and submission. A major function of 
coalitions in apes is to maintain or change the dominance structure \cite{har92,deW00};
although leveling coalitions are sometimes observed (e.g., \cite{goo86}), they are
typically of small size and short-lived. 
In sharp contrast, most known hunter-gatherer societies are egalitarian 
\cite{joh87,kna91,boe99}. 
Their weak leaders merely assist a
consensus-seeking process when the group needs to make decisions; at the band level, all main 
political actors behave as equal. It has been argued that in egalitarian societies the
pyramid of power is turned upside down with potential subordinates being able to express 
dominance because they find collective security in a large, group-wide political coalition \cite{boe99}.
One factor that may have promoted transition to an egalitarian society is the development 
of larger brains and better political/social intelligence in response to
intense within-group competition for increased social and reproductive 
success \cite{ale90,fli05,gea05,gav06a}.
Our model supports these arguments. Indeed, increased cognitive abilities would allow humans
to maintain larger group sizes, have higher  awareness of ongoing conflicts, better
abilities in attracting allies and building complex coalitions, and better memories of past events.
The changes in each of these characteristics may have shifted the group across the phase boundary
to the regime where the emergence of an  egalitarian state becomes unavoidable.  Similar effect
would follow a change in mating system that would increase father-son social bonds,
or an increase in fidelity of cultural inheritance of social networks. The fact 
that mother-daughter social bonds are often very strong suggests (everything 
else being the same) females could more easily achieve egalitarian societies.
The establishment of a stable group-wide egalitarian alliance should create conditions promoting
the origin of conscience, moralistic aggression, altruism, and other cultural norms favoring the 
group interests over those of individuals \cite{boe07}. 
Increasing within-group cohesion will also promote the group efficiency in between-group 
conflicts \cite{wra99} and intensify cultural group selection.

In humans, a secondary transition from egalitarian societies to hierarchical states took place 
as the first civilizations were emerging. 
How can it be understood in terms of the model presented here? 
One can speculate that technological and cultural advances 
made the coalition size much less important in controlling the outcome of a conflict than 
the individuals' ability to directly control and use resources (e.g., weapons, information, food)
that strongly influence conflict outcomes.
In terms of our model, this would dramatically increase the variation in individual fighting
abilities and simultaneously render the Lanchester-Osipov square law inapplicable,
making egalitarianism unstable.

Besides developing a novel and general approach for modeling coalitionary interactions and providing theoretical support to some controversial verbal arguments concerning social transitions during the origin of humans, the research presented here allows one to make a number of testable predictions. In particular, our model has identified a number of factors (such as group size, the extent to which group members are aware of within-group conflicts, cognitive abilities, aggressiveness, persuasiveness, existence of outsiders, and the strength of parent-offspring social bonds) which are predicted to increase the likelihood and size of alliances and affect in specific ways individual social success and the degree of within-group inequality. Existing data on coalitions in mammals (in particular, in dolphins and primates) and in human hunter-gatherer societies should be useful in testing these predictions and in refining our model.

{\bf Acknowledgments.}    
We thank S. Sadedine, J. Leonard, D. C. Geary, S. A. West, L. A. Bach, and E. Svensson for the comments on the manuscript and M. V. Flinn, F. B. M. de Waal, and J. Plotnick for discussions. Supported by a grant from the NIH. The funders had no role in study design, data collection and analysis, decision to publish, or preparation of the manuscript.


\bibliography{/home/sergey/PAPS/coal}

\newpage

\section*{Supporting Information}

\setcounter{figure}{0}
\renewcommand{\thefigure}{S\arabic{figure}}
\renewcommand{\theequation}{S\arabic{equation}}

Here, we present
\bi
\item some additional details on the computational methods used;
\item a set of figures (Figures~\ref{inClus}-\ref{numClusBigInterf})
illustrating the effects of individual parameters on the coalitionary structure of the model achieved within a single generation;
\item a set of figures (Figures~\ref{hypercube1} and \ref{hypercube2})
illustrating the effects of changes in multiple parameters simultaneously on the 
coalitionary structure of the model achieved within a single generation;
\item an outline of a mathematical method used to study the model analytically.
\ei


\subsection*{Some details of computational methods}

{\bf Probabilities of help}\quad
For an individual $k$ aware of a conflict between individuals $i$ and $j$, the probabilities of helping to 
$i$, to  $j$, and of no interference are set to $h_{ki}-h_{ki}h_{kj}/2, h_{kj}-h_{ki}h_{kj}/2$ 
and $(1-h_{ki})(1-h_{kj})$,
respectively. 
In numerical simulations, we set
\[
		h_{ki}  =\left[ 1+ \frac{1-\gb}{\gb} \exp(-\eta x_{ki} ) \right]^{-1},
\]
where $\gb$ and $\eta$ are scaling parameters. Note that $h_{ki}\rightarrow 1$ for $x_{ki} \rightarrow \infty$,
$h_{ki}\rightarrow 0$ for $x_{ki} \rightarrow -\infty$, and $h_{ki}=\gb$ for $x_{ki}=0$.
	
{\bf Numerical implementation}\quad
The model dynamics were simulated using Gillespie's direct method (Gillespie 1977).
That is, the next event to happen is chosen according to the corresponding rates. The time 
interval until the next event is drawn from an exponential distribution with a parameter 
equal to the sum of the rates of all possible events. All rates are recomputed after each event.\\

{\bf Reference}
\bi
\item
Gillespie, D. T. Exact stochastic simulation of coupled chemical reactions. {\em Journal of Physical
Chemistry} {\bf 81}, 2340-2361 (1977)
\ei

\subsection*{Supplementary Figures and Legends}

\begin{figure}
\centering
\subfigure[]{ 	
		\includegraphics[width=2.in]{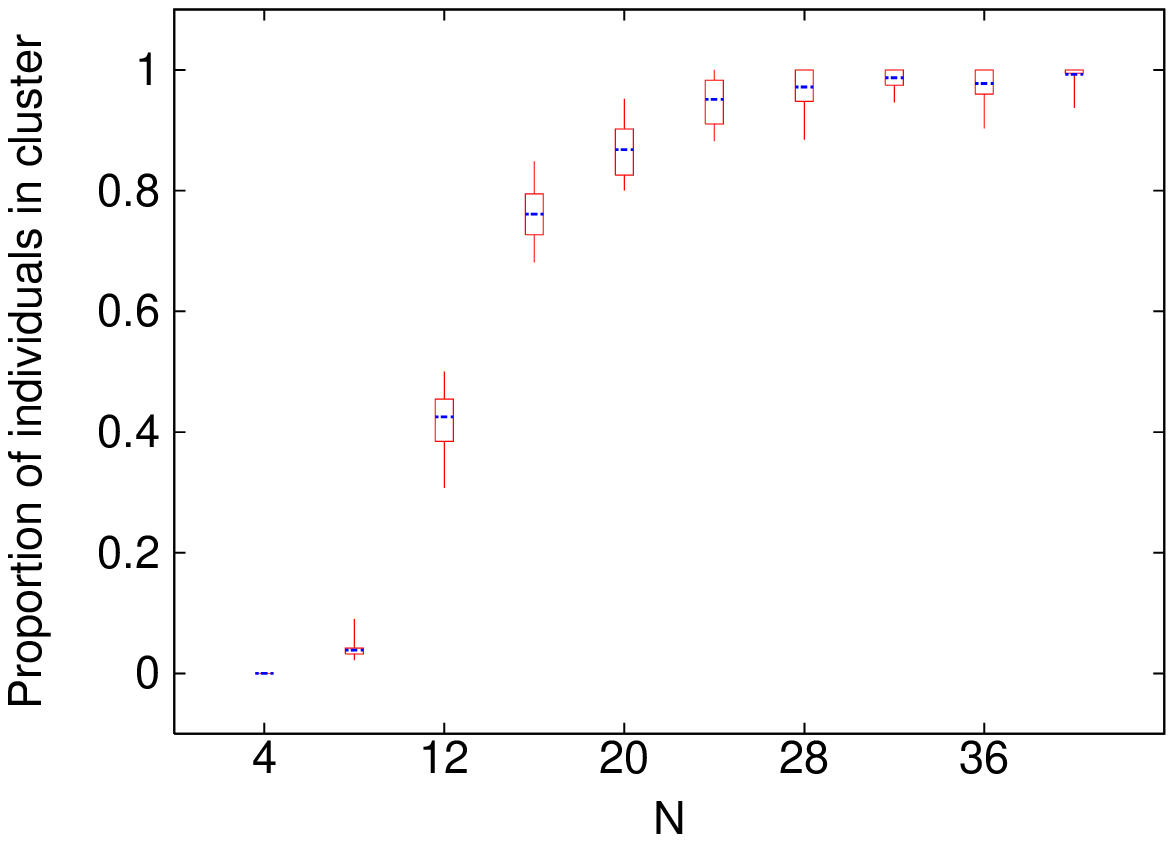}}%
 		\hspace{.5cm}%
\subfigure[]{	
		\includegraphics[width=2.in]{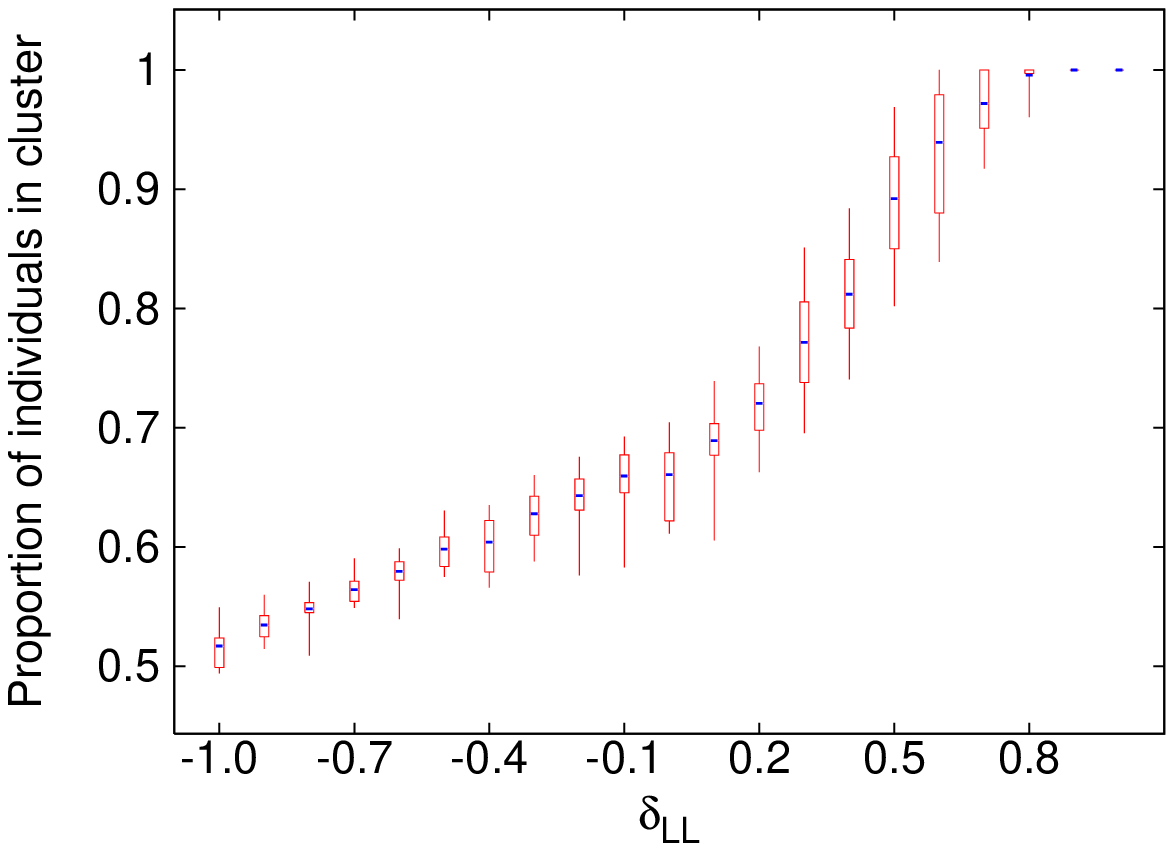}}\\
\subfigure[]{	
		\includegraphics[width=2.in]{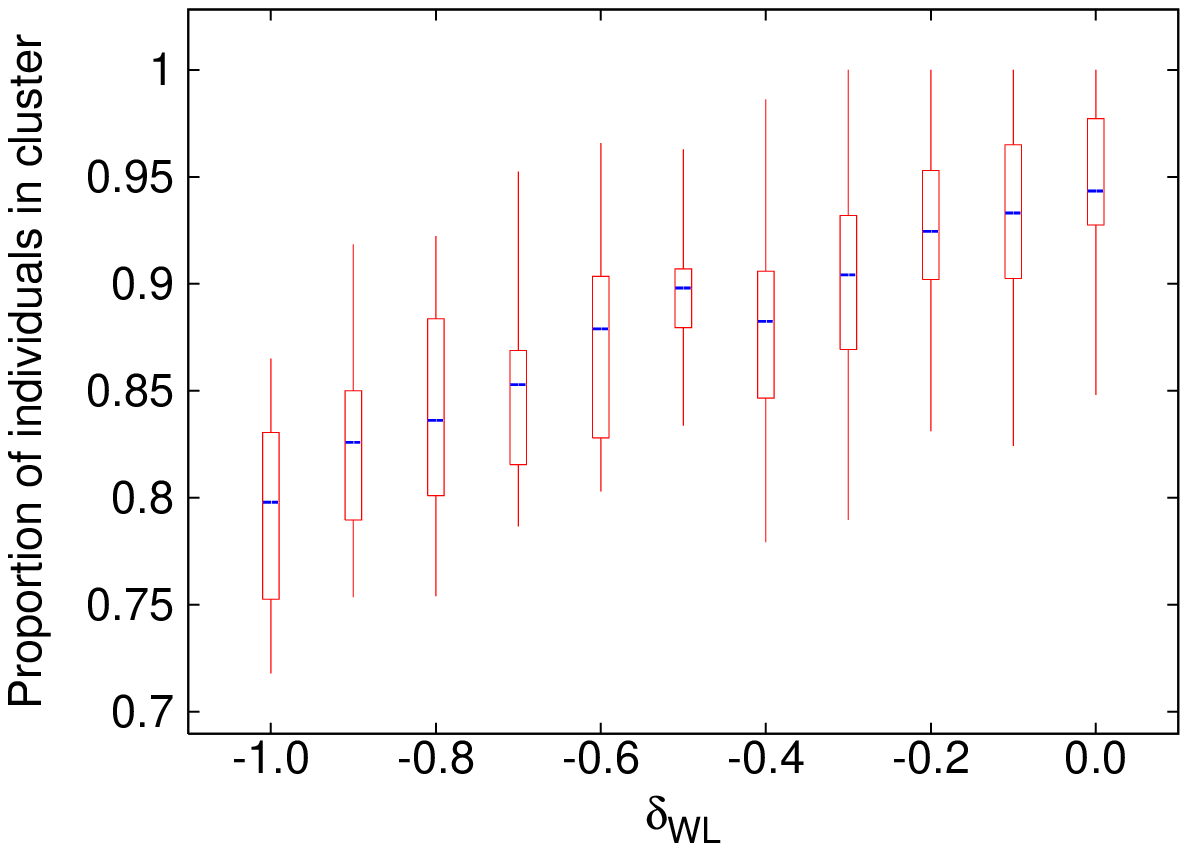}}%
 		\hspace{.5cm}%
\subfigure[]{	
		\includegraphics[width=2.in]{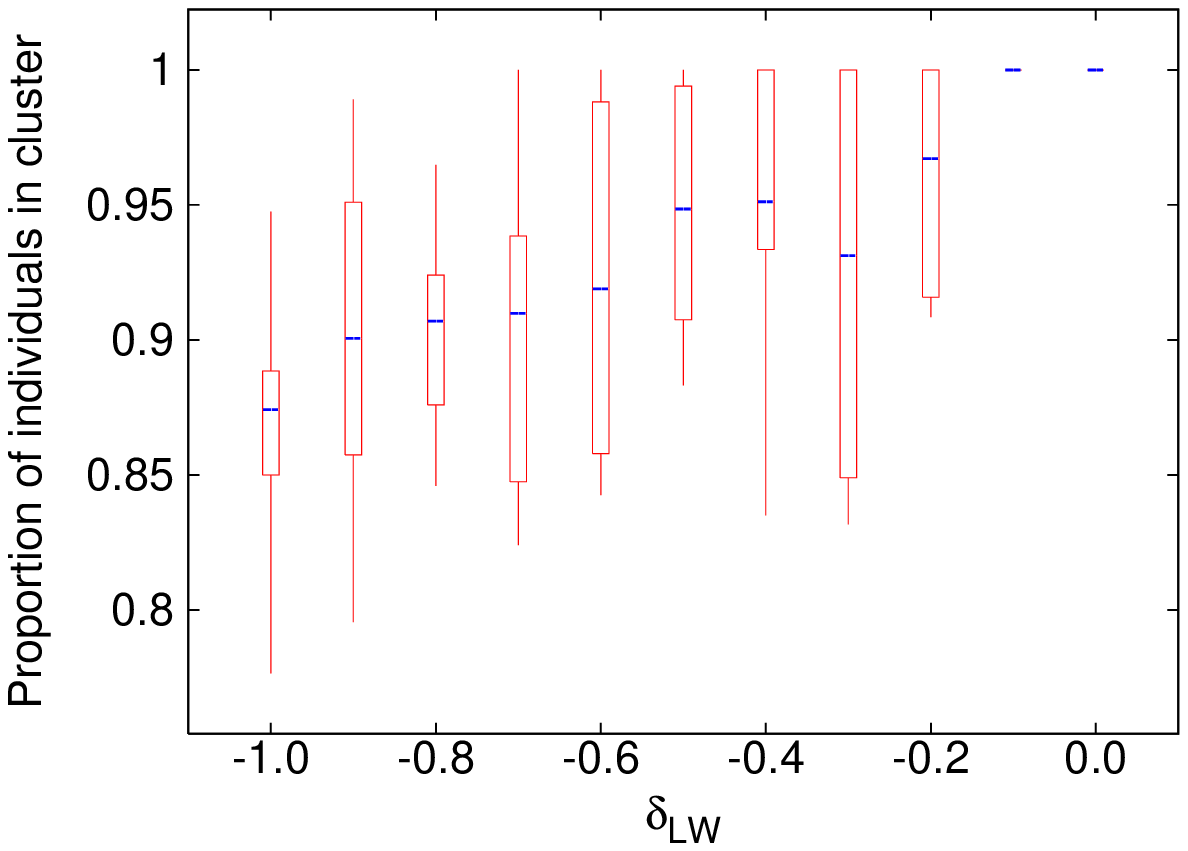}}\\
\subfigure[]{	
		\includegraphics[width=2.in]{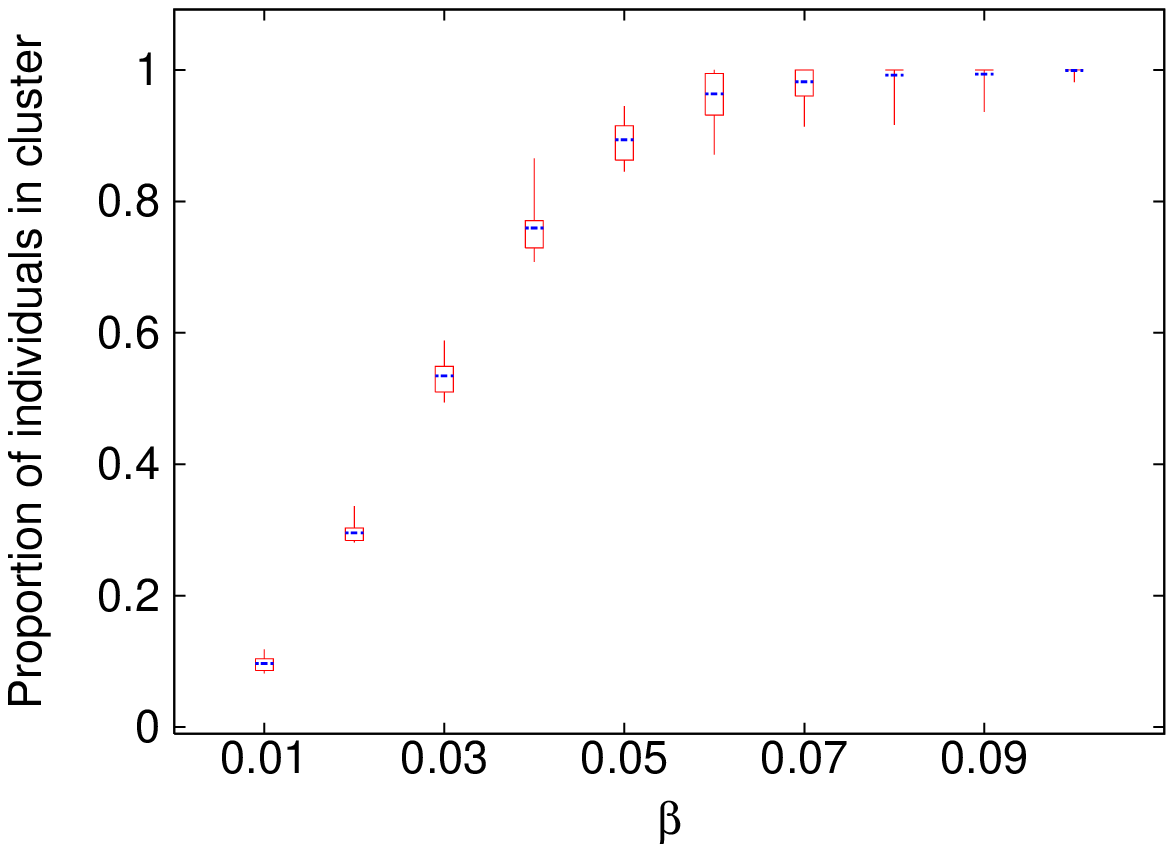}}%
 		\hspace{.5cm}
\subfigure[]{	
		\includegraphics[width=2.in]{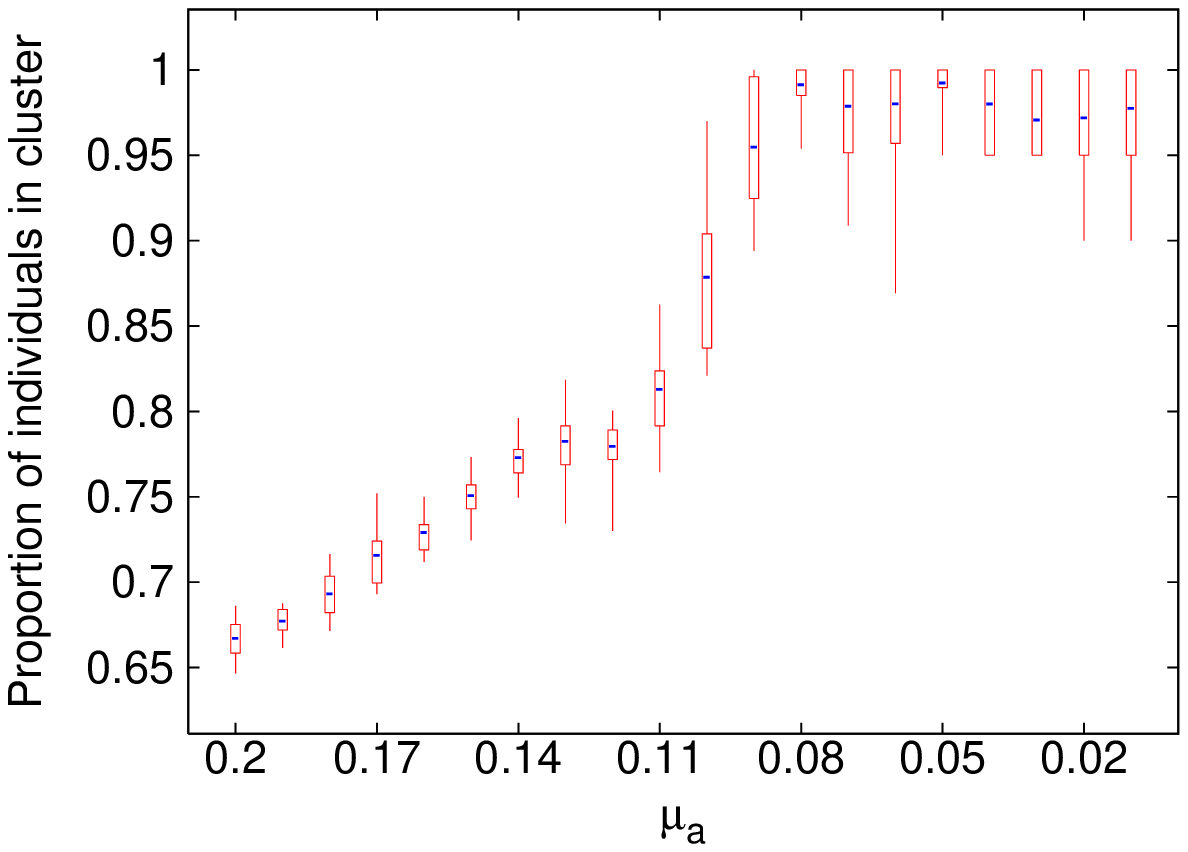}}\\		
\subfigure[]{	
		\includegraphics[width=2in]{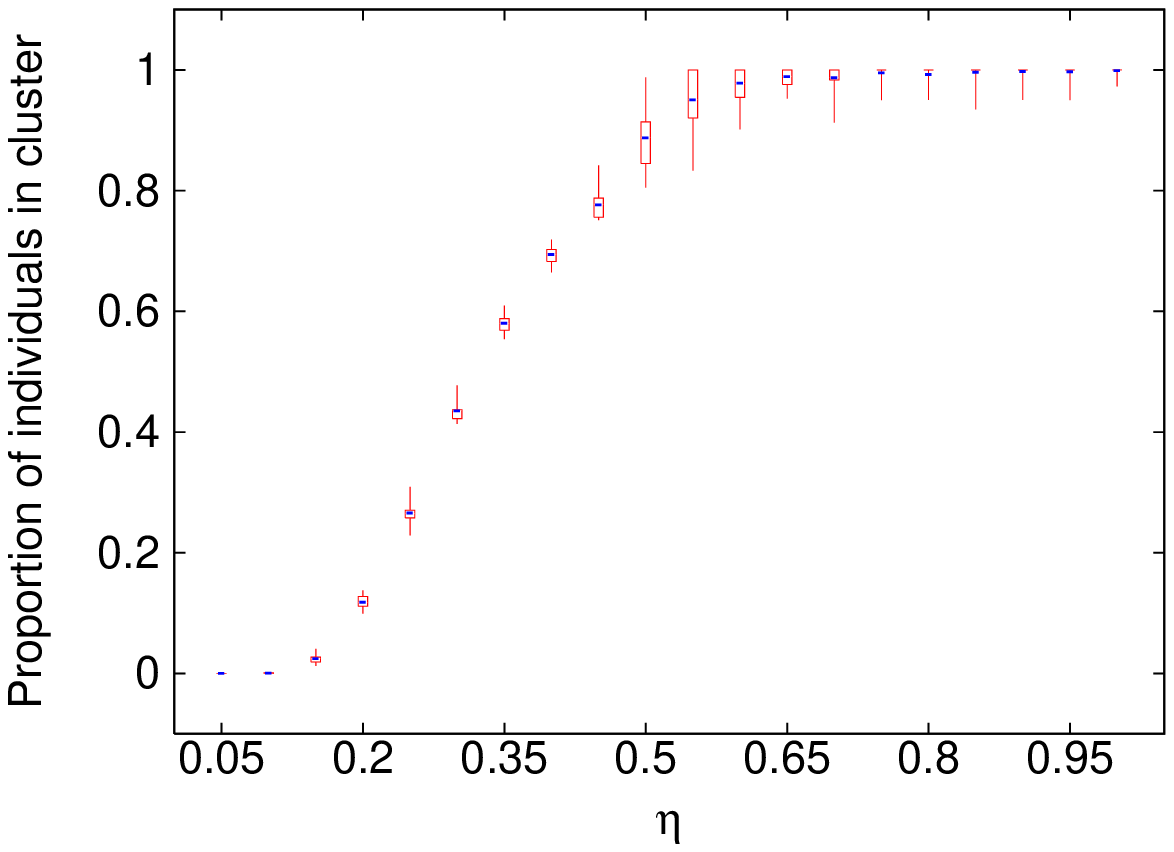}}%
		\hspace{.5cm}%
\subfigure[]{	
		\includegraphics[width=2in]{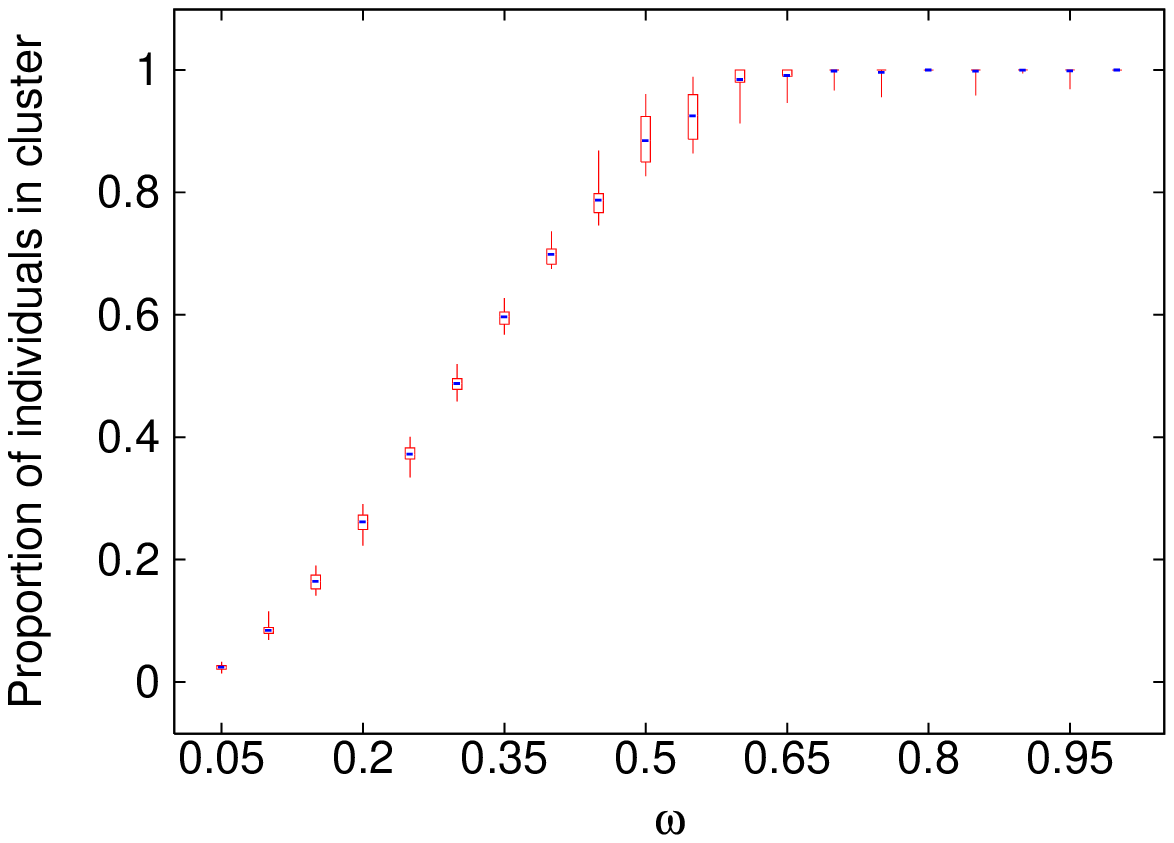}}\\
\caption{Effects of parameters $N,\gd_{ll},\gd_{wl},\gd_{lw},\gb,\mu,\eta,\go$ on
the proportion of individuals belonging to a alliance for a default set of parameter
values.
}
\label{inClus}
\end{figure}

\begin{figure}
\centering
\subfigure[]{ 	
		\includegraphics[width=2in]{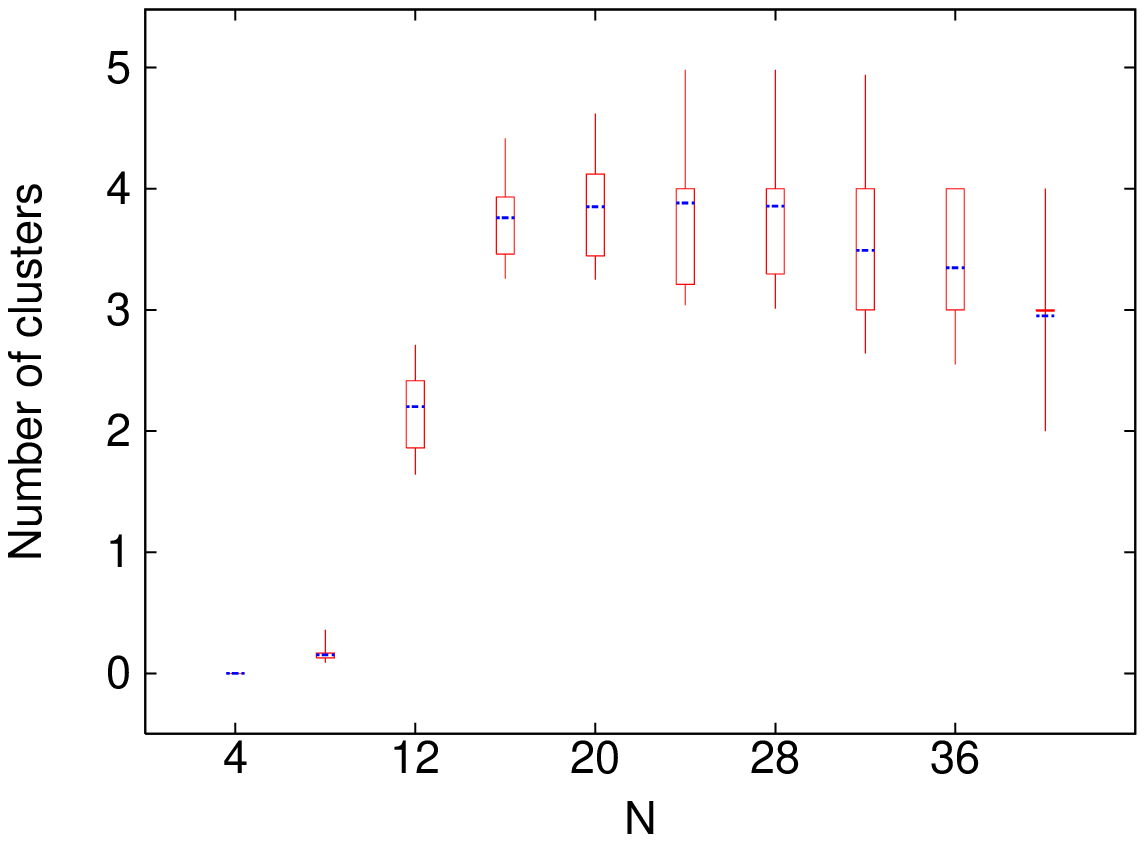}}%
 		\hspace{.5cm}%
\subfigure[]{	
		\includegraphics[width=2in]{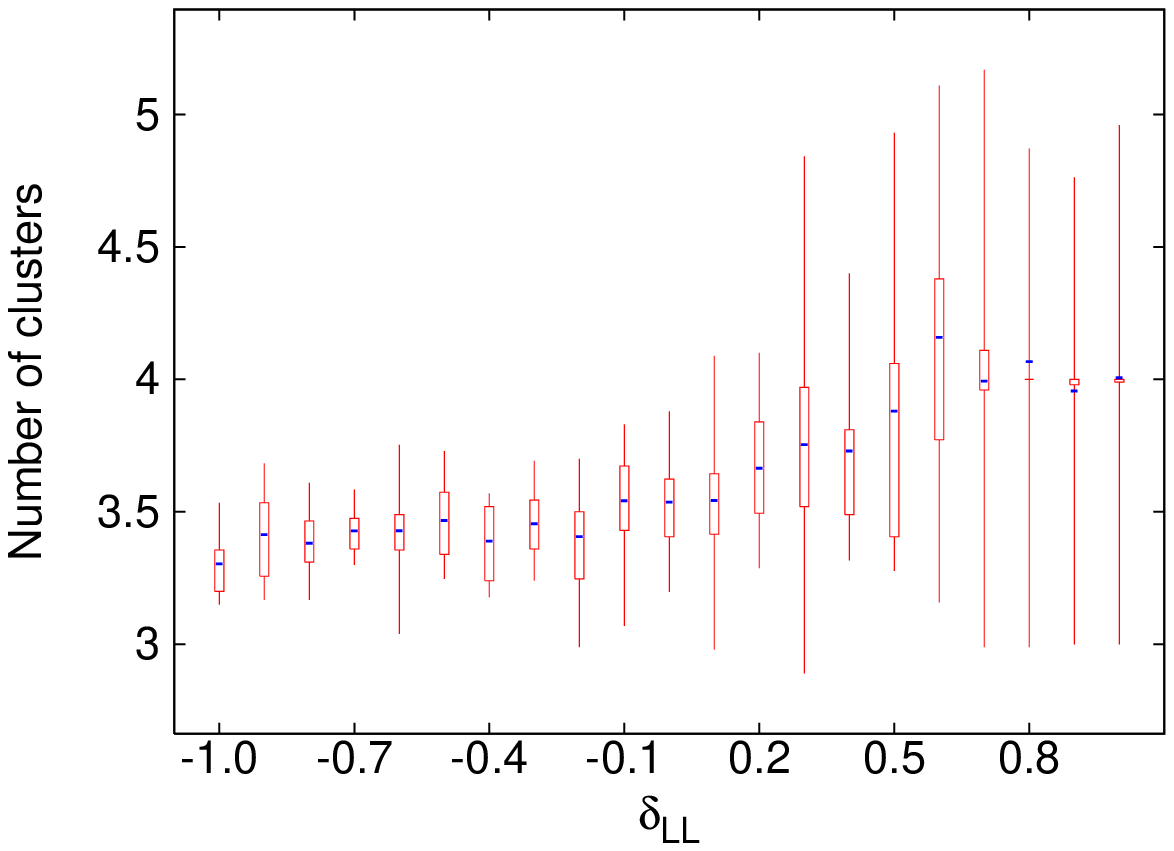}}\\
\subfigure[]{	
		\includegraphics[width=2in]{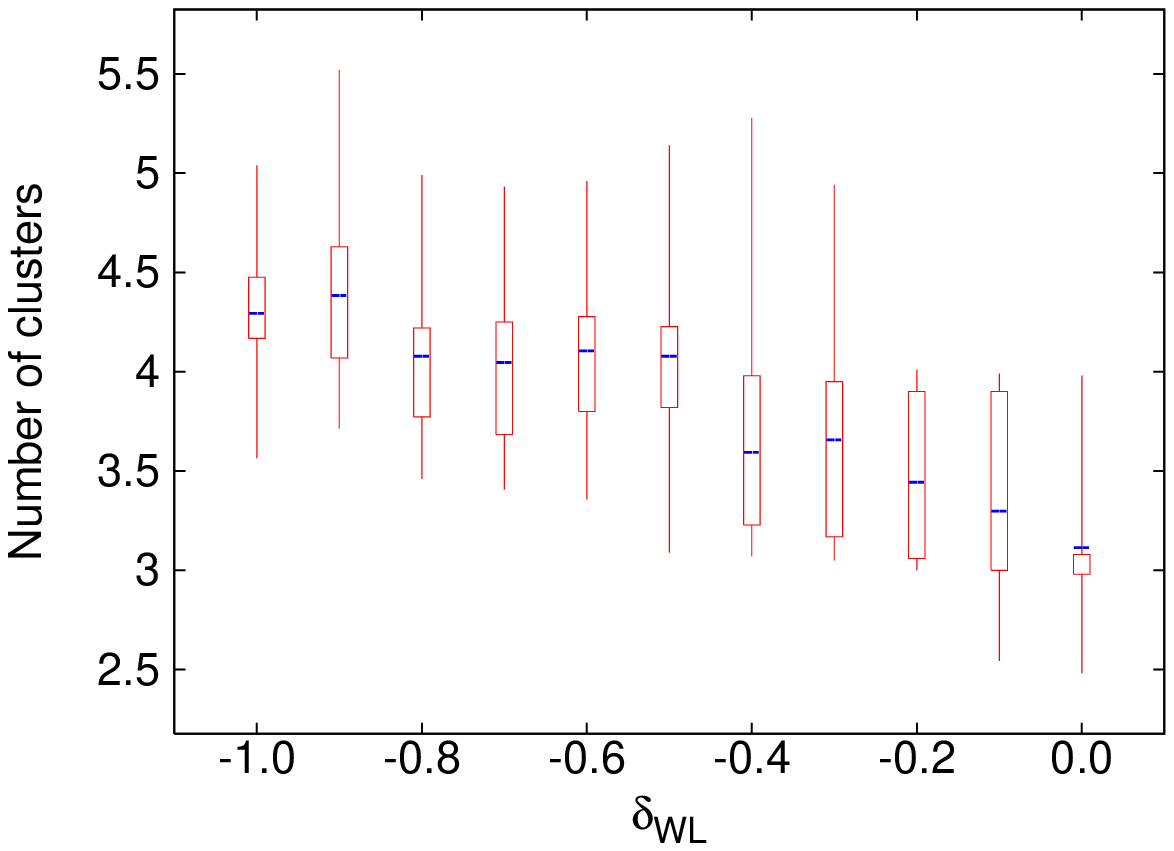}}%
 		\hspace{.5cm}%
\subfigure[]{	
		\includegraphics[width=2in]{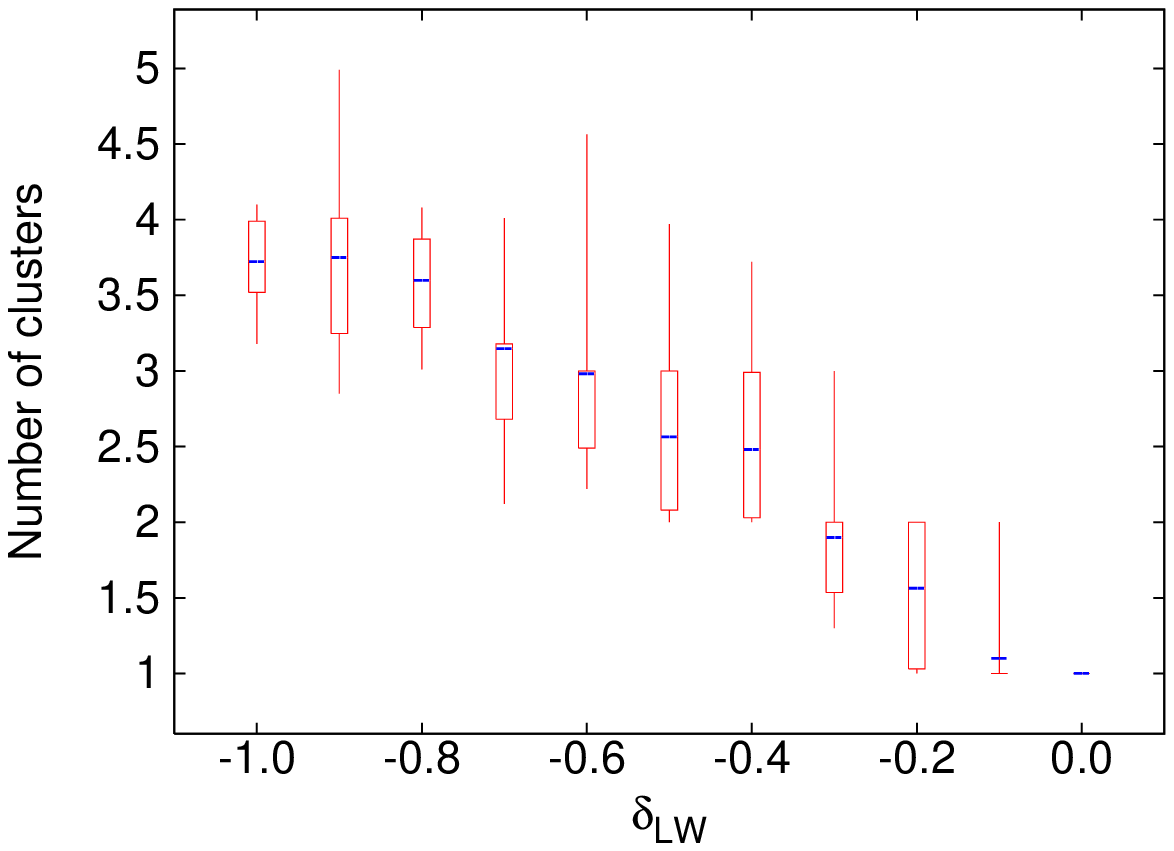}}\\
\subfigure[]{	
		\includegraphics[width=2in]{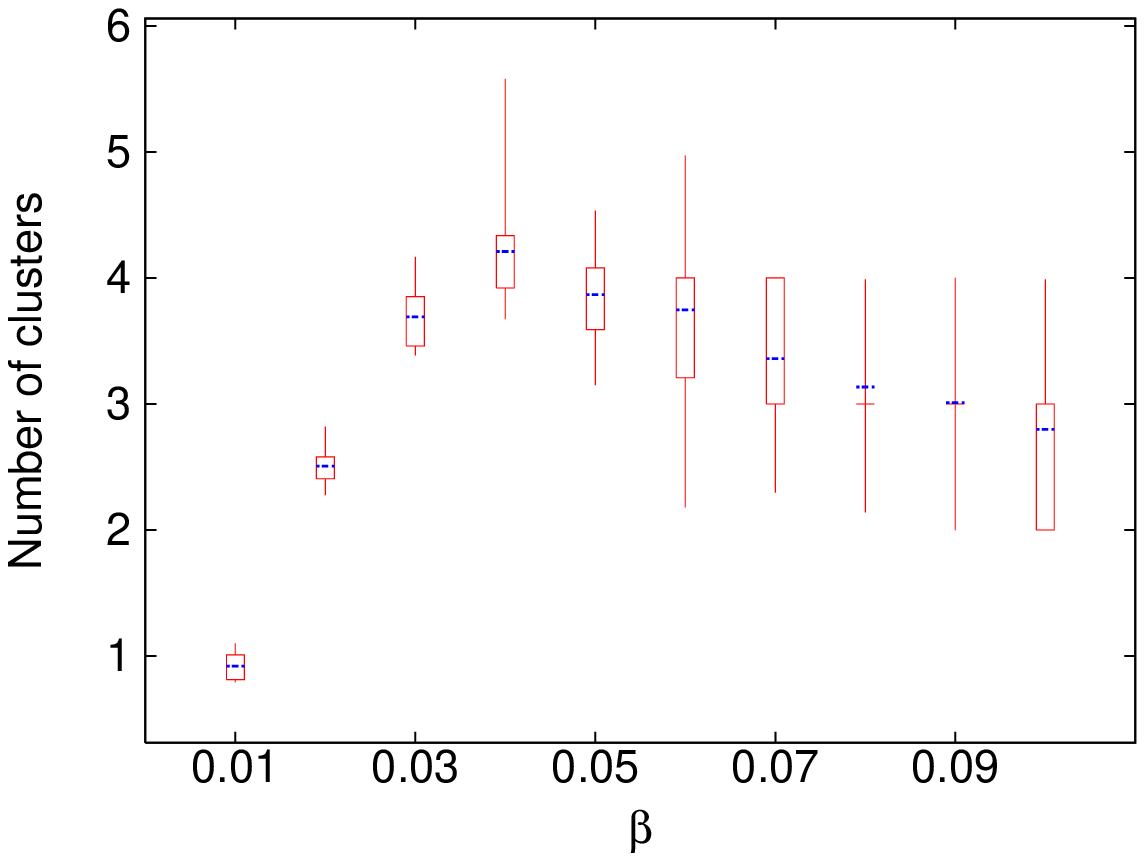}}%
 		\hspace{.5cm}
\subfigure[]{	
		\includegraphics[width=2in]{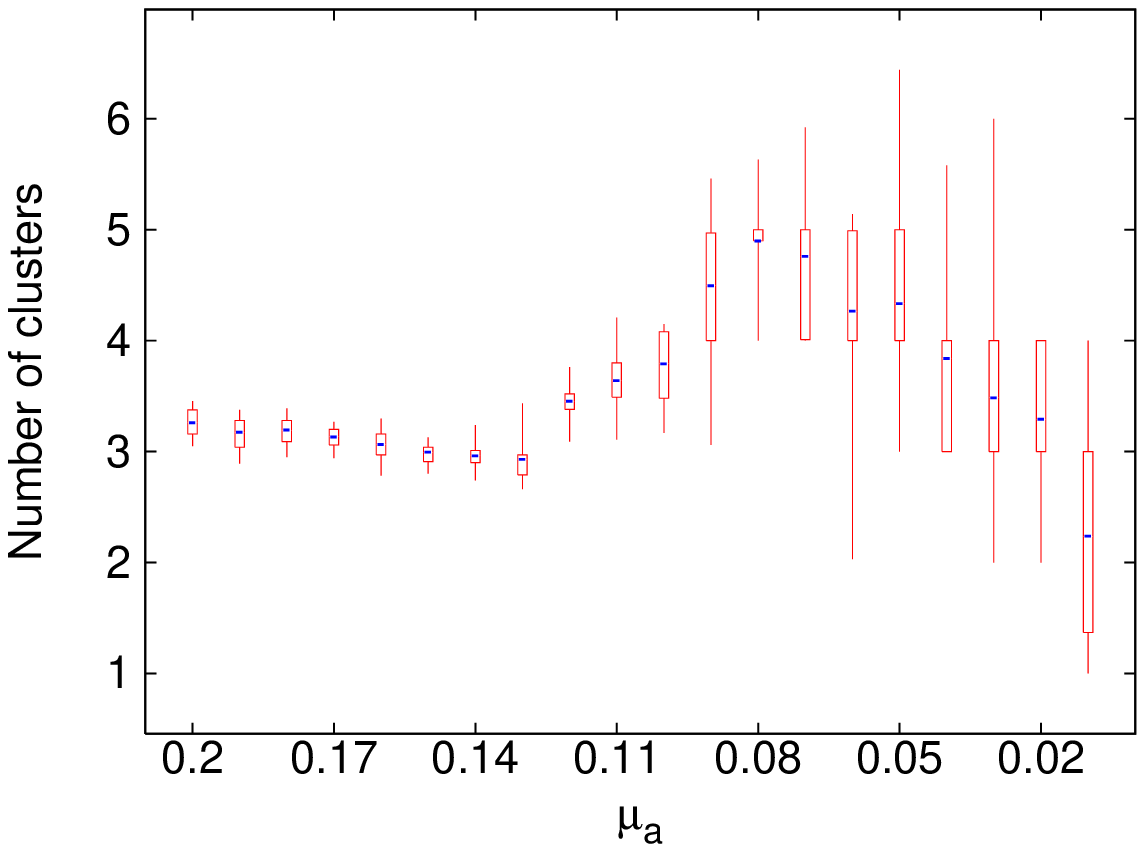}}\\		
\subfigure[]{	
		\includegraphics[width=2in]{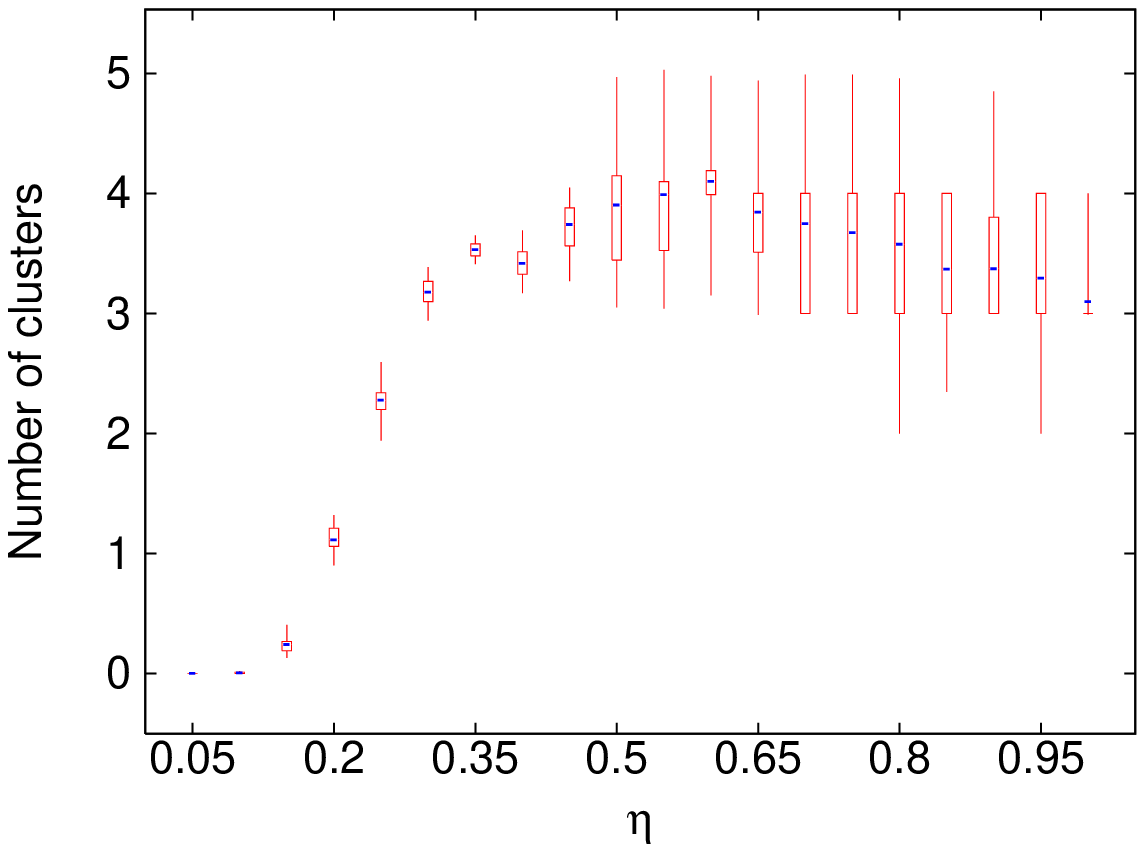}}%
		\hspace{.5cm}%
\subfigure[]{	
		\includegraphics[width=2in]{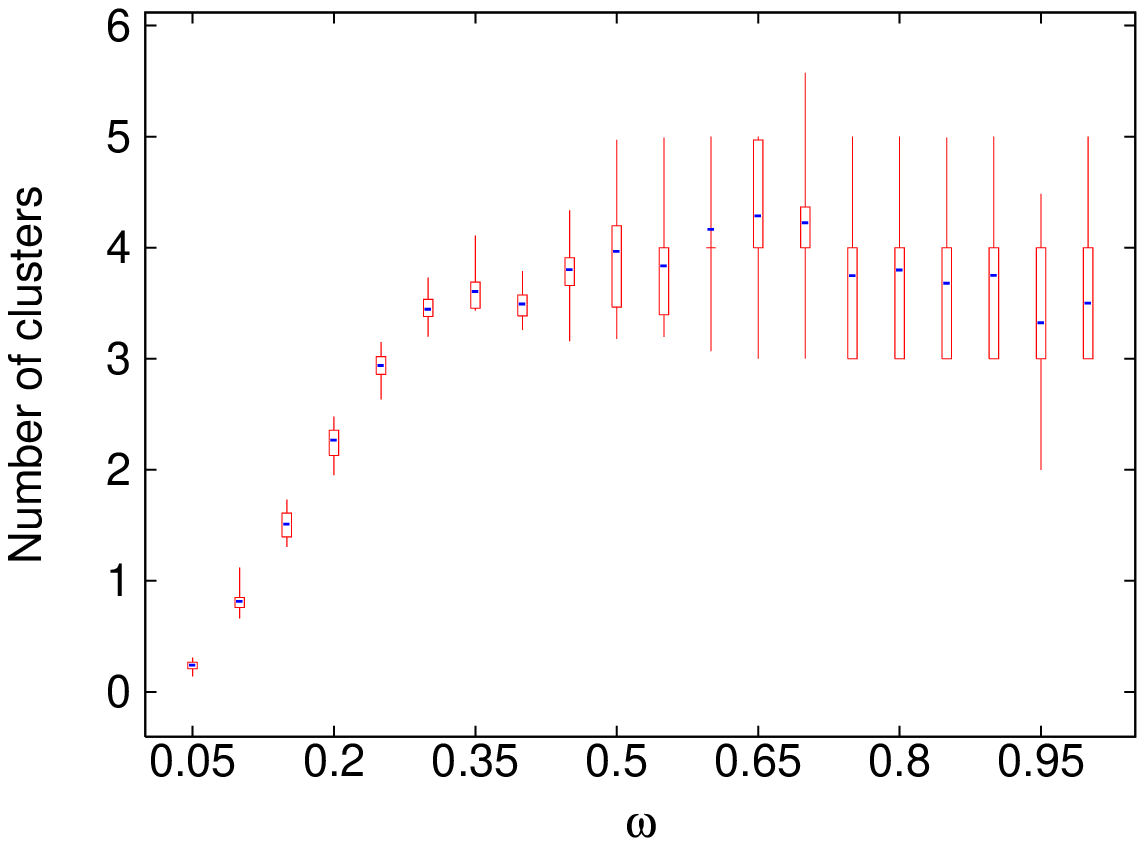}}\\
\caption{Effects of parameters $N,\gd_{ll},\gd_{wl},\gd_{lw},\gb,\mu,\eta,\go$ on
the number of alliances for a default set of parameter
values.
}
\label{numClus}
\end{figure}

\begin{figure}
\centering
\subfigure[]{ 	
		\includegraphics[width=2in]{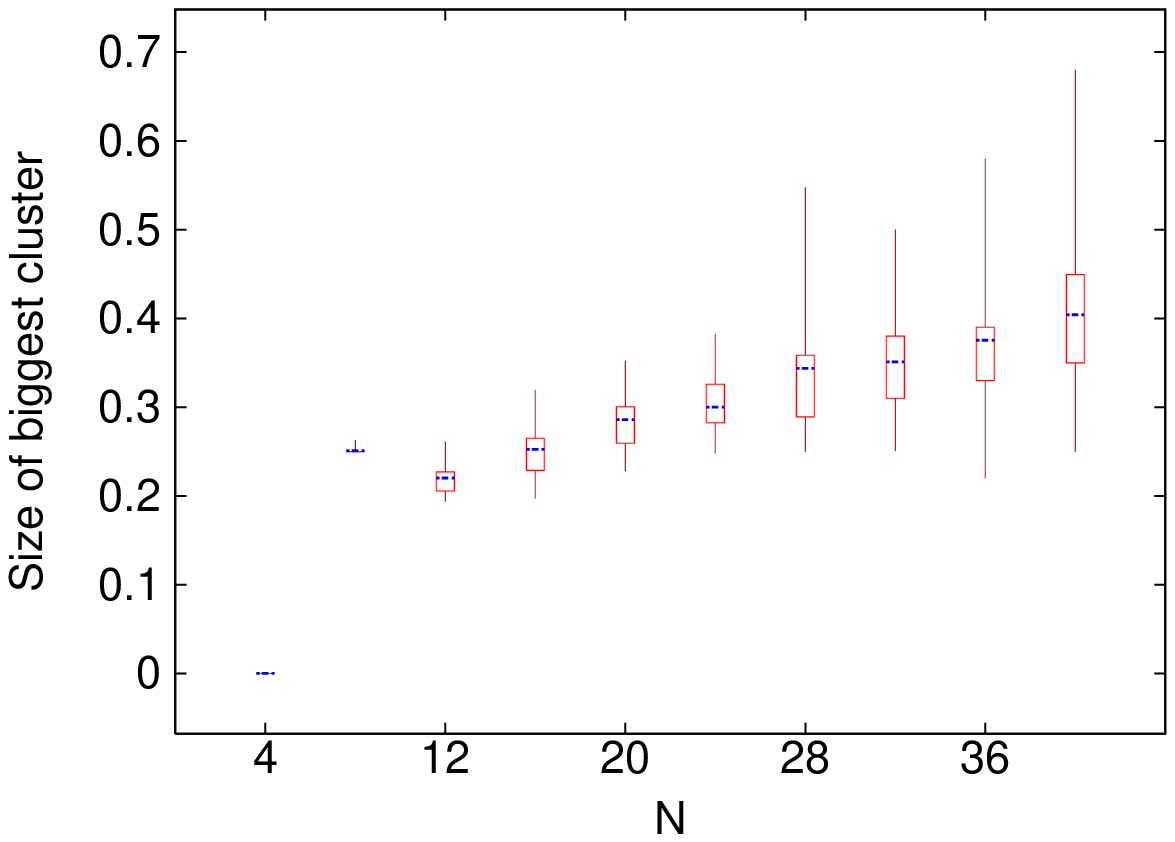}}%
 		\hspace{.5cm}%
\subfigure[]{	
		\includegraphics[width=2in]{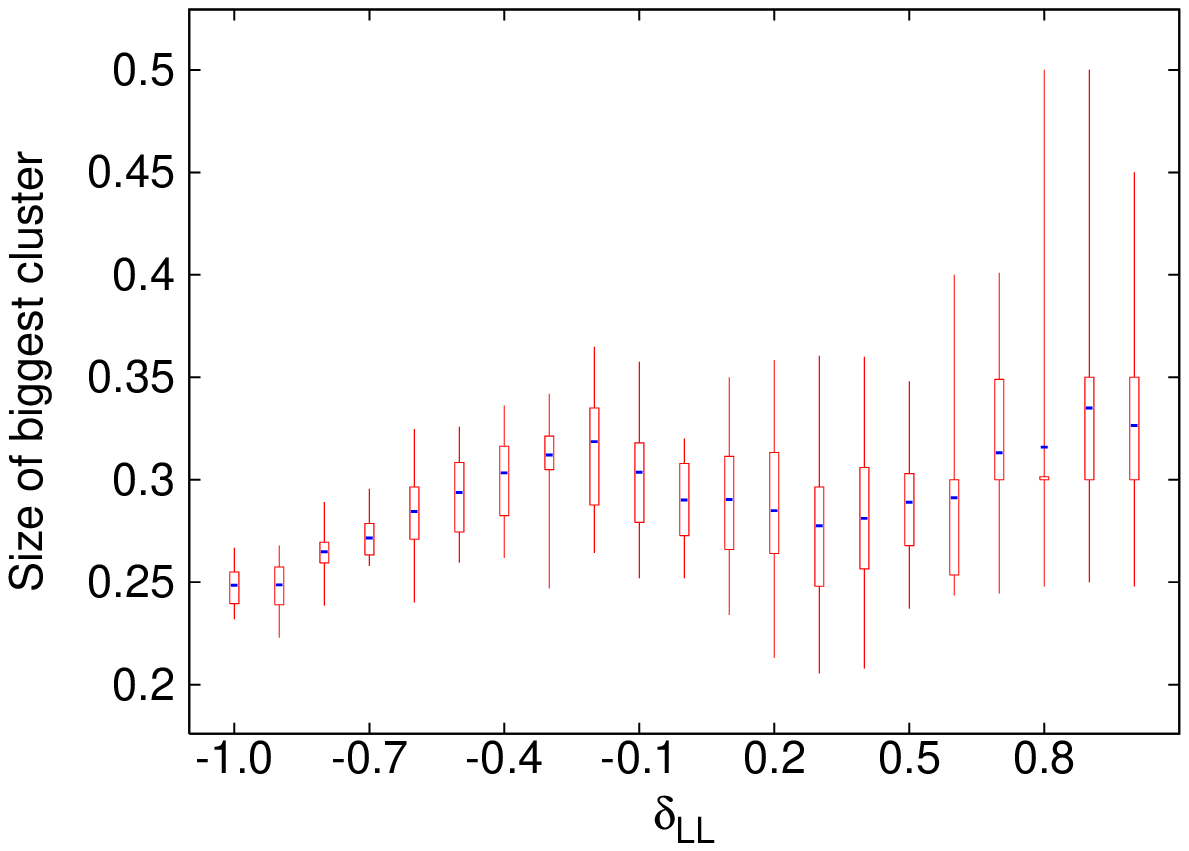}}\\
\subfigure[]{	
		\includegraphics[width=2in]{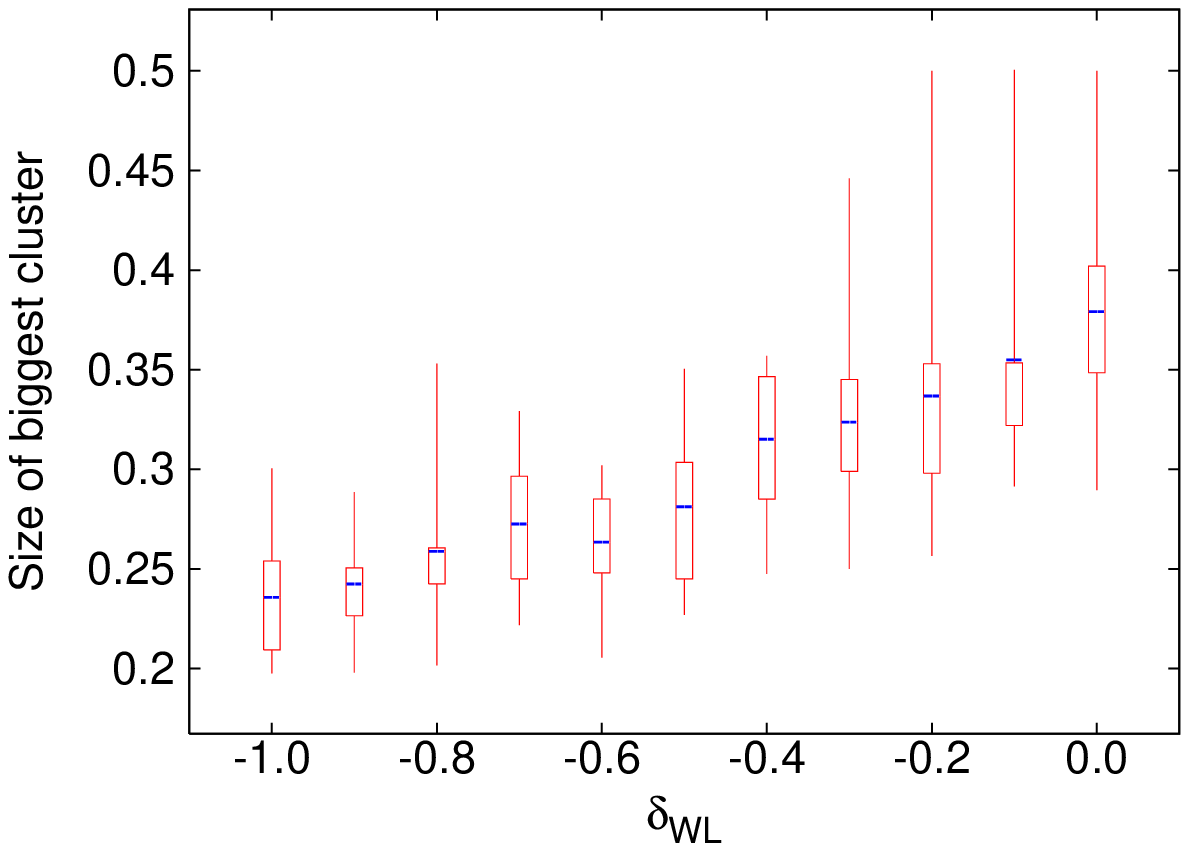}}%
 		\hspace{.5cm}%
\subfigure[]{	
		\includegraphics[width=2in]{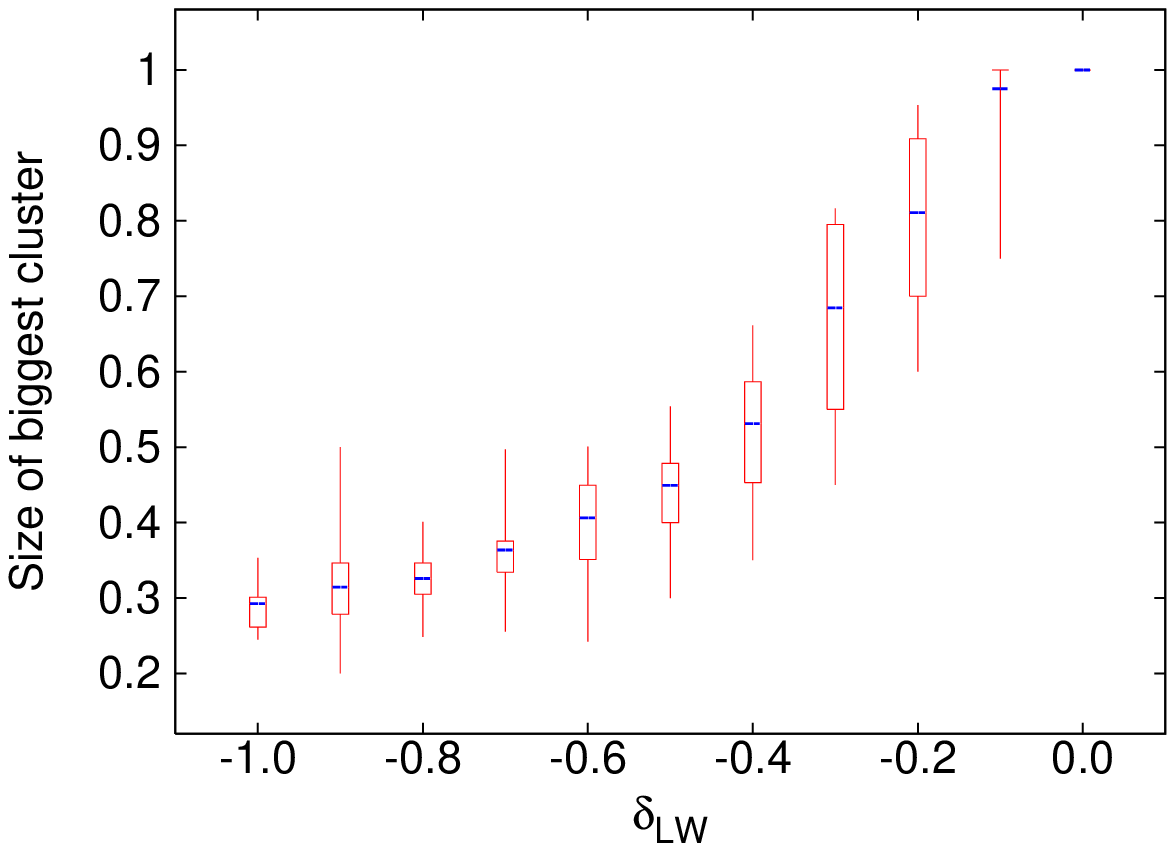}}\\
\subfigure[]{	
		\includegraphics[width=2in]{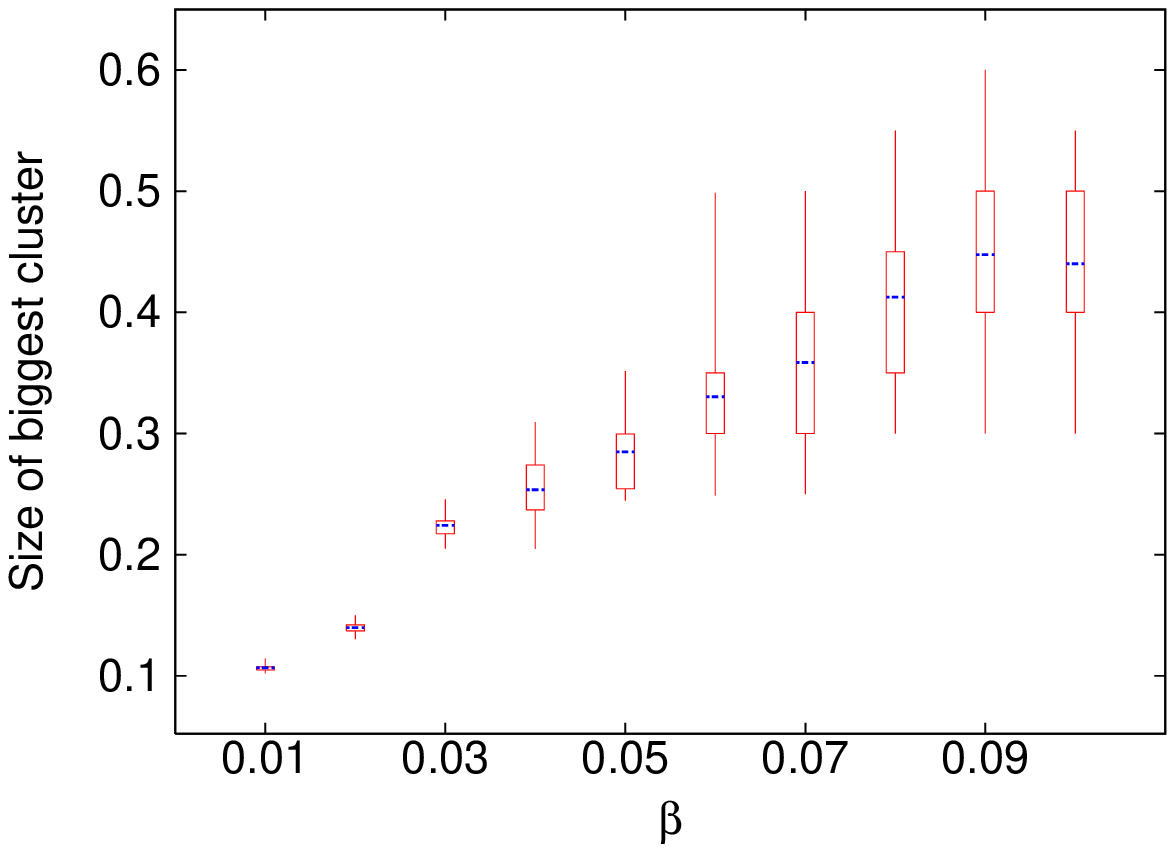}}%
 		\hspace{.5cm}
\subfigure[]{	
		\includegraphics[width=2in]{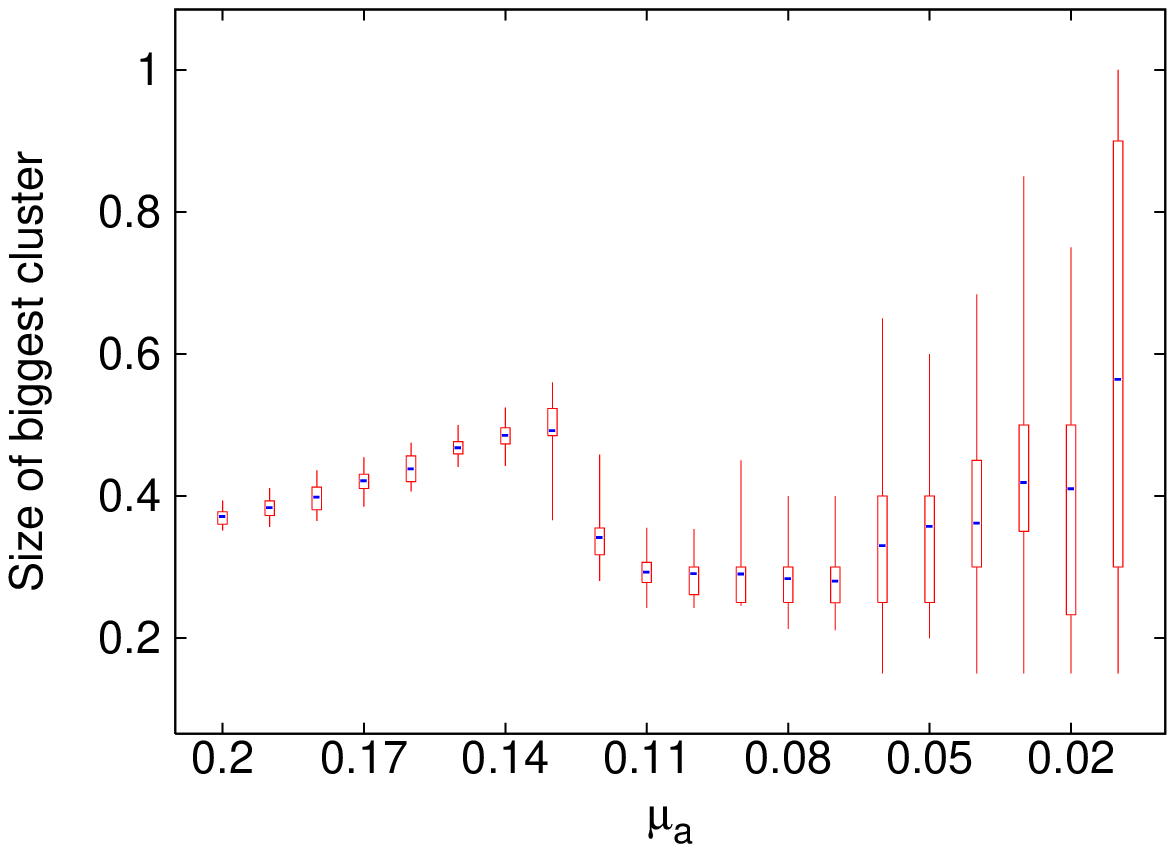}}\\		
\subfigure[]{	
		\includegraphics[width=2in]{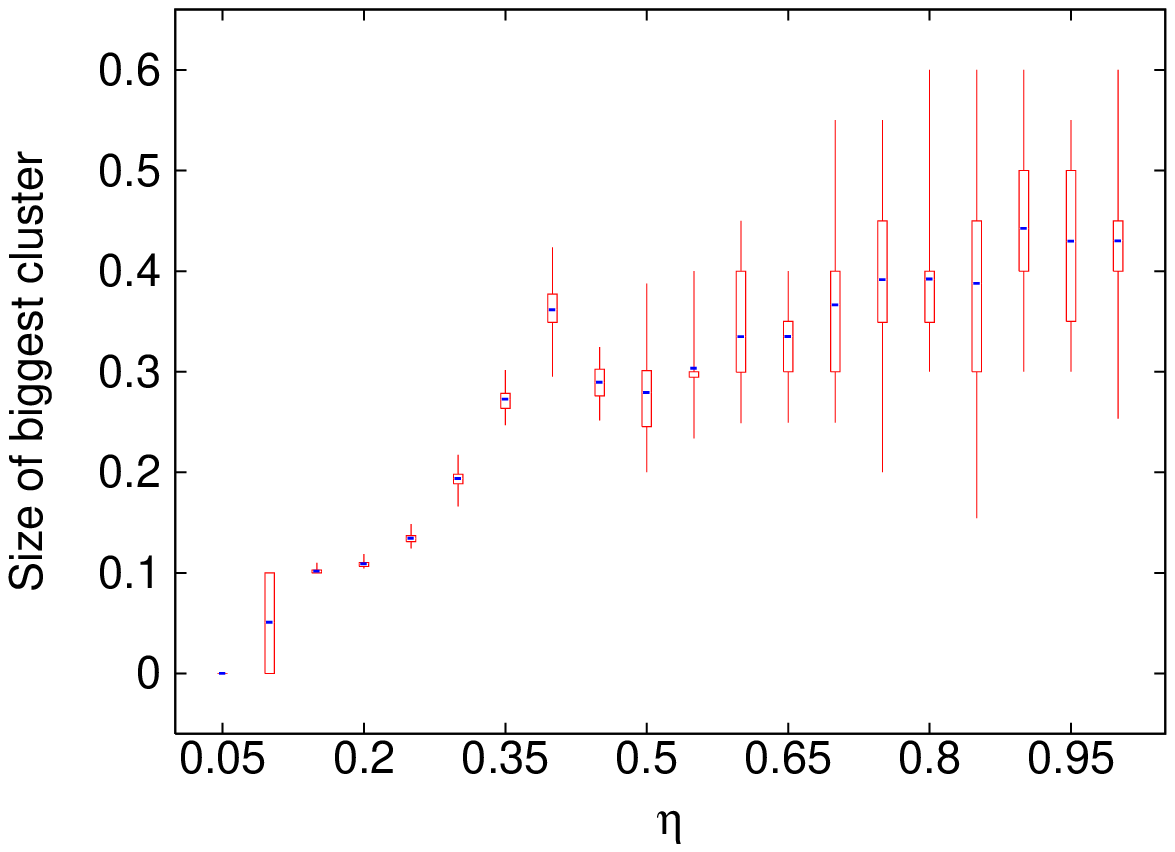}}%
		\hspace{.5cm}%
\subfigure[]{	
		\includegraphics[width=2in]{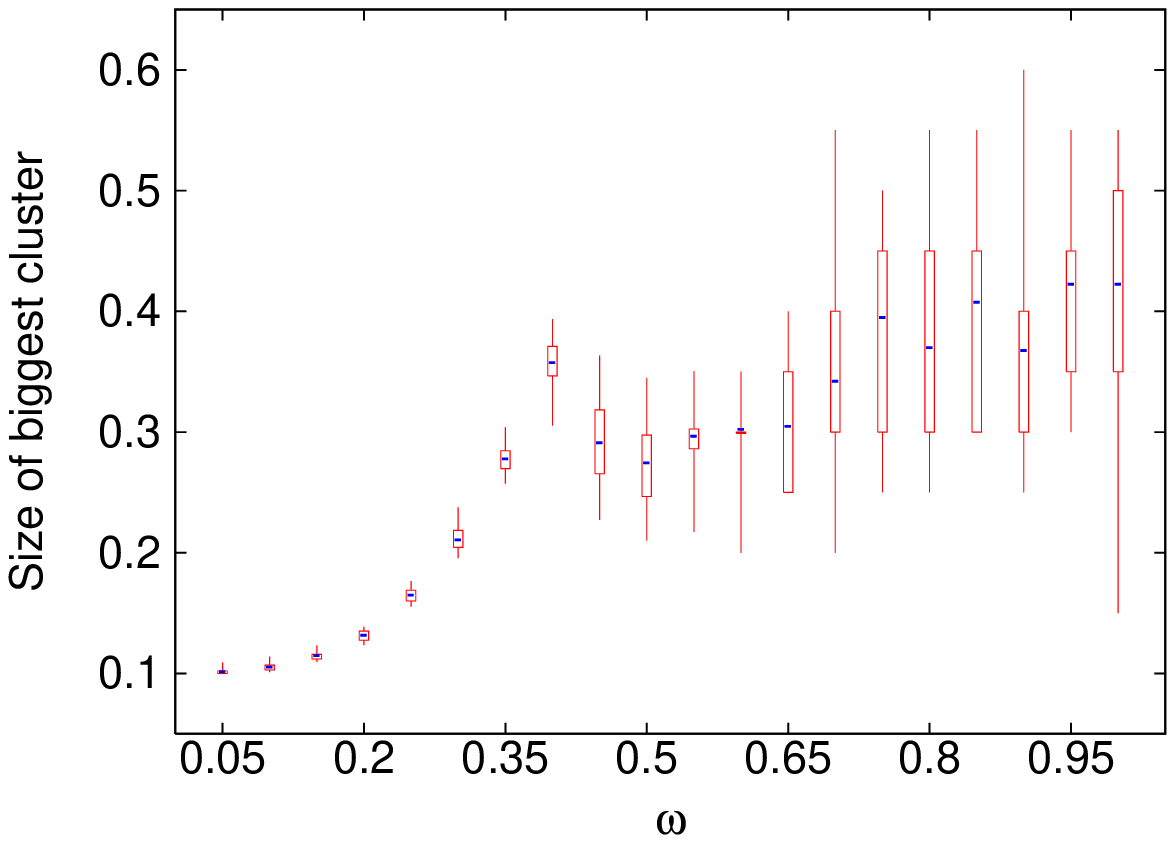}}\\
\caption{Effects of parameters $N,\gd_{ll},\gd_{wl},\gd_{lw},\gb,\mu,\eta,\go$ on
the size of the biggest alliance for a default set of parameter
values.
}
\label{szBigClus}
\end{figure}

\begin{figure}
\centering
\subfigure[]{ 	
		\includegraphics[width=2in]{./Figures/plots_C1/run_0_.eps}}%
 		\hspace{.5cm}%
\subfigure[]{	
		\includegraphics[width=2in]{./Figures/plots_C1/run_1_.eps}}\\
\subfigure[]{	
		\includegraphics[width=2in]{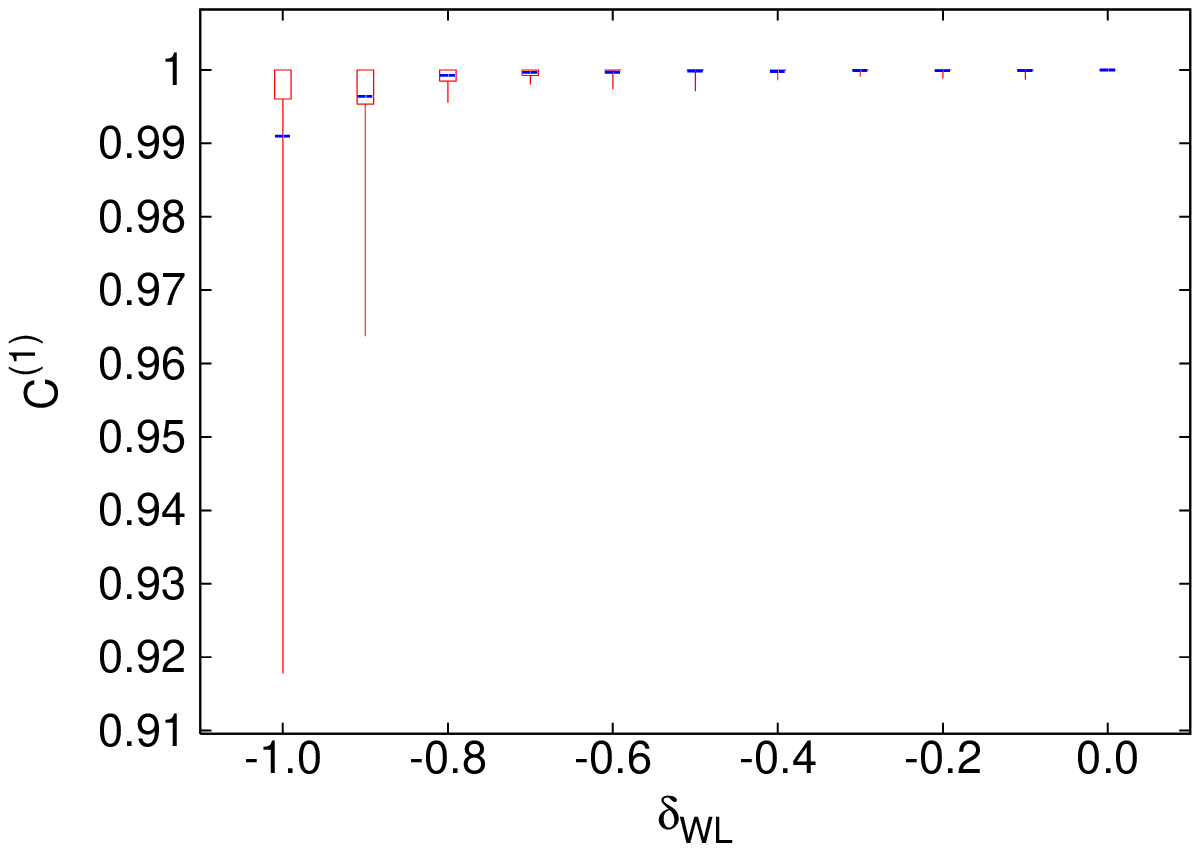}}%
 		\hspace{.5cm}%
\subfigure[]{	
		\includegraphics[width=2in]{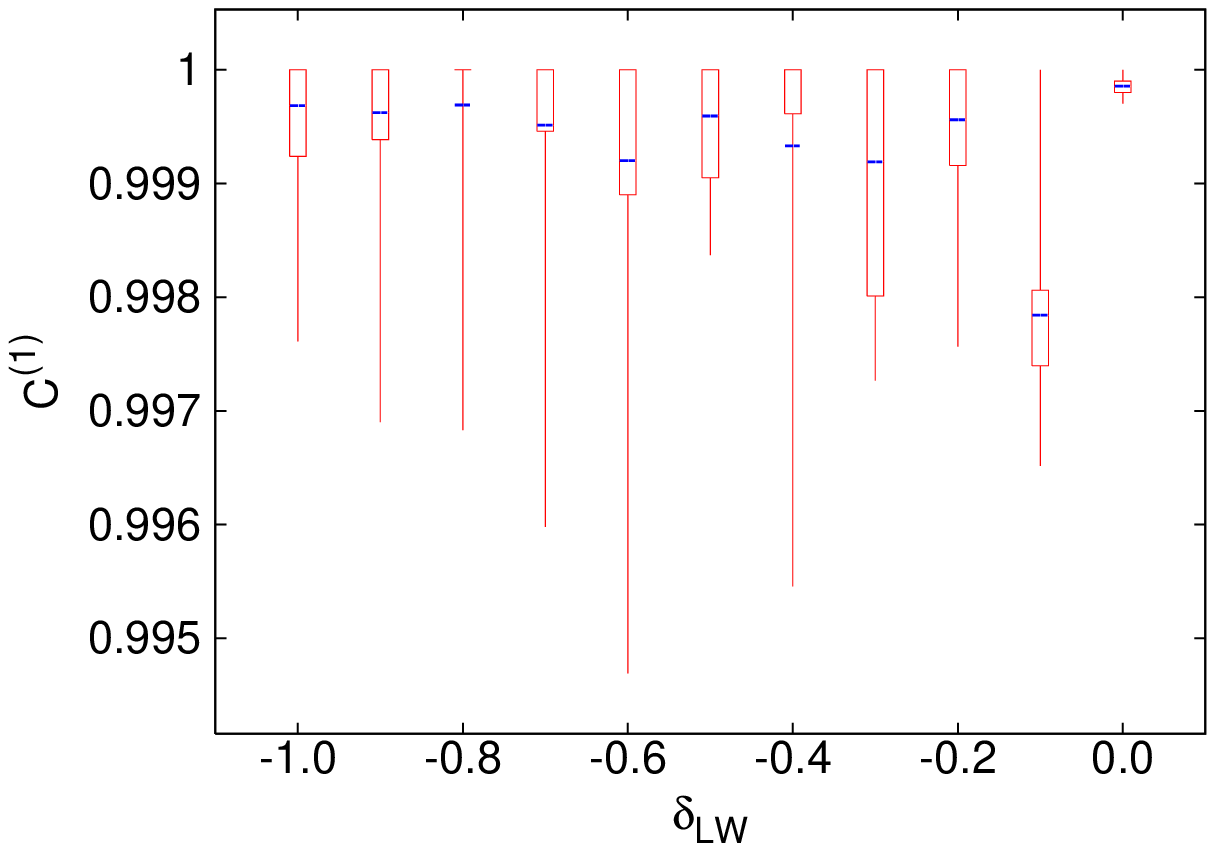}}\\
\subfigure[]{	
		\includegraphics[width=2in]{./Figures/plots_C1/run_4_.eps}}%
 		\hspace{.5cm}
\subfigure[]{	
		\includegraphics[width=2in]{./Figures/plots_C1/run_5_.eps}}\\		
\subfigure[]{	
		\includegraphics[width=2in]{./Figures/plots_C1/run_6_.eps}}%
		\hspace{.5cm}%
\subfigure[]{	
		\includegraphics[width=2in]{./Figures/plots_C1/run_7_.eps}}\\
\caption{Effects of parameters $N,\gd_{ll},\gd_{wl},\gd_{lw},\gb,\mu,\eta,\go$ on
the $C^{(1)}$ measure of the largest alliance for a default set of parameter
values.
}
\label{C1}
\end{figure}

\begin{figure}
\centering
\subfigure[]{ 	
		\includegraphics[width=2in]{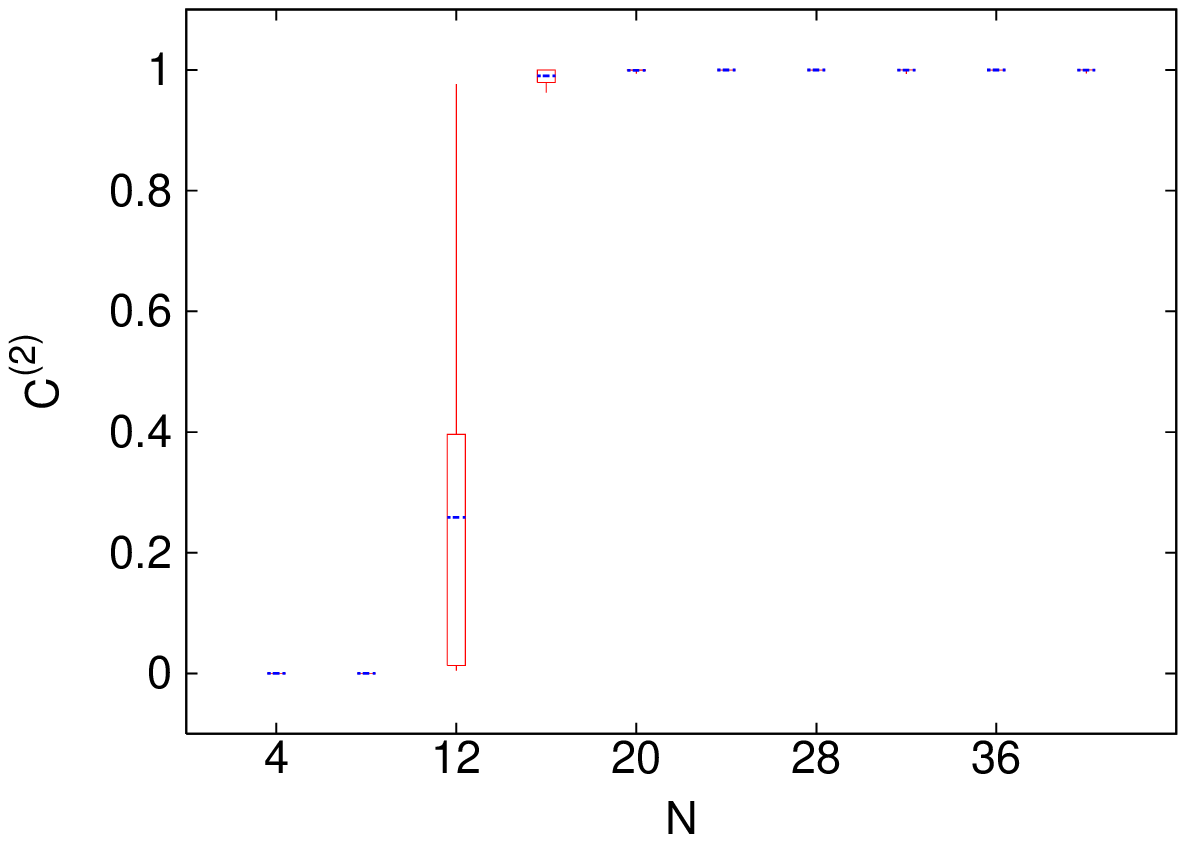}}%
 		\hspace{.5cm}%
\subfigure[]{	
		\includegraphics[width=2in]{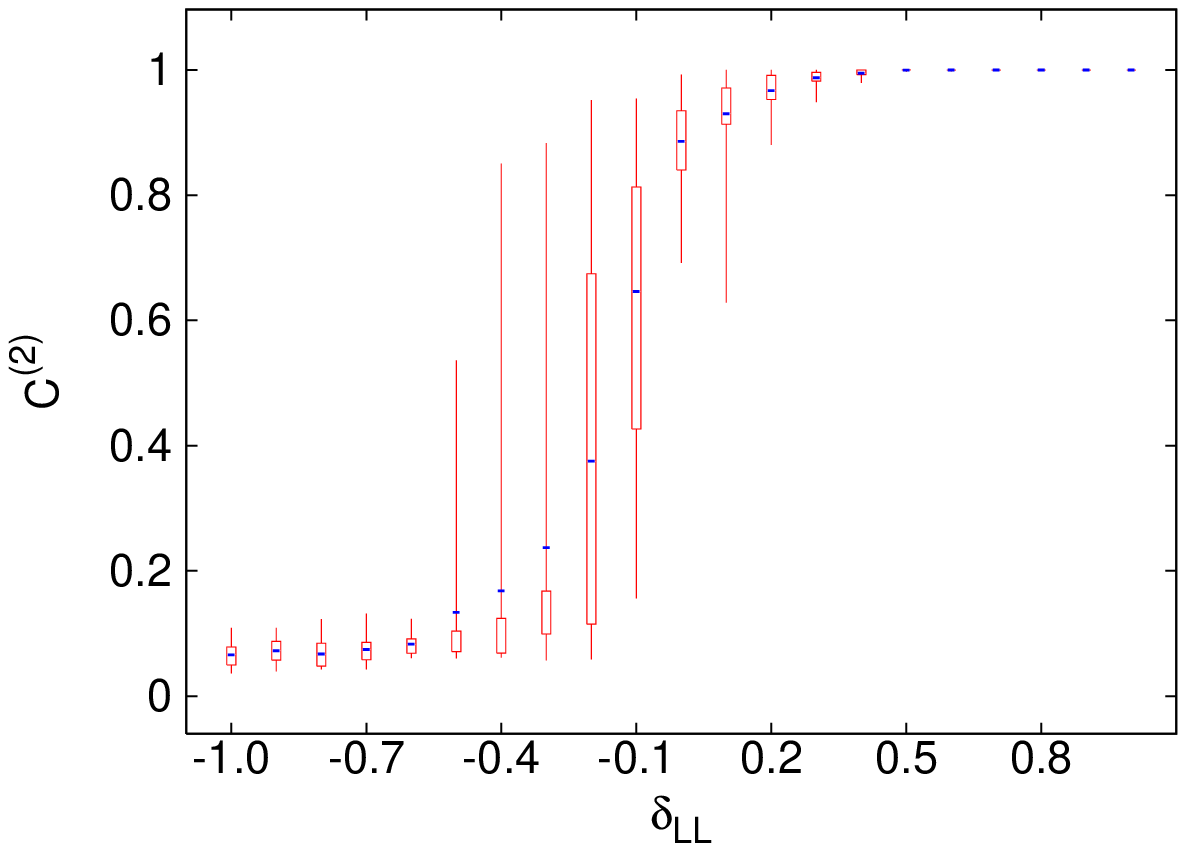}}\\
\subfigure[]{	
		\includegraphics[width=2in]{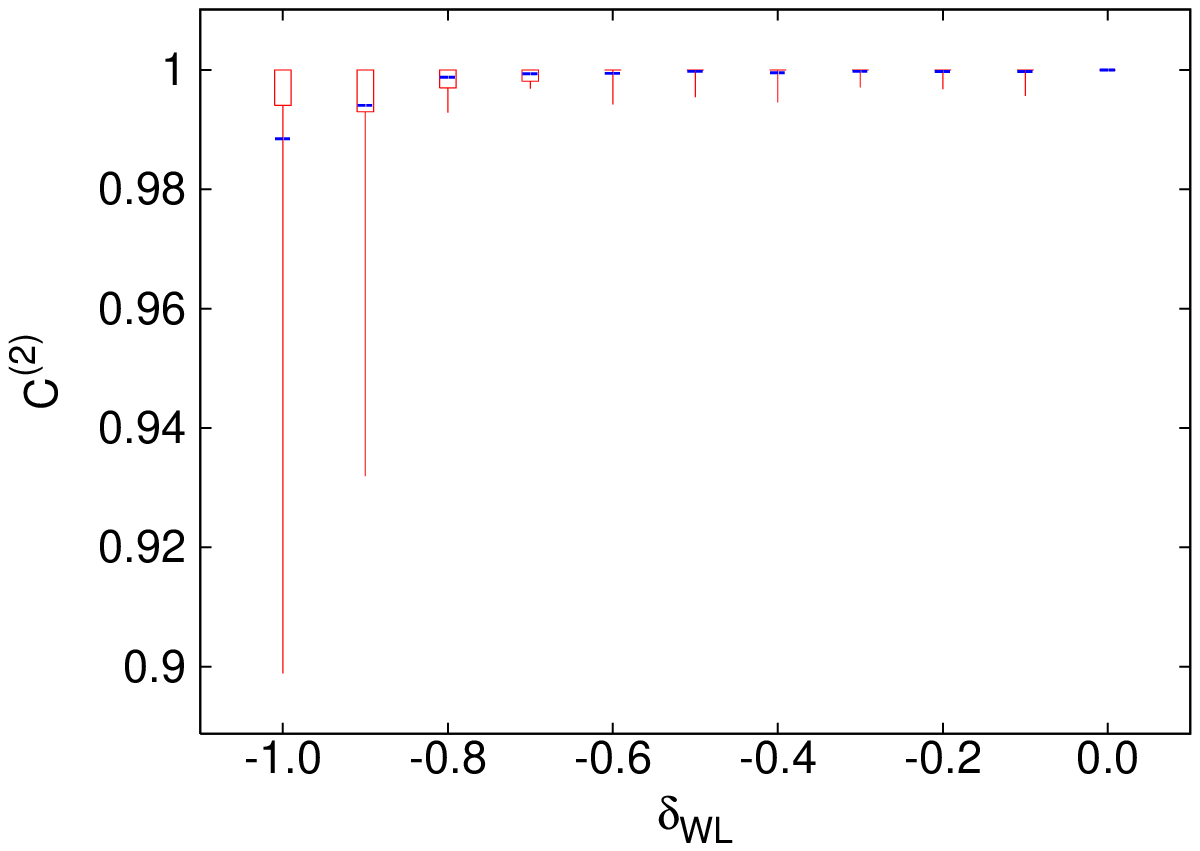}}%
 		\hspace{.5cm}%
\subfigure[]{	
		\includegraphics[width=2in]{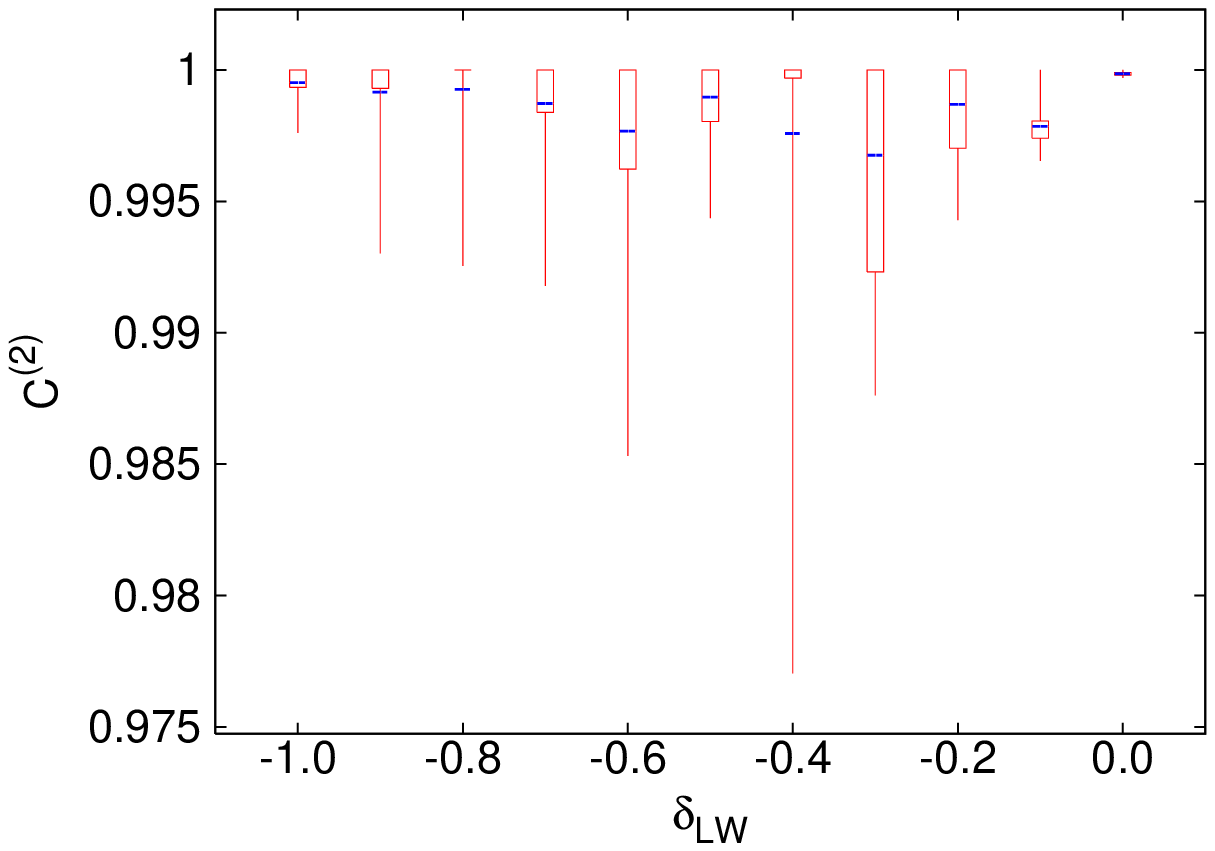}}\\
\subfigure[]{	
		\includegraphics[width=2in]{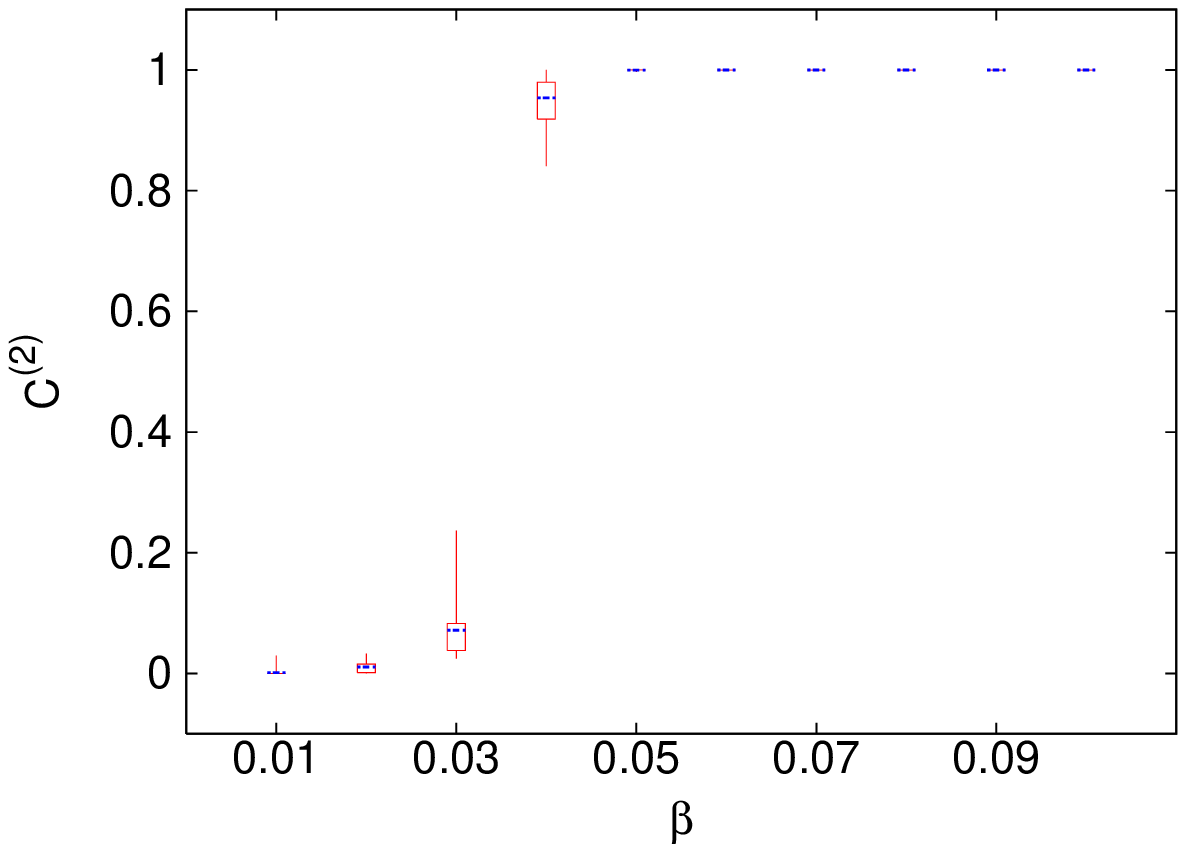}}%
 		\hspace{.5cm}
\subfigure[]{	
		\includegraphics[width=2in]{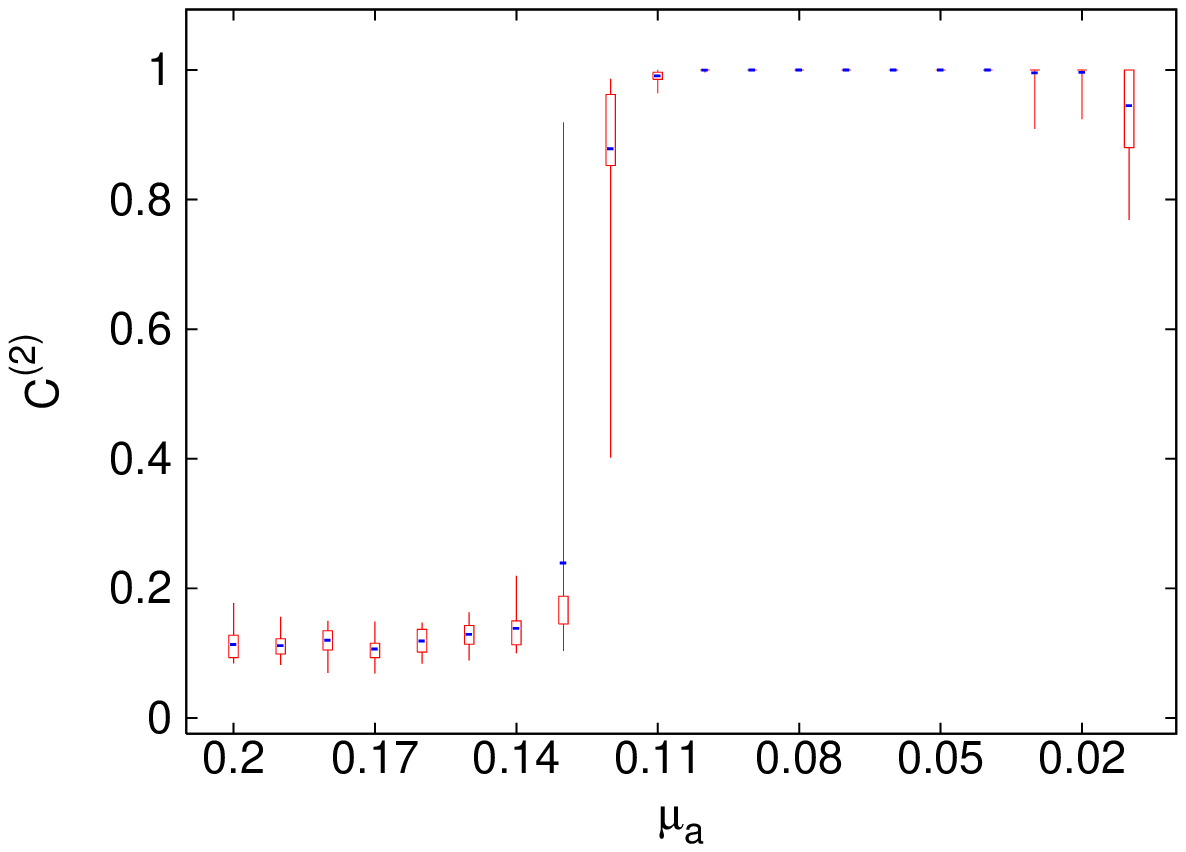}}\\		
\subfigure[]{	
		\includegraphics[width=2in]{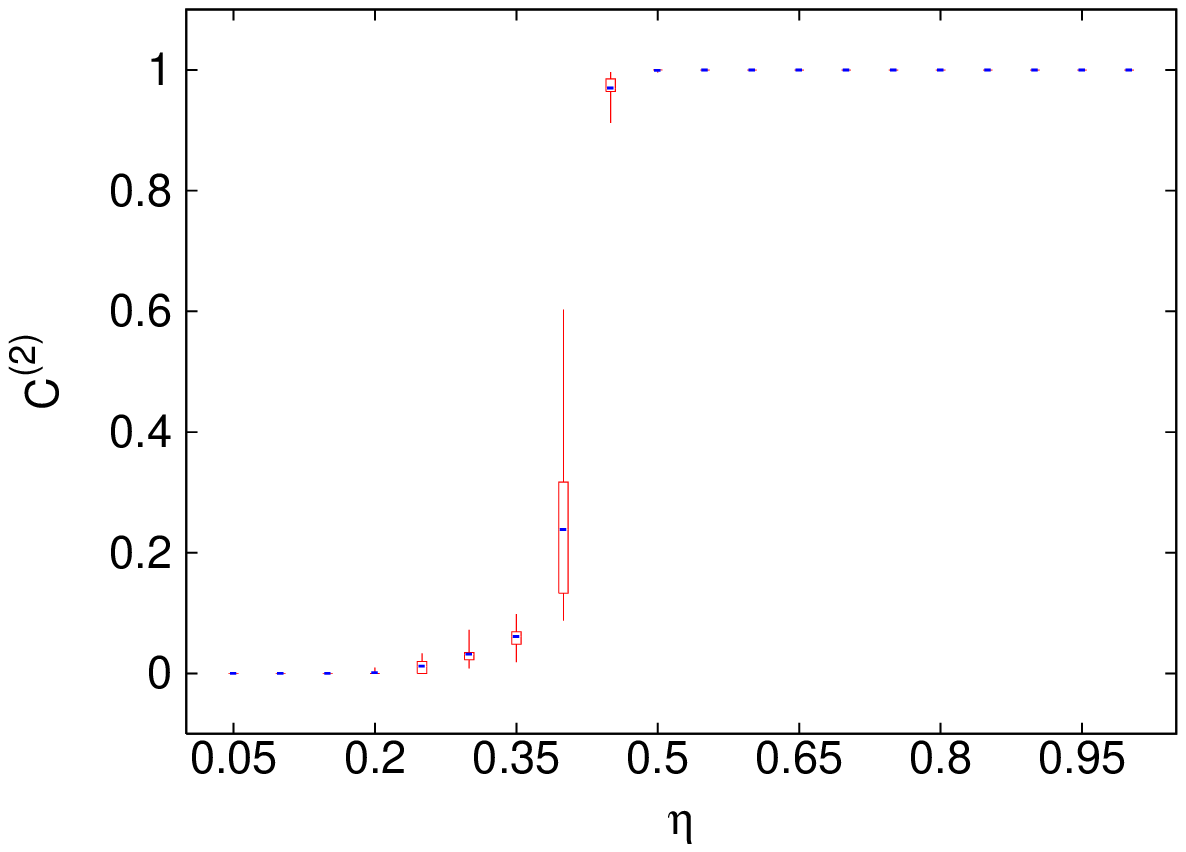}}%
		\hspace{.5cm}%
\subfigure[]{	
		\includegraphics[width=2in]{./Figures/plots_C1/run_7_.eps}}\\
\caption{Effects of parameters $N,\gd_{ll},\gd_{wl},\gd_{lw},\gb,\mu,\eta,\go$ on
the $C^{(2)}$ measure of the largest alliance for a default set of parameter
values.
}
\label{C2}
\end{figure}

\begin{figure}
\centering
\subfigure[]{ 	
		\includegraphics[width=2in]{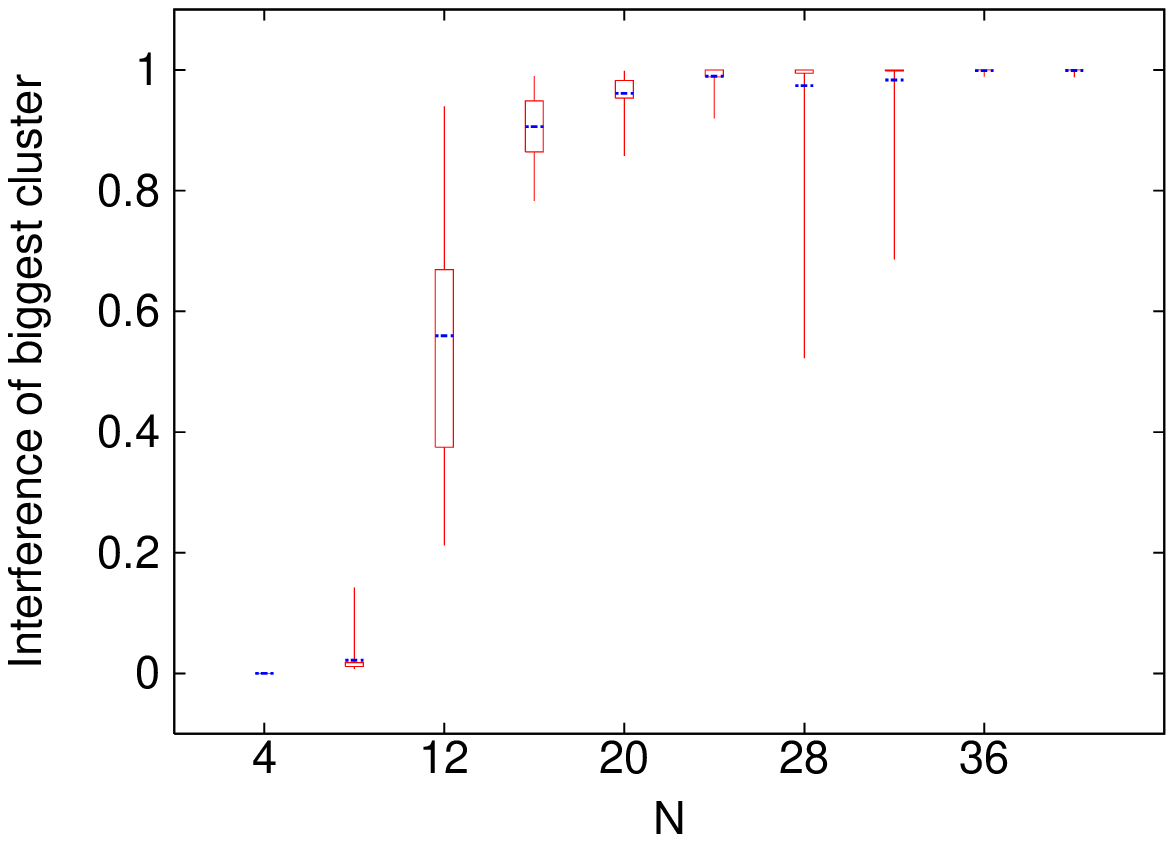}}%
 		\hspace{.5cm}%
\subfigure[]{	
		\includegraphics[width=2in]{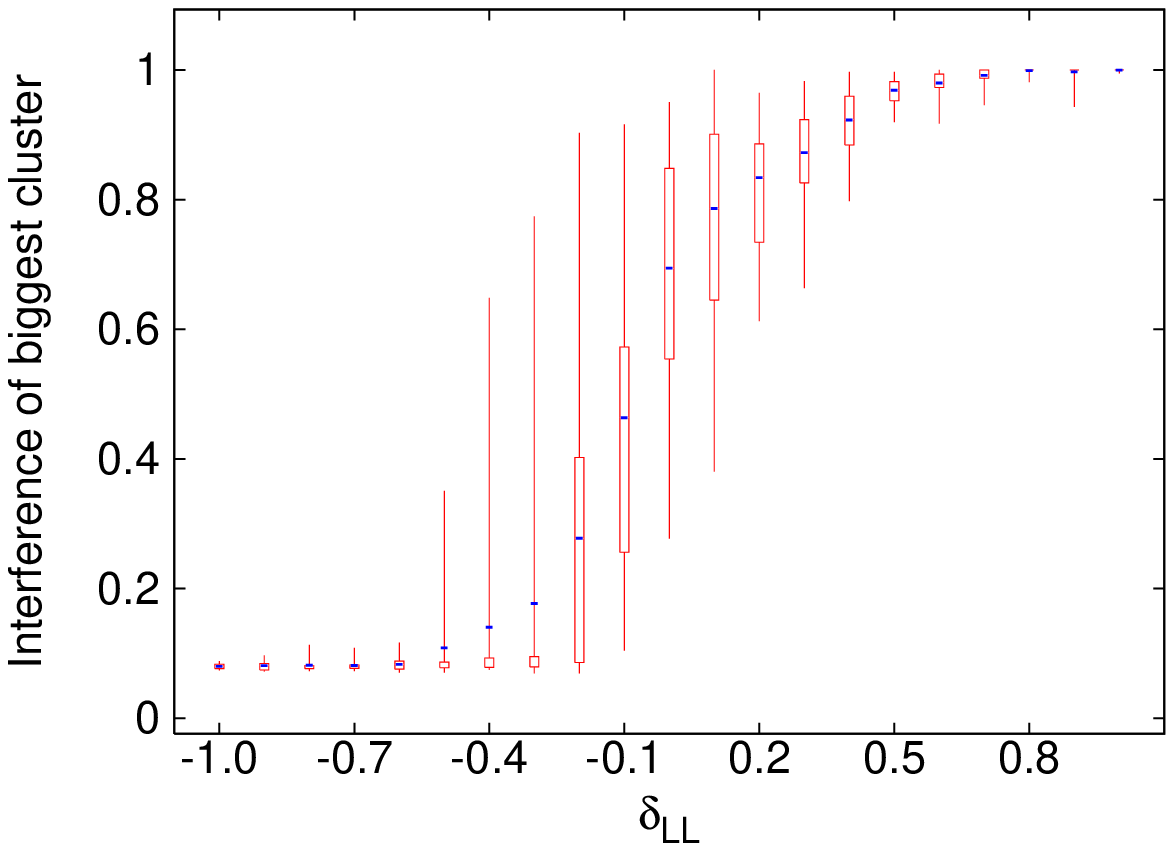}}\\
\subfigure[]{	
		\includegraphics[width=2in]{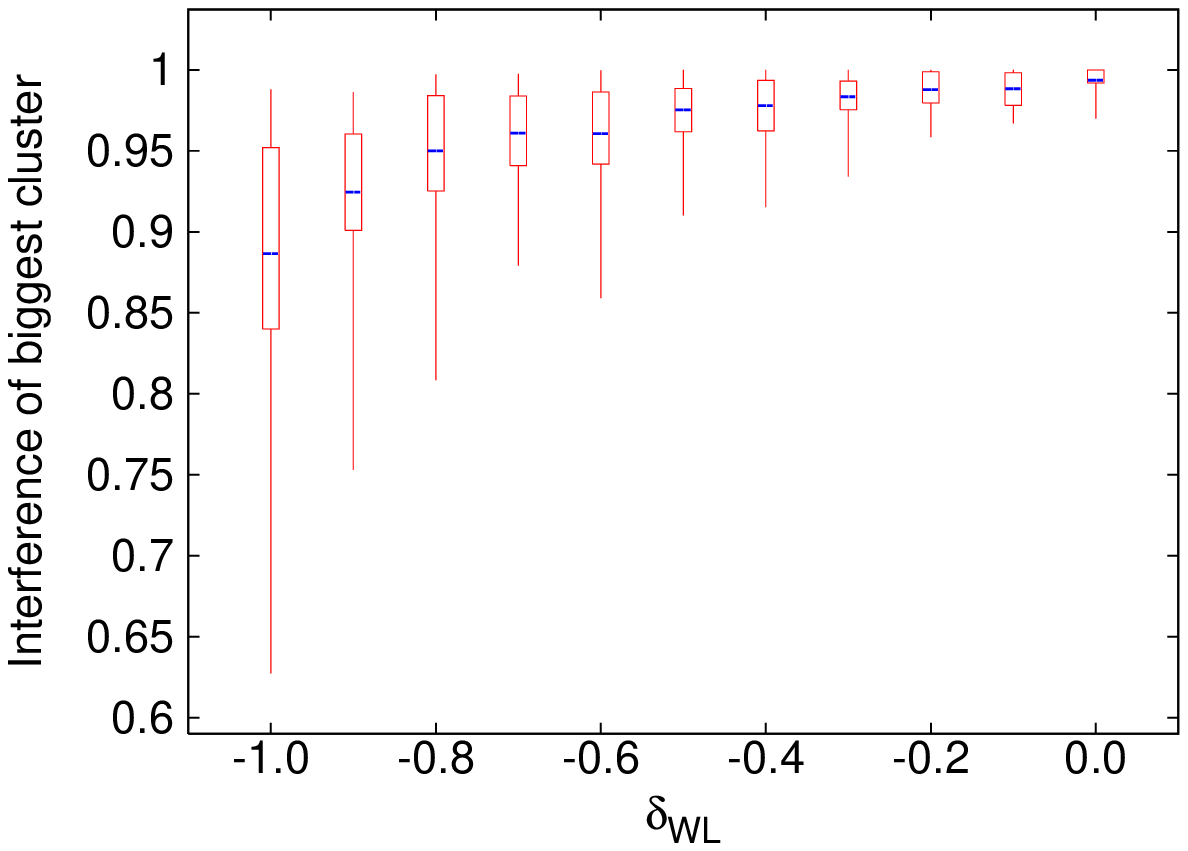}}%
 		\hspace{.5cm}%
\subfigure[]{	
		\includegraphics[width=2in]{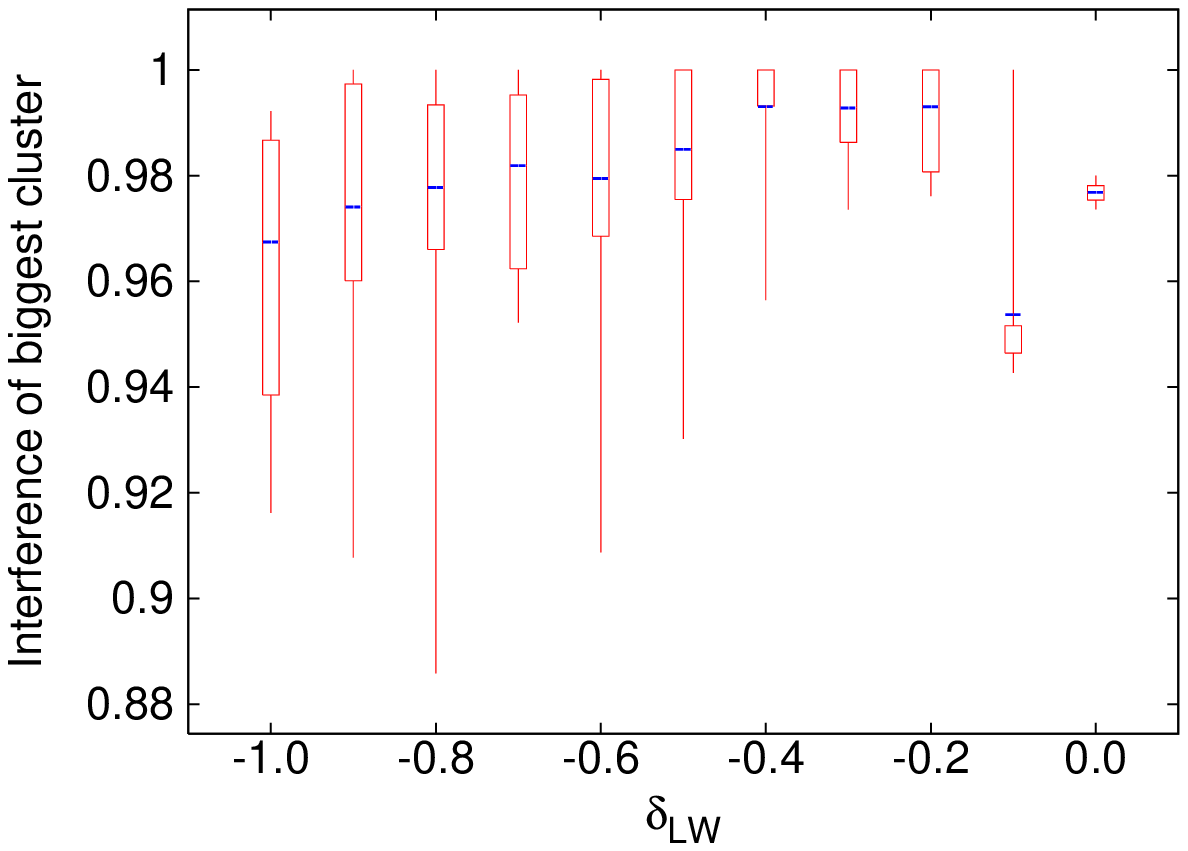}}\\
\subfigure[]{	
		\includegraphics[width=2in]{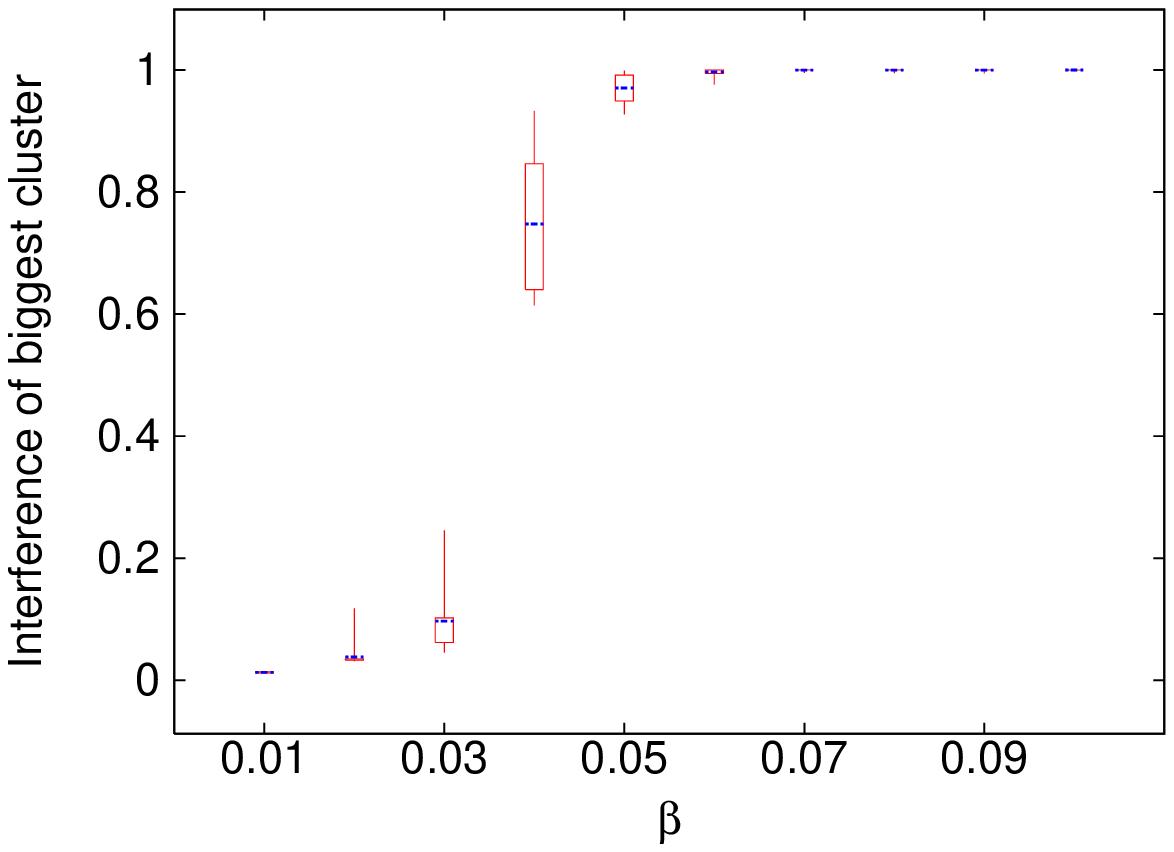}}%
 		\hspace{.5cm}
\subfigure[]{	
		\includegraphics[width=2in]{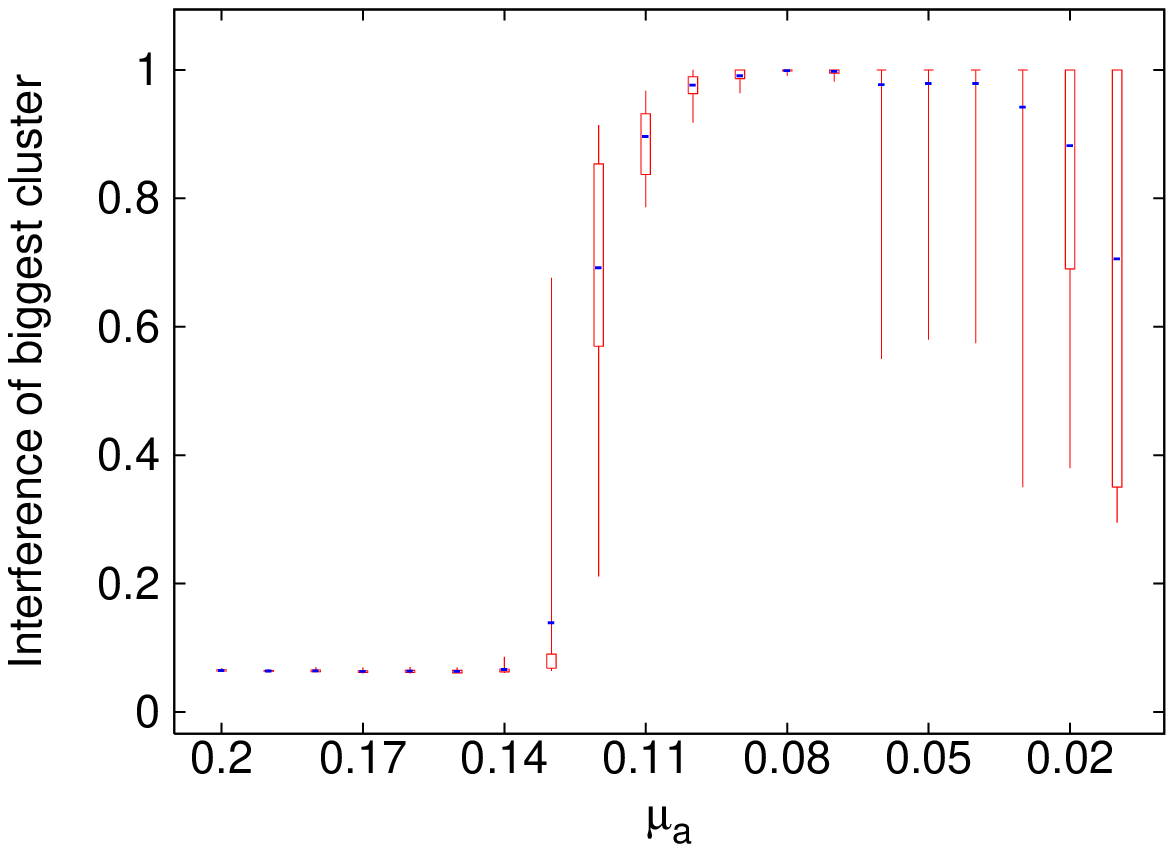}}\\		
\subfigure[]{	
		\includegraphics[width=2in]{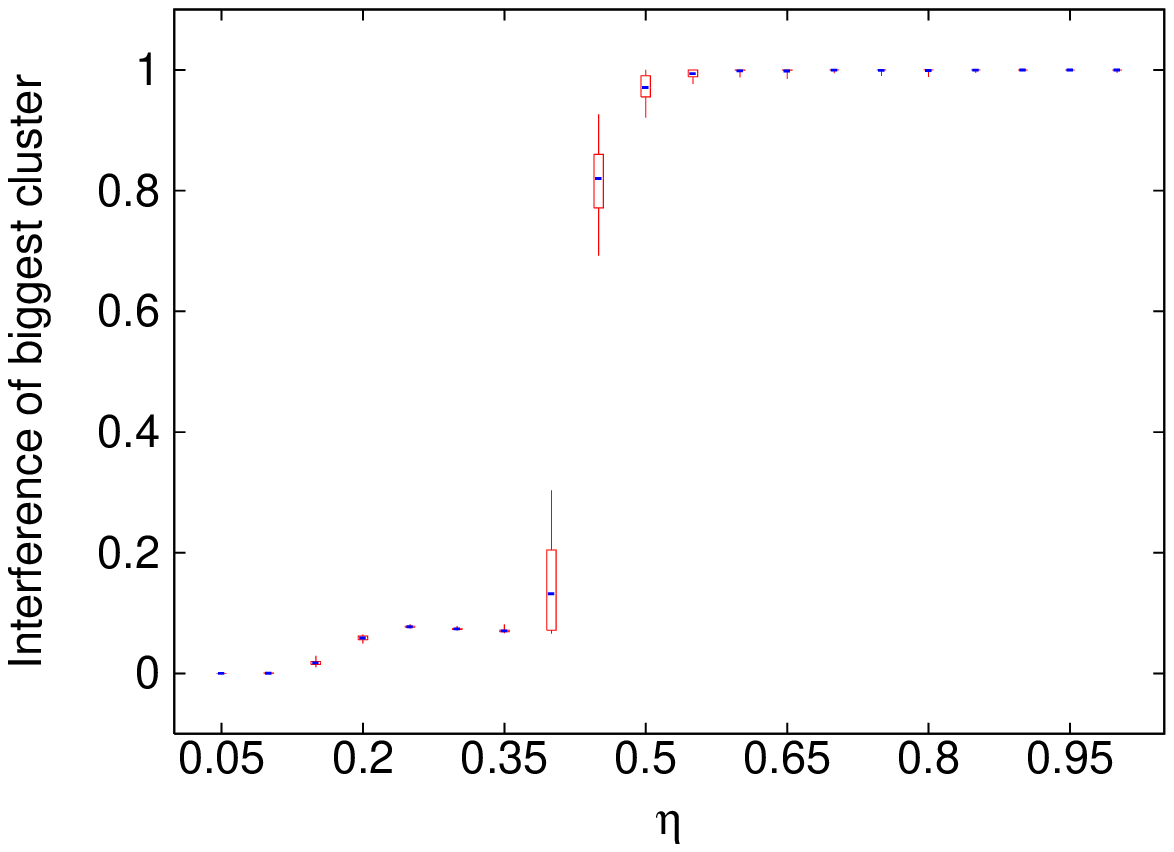}}%
		\hspace{.5cm}%
\subfigure[]{	
		\includegraphics[width=2in]{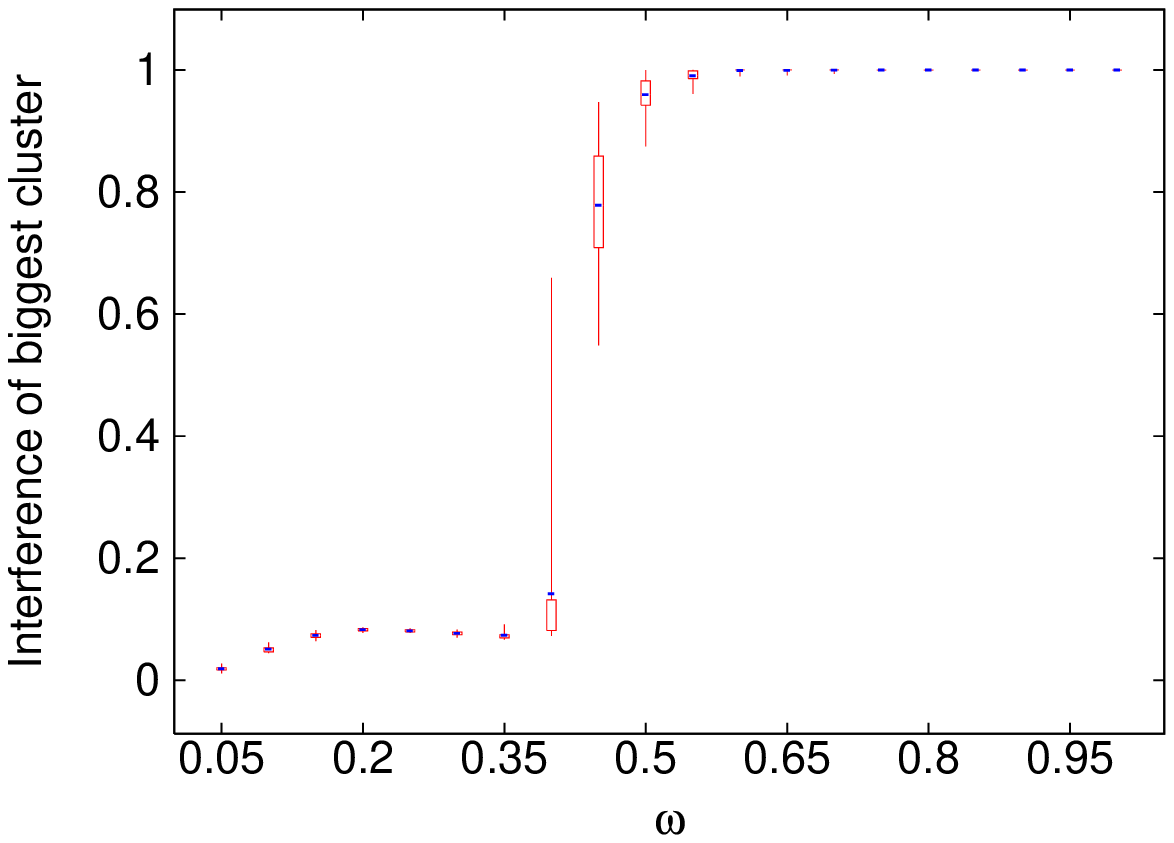}}\\
\caption{Effects of parameters $N,\gd_{ll},\gd_{wl},\gd_{lw},\gb,\mu,\eta,\go$ on
the probability of help within of the largest alliance for a default set of parameter
values.
}
\label{interfBigClus}
\end{figure}

\begin{figure}
\centering
\subfigure[]{ 	
		\includegraphics[width=2in]{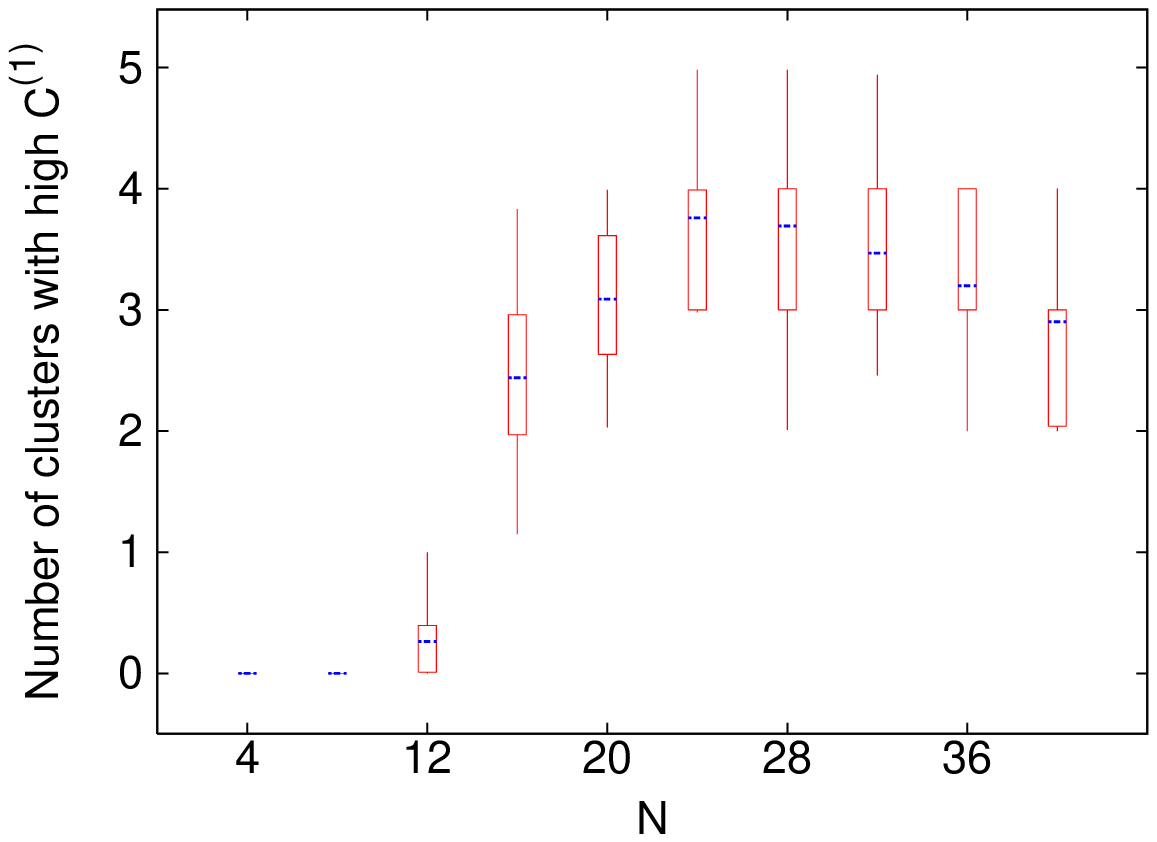}}%
 		\hspace{.5cm}%
\subfigure[]{	
		\includegraphics[width=2in]{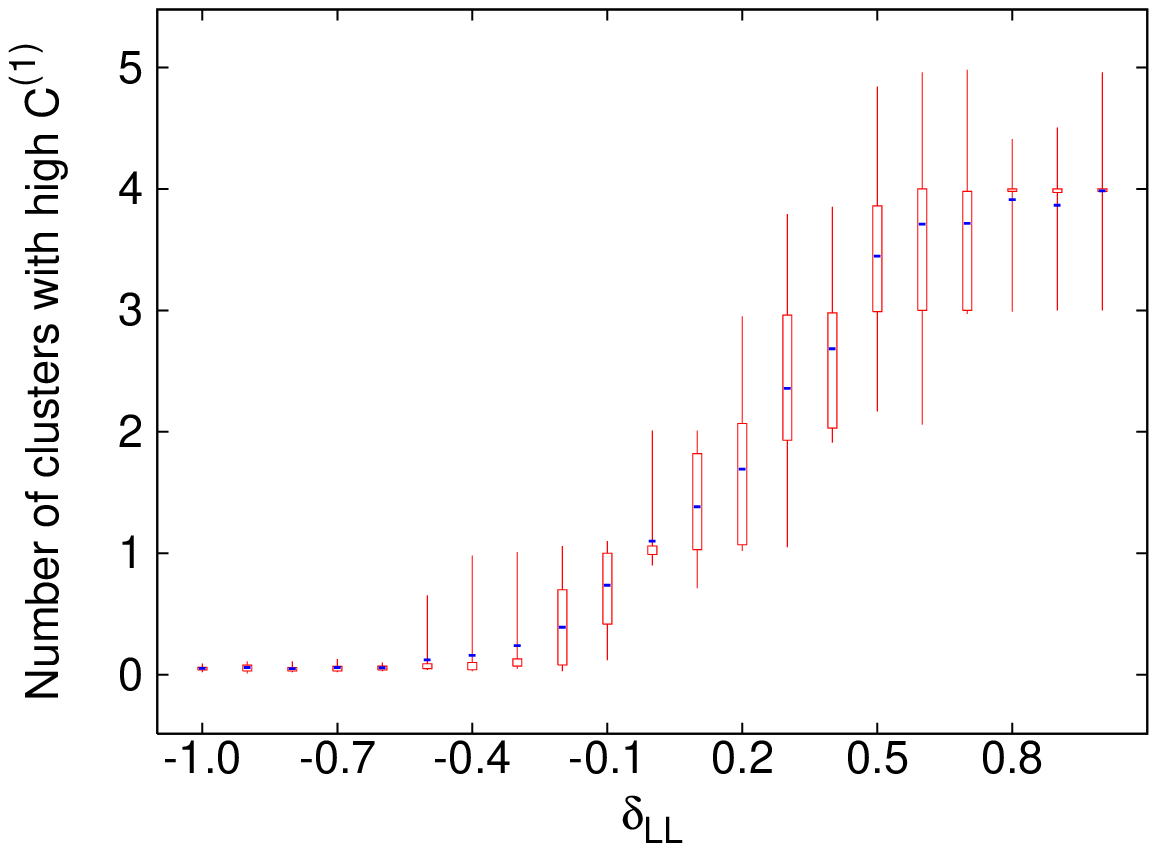}}\\
\subfigure[]{	
		\includegraphics[width=2in]{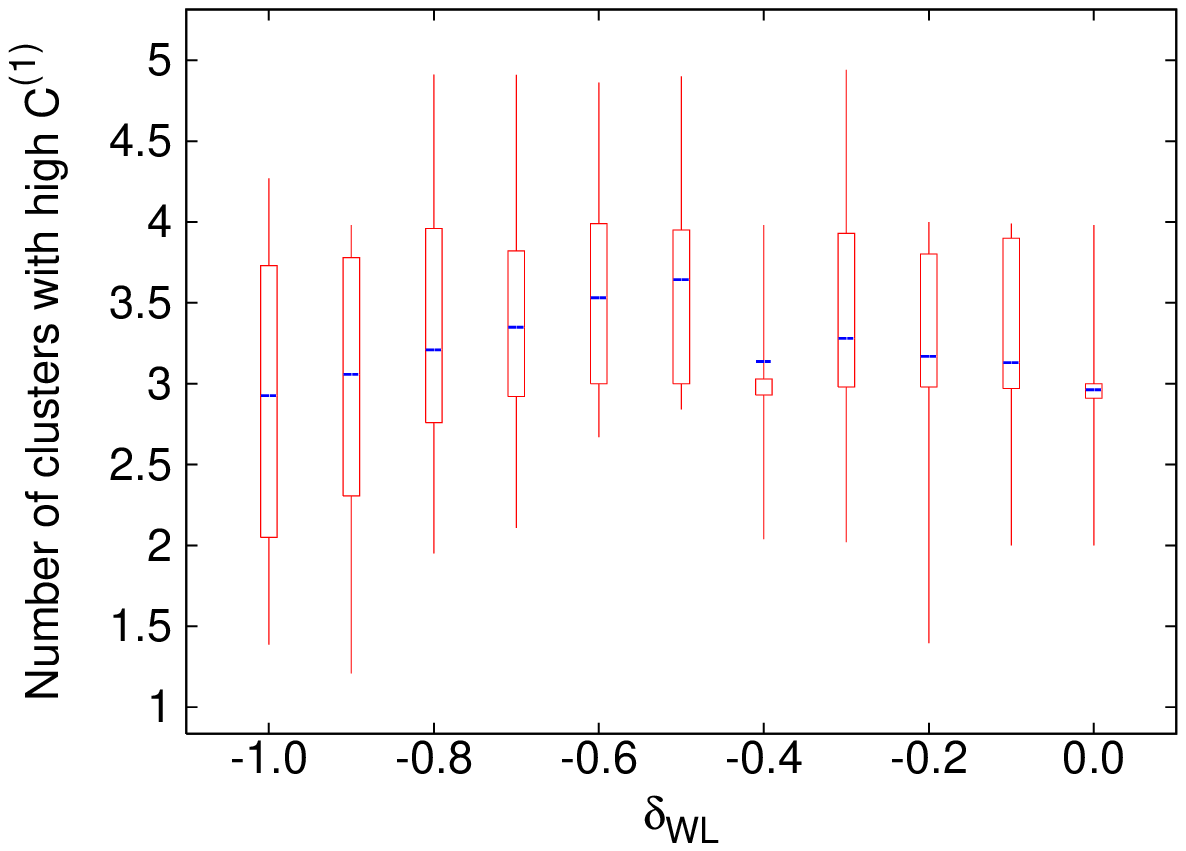}}%
 		\hspace{.5cm}%
\subfigure[]{	
		\includegraphics[width=2in]{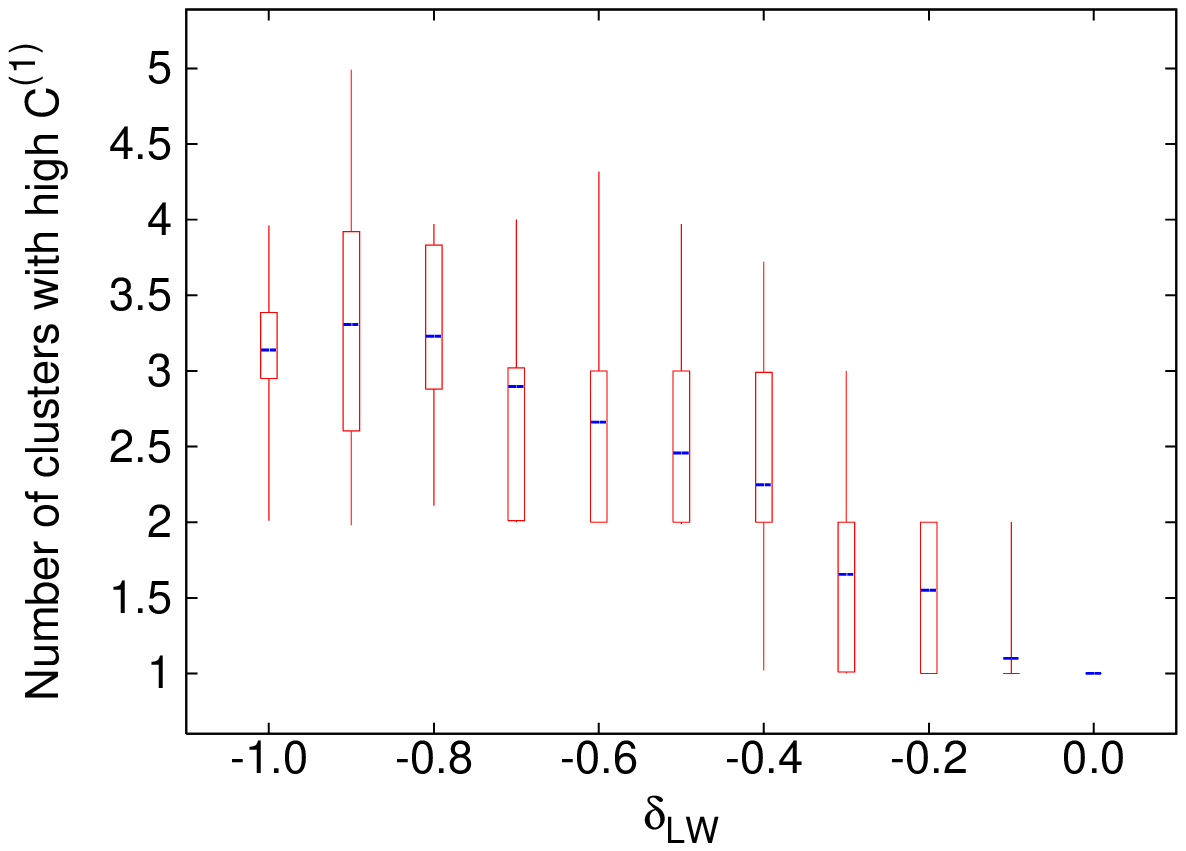}}\\
\subfigure[]{	
		\includegraphics[width=2in]{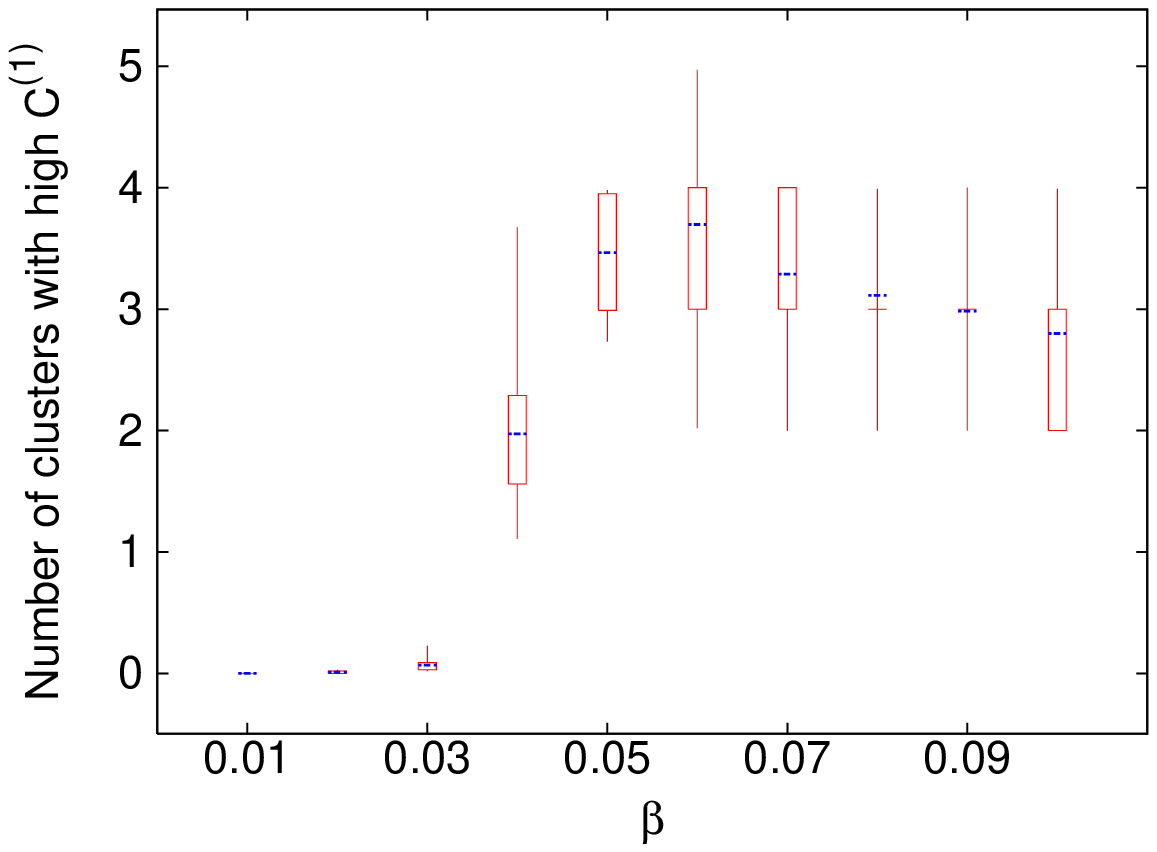}}%
 		\hspace{.5cm}
\subfigure[]{	
		\includegraphics[width=2in]{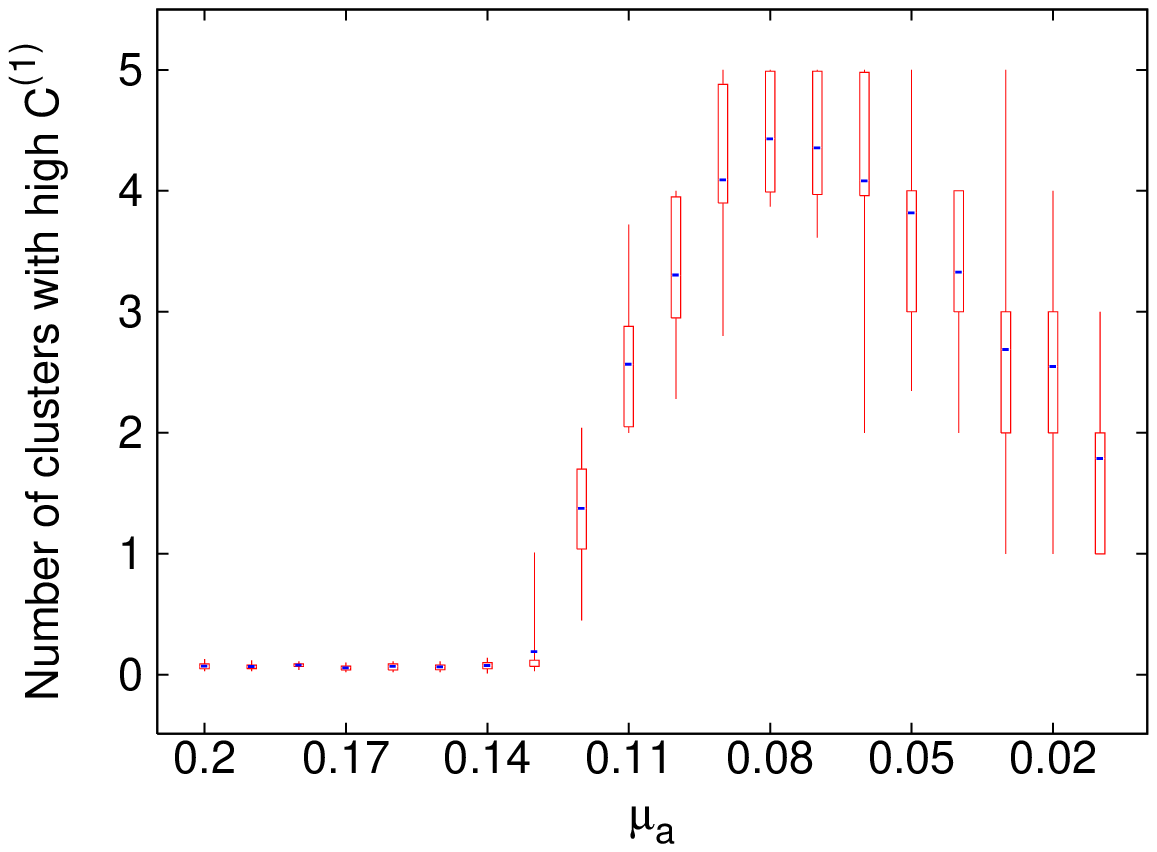}}\\		
\subfigure[]{	
		\includegraphics[width=2in]{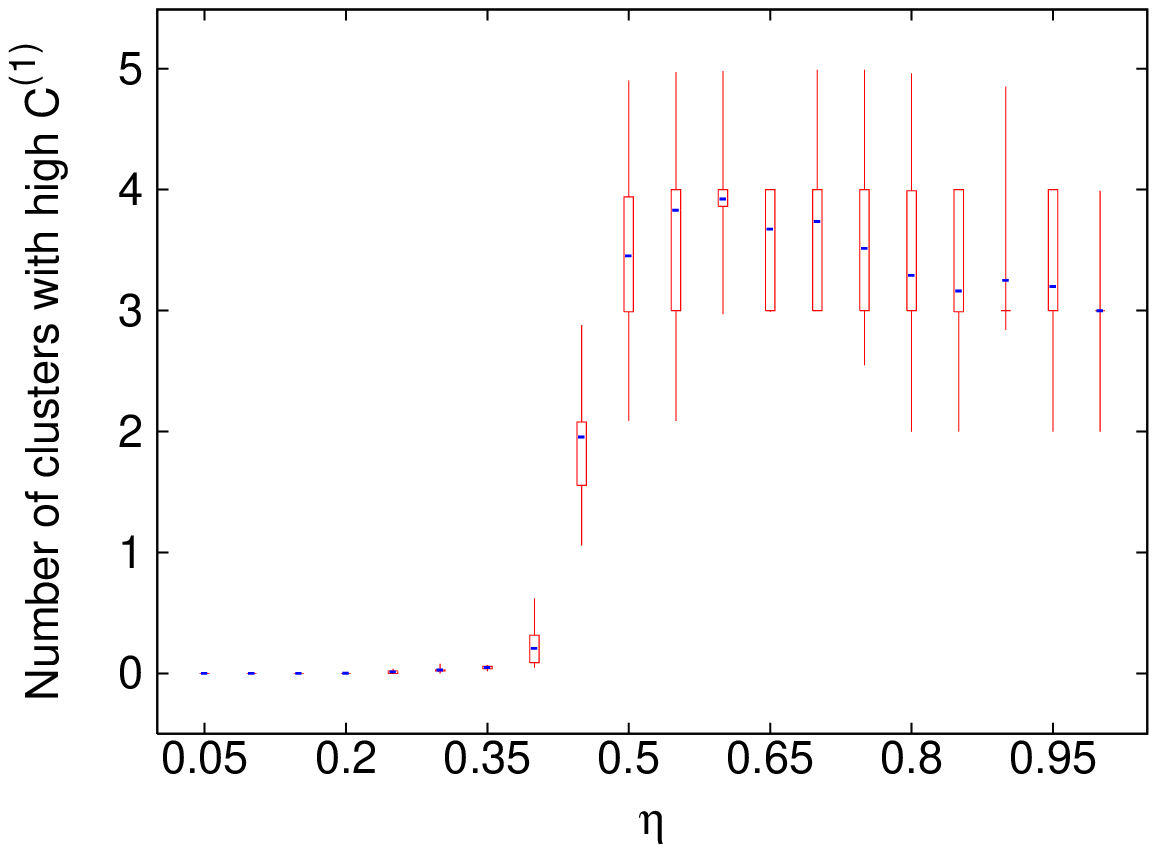}}%
		\hspace{.5cm}%
\subfigure[]{	
		\includegraphics[width=2in]{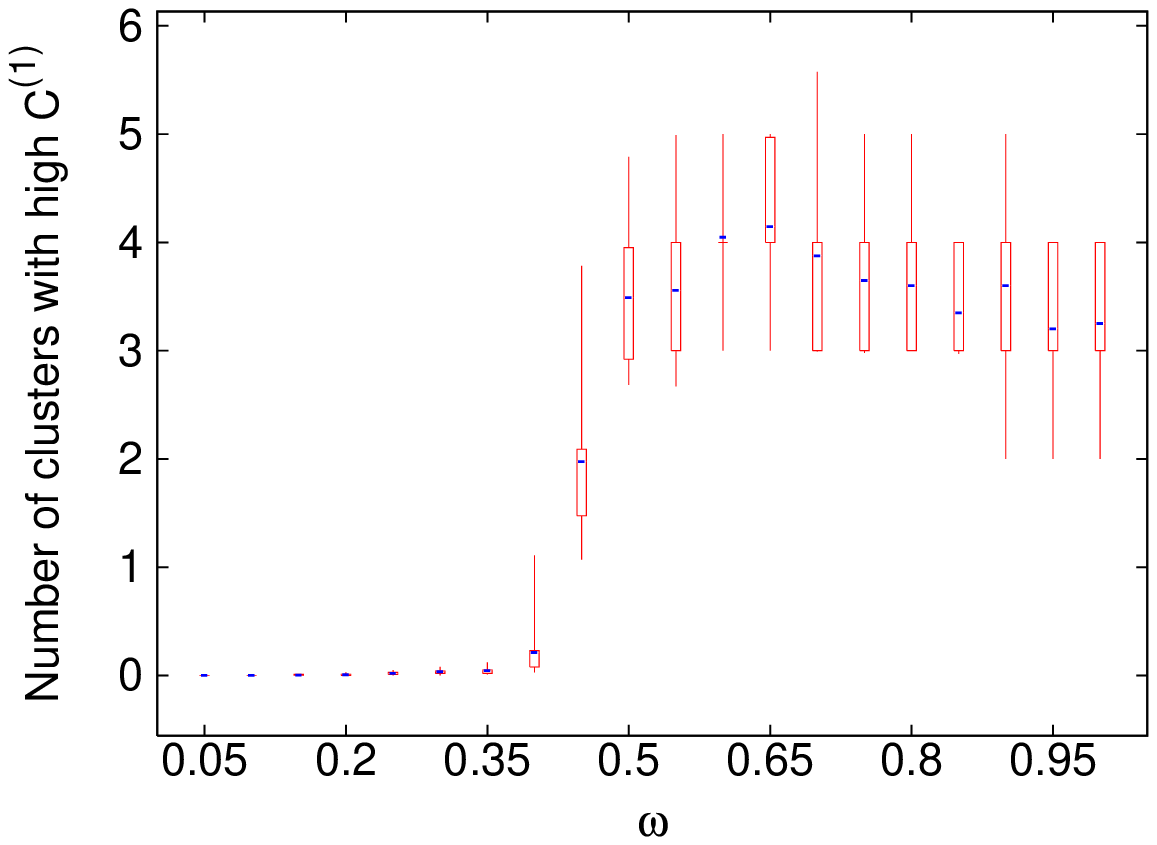}}\\
\caption{Effects of parameters $N,\gd_{ll},\gd_{wl},\gd_{lw},\gb,\mu,\eta,\go$ on
the number of alliances with $C^{1}>0.5$ for a default set of parameter values.
}
\label{numClusBigC1}
\end{figure}

\begin{figure}
\centering
\subfigure[]{ 	
		\includegraphics[width=2in]{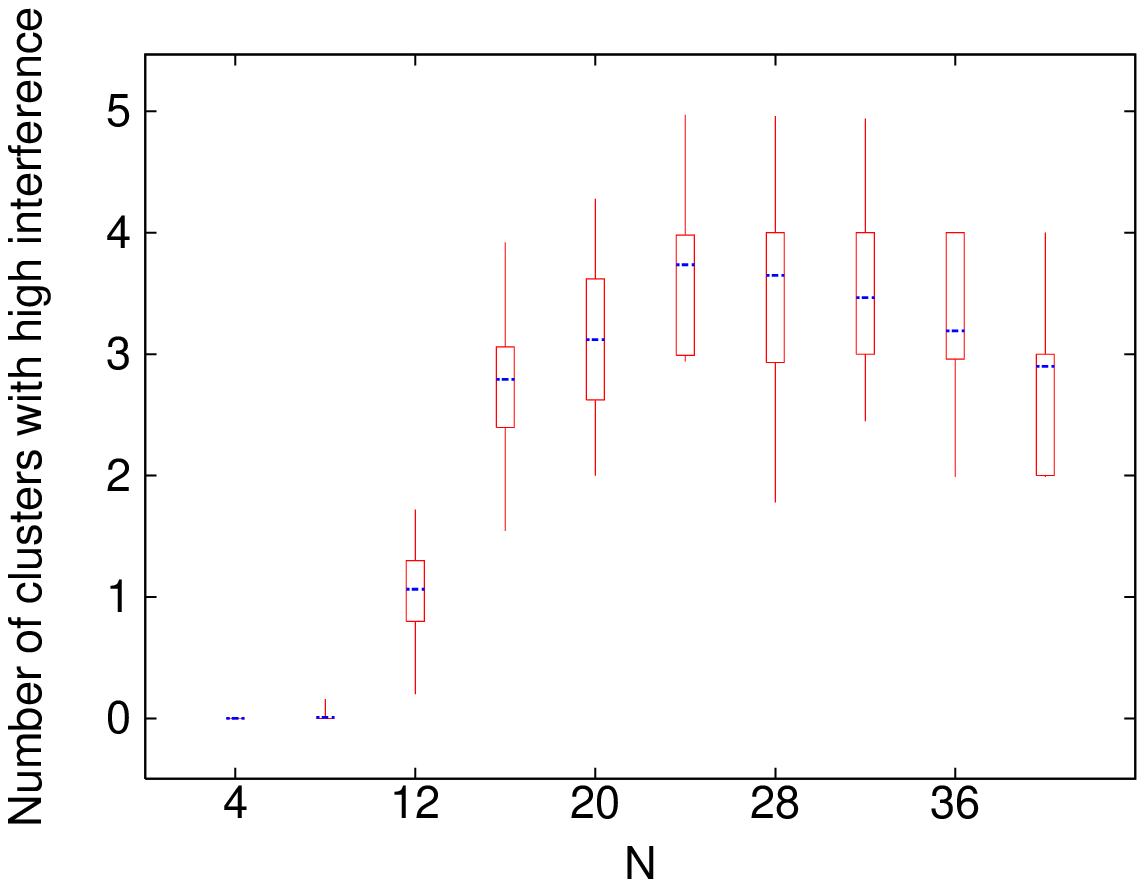}}%
 		\hspace{.5cm}%
\subfigure[]{	
		\includegraphics[width=2in]{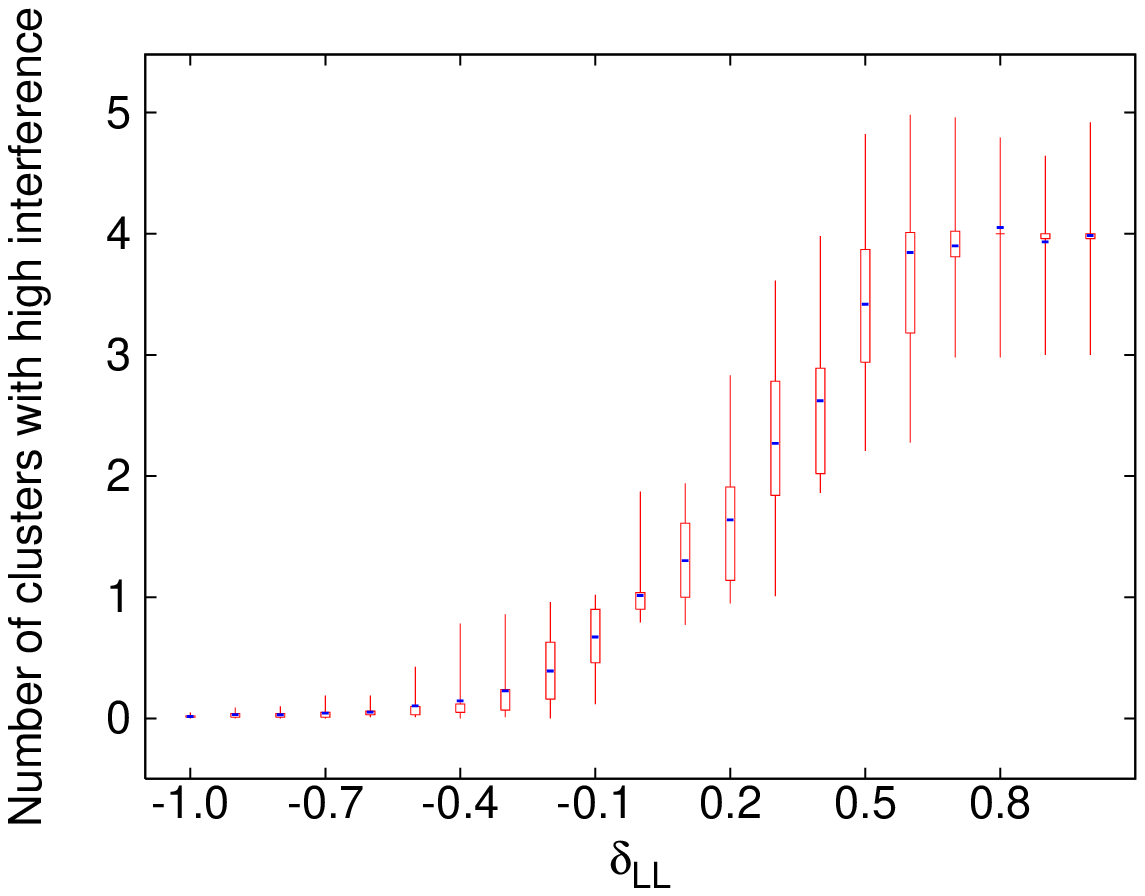}}\\
\subfigure[]{	
		\includegraphics[width=2in]{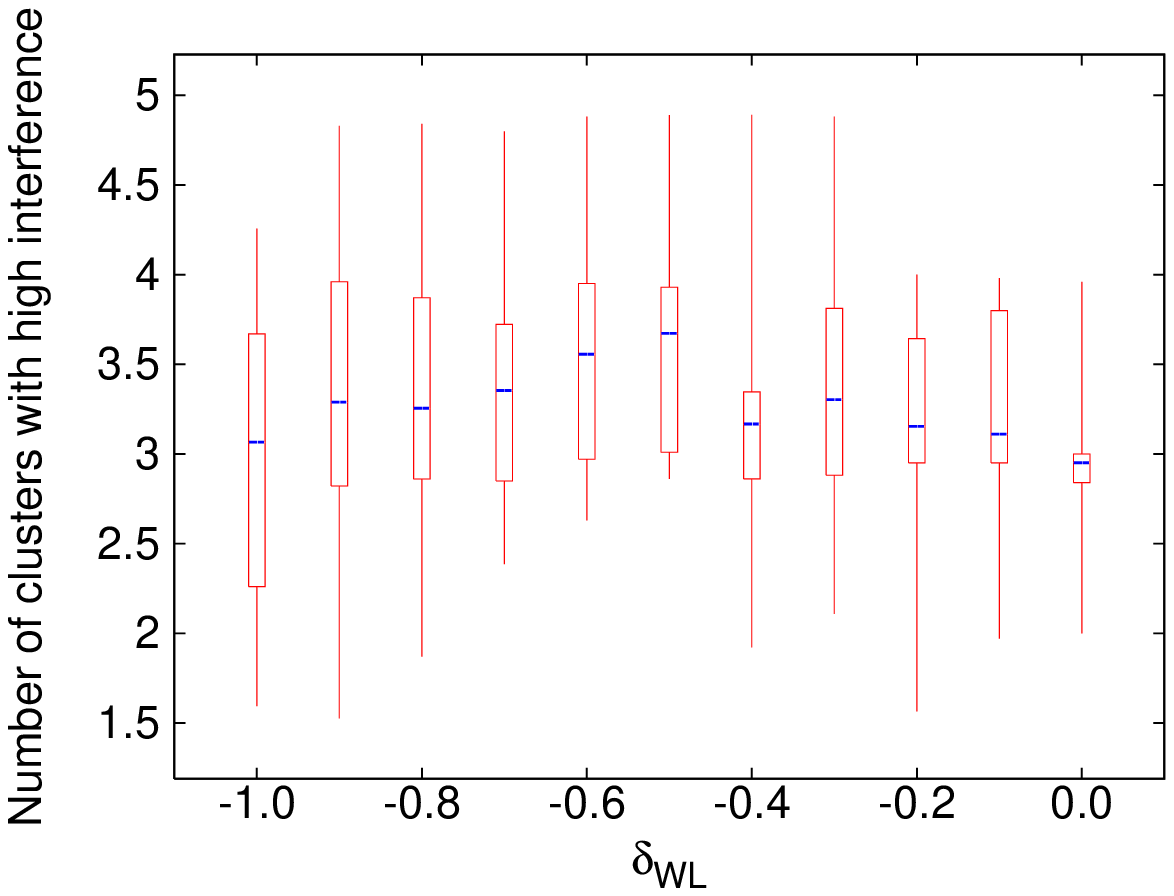}}%
 		\hspace{.5cm}%
\subfigure[]{	
		\includegraphics[width=2in]{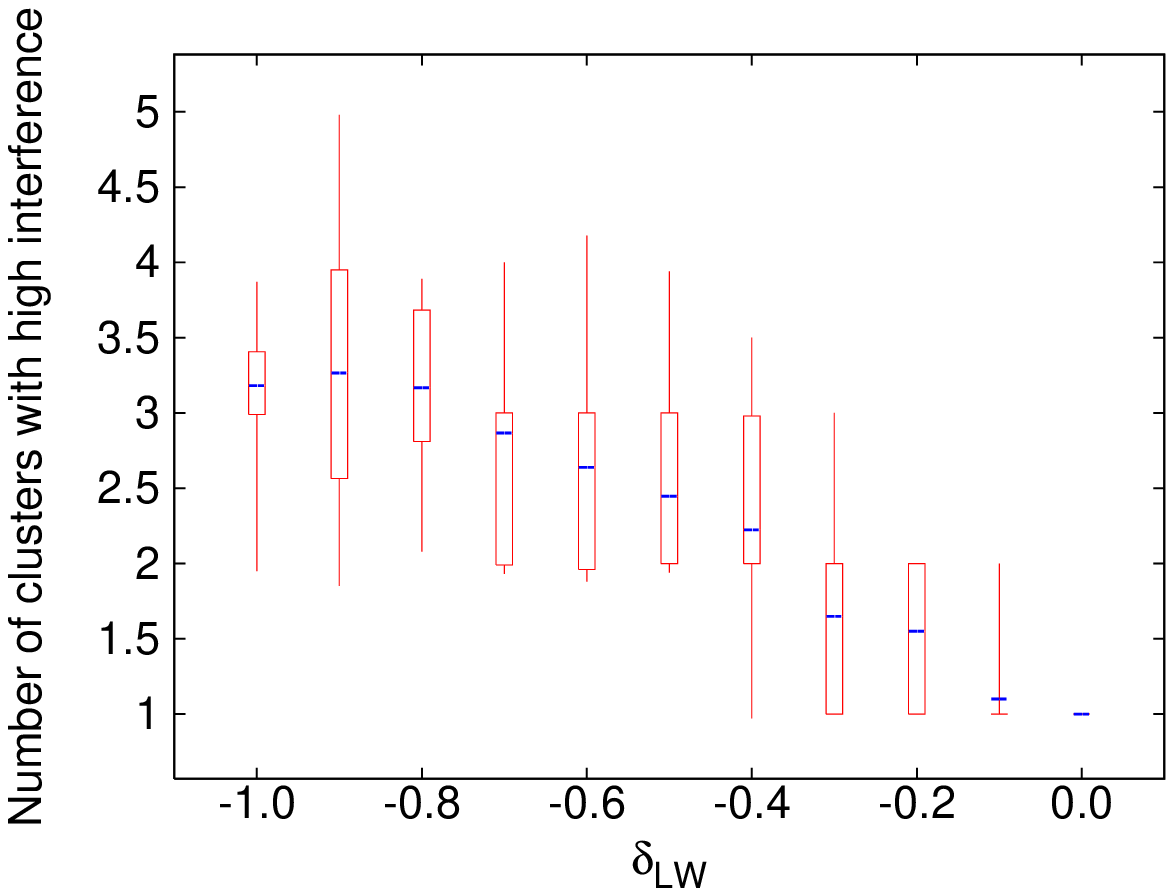}}\\
\subfigure[]{	
		\includegraphics[width=2in]{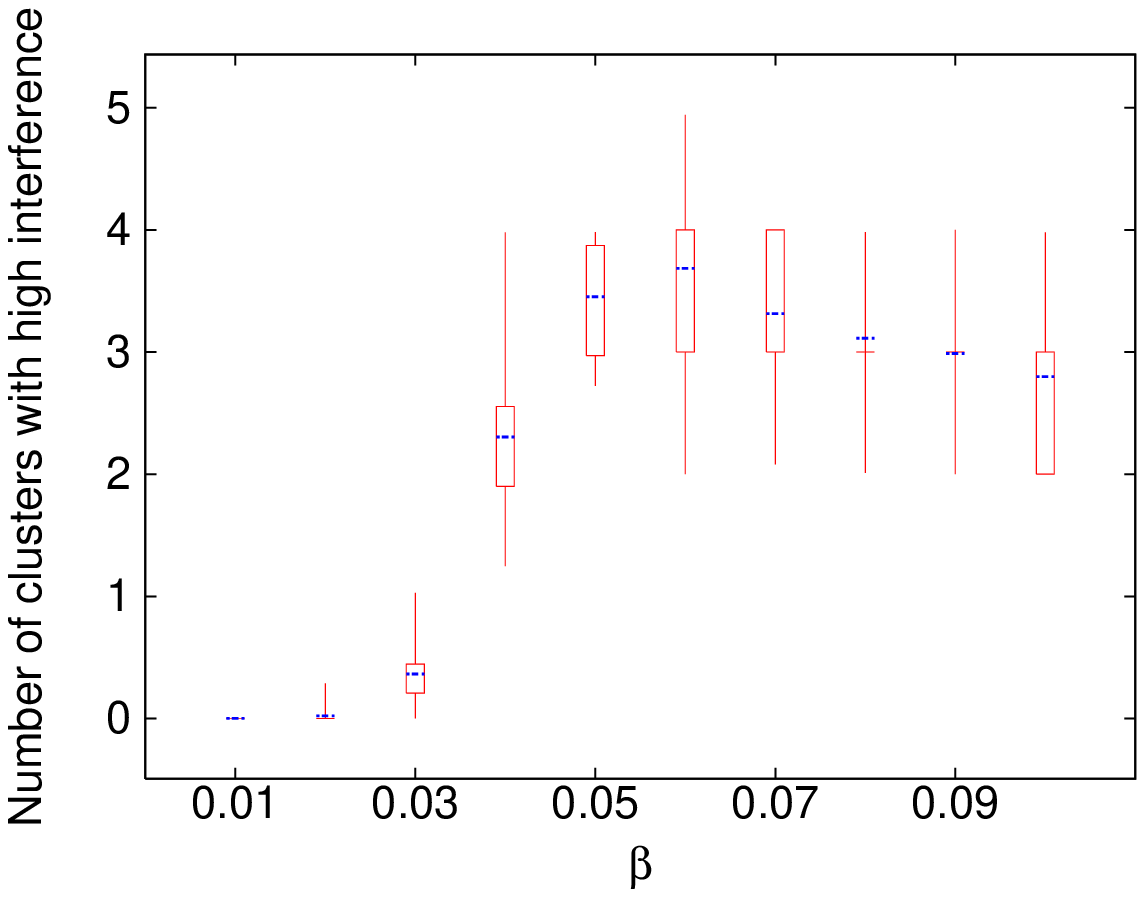}}%
 		\hspace{.5cm}
\subfigure[]{	
		\includegraphics[width=2in]{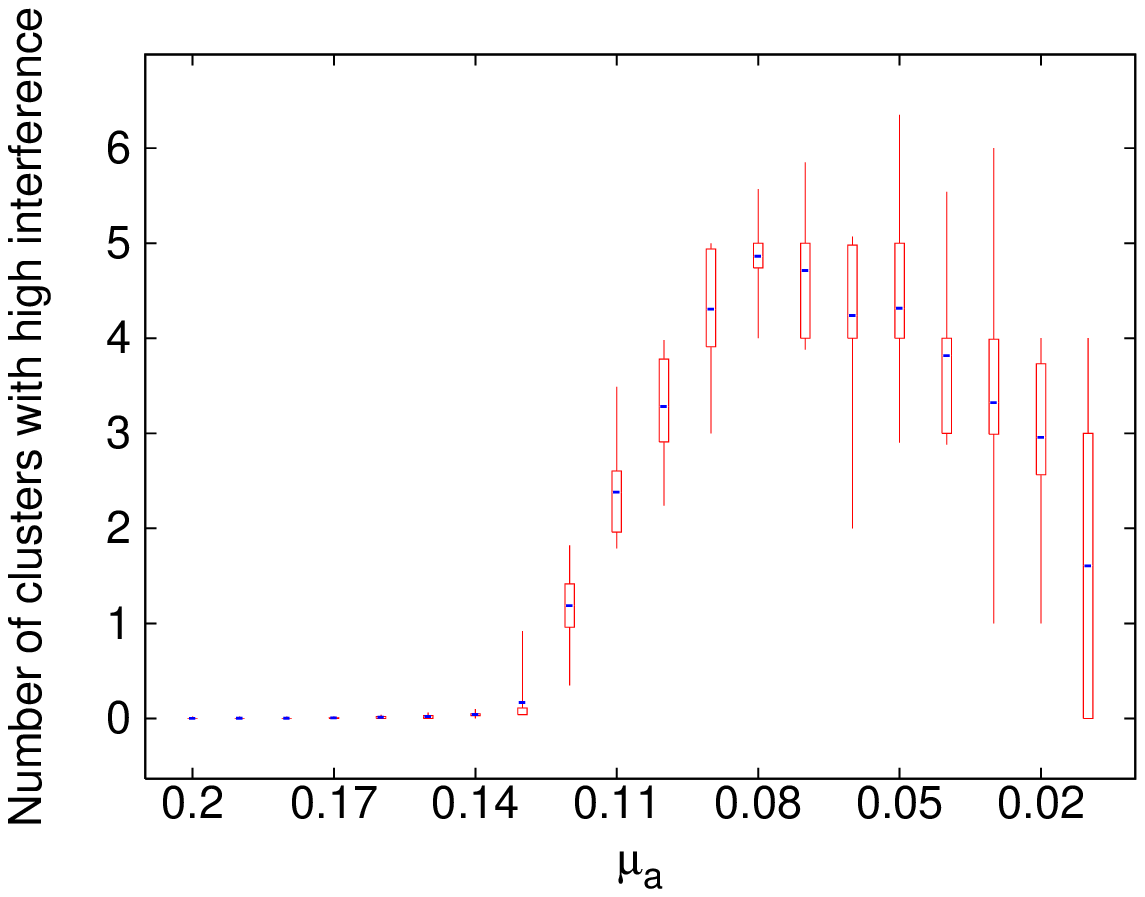}}\\		
\subfigure[]{	
		\includegraphics[width=2in]{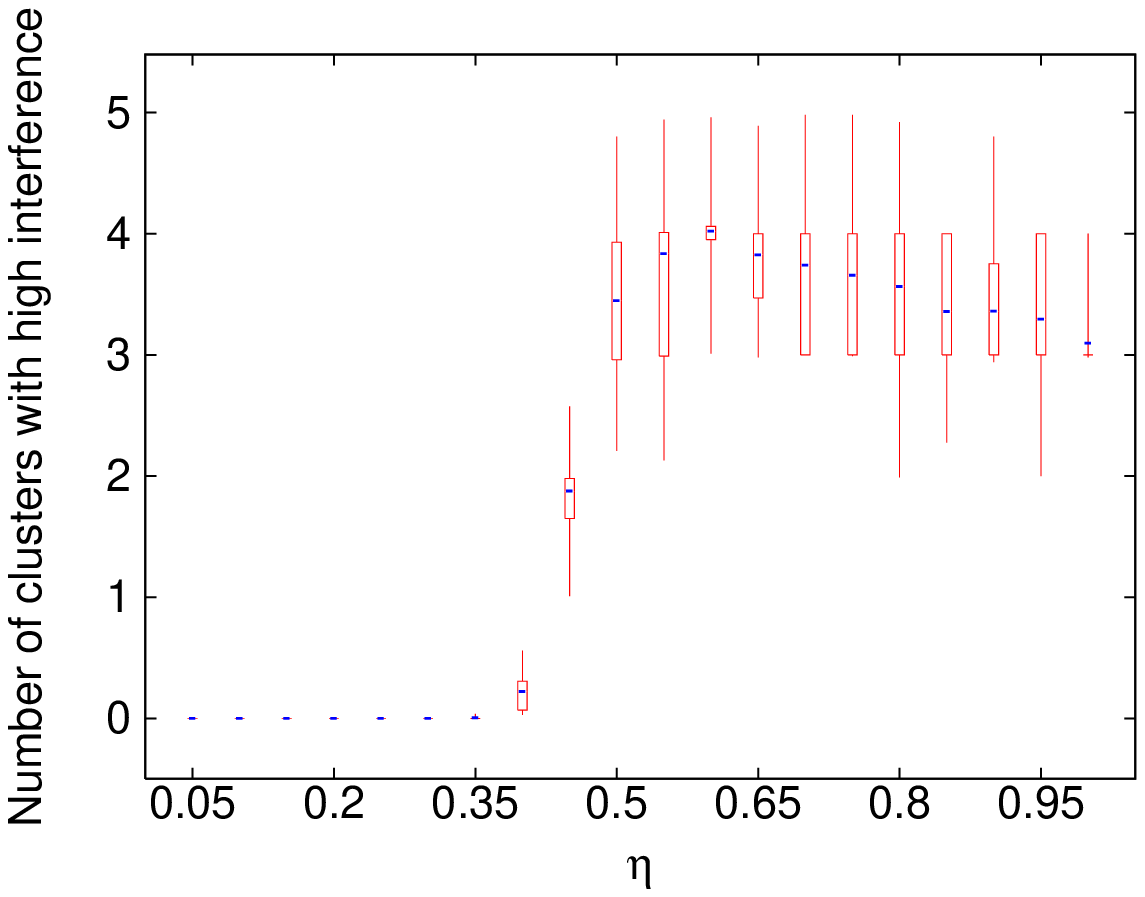}}%
		\hspace{.5cm}%
\subfigure[]{	
		\includegraphics[width=2in]{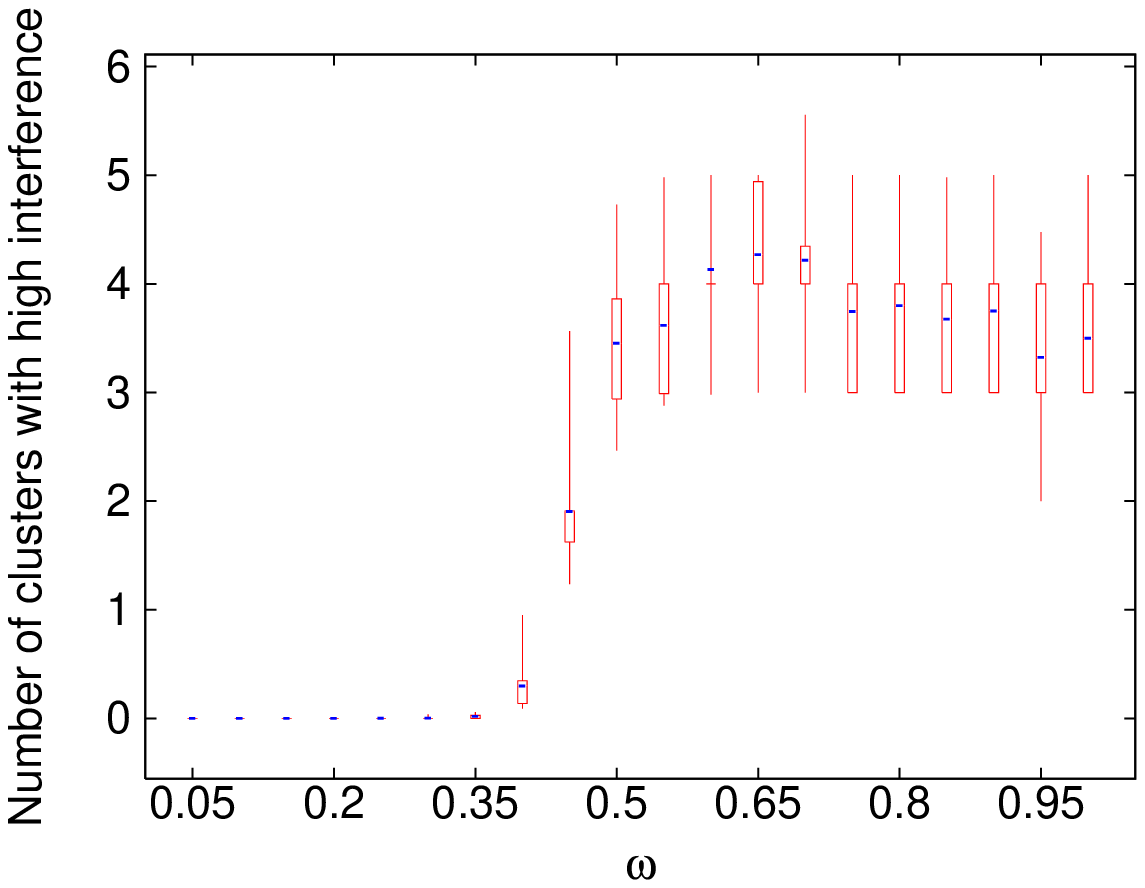}}\\
\caption{Effects of parameters $N,\gd_{ll},\gd_{wl},\gd_{lw},\gb,\mu,\eta,\go$ on
the number of alliances with within-cluster probability of interference $>0.5$ for a 
default set of parameter values.
}
\label{numClusBigInterf}
\end{figure}

{\bf Figures~\ref{inClus}-\ref{numClusBigInterf}}\quad
To obtain Figures~\ref{inClus}-\ref{numClusBigInterf} we performed 20 runs for each parameter
combination. Each of the 20 runs was characterized by a single average value (computed over 100 
observations taken between time 1000 to 2000).
All plots correspond to the Tukey Plots (i.e. show mean, min, max, quantile 1/4 and quantile 3/4), 
with 20 data points.
Other parameters were set to default values ($N=20, \gd_{ww}=1, \gd_{ll}=0.5, \gd_{wl}=-0.5,
\gd_{lw}=-1, \gb=0.05, \mu=0.1, \eta=0.5, \go=0.5$).

\begin{figure}[tbh]
\centering
\begin{tabular}{ccc}
& Number of individuals in alliances & Size of the largest alliance\\
$N=10$ 	& \subfigure{\includegraphics[width=2.5in]{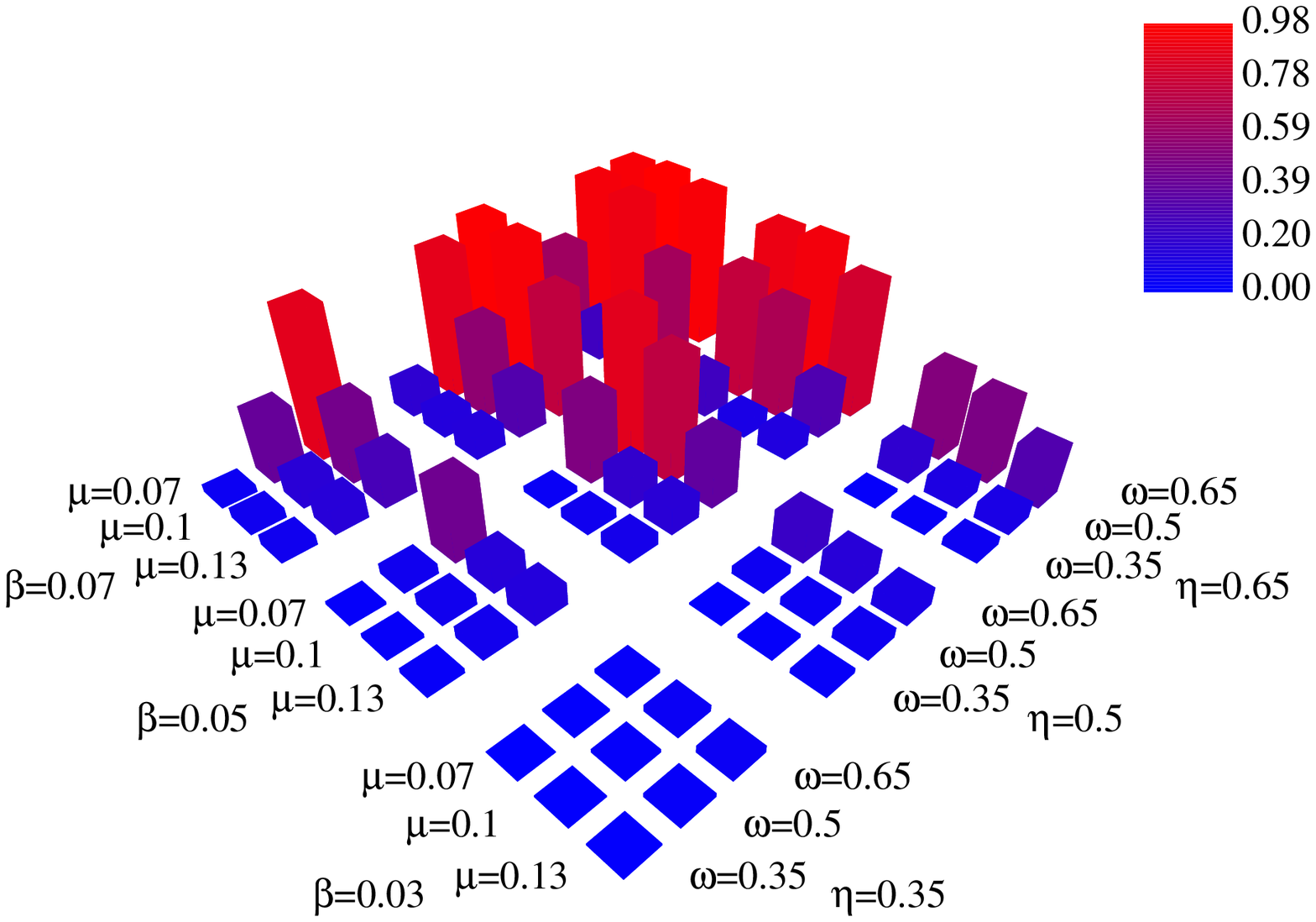}}
	& \subfigure{\includegraphics[width=2.5in]{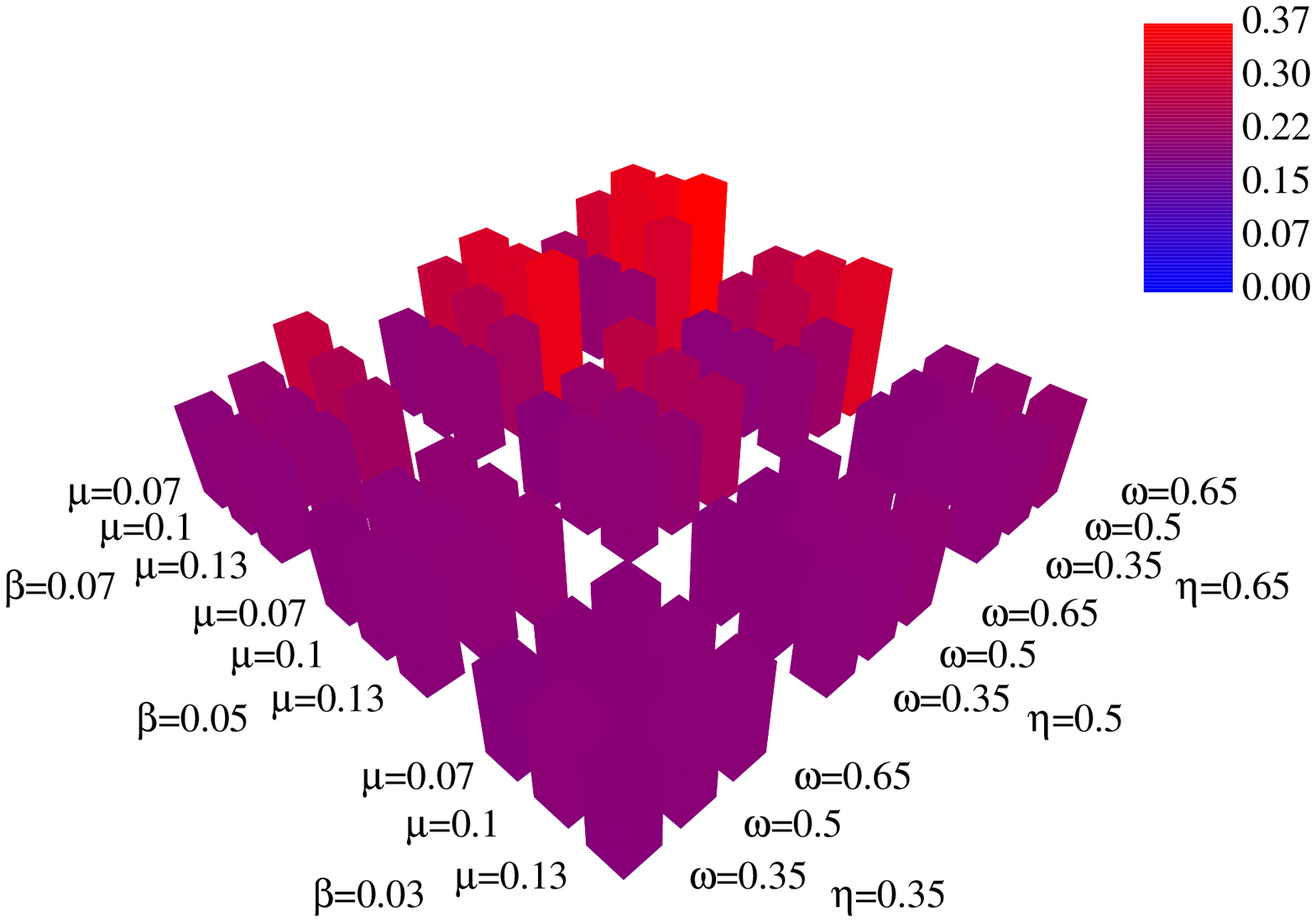}}\\
$N=20$ 	& \subfigure{\includegraphics[width=2.5in]{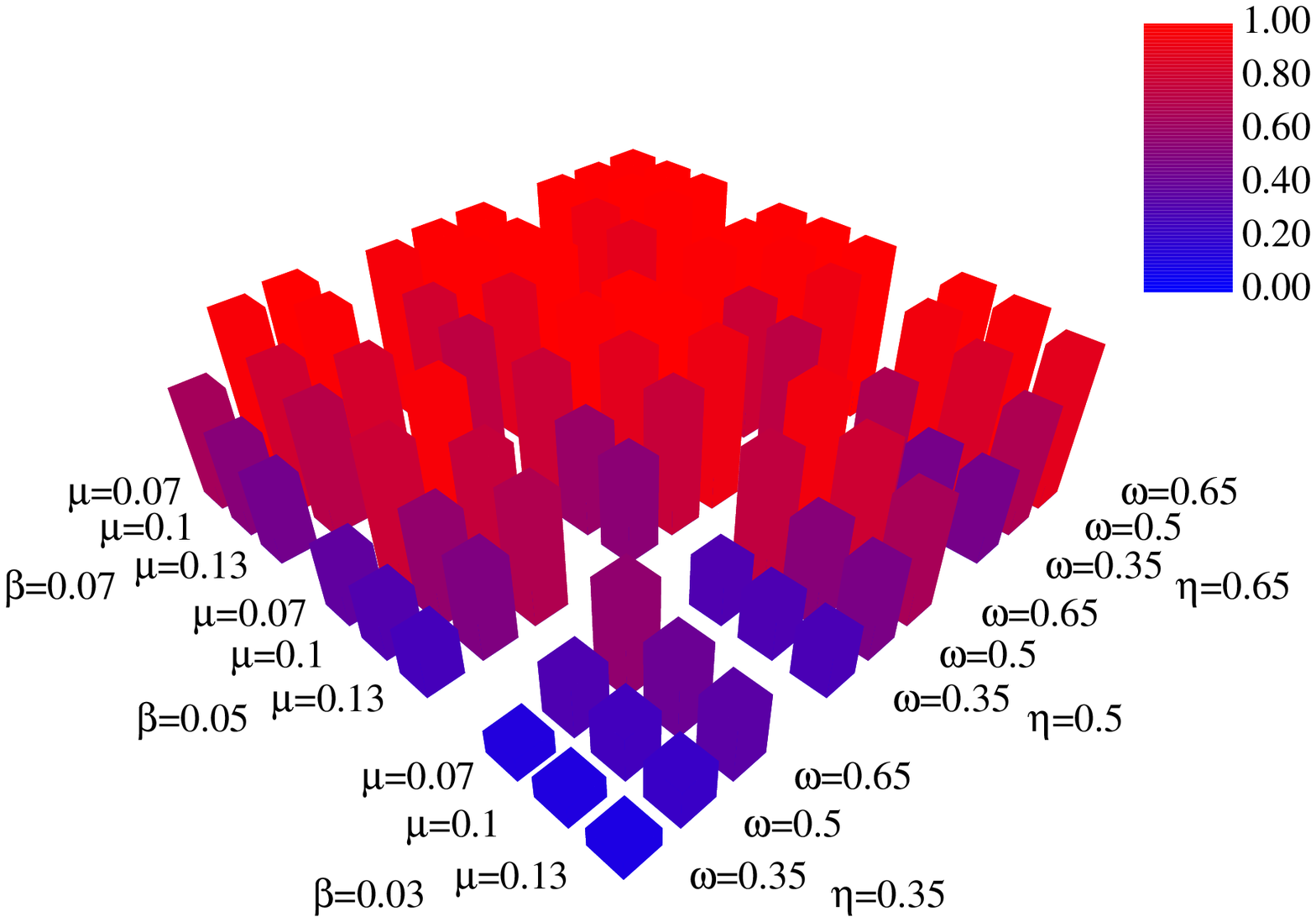}}
	& \subfigure{\includegraphics[width=2.5in]{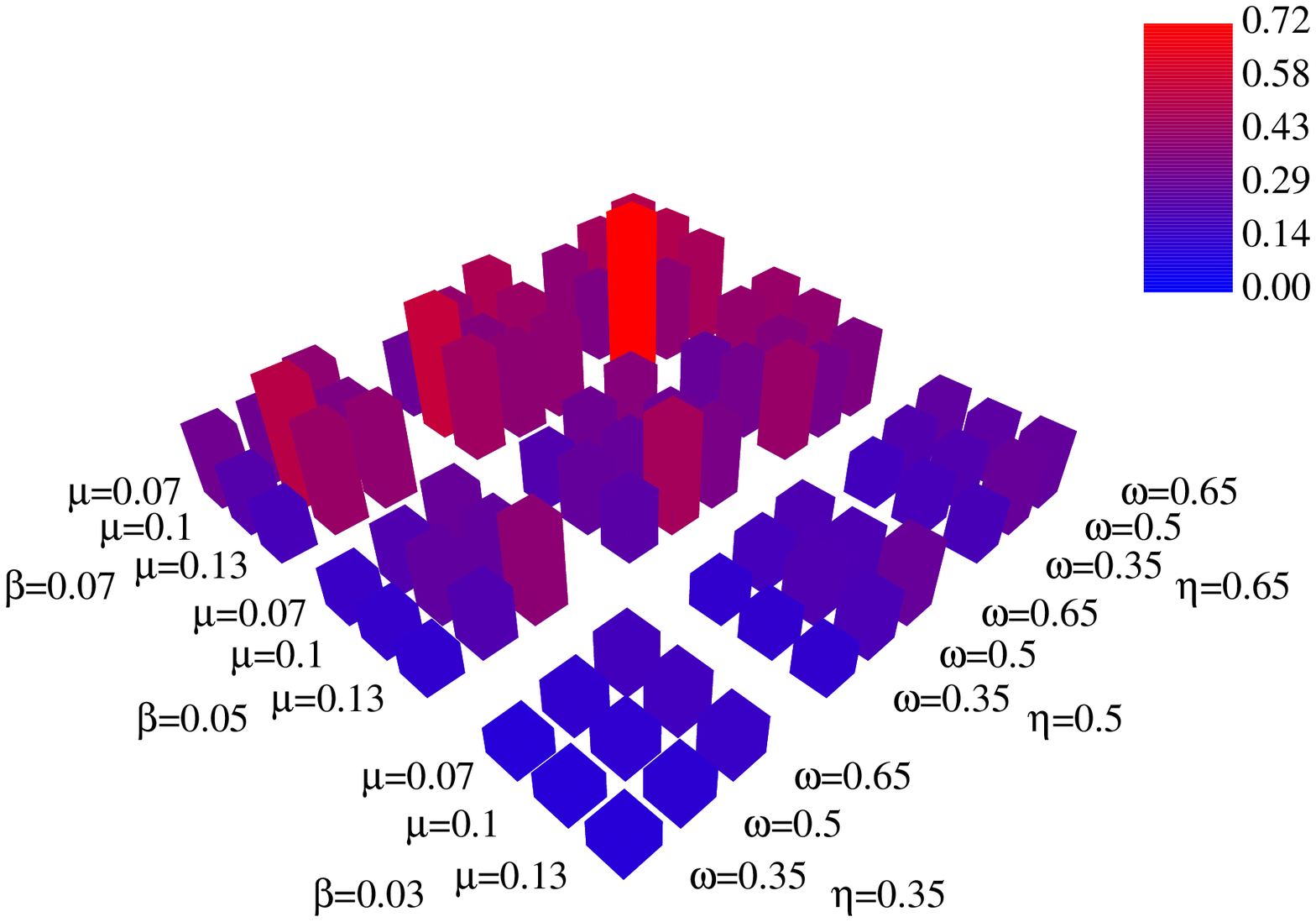}}\\
$N=30$ 	& \subfigure{\includegraphics[width=2.5in]{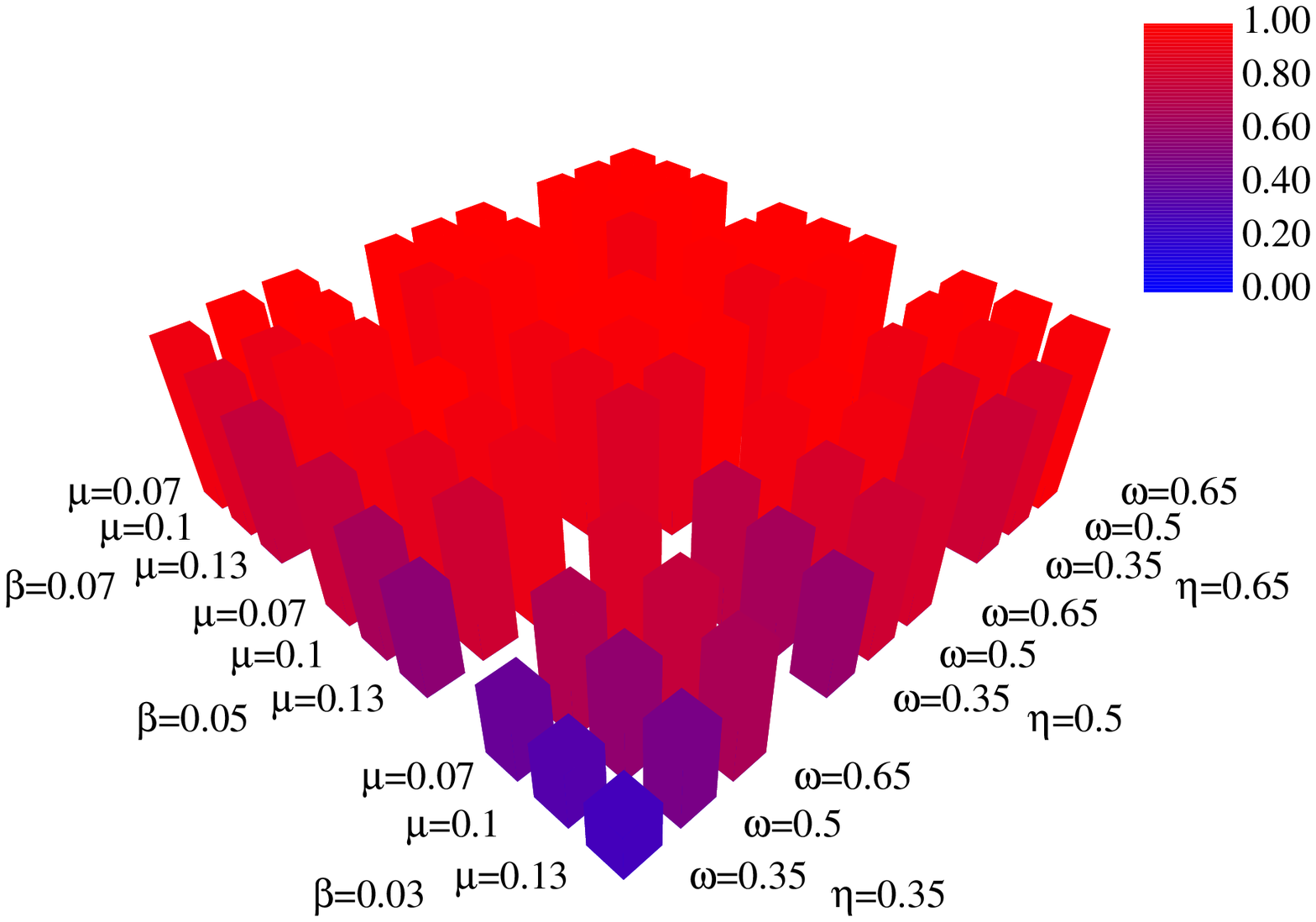}}
	& \subfigure{\includegraphics[width=2.5in]{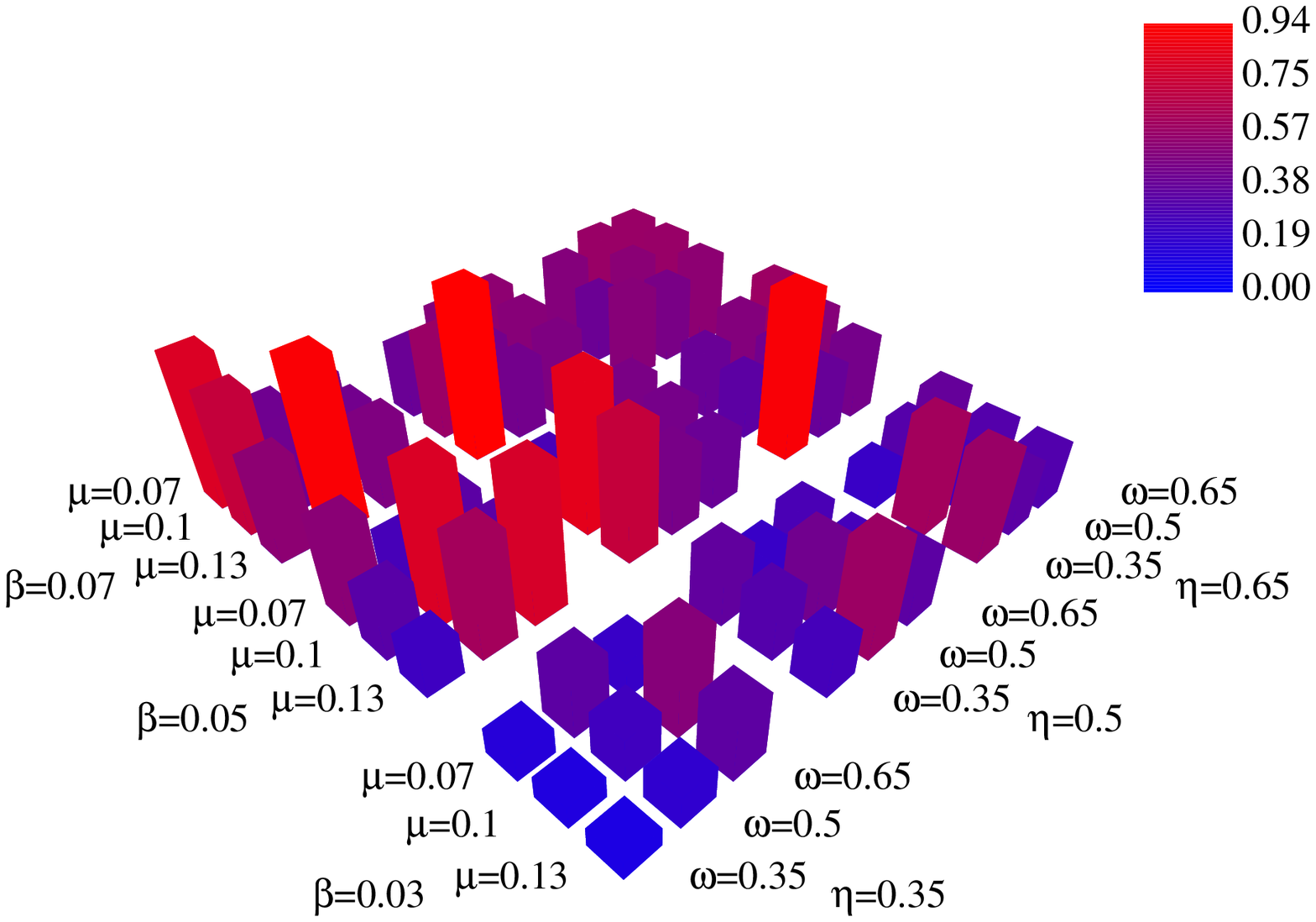}}\\
\end{tabular}
\caption{Effects of parameters $N,\gb,\mu,\eta,\go$ on
the number of individuals in alliances (first column) and the
size of the largest alliance (second column)
for $\gd_{ww}=1.0,\gd_{ll}=0.5, \gd_{wl}=-0.5, \gd_{lw}=-1.0$.
First row: $N=10$, second row: $N=20$, third row: $N=30$.
}
\label{hypercube1}
\end{figure}

\begin{figure}[tbh]
\centering
\begin{tabular}{ccc}
& $C^{(1)}$ of the largest alliance & Number of alliances with $C^{(1)}>0.5$\\
$N=10$ 	& \subfigure{\includegraphics[width=2.5in]{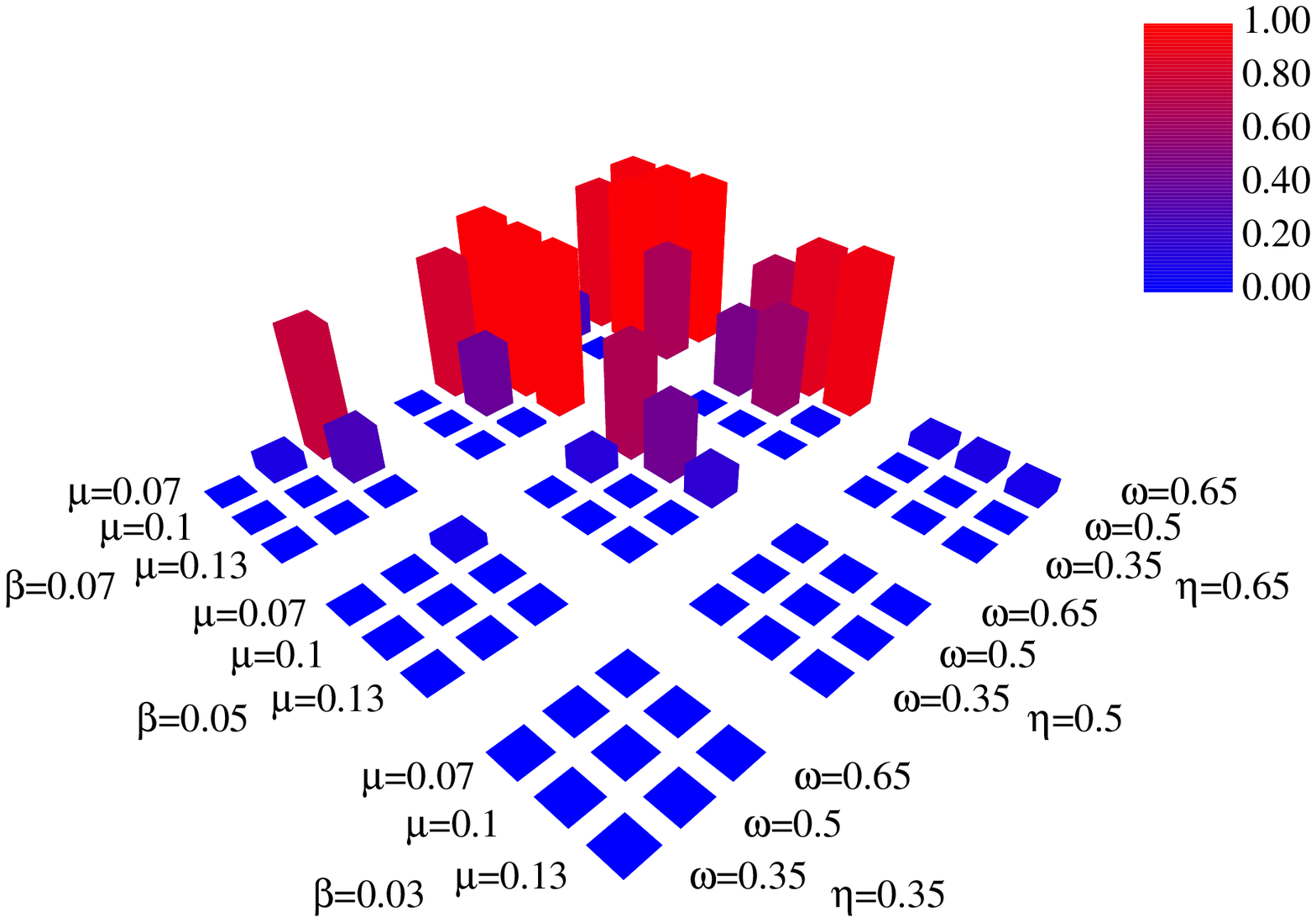}}
	& \subfigure{\includegraphics[width=2.5in]{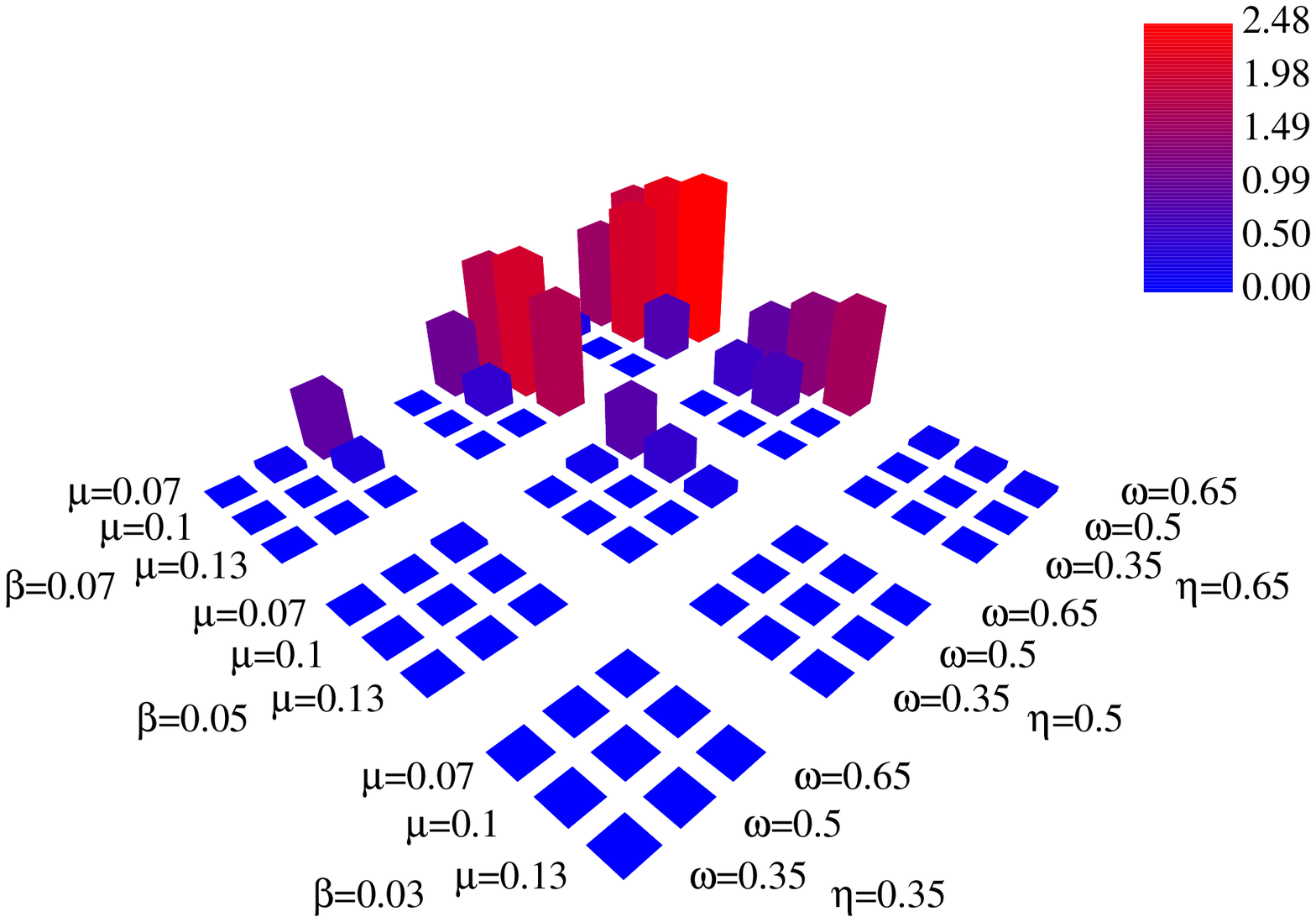}}\\
$N=20$ 	& \subfigure{\includegraphics[width=2.5in]{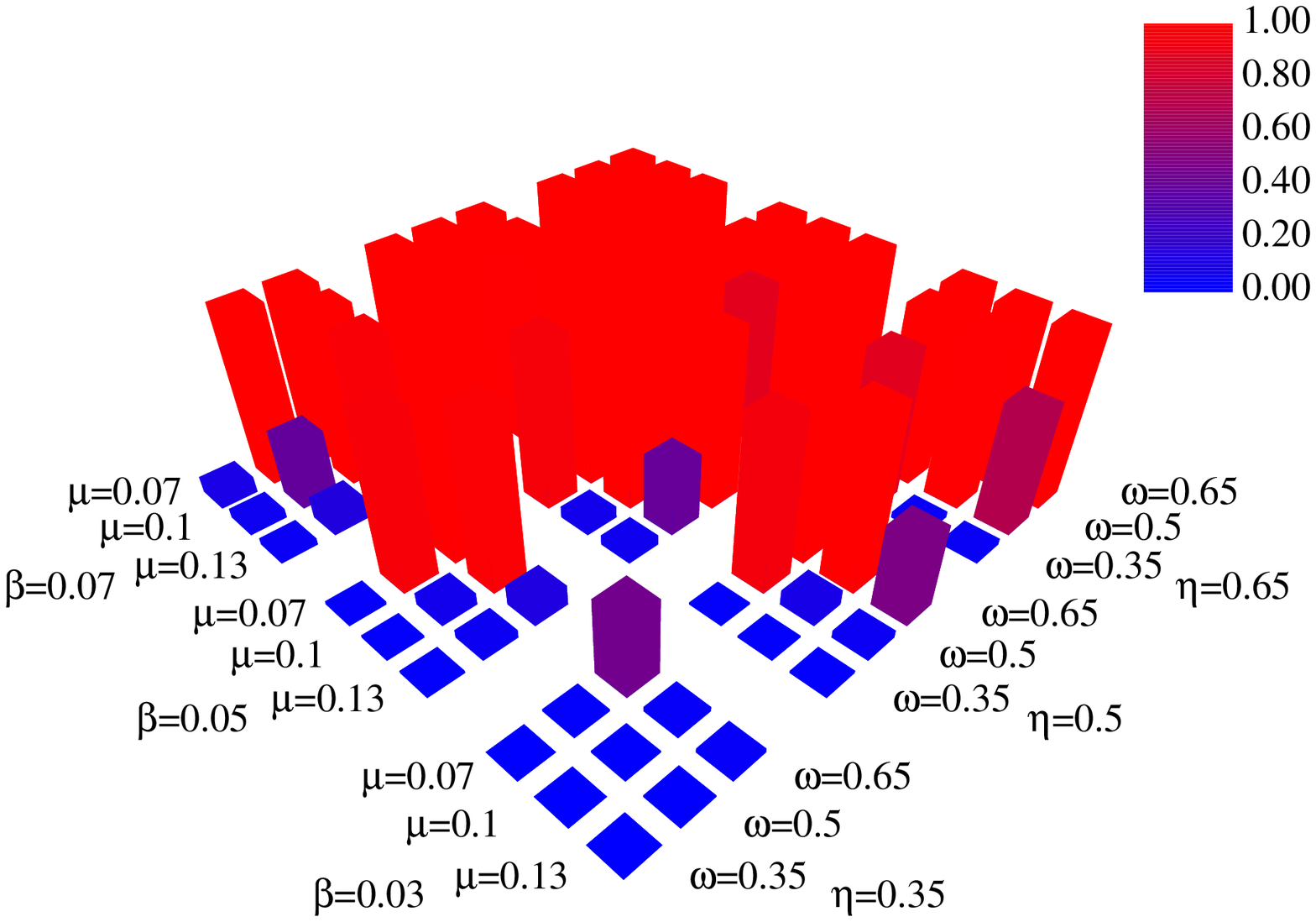}}
	& \subfigure{\includegraphics[width=2.5in]{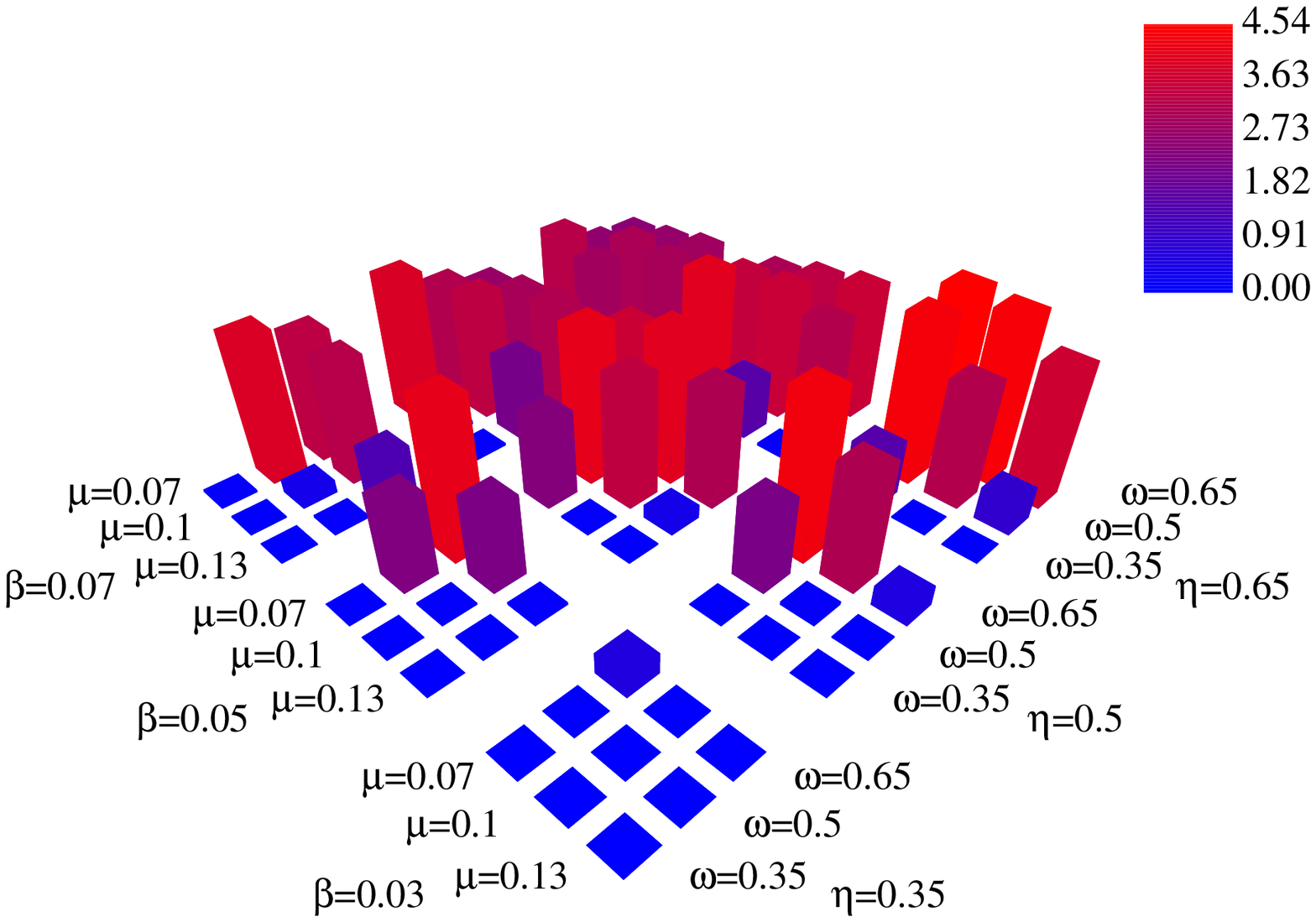}}\\
$N=30$ 	& \subfigure{\includegraphics[width=2.5in]{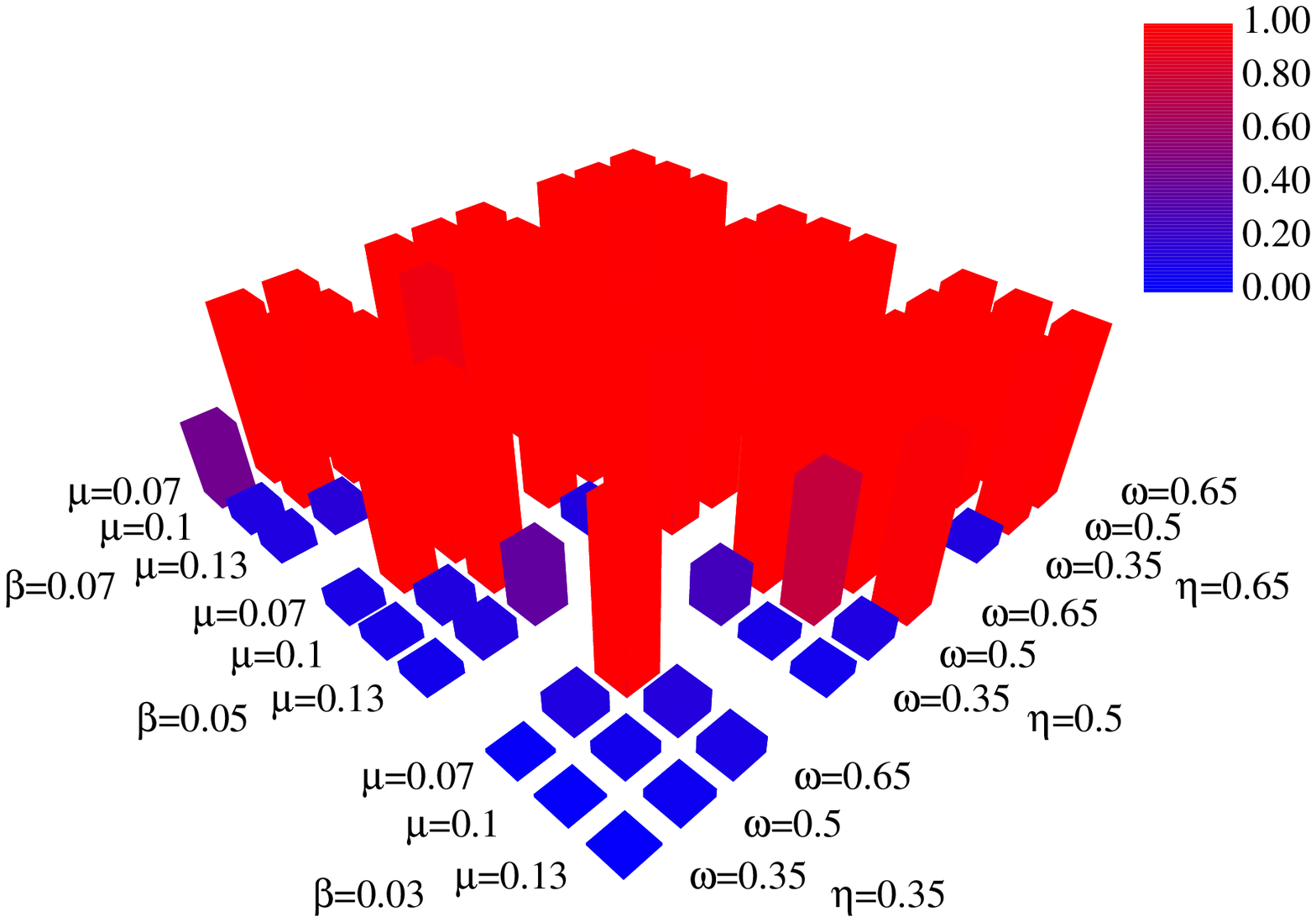}}
	& \subfigure{\includegraphics[width=2.5in]{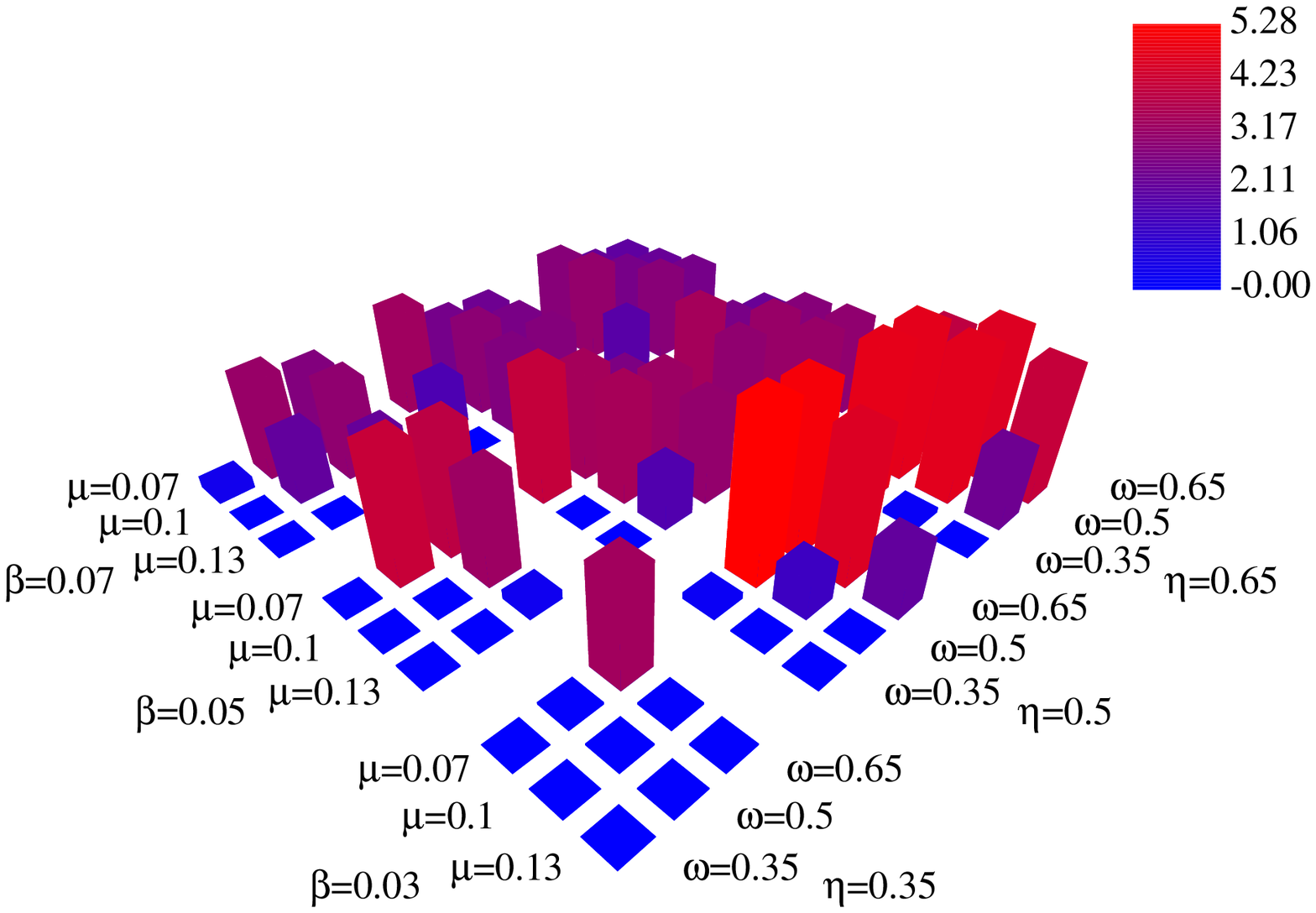}}\\
\end{tabular}
\caption{Effects of parameters $N,\gb,\mu,\eta,\go$ on
the $C^{(1)}$ measure of the largest alliance (first column), and 
the number of alliances with $C^{(1)}>0.5$ (second column) 
for $\gd_{ww}=1.0,\gd_{ll}=0.5, \gd_{wl}=-0.5, \gd_{lw}=-1.0$.
First row: $N=10$, second row: $N=20$, third row: $N=30$.
}
\label{hypercube2}
\end{figure}

{\bf Figures~\ref{hypercube1} and \ref{hypercube2}}\quad
To obtain Figures~\ref{hypercube1} and \ref{hypercube2} we performed 40 runs for each parameter
combination. 



\newpage
\clearpage
\subsection*{Supplementary Methods:
Mean field approximation for the dynamics of coalitions on the within-generation time-scale}

We consider a group of $N$ individuals in which conflicts occur at rate $\ga N$.
Below we will use two types of averages: the average over a clique (i.e., a set
of individuals who all are close allies), which
we will denote as $\langle \xi \rangle$, and the average over all possible outcomes of 
the process, which we will denote as $\ov{\xi}$ or $ E (\xi)$, where $\xi$ is a random variable.

{\bf Approximate dynamics of the mean and variance of affinities near an egalitarian state.}\quad
We assume that all $N$ individuals are close allies so that each individual aware of a conflict
interferes in it.
The average affinity of the group is 
	\[
		\langle x \rangle = \frac{1}{N(N-1)} \sum_{i \neq j} x_{ij}.
	\]
After each conflict, each affinity value changes from
$x_{ij}$ to $x_{ij}+\ge_{ij}$ where $\ge_{ij}$ is a random variable describing the change in
affinity of individual $i$ to individual $j$.
Let $a =\ov{\langle x \rangle }$ be the expected average affinity.
Since expectation and averaging are linear, 
the expected average affinity after a conflict can be written as
	\[
	a'=  \ov{\langle x +\ge \rangle}  = a + \ov{\langle \ge\rangle },
	\]
All affinities continuously decay to $0$ at a constant rate $\mu$.
Therefore, the dynamics of $a$ are described by a differential equation
	\be \label{mean}
	\frac{da}{dt}=\ga N\ \ov{\langle \ge\rangle } - \mu a.
	\ee

\vspace{0.1in} Similarly, let $v=\ov{\langle x^2 \rangle-\langle x
\rangle^2}$ be the expectation of the variance taken over all possible
outcomes of the process.  
Then the variance after a conflict is
\begin{eqnarray*}
v' & = &
E \Big( \langle \,(x + \ge)^2 \rangle - \langle x+\ge \rangle ^2 \Big) \\
& = &
E \Big( \langle x^2 \rangle + 2\langle x \,\ge \rangle + \langle \ge^2 \rangle 
- \big( \langle x \rangle^2 +2 \langle x \rangle \langle \ge \rangle +\langle \ge \rangle^2 \big)^2 \Big)\\
& = &
E \Big( \langle x^2 \rangle - \langle x \rangle^2 + \langle \ge^2 \rangle - \langle \ge \rangle^2 \Big) 
+ 2\langle x \, \ov{\ge} \rangle - 2\langle x \rangle \, \langle \ov{\ge} \rangle \\
& = & v + \ov{ \langle \ge^2 \rangle} - \ov{\langle \ge \rangle^2}.
\end{eqnarray*}
where, as an approximation, we assumed that $x$ and $\ov{\varepsilon}$ are independent
with respect to the averaging operator, i.e.,
$\langle x \, \ov{\varepsilon} \rangle = 
\langle x\rangle \, \langle \ov{\varepsilon} \rangle$.

All squares of affinities decay to $0$ at a constant rate $2\mu$.
Therefore, the dynamics of $v$ are described by a differential equation
	\be \label{variance}
	\frac{dv}{dt}=\ga N\ \big(\,\ov{ \langle \ge^2 \rangle} - \ov{\langle \ge \rangle^2}\, \big) - 2 \mu v.
	\ee

\vspace{0.1in}

First, we consider the expected change $\ov{ \langle \ge \rangle}$
in the affinity of a random pair of individuals after a conflict. There are three possibilities:
\bi
\item With probability $1/\binom{N}{2}$, the two individuals are the initiators 
of the conflict. Since either of the two initiators can be on the winning side,
the expected change in their affinity is 
	\[
	\gd_0=\frac{\gd_{WL}+\gd_{LW}}{2}.
	\]
Under our assumptions about the meaning of parameters, $\gd_0$ is negative.
\item With probability $\left[ 2(N-2)/\binom{N}{2}\right]\go$, one of the two individuals is an ``initiator'' while the 
other was aware of the conflict and interfered on behalf of one side.
Since there are four ways to distribute the two individuals over the winning and losing coalitions
and each occurs with equal probability,
the expected change in their affinity is  
	\[
	\ov{\gd}= \frac{\gd_{WW}+\gd_{WL}+\gd_{LW}+\gd_{LL}}{4}.
	\]
\item With probability
$\left[1-1/\binom{N}{2}-2(N-2)/\binom{N}{2}\right]\go^2$, neither individual is the initiator
of the conflict but both are aware of it and 
interfere in the conflict. The expected change in their affinity is $\ov{\gd}$.
\ei
Therefore,  
\bs
\begin{eqnarray}
	\ov{\langle \ge\rangle } 
	& = & \frac{1}{\binom{N}{2}}\ \gd_0
			+  \frac{2(N-2)}{\binom{N}{2}}\ \go\ \ov{\gd}
			+ \frac{ \binom{N}{2} - 2(N-2) -1}{\binom{N}{2}}\ \go^2\ \ov{\gd}\\ \label{single_clique-a}
	& = & \go^2\ \ov{\gd} + 
	    \frac{4(N-2)}{N(N-1)}\ \go(1-\go)\ov{\gd}
			+ \frac{2}{N(N-1)} (\gd_0-\go^2\ \ov{\gd}). \label{single_clique-b}
\end{eqnarray}
\es
Then, equations~(\ref{mean},\ref{single_clique-b}) predict that the average affinity 
in the egalitarian state evolves to an equilibrium value
	\be \label{egal}
	    a^* = \frac{\ga N}{\gm}  \left[ \go^2\ \ov{\gd} + 
	    \frac{4(N-2)}{N(N-1)}\ \go(1-\go)\ov{\gd}
			+ \frac{2}{N(N-1)} (\gd_0-\go^2\ \ov{\gd}) \right]. 
	\ee
The average affinity is positive only if $\ov{\gd}>0$. The last term in the brackets can be
neglected relative to the first term even for small groups (e.g., $N \geq 5$). 
The second term in the brackets 
can be neglected for larger groups (e.g., $N \geq 40$) if $\go$ is not too small.
Under these conditions, $a^* \approx \frac{\ga N}{\gm} \go^2\ \ov{\gd}$.


In a similar way and using  the results above,
	\be \label{1st}
	   \ov{ \langle \ge^2 \rangle } = \frac{1}{\binom{N}{2}}\ \gd_1
			+  \frac{2(N-2)}{\binom{N}{2}}\ \go\ \gd_2
			+ \frac{ \binom{N}{2} - 2(N-2) -1}{\binom{N}{2}}\ \go^2\ \gd_2,
	\ee
where
	\[
	    \gd_1 = \frac{\gd_{WL}^2+\gd_{LW}^2}{2},\ \ \gd_2 =\frac{\gd_{WW}^2+\gd_{WL}^2+\gd_{LW}^2+\gd_{LL}^2}{4}.
	\]

\vspace{0.1in}

\noindent More involved calculations show that 
\bs
\begin{eqnarray}
\ov{\langle \ge \rangle^2 } & = &\frac{1}{N^2(N-1)^2} \sum_{i\neq j} \sum_{k \neq l} \ov{\ge_{ij}\ge_{kl}}\\
& = &\frac{1}{N^2(N-1)^2} \sum_{i\neq j}  (\ov{\ge_{ij}\ge_{ij}}+\ov{\ge_{ij}\ge_{ji}})\\
& & + \frac{1}{N^2(N-1)^2} \sum_{i\neq j} \sum_{k\neq i,j} (\ov{\ge_{ij}\ge_{ik} }+\ov{\ge_{ij}\ge_{kj}}+ \ov{\ge_{ij}\ge_{jk}}+\ov{\ge_{ij}\ge_{ki}})\\
& & + \frac{1}{N^2(N-1)^2} \sum_{i\neq j} \sum_{k,l\neq i,j} \ov{\ge_{ij}\ge_{kl} }\\
	 & \equiv & \frac{ 1 }{N(N-1)} A_1+ \frac{4(N-2)}{N(N-1)} A_2+ \frac{(N-2)(N-3)}{N(N-1)} A_3,
\end{eqnarray}
where
	\begin{eqnarray}
	A_1 & = & \frac{1}{N(N-1)} \sum_{i\neq j}  (\ov{\ge_{ij}\ge_{ij}}+\ov{\ge_{ij}\ge_{ji}}) ,\\
	A_2 & = & \frac{1}{4N(N-1)(N-2)} \sum_{i\neq j} \sum_{k\neq i,j} (\ov{\ge_{ij}\ge_{ik} }+\ov{\ge_{ij}\ge_{kj}}+ \ov{\ge_{ij}\ge_{jk}}+\ov{\ge_{ij}\ge_{ki}}),\\
	A_3 & = & \frac{1}{N(N-1)(N-2)(N-3)} \sum_{i\neq j} \sum_{k,l\neq i,j} \ov{\ge_{ij}\ge_{kl} }.
	\end{eqnarray}
\es

The term $A_1$ can be interpreted as the expected value of 
$\gD=\ge_{ij}\ge_{ij}+\ge_{ij}\ge_{ji}$ for a random pair of individuals ($i$ and $j$).
There are three cases to consider.\bi
\item With probability $1/\binom{N}{2}$, the focal individuals
are the initiators of the conflict. In this case, $\gD=2\gd_0^2$.
\item With probability $\left[2(N-2)/\binom{N}{2} \right]\go$,
one of the two focal individuals is the initiator of the
conflict while the other is aware of it.
\item With probability $\left[1-1/\binom{N}{2}-2(N-2)/\binom{N}{2} \right]\go^2$,
both focal individuals are aware of the conflict. In the last two cases,
$\gD=\gd_0^2 + 2 \gd_2 -  \gd_1$.
\ei
Therefore,
	\be
	A_1=\frac{1}{\binom{N}{2}}\ 2 \gd_0^2
			+  \left[ \frac{2(N-2)}{\binom{N}{2}}\ \go
			+ \frac{ \binom{N}{2} - 2(N-2) -1}{\binom{N}{2}}\ \go^2\right] (\gd_0^2 + 2 \gd_2 -  \gd_1 )
	\ee

The term $A_2$ can be interpreted as the expected value of $\gD=\ge_{ij}\ge_{ik} +\ge_{ij}\ge_{kj}+ \ge_{ij}\ge_{jk}+\ge_{ij}\ge_{ki}$ for a random triple of individuals ($i,j$ and $k$).
There are three cases to consider.
\bi
\item With probability $\left[3/\binom{N}{2} \right]\go$,
two of the three focal individuals are the initiators of the
conflict while the third is aware of it. In this case, $\gD=(8\gd_o \ov{\gd}+4\ov{\gd}^2)/3$.
\item With probability $\left[3(N-3)/\binom{N}{2} \right]\go^2$,
one of the three focal individuals is the initiator of the
conflict while the two others are aware of it. 
\item With probability $\left[1-3(N-3)/\binom{N}{2}-3/\binom{N}{2}\right] \go^3$,
none of the three focal individuals are the initiators of the
conflict but all are aware of it. 
\ei
To evaluate $\gD$ in the last  two cases, one needs to
consider changes in affinities corresponding to all possible ways to assign three 
individuals to the winning and losing coalitions. This is done in the table below:
%
	\begin{center}
	\begin{tabular}{c|c|cccc}                                      
  winners & losers  & $\ge_{ij}\ge_{ik}$  & $\ge_{ij}\ge_{kj}$   &  $\ge_{ij}\ge_{jk}$ & $\ge_{ij}\ge_{ki}$\\  \hline
$ijk$ & -          & $\gd_{WW}^2 $ &  $\gd_{WW}^2 $  &  $\gd_{WW}^2$  &  $\gd_{WW}^2 $    \\
$ij$  & $k$        & $\gd_{WW}\gd_{WL} $ &  $\gd_{WW}\gd_{LW} $  &  $\gd_{WW}\gd_{WL} $  &  $\gd_{WW}\gd_{LW} $  \\
$ik$  & $j$        & $\gd_{WL}\gd_{WW} $ &  $\gd_{WL}^2 $  &  $\gd_{WL}\gd_{LW} $  &  $\gd_{WL}\gd_{WW} $  \\
$jk$  & $i$        & $\gd_{LW}^2 $ &  $\gd_{LW}\gd_{WW} $  &  $\gd_{LW}\gd_{WW} $  &  $\gd_{LW}\gd_{WL} $  \\ 
$i$  & $jk$        & $\gd_{WL}^2 $ &  $\gd_{WL}\gd_{LL} $  &  $\gd_{WL}\gd_{LL} $  &  $\gd_{WL}\gd_{LW} $  \\ 
$j$  & $ik$        & $\gd_{LW}\gd_{LL} $ &  $\gd_{LW}^2 $  &  $\gd_{LW}\gd_{WL} $  &  $\gd_{LW}\gd_{LL} $  \\ 
$k$  & $ij$        & $\gd_{LL}\gd_{LW} $ &  $\gd_{LL}\gd_{WL} $  &  $\gd_{LL}\gd_{LW} $  &  $\gd_{LL}\gd_{WL} $  \\
$-$  & $ijk$       & $\gd_{LL}^2 $ &  $\gd_{LL}^2 $  &  $\gd_{LL}^2 $  &  $\gd_{LL}^2 $
	\end{tabular} 
	\end{center}
Using this table, 
\[
	\begin{aligned}
	\gD = &  \frac{1}{8}(4\gd_{WW}^2+4\gd_{LL}^2+2\gd_{LW}^2 +2\gd_{WL}^2 +4\gd_{WW}\gd_{WL} +4\gd_{WW}\gd_{LW}  +4\gd_{LL}\gd_{LW} + 4\gd_{LL}\gd_{WL}+4\gd_{LW}\gd_{WL}) \\
	=&  \frac{1}{8}[2(\gd_{WW}+\gd_{LL}+\gd_{LW}+\gd_{WL})^2+2(\gd_{WW}-\gd_{LL})^2]\\ 
	=&   4 \ov{\gd}^2+\frac{1}{4} (\gd_{WW}-\gd_{LL})^2.
	\end{aligned}
\]
Therefore, 
\be
A_2=\frac{4}{\binom{N}{2}}\go (2\gd_o \ov{\gd}+\ov{\gd}^2)+
\left[ \frac{3(N-3)}{\binom{N}{2}} \go^2+\frac{\binom{N}{2}-3(N-3)-3}{\binom{N}{2}} \go^3\right] (4 \ov{\gd}^2+\frac{1}{4} \left[\gd_{WW}-\gd_{LL})^2\right].
\ee

The term $A_3$ can be interpreted as the expected value of $\gD=\ge_{ij}\ge_{kl}$
for a random quartet of individuals ($i,j,k$ and $l$). There are three cases to
consider:
\bi
\item With probability $\left[6/\binom{N}{2} \right]\go^2$,
two of the four focal individuals are the initiators of the
conflict while the two others are aware of it. In this case,
$\gD = \gd_{3} \equiv \left[ 4\gd_0^2 + 3\gd_0(\gd_{WW}+\gd_{LL}) + 2\gd_{LL}\gd_{WW}\right]/12$
\item With probability $\left[4(N-4)/\binom{N}{2} \right]\go^3$,
one of the four focal individuals is the initiator of the
conflict while the three others are aware of it. In this case,
$\gD=\ov{\gd}^2$.
\item With probability $\left[1-4(N-4)/\binom{N}{2}-6/\binom{N}{2}\right] \go^4$,
none of the three focal individuals are the initiators of the
conflict but all are aware of it. In this case,
$\gD=\ov{\gd}^2$.
\ei
%

Therefore, 
\be
A_3=\frac{1}{\binom{N}{2}}\go^2 \gd_3+
\left[ \frac{4(N-4)}{\binom{N}{2}} \go^3+\frac{\binom{N}{2}-4(N-4)-6}{\binom{N}{2}}   \go^4 \right] \ov{\gd}^2.
\ee

Keeping only the leading terms in $1/N$, $\ov{\langle \ge_{ij}^2 \rangle}=\go^2 \ov{\gd^2}, 
\ov{\langle \ge_{ij}\ge_{kl} \rangle }
=  (\go^2 \ov{\gd})^2$, which results in an equation for $v$:
\be
	\frac{dv}{dt}= \ga N \go^2 \left[ \var \gd +(1-\go^2) \ov{\gd}^2 \right] - 2 \mu v,
\ee
where  $\var \gd = \ov{\gd^2}-\ov{\gd}^2$.  Higher order corrections (in $1/N$) can be found
in a straightforward way from the formula given above.

Keeping only the leading terms in $1/N$, the mean field approximation predicts the following 
equilibrium values at the egalitarian regime
%
\[
	\begin{aligned}
	a^*= & \frac{\ga N \go^2  \ov{\gd}}{ \mu},\\
	v^*= & \frac{\ga N \go^2 \left[ \var \gd +(1-\go^2) \ov{\gd}^2 \right]}{ 2 \mu},
	\end{aligned}
\]

The egalitarian state is stable if the fluctuations of pairwise affinities
around $a^*$ do not result in negative affinities.
We conjecture that the egalitarian state is stable if $a^*>3 \sqrt{v^*}$, which is roughly
equivalent to $(a^*)^2>10 v^*$, which in turn can be rewritten as
	\[
	   \frac{2 \ga N \go^2}{\mu}>10 \left( \frac{\var \gd}{\ov{\gd}^2}+1-\go^2 \right).
	\]

{\bf The strongest clique comprising $N_1$ individuals; other $N_2=N-N_1$ individuals belong to weaker cliques.}\quad
We assume that all $N_1$ individuals in the clique are close allies that always help each 
other and never help outsiders. 
To evaluate the expected average over the clique $\ov{\langle \ge \rangle}$,
we need to find the expected value of $\gD=\ge_{ij}$ for a random pair from 
the strongest clique.
One needs to consider five possibilities:
\bi
	\item With probability $1/\binom{N}{2}$, the focal individuals are the initiators 
of the conflict. In this case, $\gD=\gd_0$.
	\item With probability $2(N_1-2)\go/\binom{N}{2}$, one of the focal individuals is
an initiator of a conflict involving another member of the clique while the 
other is aware of the conflict and interferes on behalf of one side.
In this case, $\gD=\ov{\gd}$.
  	\item With probability
$\left[\binom{N_1-2}{2}/\binom{N}{2}\right]\go^2$,
both focal individuals are aware 
of and interfere in a conflict between two other members of the clique. 
In this case, $\gD=\ov{\gd}$.
	\item With probability $\left[2N_2/\binom{N}{2}\right]\go$, one of the focal individuals is
an initiator of a conflict involving an outsider while the 
other is aware of the conflict and interferes on behalf of the clique member.
Assuming that the clique always wins, $\gD=\gd_{WW}$.
  	\item With probability
$\left[(N_1-2)N_2/\binom{N}{2}\right]\go^2$, both focal individuals are aware of and 
interfere in a conflict between a member of the clique and an outsider. 
Assuming that the clique always wins, $\gD=\gd_{WW}$.

\ei
Therefore, 
	\be  \label{single}
	\ov{\langle \ge \rangle} = \frac{1}{\binom{N}{2}} \gd_0
			+ \frac{2(N_1-2)}{\binom{N}{2}} \go \ov{\gd}
			+ \frac{ \binom{N_1-2}{2}}{\binom{N}{2}} \go^2 \ov{\gd}
			+ \frac{2N_2}{\binom{N}{2}} \go \gd_{WW}
			+ \frac{(N_1-2)N_2}{\binom{N}{2}} \go^2 \gd_{WW}.
	\ee
Assume that $N_1=N-1,N_2=1$ (i.e., the single outsider case).
Then the dynamics of the average within-clique affinity $a$ are described by equation
	\[
	\frac{da}{dt}=\ga N \left[  
	\frac{1}{\binom{N}{2}} \gd_0
			+ \frac{2(N-3)}{\binom{N}{2}} \go \ov{\gd}
			+ \frac{ \binom{N-3}{2}}{\binom{N}{2}} \go^2 \ov{\gd}
			+ \frac{2}{\binom{N}{2}} \go \gd_{WW}
			+ \frac{(N-3)}{\binom{N}{2}}\go^2 \gd_{WW}  \right] -\mu a.
	\]
Thus, the average affinity under the single outsider regime is predicted to evolve to
	\[
	    a_s^*= \frac{\ga N}{\gm}  \left[  \go^2\ \ov{\gd} -\frac{6(N-2)}{N(N-1)}\go^2\ \ov{\gd}
	    +\frac{2(N-3)}{N(N-1)} \left(2 \go \ov{\gd}+\go^2 \gd_{WW}\right)
	    + \frac{2}{N(N-1)} (\gd_0 + 2 \go \gd_{ww}) \right].
	\]

Keeping only terms of order $O(1/N)$ and larger in the brackets, 
	\be
a_s^*= \frac{\ga N}{\gm}  \left[ \go^2\ \ov{\gd} + 
		\frac{4}{N}  \go(1-\go)\ov{\gd}+\frac{2}{N}\go^2 (\gd_{WW}-\ov{\gd}) \right]
	\ee
It is illuminating to compare this expression with expression~(\ref{egal}) approximating
the average affinity under egalitarian regime.
Under the same assumptions, expression~(\ref{egal}) simplifies to
 	\be 
	    a^* = \frac{\ga N}{\gm}  \left[ \go^2\ \ov{\gd} + 
	    \frac{4}{N}\ \go(1-\go)\ov{\gd} \right]. 
	\ee

If $N$ is not too large, $a^*$ can be substantially smaller than $a_s^*$. It is in this
situation when a single outsider can have a strong stabilizing effect on a small coalition.
For example, let $\ga=1, N=20, \go=0.5, \gm=0.05$ and $\gd_{WW}=1, \gd_{LL}=0.5, \gd_{LW}=-0.5,
\gd_{WL}=-0.5$ so that $\ov{\gd}=0.125$. Then $a^*=15.00$ but
$a^*_s=23.75$, so that a single outsider significantly increases the average affinity of
the clique. A single outsider will also reduce variance $v$, the effect of which will further
strengthen the stability of the coalition.\\	

%
%
%
%

\end{document}